\documentclass{aastex631}

\usepackage{CJK}
\usepackage{caption2}
\usepackage{color}
\usepackage{ulem}
\usepackage{multirow}
\usepackage{booktabs}
\usepackage{threeparttable}
\usepackage{color}
\usepackage{multirow}
\usepackage{rotating}
\usepackage{cancel,soul,ulem,amsmath} 
\usepackage{booktabs}
\shorttitle{Dense outflowing molecular gas}
\shortauthors{Yani Xu et al.}
\graphicspath{{./}{figures/}}
\usepackage{graphicx}
\usepackage{epstopdf}
\usepackage{subfigure}
\usepackage{amssymb}
\makeatletter

\newcommand{\Rmnum}[1]{\expandafter\@slowromancap\romannumeral #1@} 
\makeatother

\begin{document}
\begin{CJK*}{UTF8}{gbsn}
\title{Dense Outflowing Molecular Gas in Massive Star-forming Regions}
\date{\today}

\author[0009-0009-6136-0417]{Yani Xu}
\author[0000-0001-6106-1171]{Junzhi Wang}\thanks{E-mail: junzhiwang@gxu.edu.cn}
\affiliation{Guangxi Key Laboratory for Relativistic Astrophysics, School of Physical Science and Technology, Guangxi University, Nanning 530004, China}
\author[0000-0001-6016-5550]{Shu Liu}
\affiliation{National Astronomical Observatories, Chinese Academy of Sciences, Beijing 100101, People's Republic of China}
\author[0000-0003-3520-6191]{Juan Li}
\affiliation{Shanghai Astronomical Observatory, Chinese Academy of Sciences, No. 80 Nandan Road, Shanghai, 200030, China}
\affiliation{Key Laboratory of Radio Astronomy, Chinese Academy of Sciences, Nanjing 210033, China}
\author[0000-0002-2243-6038]{Yuqiang Li}
\affiliation{Shanghai Astronomical Observatory, Chinese Academy of Sciences, No. 80 Nandan Road, Shanghai, 200030, China}
\affiliation{School of Astronomy and Space Sciences, University of Chinese Academy of Sciences, No. 19A Yuquan Road, Beijing 100049, People's Republic of China}
\author[0009-0001-2414-7221]{Rui Luo}
\affiliation{Guangxi Key Laboratory for Relativistic Astrophysics, School of Physical Science and Technology, Guangxi University, Nanning 530004, China}
\author[0000-0003-0128-4570]{Chao Ou}
\affiliation{Guangxi Key Laboratory for Relativistic Astrophysics, School of Physical Science and Technology, Guangxi University, Nanning 530004, China}
\author[0000-0001-9047-846X]{Siqi Zheng}
\affiliation{Shanghai Astronomical Observatory, Chinese Academy of Sciences, No. 80 Nandan Road, Shanghai, 200030, China}
\affiliation{Key Laboratory of Radio Astronomy, Chinese Academy of Sciences, Nanjing 210033, China}
\affiliation{School of Astronomy and Space Sciences, University of Chinese Academy of Sciences, No. 19A Yuquan Road, Beijing 100049, People's Republic of China}
\author[0009-0001-0943-1195]{Yijia Liu}
\affiliation{Department of Physics, Anhui Normal University, Wuhu, Anhui 241002, People's Republic of China}

\begin{abstract}

Dense outflowing gas, traced by transitions of molecules with large dipole moment, is important for understanding mass loss and feedback of massive star formation. HCN 3-2 and HCO$^+$ 3-2 are good tracers of dense outflowing molecular gas, which are closely related to active star formation. 
In this study, we present on-the-fly (OTF) mapping observations of HCN 3-2 and HCO$^+$ 3-2 toward a sample of 33 massive star-forming regions using the 10-m Submillimeter Telescope (SMT). With the spatial distribution of line wings of HCO$^+$ 3-2 and HCN 3-2, outflows are detected in 25 sources, resulting in a detection rate of 76$\%$. The optically thin H$^{13}$CN and H$^{13}$CO$^+$ 3-2 lines are used to identify line wings as outflows and estimate core mass. The mass $M_{out}$, momentum $P_{out}$, kinetic energy $E_{K}$, force $F_{out}$ and mass loss rate $\dot M_{out}$ of outflow and core mass, are obtained for each source. A sublinear tight correlation is found between the mass of dense molecular outflow and core mass, with an index of $\sim$ 0.8 and a correlation coefficient of 0.88.

\end{abstract}


\newpage

\section{Introduction} \label{sec:intro}

Molecular outflows were first detected within the core of the Orion molecular cloud \citep{1976ApJ...209L.137Z, 1976ApJ...210L..39K} and well studied in past decades \citep{1996ApJ...472..225S, 2004A&A...426..503W, 2005ApJ...625..864Z}.  Numerous studies have shown the significant relevance of outflows to the evolution and feedback of star formation \citep{1987ARA&A..25...23S, 2005ApJ...625..864Z, 2016ARA&A..54..491B}. Outflows form during the major accretion phase of protostellar evolution, serving as one of the earliest observable signatures of star formation \citep{1997IAUS..182P..63B, 2019MNRAS.488.4638L, 2020ApJ...901...31L}. As stars originate from the gravitational collapse of molecular clouds, accretion disks form around the protostars. Molecular outflows play a crucial role in removing excess angular momentum from the accreting material, thus facilitating accretion and increasing the mass of protostellar cores \citep{1987ARA&A..25...23S, 2000prpl.conf..789S, 2000prpl.conf..759K, 2010A&A...518L..77M}. 

The emission lines of CO and SiO are tracers for detecting molecular outflows \citep{2001ApJ...552L.167Z, 2005ApJ...625..864Z,  2023A&A...677A.148G}. Low-$J$ CO lines are widely used to trace total molecular outflows in star-forming regions  \citep{1976ApJ...209L.137Z, 1976ApJ...210L..39K, 2004A&A...426..503W}. SiO serves as a well-known tracer for molecular outflows, specifically indicating the presence of shocked gas resulting from outflow collisions with the surrounding environment \citep{1992A&A...254..315M, 2008A&A...482..809G, 2022MNRAS.512.5214D}. However, the abundance of SiO emission is enhanced due to grain sputtering \citep{1995ApJ...441..689B, 2021anms.confE..15R}, leading to significant uncertainties in obtaining outflow parameters with SiO lines. 

Dense outflowing molecular gas, situated closer to the central source than outflows traced by low-$J$ CO \citep{1991ApJ...373..137Y, 2006ApJ...646.1070A, 2015ApJ...810...10Z}, is more reflective of the evolution and feedback of active massive star formation, making it a powerful tool for studying the  massive star-formation process \citep{2010ApJ...724..687L, 2010ApJ...723.1019H}. Transitions of molecules with large dipole moments, such as lines of HCN, HCO$^{+}$, HNC, and CS, etc, require high densities ($n>10^4\ \rm cm^{-3}$) for efficient excitation. They are almost unaffected by contamination from environmental gas emissions, making them well suited for tracing dense outflowing molecular gas. 

Molecular outflows had been revealed with dense gas tracers and CO lines in past years.
The mapping of the outflows toward the dark cloud L1287, traced by CO 1-0 and HCO$^+$ 1-0, implies that the outflows are embedded in high-density gas and driven by a cold-type infrared source \citep{1991ApJ...373..137Y}.
CO 1-0 and HCO$^+$ 1-0 outflows in low-mass protostars were imaged using OVRO millimeter array of six 10.4 m telescopes \citep{2006ApJ...646.1070A}, indicating that dense gas from the outer regions of the circumstellar envelope, entrained by high-velocity flow, erodes the envelope and widens the outflow cavities. 
\citet{2015ApJ...810...10Z} found a more extended CO 3-2 outflow than that of high-density tracers in the massive star-forming region S255IR, indicating it is supposed to be driven by the jet bow shock mechanism. 
The mechanical luminosities, mass, mass-loss rates, and forces of outflows ($^{12}$CO, $^{13}$CO, HCO$^+$ 1-0, CS, SiO 2-1) toward nine nearby high-mass star-forming regions were studied with the 14 m millimeter telescope of the Purple Mountain Observatory \citep{2021ApJS..253...15L}. 
The morphology, kinematics, and energetics of the DR21 Main outflow traced by HCN, HCO$^+$ 1-0 were studied with IRAM 30 m and NOEMA telescopes \citep{2023A&A...679A..66S}, revealing it resemble a typical bipolar outflow. However, there still lacks sufficient studies on dense outflowing molecular gas in massive star-forming regions, particularly with large sample in the literature \citep{2021ApJS..253...15L, 2023A&A...679A..66S}, which could provide different informations compared to those obtained from CO observations. Furthermore, acquiring such information is crucial for enhancing our understanding of massive star-formation processes.


In this paper, we present HCO$^+$, HCN, H$^{13}$CO$^+$ and H$^{13}$CN 3-2 on-the-fly (OTF) observations toward a sample of 33 massive star forming regions with strong ($>$0.5K) H$^{13}$CN 2-1 emissions in the Milky Way from \citet{2014ApJ...783..130R}. The parallax measurements of this sample \citep{2014ApJ...783..130R} were obtained using Very Long Baseline Interferometry techniques, providing accurate distances, which are crucial for reducing uncertainties in calculating parameters such as mass, mass-loss rate and force of molecular outflow in each source. The observations and data reductions are described in Sect. 2. Section 3 presents the calculation and results on outflows parameters. In Sect.4, we discuss the relationship between outflows and cores.

\section{OBSERVATIONS AND DATA REDUCTION} \label{sec:obs}
The present study selects 33 massive star-forming regions from \citet{2014ApJ...783..130R} with parallax distances. The observations were carried out using the 10-m Submillimeter Telescope (SMT; Mt. Graham, Arizona) in the OTF mapping mode with 1.3 mm ALMA band 6 receiver. The OTF maps have sizes of 2$^{\prime}$×2$^{\prime}$ regions for HCO$^{+}$ 3-2 and HCN 3-2, 1.5$^{\prime}$×1.5$^{\prime}$ for H$^{13}$CO$^{+}$ 3-2 and H$^{13}$CN 3-2, respectively. The mapping center of each source is maser position from \citet{2014ApJ...783..130R} and presented in Table \ref{tab1}. The maps of HCO$^+$ 3-2 at rest frequency of 267.557633 GHz and HCN 3-2 at rest frequency of 265.886431 GHz were obtained in the upper sideband (USB) simultaneously. The isotopic molecules, H$^{13}$CO$^+$ 3-2 with rest frequency of 260.255342 GHz and H$^{13}$CN 3-2 with rest frequency of 259.011787 GHz were observed simultaneously also in the USB with the second tuning setup. The resulting maps have spatial resolution of $\sim$27.8$^{\prime \prime}$ at 260 GHz and step size of 15$^{\prime \prime}$ after re-gridding the OTF data of HCO$^{+}$ 3-2 and HCN 3-2 for each source. The map of H$^{13}$CO$^{+}$ or H$^{13}$CN 3-2 is gridded to match the center and each position of the spatial resolved HCO$^{+}$ or HCN 3-2 data. The more detailed information can be found in \citet{2023MNRAS.525.4761L}. The antenna temperature $T^{*}_{A}$ is converted to main beam brightness temperature $T_{mb}$ using the equation $T_{mb}= T^{*}_{A}/ \eta_{b}$, where the main beam efficiency $\eta_b$ is 0.77.

The data are reduced using the CLASS in GILDAS software package, following a specific process for each line of each source. 
At first, a refined average spectral line is obtained by averaging the spectral lines of all pixels in the mapping area to select line-free channels for baseline fitting. A first-order linear baseline fit is then applied to spectral line in each pixel. The spatially averaged H$^{13}$CO$^+$ and H$^{13}$CN 3-2 lines are well fitted with single component Gaussian profile, which are used to define the velocity ranges of HCO$^+$ and HCN 3-2 line wings.  The `print area' command is employed to obtain velocity integrated intensity for line wings in each pixel, as well as for the dense cores of H$^{13}$CN and H$^{13}$CO$^+$ 3-2 lines. 


\section{RESULTS}\label{sec:res}
\subsection{Line profiles of HCN 3-2 and HCO$^+$ 3-2}

We identify outflows by inspecting the line wing emission in the HCO$^+$ 3-2 \& HCN 3-2 spectra. HCO$^+$ 3-2 and HCN 3-2 are optically thin in line wing and optically thick in line core. The isotopic lines H$^{13}$CO$^+$ 3-2 and H$^{13}$CN 3-2, which are normally optically thin, are used to identify the velocity ranges of line cores. The line wings of HCO$^+$ 3-2 \& HCN 3-2 are defined as the velocity ranges with HCO$^+$ 3-2 \& HCN 3-2 and without H$^{13}$CO$^+$ 3-2 \& H$^{13}$CN 3-2 emission. Such emission of HCO$^+$ 3-2 and HCN 3-2 at lines wings are optically thin. The line wings of HCO$^+$ 3-2 and HCN 3-2 obtained in this way for each source are considered as dense molecular outflows in subsequent analyses.

Fig. \ref{fig1} presents the spatially averaged spectra of HCO$^+$ 3-2 and HCN 3-2, as well as their isotopic lines, H$^{13}$CO$^+$ 3-2 and H$^{13}$CN 3-2, for the 33 sources. In terms of spectral characteristics, the intensities between blue and red wings in most sources are similar, although certain sources exhibit significant differences, such as G005.88-00.39 and G043.16+00.01. Furthermore, the velocity ranges of the line wings for HCN 3-2 are generally wider than those of HCO$^+$ 3-2. Among the 33 sources, a total of 25 were found to exhibit outflows, while the remaining eight showed no evidence of outflows. Within the subset of sources with outflows, 18 exhibited bipolar lobes traced by both HCO$^+$ 3-2 and HCN 3-2, while 3 sources had a single lobe traced by both lines. Additionally, bipolar lobes were identified solely with HCO$^+$ 3-2 in 2 sources, whereas in another 2 sources, the bipolar lobes were solely identified with HCN 3-2. The parameters for all sources are presented in Table \ref{tab1}.

\subsection{Integrated intensity maps}
Fig. \ref{fig2} shows the velocity integrated maps of blue and/or red wings of HCO$^+$ 3-2, overlaid on greyscale of H$^{13}$CO$^+$ 3-2. In Fig. \ref{fig3}, the velocity integrated maps of blue and/or red wings of HCN 3-2 are presented, with H$^{13}$CN 3-2 as greyscale background. 

Regarding HCO$^+$ 3-2, clear bipolar morphology is found in 19 outflows, while 3 outflows exhibit only red lobes. In 7 sources, the locations between blue and red lobes can be well separated, with the largest observed separation of 54$''$ in G015.03-00.67, corresponding to a physical distance of 0.026 pc. In the remaining 12 sources, the lobes cannot be distinctly separated under the current spatial resolution. Furthermore, there are 11 sources where outflows are offset from the core centers. 
Similarly, for HCN 3-2, 19 outflows exhibit clear bipolar morphology, while 3 outflows exhibit only red lobes. The separation between blue and red lobes can be well resolved in 10 sources, including G015.03-00.67 where the largest observed separation of 54$''$ is noted. In the remaining 9 sources, the lobes cannot be clearly separated under the current spatial resolution. 9 sources show outflows offset from the core centers. 
The HCN 3-2 and HCO$^+$ 3-2 outflows present different morphology in each source for 9 sources with blue and red lobes well separated, indicating that these outflows do have varying physical and/or chemical conditions at different regions. The observational parameters for all sources are listed in Table \ref{tab2}.

\subsection{Parameters of outflows and cloud cores}
\subsubsection{The parameters with fixed abundance ratios}
With the reasonable assumptions that HCO$^+$ 3-2, HCN 3-2 at line wings and H$^{13}$CO$^+$ 3-2, H$^{13}$CN 3-2 in cloud cores are optically thin under local thermodynamic equilibrium (LTE) conditions, the column densities of the molecules can be derived with
\begin{equation}
N_{tot}=\frac{8\pi k\nu^2}{hc^3A_{ul}}\frac{Q(T_{ex})}{g_u}e^{E_u/kT_{ex}}\int_{}^{}T_{mb}{\rm d}v \qquad (\rm cm^{-2}),\label{equation1}
\end{equation}

where $k$ is the Boltzmann constant, $h$ is the the Planck constant, $c$ is the speed of light, $\nu$ is the frequency of the molecular emission line, $A_{ul}$ the Einstein emission coefficient, $T_{ex}$ is the excitation temperature, and $Q(T_{ex})$ is the partition function, $g_u$ is the upper-level degeneracy, $E_u$ is the upper-level energy, and $T_{mb}$ is the main beam temperature. The values of $A_{ul}$, $Q(T_{ex})$, $g_u$ and $E_u$ for the four molecular lines are from the Cologne Database for Molecular Spectroscopy \footnote{\url{https://cdms.astro.uni-koeln.de/classic/predictions/catalog/partition_function.html}}  \citep{2001A&A...370L..49M, 2005JMoSt.742..215M}. The values of $\int_{}^{}T_{mb}{\rm d}v$ are the sum over the entire region condensed into a single beam. The assumption of $T_{ex}=18.75\ \rm K$ is used in the calculations of column density, as the impact of $T_{ex}$ on results is not that significant. For example, when $T_{ex}=37.5\ \rm K$, the calculated outflow mass is $\sim$1.9 times of that obtained at 18.75 K, as also discussed for CO 3-2 with different $T_{ex}$ \citep{2022A&A...660A..39S}.

We first derive the outflow mass by using the assumed relative abundance $\rm [HCN/H_2]=5 \times10^{-9}$ and $\rm [HCO^+/H_2]=2.3 \times10^{-9}$ obtained in Orion \citep{1987ApJ...315..621B}, with 
\begin{equation}
M=\frac{N_{tot}\times S_{beam}\times D^2 \times m_{H_{2}}\times H}{M_\odot},\label{equation2}
\end{equation}
where $N_{tot}$ is the column densities obtained before, $S_{beam}$ denotes the area of the beam size in square radian, $m_{H_{2}}=3.35\times10^{-24}\ \rm g$ is the mass of a hydrogen molecule, $H $ represent the relative abundance of HCN and HCO$^+$ with hydrogen, $D$ is the distance of the source to the sun in cm, and $M_\odot$ is the solar mass in $\rm g$. It can also be used for calculating the mass of cloud cores with H$^{13}$CN 3-2 and H$^{13}$CO$^+$ 3-2 emission. 

The mass loss rate is defined as
\begin{equation}
\dot{M}_{out}= M/t_{dyn}, \label{equation3}
\end{equation}
Subsequently, the outflow momentum is calculated as:
\begin{equation}
P_{out}=M_{out}v_{out}, \label{equation4}
\end{equation}
and the time-averaged kinetic energy of the outflow, is taken as
\begin{equation}
E_k=\frac1{2}M_{out}v_{out}^2\label{equation5}
\end{equation}
also, the outflow force is obtained as
\begin{equation}
F_{out}=\dot{M}_{out}v_{out}, \label{equation6}
\end{equation}

where $t_{dyn}$ represents the dynamical timescale of the outflow and is determined by the ratio between the linear size of the outflow along the flow axis ($R_{out}$) and its flow velocity ($v_{out}$). The method for estimating $t_{dyn}$ involves calculating the ratio of the separation between the peak of the blue and red lobes to the mean outflow velocity \citep{1990ApJ...354..687C, 2005ApJ...625..864Z}. Given the inherent difficulty in determining the orientation of the outflow, the inclination angle $c_1$ of $45^\circ$ was used for the calculations. The impact of inclination angle was discussed in \citet{1990ApJ...354..687C}. The resulting HCN outflow properties are presented in Table \ref{tab2} and \ref{tab3}.
\subsubsection{The parameters obtained without fixed HCO$^+$ abundance}
The abundance [HCN/H$_2$] is generally assumed to be similar in different Galactic molecular clouds. However, the abundance [HCO$^+$/H$_2$] varies among different sources. Therefore, a crude method is used to obtain the abundance [HCO$^+$/H$_2$] for each source with following assumptions, to calculate outflow mass traced by HCO$^+$ 3-2. First, the abundance ratio of $^{12}$C to $^{13}$C is equal to that of HCN \& HCO$^+$ to H$^{13}$CN \& H$^{13}$CO$^+$. Second, the abundance ratio of $^{12}$C to $^{13}$C remains consistent between the outflow and cloud cores. Third, the cloud cores traced by H$^{13}$CN and H$^{13}$CO$^+$ are with identical mass.
The abundance ratio of $^{12}$C to $^{13}$C in each source can be determined by \citet{2023A&A...670A..98Y} based on the Galactocentric distance, with the following calculation formula:
\begin{equation}
^{12}C/^{13}C=(3.77\pm 0.81)R_{GC}+(20.76\pm 4.61).
\end{equation}

The abundance [H$^{13}$CN/H$_2$] in each source can be determined using the abundance [HCN/H$_2$] and $^{12}$C/$^{13}$C ratio. The mass of each cloud core can be calculated using equations (\ref{equation1}) and (\ref{equation2}) with H$^{13}$CN 3-2 line. Assuming that both H$^{13}$CN 3-2 and H$^{13}$CO$^+$ 3-2 trace the same dense core, the abundance [H$^{13}$CO$^+$/H$_2$] can be calculated. Subsequently, the abundance [HCO$^+$/H$_2$] in each source can be obtained with $^{12}$C/$^{13}$C ratio. Note that  inaccuracies in the $^{12}$C to $^{13}$C ratio in each source only affect the estimation of core mass, while the derived [HCO$^+$/H$_2$] abundance and the outflow mass traced by HCO$^+$ 3-2 obtained in this way remain unaffected, as the $^{12}$C to $^{13}$C ratio cancels out during these calculations. Consequently, the outflow mass in each source, traced by HCO$^+$ 3-2, can be determined with the derived [HCO$^+$/H$_2$] abundance. Furthermore, the mass-loss rate, momentum, kinetic energy, and force of outflow can be obtained. The calculated results with unfixed HCO$^+$ abundance are presented in Table \ref{tab2} and Table \ref{tab3}.

The outflow masses in 25 of 33 massive star forming regions, as the summation of blue and red ones if both are detected, range from 0.28 M$_\odot$ in G012.80-00.20 to 165.4 M$_\odot$ in G043.16+00.01, with typical values in the tens of M$_\odot$ range. The core masses vary widely from 3.95 M$_\odot$ in G183.72-03.60 to 8730 M$_\odot$ in G043.16+00.01. The dynamical timescales for dense outflowing molecular gas span from $2.6 \times 10^3$ to $6.6 \times 10^4$ yr. The mass-loss rates range from $2.1 \times 10^{-5}$ to $5.2 \times 10^{-3} \ \rm M_\odot \ yr^{-1}$, with typical values in the range of $10^{-4}$ to $10^{-3}\ \rm M_\odot\ yr^{-1}$. The outflow momentum spans from a few to several thousand $\rm M_\odot\ km\ s^{-1}$, while the outflow kinetic energy ranges from a few to tens of thousands $\rm M_\odot\ km^2s^{-2}$. Additionally, outflow forces range from $8.1 \times 10^{-5}$ to $\rm 3.44 \times 10^{-2}\ M_\odot\ km\ s^{-1}  yr^{-1}$, with typical values between $10^{-3}$ and $10^{-2}$ $\rm M_\odot\ km\ s^{-1}yr^{-1}$. 

Fig. \ref{fig4} presents the plots of outflow mass from 25 sources. The abscissa indicates the outflow mass traced by HCN 3-2 and the ordinate indicates that traced by HCO$^+$ 3-2. In the left of Fig. \ref{fig4}, the outflow mass in each source traced by HCO$^+$ 3-2 was obtained with fixed abundance ratio of $[\rm HCO^+/H_2]=2.3 \times10^{-9}$, while in the right of Fig. \ref{fig4} it was obtained with HCO$^+$ abundance calculated with the method described before. Least squares fitting was employed for both plots, both of which presented tight correlations between the outflow  mass traced by HCN 3-2 and HCO$^+$ 3-2.

\section{DISCUSSION}\label{sec:dis}

\subsection{Relationship between outflows and cores}

A positive correlation between the outflow mass and the core mass was found in our sample with 23 sources, which was shown in the left pane Fig. \ref{fig5}. The mass of outflows ($M_{out}$) were traced by HCN 3-2, while the core mass were traced by H$^{13}$CN 3-2. The solid line is the result of least squares fitting, which is $M_{out}\sim 0.18\times M_{core}^{0.77}$ with the correlation coefficient ($R$) of 0.92. Due to the inability to resolve distances along the line of sight, t$_{dyn}$ of eight sources cannot be calculated. The right panel in Fig. \ref{fig5} is the plot of the mass-loss rate ${\dot M_{out}}$ traced by HCN 3-2 versus the core mass $M_{core}$ traced by H$^{13}$CN 3-2 for 15 sources. The result of least squares fitting for mass-lose rate and core mass is $\dot M_{out}\sim 0.16\times M_{core}^{0.75}$, with correlation coefficient ($R$) of 0.88.
 
 The trend of increment of outflow mass with increasing core mass had been found with other tracers. 
 \citet {2002ASPC..267..341B} used CO 2-1 to trace the outflow toward 26 massive star forming regions and found the correlation between the outflow mass and the core mass derived from the 1.2 mm dust emission. The best fit, spanning over three orders of magnitude in core mass, was $M_{out} \sim 0.1\times M_{core}^{0.8}$. \citet{2015MNRAS.453..645M} mapped the outflows and cores of 65 Massive Young Stellar Objects (MYSOs), and compact $\rm H_{\Rmnum{2}}$ regions, resulting for core mass and outflow mass as $M_{out} \propto M_{core}^{0.8}$. C$^{18}$O 3-2 was used to establish the core mass and CO 3-2 was used to calculate the outflow mass in \cite{2015MNRAS.452..637M}. Comparing the indices of the three power functions, it can be found that there is a similar trend between the outflows and cores in dense gas and those in large-scale gas.
 
 The relationship between the mass of outflows and the core mass is sublinear with the index of $\sim$0.8, based on our results for dense outflowing molecular gas, as well as for total outflowing gas traced by CO lines \citep{2002ASPC..267..341B,2015MNRAS.453..645M}. Such results indicate that dense gas fraction in such outflows may be similar and the ratio of outflowing gas and the gas in the cores decreases with increasing core mass. Further theoretical studies and large sample high-resolution observations are needed for understanding such sublinear relation. 

 \subsection{Different outflow properties traced by different molecules}
Outflowing molecular gas traced by different tracers, can reveal multiple phase of molecular outflows in massive star forming regions. Normally, molecular outflows traced by low-$J$ CO are more extended than that traced by dense gas tracers, such as seen in G121.29+00.65 (L1287) with the SMA \citep{2019A&A...621A.140J} and in a sample of outflows with ALMA \citep{2021MNRAS.507.4316B}. In L1287 as an example, the blue wings of HCO$^+$ 1-0, CO 1-0 \citep{1991ApJ...373..137Y} and HCN 3-2 (see Fig. \ref{fig3} ) exhibit higher intensity and larger spatial distribution compared to their corresponding red wings, while the HCO$^+$ 3-2 emission shows reversed wing characteristics. Additionally,  similar spatial distribution of the blue and red wings of HCO$^+$ 1-0 and CO 1-0 emissions was reported in  \cite{1991ApJ...373..137Y}, while  the wings of HCO$^+$ 3-2 and HCN 3-2 present different seperations to that of HCO$^+$ 1-0 and CO 1-0, with red lobe located in the northeast and blue lobe in the southwest (see Fig. \ref{fig2} ). These differences may arise from variations in volume density and kinetic temperature of different outflowing gas. Line wings of dense gas tracers HCN 3-2 and HCO$^+$ 3-2 in this sample of massive star forming regions provide good candidates for further high-resolution observations to understand outflow properties and the feedback of massive formation in the future.

\section{SUMMARY}\label{sec:dis}
With OTF mapping observations toward a sample of 33 massive star forming regions using 10-m SMT for HCN 3-2, HCO$^+$ 3-2, H$^{13}$CN 3-2, H$^{13}$CO$^+$ 3-2, we detected dense outflowing gas in 25 sources, with bipolar lobes in 18 sources and a single lobe in 3 sources seen with both HCN 3-2 and HCO$^+$ 3-2, while bipolar outflows only seen with HCO$^+$ 3-2 in two sources and only seen with HCN 3-2 in another sources. The outflow mass ranges from 0.28 to 165.4 M$_\odot$, while the core mass spans from 3.95 to 8730 M$_\odot$ and the typical mass-loss rate is a few times $10^{-4}$ $\rm M_\odot yr^{-1}$. 
A sublinear tight correlation, with index of $\sim$0.8 and correlation coefficient of 0.88, between the mass of dense molecular outflow and core mass, is found. The results of detected dense gas outflows provide a good sample for further high-resolution observations for studying outflows and star formation feedback during massive star forming process. 
 
\section*{Acknowledgements}
We would like to acknowledge the help of the staff of the 10-m SMT  for assistance with the observations. This work is supported by National Key R$\&$D Program of China under grant 2023YFA1608204 and the National Natural Science Foundation of China grant 12173067.

\clearpage
\renewcommand\thetable{\arabic{table}} 
\setcounter{table}{0}
\renewcommand\arraystretch{0.94}
\setlength{\tabcolsep}{2.pt}
\centerwidetable
\small
\begin{threeparttable}
\begin{longtable}[c]{llllllllllll}
	\caption{Basic information and data parameters of 33 sources}
	\label{tab1}\\
	\hline 
	\hline
	Source      & R.A.(J2000)          & Decl.(J2000)             & $D$       & $D_{GC}$   & Line                 & $\int T_{\rm mb}\rm  d\rm{v}_{Core}^a$           &  $\int T_{\rm mb}\rm  d\rm{v}_{Blue}^b$                    & $\int T_{\rm mb}\rm  d\rm{v}_{Red}$   & $v_{LSR}$   & Blue Range    & Red Range     \\
	Alias           & (hh:mm:ss)  & (dd:mm:ss)   & (kpc) & (kpc)            & (3-2)            & (K·km\,s$^{-1}$)           & (K·km\,s$^{-1}$)               & (K·km\,s$^{-1}$)  &(km\,s$^{-1}$)   &(km\,s$^{-1}$)    &(km\,s$^{-1}$)    \\	
\hline
\endfirsthead
\caption[]{(continued)}\\	         
	\hline 
	\hline
	Source          & R.A.(J2000)         & Decl.(J2000)        & D       & D$_{GC}$   & Line                 & $\int T_{\rm mb}\rm  d\rm{v}$$_{Core}$           &  $\int T_{\rm mb}\rm  d\rm{v}$$_{Blue}$                     & $\int T_{\rm mb}\rm  d\rm{v}$$_{Red}$        & $v_{LSR}$      & Blue Range    & Red Range     \\
	Alias           & (hh:mm:ss)  & (dd:mm:ss)   & (kpc) & (kpc)            & (3-2)            & (K·km\,s$^{-1}$)           & (K·km\,s$^{-1}$)               & (K·km\,s$^{-1}$)   &(km\,s$^{-1}$)  &(km\,s$^{-1}$)    &(km\,s$^{-1}$)  \\
  \hline
 \endhead
\hline
\endfoot
G121.29+00.65 & 00:36:47.35 & 63:29:02.2 & 0.93 & 8.8 & HCN & 5.28(0.33) & 14.79(0.79) & 6.71(0.56) &-18 & -32 -22 & -13 -8 \\
~ & ~ & ~ & ~ & ~ & HCO$^+$ & 9.97(0.27) & 8.39(0.81) & 10.56(0.81) &-18 & -30 -22 & -13 -5 \\
G123.06-06.30 & 00:52:24.70 & 56:33:50.5 & 2.82 & 10.1 & HCN & 2.98(0.16) & 7.49(0.46) & 4.06(0.4) &-31 & -44 -35 & -25 -18 \\
~ & ~ & ~ & ~ & ~ & HCO$^+$ & 6.11(0.29) & 5.00(0.41) & 4.36(0.41) &-30 & -43 -35 & -25 -17 \\
G183.72-03.60 & 05:40:24.23 & 23:50:54.7 & 1.75 & 10.1 & HCN & 0.48(0.12) & 6.10(0.56) & 7.25(0.6) &2   & -7 0 & 4 12 \\
~ & ~ & ~ & ~ & ~ & HCO$^+$ & 1.73(0.19) & 6.99(0.37) & 8.36(0.34) &2  & -6 0 & 4 9 \\
G005.88-00.39 & 18:00:30.31 & -24:04:04.5 & 2.99 & 5.3 & HCN & 57.50(1.17) & 49.91(1.25) & 86.62(1.32) &11 & -25 1 & 18 47 \\
~ & ~ & ~ & ~ & ~ & HCO$^+$ & 49.90(0.60) & 43.37(1.24) & 91.94(1.49) &11 & -24 1 & 18 54 \\
G010.62-00.38 & 18:10:28.55 & -19:55:48.6 & 4.95 & 3.6 & HCN & 53.41(0.54) & 10.10(1.12) & 19.03(1.17) &-3  & -23 -13 & 6 17 \\
~ & ~ & ~ & ~ & ~ & HCO$^+$ & 63.18(0.49) & 5.03(0.81) & 13.20(0.92) &-3  & -20 -13 & 6 15 \\
G011.91-00.61 & 18:13:58.12 & -18:54:20.3 & 1.25 & 5.1 & HCN & 5.99(0.27) & 12.87(1.08) & 8.46(0.99) &37 & 16 29 & 44 55 \\
~ & ~ & ~ & ~ & ~ & HCO$^+$ & 5.49(0.28) & 3.21(0.91) & 2.09(0.91) &36  & 22 29 & 44 51 \\
G012.80-00.20 & 18:14:14.23 & -17:55:40.5 & 2.92 & 5.5 & HCN & 74.06(0.96) & 71.03(1.86) & 102.29(2.38) &36   & 18 29 & 42 60 \\
~ & ~ & ~ & ~ & ~ & HCO$^+$ & 108.30(0.69) & 31.95(1.2) & 49.62(1.47) &36   & 23 29 & 42 51 \\
G015.03-00.67 & 18:20:24.81 & -16:11:35.3 & 1.98 & 6.4 & HCN & 38.54(0.42) & 31.65(0.75) & 36.70(0.81) &20  & 7 13 & 26 33 \\
~ & ~ & ~ & ~ & ~ & HCO$^+$ & 82.64(0.54) & 31.02(1.29) & 41.35(1.29) &20 & 5 13 & 26 34 \\
G023.43-00.10 & 18:34:39.29 & -08:31:25.4 & 5.88 & 3.7 & HCN & 8.40(0.47) & 13.64(0.79) & 14.60(0.97) &102 & 78 90 & 110 128 \\
~ & ~ & ~ & ~ & ~ & HCO$^+$ & 10.16(0.52) & 6.63(0.97) & 5.59(0.91) &101  & 81 90 & 110 118 \\
G023.00-00.41 & 18:34:40.20 & -09:00:37.0 & 4.59 & 4.5 & HCN & 4.56(0.22) & 11.39(0.67) & 8.49(0.61) &76  & 57 70 & 85 96 \\
~ & ~ & ~ & ~ & ~ & HCO$^+$ & 5.11(0.32) & 13.87(0.75) & 4.57(0.71) &75 & 59 70 & 85 95 \\
G029.95-00.01 & 18:46:03.74 & -02:39:22.3 & 5.26 & 4.6 & HCN & 13.29(0.26) & 17.72(1.61) & 10.94(1.53) &97 & 81 92 & 104 114 \\
~ & ~ & ~ & ~ & ~ & HCO$^+$ & 9.76(0.24) & 6.94(1.25) & 4.70(1.25) &98 & 84 92 & 104 112 \\
G032.04+00.00 & 18:49:36.58 & -00:45:46.9 & 5.18 & 4.2 & HCN & 4.75(0.21) & 14.39(1.23) & 15.23(1.14) &95 & 73 88 & 102 115 \\
~ & ~ & ~ & ~ & ~ & HCO$^+$ & 5.05(0.24) & 4.54(0.68) & 5.84(0.64) &94 & 79 88 & 102 110 \\
G043.16+00.01 & 19:10:13.41 & 09:06:12.8 & 11.11 & 7.6 & HCN & 31.97(0.55) & 7.96(1.17) & 17.73(1.21) &8  & -25 -11 & 25 40 \\
~ & ~ & ~ & ~ & ~ & HCO$^+$ & 50.52(0.62) & 15.95(0.91) & 52.64(1.22) &8 & -25 -11 & 25 50 \\
G069.54-00.97 & 20:10:09.07 & 31:31:36.0 & 2.46 & 7.8 & HCN & 7.18(0.29) & 11.04(0.91) & 10.84(0.94) &12 & -10 5 & 18 34 \\
~ & ~ & ~ & ~ & ~ & HCO$^+$ & 17.76(0.43) & 17.26(1.07) & 16.93(1.04) &12  & -11 5 & 18 33 \\
G075.76+00.33 & 20:21:41.09 & 37:25:29.3 & 3.51 & 8.2 & HCN & 8.15(0.33) & 14.53(0.73) & 23.82(0.8) &-1 & -19 -8 & 5 18 \\
~ & ~ & ~ & ~ & ~ & HCO$^+$ & 8.50(0.36) & 10.36(0.53) & 17.41(0.59) &-1 & -17 -8 & 5 16 \\
G081.87+00.78 & 20:38:36.43 & 42:37:34.8 & 1.30 & 8.2 & HCN & 27.54(0.40) & 18.51(1.2) & 24.22(1.31) &10 & -7 3 & 16 28 \\
~ & ~ & ~ & ~ & ~ & HCO$^+$ & 38.70(0.38) & 22.19(1.23) & 25.91(1.23) &10 & -6 3 & 16 25 \\
G081.75+00.50 & 20:39:01.99 & 42:24:59.3 & 1.50 & 8.2 & HCN & 8.76(0.23) & 26.05(0.42) & 21.77(0.45) &-3 & -15 -9 & 1 8 \\
~ & ~ & ~ & ~ & ~ & HCO$^+$ & 18.94(0.33) & 10.92(1.09) & 16.32(1.18)&-4  & -15 -9 & 1 8 \\
G109.87+02.11 & 22:56:18.10 & 62:01:49.5 & 0.70 & 8.6 & HCN & 20.59(0.38) & 16.27(0.91) & 23.13(0.91) &-10 & -28 -17 & -4 7 \\
~ & ~ & ~ & ~ & ~ & HCO$^+$ & 45.42(0.40) & 58.56(2.48) & 19.81(2.07) &-10 & -40 -17 & -4 20 \\
G028.86+00.00 & 18:43:46.22 & -03:35:29.6 & 7.41 & 4.0 & HCN & 3.57(0.06) & … & …  &105 & … & … \\
~ & ~ & ~ & ~ & ~ & HCO$^+$ & 4.25(0.08) & 3.96(0.56) & 5.26(0.52) &104  & 91 99  & 107 114 \\
G043.79-00.12 & 19:11:53.99 & 09:35:50.3 & 6.02 & 5.7 & HCN & 3.49(0.04) & … & … &44 & … & … \\
~ & ~ & ~ & ~ & ~ & HCO$^+$ & 4.21(0.04) & 5.92(0.62) & 5.09(0.52) &44 & 31 38 & 50 55 \\
G031.28+00.00 & 18:48:12.39 & -01:26:30.7 & 4.27 & 5.2 & HCN & 4.70(0.08) & 6.54(1.01) & 7.04(1.15) &109 & 92 102 & 115 128 \\
G031.58+00.00 & 18:48:41.68 & -01:09:59.0 & 4.90 & 4.9 & HCN & 2.74(0.06) & 7.25(0.41) & 10.86(0.67) &96  & 86 92 & 101 117 \\
G188.94+00.88 & 06:08:53.53 & 21:38:28.7 & 2.10 & 10.4 & HCN & 4.85(0.11) & … & 18.78(0.91) & 3 & … & 7 15 \\
~ & ~ & ~ & ~ & ~ & HCO$^+$ & 9.66(0.09) & … & 20.37(0.65) &3 & … & 7 17 \\
G009.62+00.19 & 18:06:14.66 & -20:31:31.7 & 5.15 & 3.3 & HCN & 18.77(0.17) & … & 20.22(0.81) &6 & … & 14 26 \\
~ & ~ & ~ & ~ & ~ & HCO$^+$ & 14.01(0.11) & … & 9.93(0.99) &6 & … & 14 30 \\
G011.49-03.60 & 18:16:22.13 & -19:41:27.2 & 1.25 & 7.1 & HCN & 2.44(0.05) & … & 8.54(0.64) &11 & … & 13 23 \\
~ & ~ & ~ & ~ & ~ & HCO$^+$ & 7.23(0.08) & … & 5.83(0.44) &11 & … & 13 20 \\
\hline
 \end{longtable}	
  \footnotesize   {\textbf{Notes.}  Columns are: the source names/coordinates, right ascension (R.A.), declination (Decl.), trigonometric parallax distance ($D$) \citep{2014ApJ...783..130R}, Galactocentric distance ($D_{GC}$), line, flux integral intensity of cores ($\int T_{\rm mb}\rm  d\rm{v}_{Core}$), flux integral intensity of blue wing ($\int T_{\rm mb}\rm  d\rm{v}_{Blue}$), flux integral intensity of red wing ($\int T_{\rm mb}\rm  d\rm{v}_{Red}$), velocity range of blue wing, velocity range of red wing. $a.$ Calculations are conducted using H$^{13}$CO$^+$ or H$^{13}$CN 3-2. $b.$ Calculations are conducted using HCO$^+$ or HCN 3-2.  }
                         \end{threeparttable}

 \clearpage
\begin{table*}
\centering
\setlength{\tabcolsep}{0.1in}
\centering
\caption{Observational parameters for each source of HCN 3-2, HCO$^+$ 3-2 \label{tab2}}

\vspace{-1mm}
\begin{tabular}{lllllllll}
  \hline
    \hline
             
  \multirow{2}{*}{Source name}          
    &   \multicolumn{4}{c}{HCN 3-2 }       &\multicolumn{4}{c}{HCO$^+$ 3-2 }    \\
 &  1 $\sigma _{blue}$    & 1 $\sigma _{red}$  & starting                  &  step                    &  1 $\sigma _{blue}$    & 1 $\sigma _{red}$  & starting                 &  step  \\
 &  K km\,s$^{-1}$          &  K km\,s$^{-1}$       &  K km\,s$^{-1}$     &  K km\,s$^{-1}$    &  K km\,s$^{-1}$          &  K km\,s$^{-1}$      &  K km\,s$^{-1}$     &  K km\,s$^{-1}$   \\
    
\hline
G121.29+00.65  & 0.22 & 0.16 & 0.95 & 0.36 & 0.17 & 0.17 & 1.00 & 0.18\\
G123.06-06.30  & 0.14 & 0.12 & 0.70 & 0.26 & 0.14 & 0.14 & 0.50 & 0.20\\
G183.72-03.60  & 0.13 & 0.14 & 0.60 & 0.35 & 0.12 & 0.11 & 0.65 & 0.25\\
G005.88-00.39  & 0.40 & 0.43 & 2.00 & 6.00 & 0.35 & 0.42 & 2.00 & 6.00\\
G010.62-00.38  & 0.38 & 0.40 & 1.65 & 1.00 & 0.29 & 0.33 & 1.50 & 0.80\\
G011.91-00.61  & 0.35 & 0.32 & 1.30 & 0.45 & 0.24 & 0.24 & 0.80 & 0.25\\
G012.80-00.20  & 0.24 & 0.31 & 2.40 & 1.80 & 0.17 & 0.21 & 1.00 & 1.20\\
G015.03-00.67  & 0.14 & 0.15 & 1.10 & 0.50 & 0.16 & 0.16 & 1.10 & 0.60\\
G023.43-00.10  & 0.26 & 0.32 & 1.00 & 1.00 & 0.21 & 0.19 & 0.90 & 0.30\\
G023.00-00.41  & 0.21 & 0.19 & 1.15 & 0.40 & 0.21 & 0.20 & 0.75 & 0.50\\
G029.95-00.01  & 0.32 & 0.31 & 1.00 & 0.66 & 0.25 & 0.25 & 0.90 & 0.30\\
G032.04+00.00  & 0.27 & 0.25 & 1.20 & 0.80 & 0.21 & 0.20 & 0.75 & 0.35\\
G043.16+00.01  & 0.20 & 0.21 & 1.00 & 0.80 & 0.21 & 0.28 & 0.90 & 3.20\\
G069.54-00.97  & 0.27 & 0.28 & 0.90 & 0.80 & 0.24 & 0.23 & 0.75 & 1.10\\
G075.76+00.33  & 0.24 & 0.26 & 1.50 & 0.60 & 0.18 & 0.20 & 1.00 & 0.70\\
G081.87+00.78  & 0.26 & 0.28 & 1.50 & 0.80 & 0.22 & 0.22 & 1.20 & 1.00\\
G081.75+00.50  & 0.07 & 0.08 & 0.50 & 0.80 & 0.17 & 0.18 & 0.80 & 0.60\\
G109.87+02.11 & 0.27 & 0.27 & 2.48 & 0.50 & 0.38 & 0.32 & 3.00 & 2.00\\
G028.86+00.00 & ... & ... & ... & ... & 0.19 & 0.18 & 0.70 & 0.20 \\
G043.79-00.12 & ... & ... & ... & ... & 0.17 & 0.15 & 0.47  & 0.15 \\
G031.28+00.00 & 0.23 & 0.26 & 0.85 & 0.27 & ... & ... & ... & ... \\
G031.58+00.00 & 0.14 & 0.23 & 0.93 & 0.30 & ... & ... & ... & ... \\
G188.94+00.88 & ... & 0.25 & 0.90 & 0.30 & ... & 0.27 & 1.00 & 0.30 \\
G009.62+00.19 & ... & 0.31 & 1.00 & 1.10 & ... & 0.32 & 0.96 & 1.00 \\
G011.49-03.60 & ... & 0.24 & 1.00 & 0.40 & ... & 0.18 & 0.90 & 0.25 \\
\hline
\end{tabular}
\\
\end{table*}

 \clearpage
\renewcommand\thetable{\arabic{table}}    
\setcounter{table}{2}
\renewcommand\arraystretch{0.9}
\setlength{\tabcolsep}{7pt}
\centerwidetable
\small
\begin{threeparttable}
\begin{longtable}{lllllllll}
	\caption{Source characteristics of 33 sources}
	\label{tab3}\\
	\hline 
	\hline
	Source        &$M$ $_{Core}$    & Line          &$ f(X)$            & $t_{dyn}$                      & $M$ $_{Blue}$                        &$M$ $_{Red}$                    &$\dot{M}$ $_{Blue}$                       & $\dot{M}$ $_{Red}$  \\
	Alias           & (M$_{\odot}$)     &                 & (10$^{-9}$ )    & (10$^{4}$ yr)          & (M$_{\odot}$)                   &(M$_{\odot}$)              & (10$^{-4}$ )                    & (10$^{-4}$ )  \\	
	                &     &                 &                       &                                 &                                          &                                   & (M$_{\odot}$ yr$^{-1}$  )       & (M$_{\odot}$ yr$^{-1}$  )          \\
\hline
\endfirsthead
\caption[]{(continued)}\\	         
	\hline 
	\hline
	Source        &$M$ $_{Core}$    & Line          &$ f(X)$            & $t_{dyn}$                          & $M$ $_{Blue}$                        &$M$ $_{Red}$                    &$\dot{M}$ $_{Blue}$                       & $\dot{M}$ $_{Red}$  \\
	Alias           & (M$_{\odot}$)     &                 & (10$^{-9}$ )    & (10$^{4}$ yr)          & (M$_{\odot}$)                   &(M$_{\odot}$)              & (10$^{-4}$ )                    & (10$^{-4}$ )  \\	
	                &     &                 &                       &                                 &                                          &                                   & (M$_{\odot}$ yr$^{-1}$ )        & (M$_{\odot}$ yr$^{-1}$   )         \\
\hline
 \endhead
\hline
\endfoot
G121.29+00.65 & 11.12(0.69) & HCN & 5.00 & 1.44  & 0.49(0.03) & 0.22(0.02) & 0.34(0.02) & 0.15(0.01) \\
 ~ & ~ & HCO$^+$ & 5.52 & 1.14  & 0.14(0.01) & 0.18(0.01) & 0.13(0.01) & 0.16(0.01) \\
 G123.06-06.30 & 63.63(3.42) & HCN & 5.00 & 1.49  & 2.28(0.14) & 1.24(0.12) & 1.53(0.09) & 0.83(0.08) \\
 ~ & ~ & HCO$^+$ & 5.99 & 1.39  & 0.46(0.04) & 0.40(0.04) & 0.54(0.04) & 0.47(0.04) \\
 G183.72-03.60 & 3.95(0.99) & HCN & 5.00 & 1.92  & 0.72(0.07) & 0.85(0.07) & 0.37(0.03) & 0.44(0.04) \\
 ~ & ~ & HCO$^+$ & 10.53 & 1.97  & 0.23(0.01) & 0.28(0.01) & 0.12(0.01) & 0.14(0.01) \\
 G005.88-00.39 & 918.75(18.69) & HCN & 5.00 & ...  & 17.00(0.43) & 29.60(0.45) & … & … \\
 ~ & ~ & HCO$^+$ & 2.53 & ...  & 16.90(0.48) & 35.90(0.58) & … & … \\
 G010.62-00.38 & 1929.15(19.5) & HCN & 5.00 & ...  & 9.46(1.05) & 17.80(1.10) & … & … \\
 ~ & ~ & HCO$^+$ & 3.45 & ...  & 5.34(0.86) & 14.00(0.98) & … & … \\
 G011.91-00.61 & 16.42(0.74) & HCN & 5.00 & ...  & 0.77(0.06) & 0.51(0.06) & … & … \\
 ~ & ~ & HCO$^+$ & 2.68 & 0.77  & 0.17(0.05) & 0.11(0.05) & 0.27(0.08) & 0.18(0.08) \\
 G012.80-00.20 & 1157.96(15.01) & HCN & 5.00 & 1.09  & 23.30(0.61) & 33.50(0.78) & 21.40(0.56) & 30.90(0.72) \\
 ~ & ~ & HCO$^+$ & 4.26 & 1.77  & 9.83(0.37) & 15.30(0.45) & 4.04(0.15) & 6.28(0.19) \\
 G015.03-00.67 & 301.61(3.29) & HCN & 5.00 & ...  & 4.75(0.11) & 5.51(0.12) & … & … \\
 ~ & ~ & HCO$^+$ & 6.25 & 3.15  & 3.52(0.15) & 4.69(0.15) & 0.69(0.03) & 0.92(0.03) \\
 G023.43-00.10 & 434(24.28) & HCN & 5.00 & 1.51  & 18.10(1.05) & 19.30(1.28) & 12.00(0.69) & 12.80(0.85) \\
 ~ & ~ & HCO$^+$ & 3.53 & ...  & 9.01(1.32) & 7.60(1.24) & … & … \\
 G023.00-00.41 & 157.94(7.62) & HCN & 5.00 & 2.1  & 9.19(0.54) & 6.85(0.49) & 4.38(0.26) & 3.27(0.24) \\
 ~ & ~ & HCO$^+$ & 3.27 & 1.55  & 6.60(0.36) & 2.17(0.34) & 6.45(0.35) & 2.13(0.33) \\
 G029.95-00.01 & 610.39(11.94) & HCN & 5.00 & ...  & 18.80(1.70) & 11.60(1.62) & … & … \\
 ~ & ~ & HCO$^+$ & 2.15 & ...  & 8.85(1.59) & 6.00(1.59) & … & … \\
 G032.04+00.00 & 202.3(8.94) & HCN & 5.00 & ...  & 14.80(1.26) & 15.60(1.17) & … & … \\
 ~ & ~ & HCO$^+$ & 3.11 & ...  & 2.68(0.40) & 3.45(0.38) & … & … \\
 G043.16+00.01 & 8730.22(150.19) & HCN & 5.00 & ...  & 37.50(5.51) & 83.60(5.70) & … & … \\
 ~ & ~ & HCO$^+$ & 4.61 & ...  & 38.40(2.19) & 127.00(2.94) & … & … \\
 G069.54-00.97 & 98(3.96) & HCN & 5.00 & 0.79  & 2.56(0.21) & 2.51(0.22) & 3.22(0.27) & 3.16(0.27) \\
 ~ & ~ & HCO$^+$ & 7.25 & ...  & 1.08(0.07) & 1.06(0.07) & … & … \\
 G075.76+00.33 & 233.73(9.46) & HCN & 5.00 & 3.03  & 6.85(0.34) & 11.20(0.38) & 2.26(0.11) & 3.71(0.13) \\
 ~ & ~ & HCO$^+$ & 3.04 & 3.06  & 4.40(0.23) & 7.40(0.25) & 1.53(0.08) & 2.57(0.09) \\
 G081.87+00.78 & 108.3(1.57) & HCN & 5.00 & 0.48  & 1.20(0.08) & 1.57(0.08) & 2.47(0.16) & 3.23(0.18) \\
 ~ & ~ & HCO$^+$ & 4.12 & 0.48  & 1.17(0.06) & 1.36(0.06) & 2.10(0.12) & 2.45(0.12) \\
 G081.75+00.50 & 45.88(1.2) & HCN & 5.00 & 2.65  & 2.24(0.04) & 1.87(0.04) & 0.85(0.01) & 0.71(0.01) \\
 ~ & ~ & HCO$^+$ & 6.33 & ...  & 0.39(0.04) & 0.59(0.04) & … & … \\
 G109.87+02.11 & 24.26(0.45) & HCN & 5.00 & 0.38  & 0.31(0.02) & 0.43(0.02) & 0.81(0.05) & 1.15(0.05) \\
 ~ & ~ & HCO$^+$ & 6.45 & 0.22 & 0.49(0.02) & 0.17(0.02) & 2.26(0.10) & 0.77(0.08) \\
G028.86+00.00 & 303.91(5.11) & HCN & 5.00 & … & ... & ... & ... & ... \\
~ & ~ & HCO$^+$ & 3.81 & 4.07 & 6.33(0.90) & 8.49(0.84) & 1.55(0.22) & 2.08(0.21) \\
G043.79-00.12 & 236.15(2.71) & HCN & 5.00 & … & ... & ... & ... & ... \\
~ & ~ & HCO$^+$ & 3.60 & … & 6.67(0.70) & 5.73(0.59) & ... & ... \\
G031.28+00.00 & 151.70(2.58) & HCN & 5.00 & 6.63 & 4.56(0.70) & 4.91(0.80) & 0.69(0.11) & 0.74(0.12) \\
G031.58+00.00 & 113.06(2.48) & HCN & 5.00 & 1.57 & 6.67(0.38) & 9.99(0.62) & 4.25(0.24) & 6.37(0.39) \\
G188.94+00.88 & 58.54(1.33) & HCN & 5.00 & 0.55 & ... & 2.20(0.11) & ... & 3.98(0.19) \\
~ & ~ & HCO$^+$ & 6.07 & 0.57 & ... & 1.15(0.04) & ... & 2.01(0.06) \\
G009.62+00.19 & 707.39(6.41) & HCN & 5.00 & … & ... & 20.50(0.82) & ... & ... \\
~ & ~ & HCO$^+$ & 2.20 & … & ... & 13.40(1.34) & ... & ... \\
G011.49-03.60 & 8.11(0.17) & HCN & 5.00 & 0.46 & ... & 0.51(0.04) & ... & 1.10(0.08) \\
~ & ~ & HCO$^+$ & 9.27 & 0.54 & ... & 0.11(0.01) & ... & 0.21(0.02) \\
        \hline
               \end{longtable}
\footnotesize   {\textbf{Notes.}  Columns are: the source names/coordinates, core mass calculated by H$^{13}$CN, line, absolute fractional abundance $f(X$) of a species X relative to the most abundant interstellar molecule, H2, defined by $f(X)=N(X)/N(H_2)$, dynamical timescale of outflow ($t_{dyn}$), flux integral intensity of cores ($\int T_{\rm mb}\rm  d\rm{v}_{Core}$), flux integral intensity of blue wing ($\int T_{\rm mb}\rm  d\rm{v}_{Blue}$), flux integral intensity of red wing ($\int T_{\rm mb}\rm  d\rm{v}_{Red}$), velocity range of blue wing, velocity range of red wing. }
 \end{threeparttable}
 
\clearpage
\renewcommand\thetable{\arabic{table}}    
\setcounter{table}{3}
\renewcommand\arraystretch{0.9}
\setlength{\tabcolsep}{3pt}
\centerwidetable
\small
\begin{threeparttable}
\begin{longtable}{llllllll}
	\caption{Outflow properties of 33 sources}
	\label{tab4}\\
	\hline 
	\hline
	Source     & Line                                & $P$ $_{Blue}$                        &$P$ $_{Red}$                & $E_k$ $_{Blue}$                        &$E_k$ $_{Red}$        &  $F_{Blue}$     &  $F_{Red}$ \\
	Alias        &                         & (M$_{\odot}$km s$^{-1}$)     &(M$_{\odot}$km s$^{-1}$)        &$\rm (M_{\odot}km^2 s^{-2}) $    & $\rm (M_{\odot}km^2 s^{-2}) $       & (10$^{-4}M_{\odot}$ )                    &(10$^{-4}M_{\odot}$ )  \\	
	                  &                                                 &                                          &                                   &                                                   &                                     & ( $\rm km\ s^{-1}yr^{-1}$  )       & ( $\rm km\ s^{-1}yr^{-1}$  )           \\
\hline
\endfirsthead
\caption[]{(continued)}\\	         
	\hline 
	\hline
	Source     & Line                                & $P$ $_{Blue}$                        &$P$ $_{Red}$                & $E_k$ $_{Blue}$                        &$E_k$ $_{Red}$        &  $F_{Blue}$     &  $F_{Red}$ \\
	Alias        &                         & (M$_{\odot}$km s$^{-1}$)     &(M$_{\odot}$km s$^{-1}$)        &$\rm (M_{\odot}km^2 s^{-2}) $    & $\rm (M_{\odot}km^2 s^{-2}) $       & (10$^{-4}M_{\odot}$ )                    &(10$^{-4}M_{\odot}$ )  \\	
	                  &                                                 &                                          &                                   &                                                   &                                     & ( $\rm km\ s^{-1}yr^{-1}$  )       & ( $\rm km\ s^{-1}yr^{-1}$  )           \\
\hline
 \endhead
\hline
\endfoot
G121.29+00.65 & HCN & 4.48(0.24) & 2.03(0.17) & 20.57(1.10) & 9.33(0.78) & 3.11(0.17) & 1.41(0.12) \\ 
        ~ & HCO+ & 1.34(0.13) & 1.69(0.13) & 6.13(0.59) & 6.13(0.59) & 1.17(0.11) & 1.47(0.11) \\ 
        G123.06-06.30 & HCN & 21.66(1.33) & 11.74(1.16) & 102.79(6.31) & 102.79(5.49) & 14.49(0.89) & 7.86(0.77) \\ 
        ~ & HCO+ & 7.60(0.62) & 6.63(0.62) & 38.93(3.19) & 38.93(3.19) & 5.49(0.45) & 4.79(0.45) \\ 
        G183.72-03.60 & HCN & 3.28(0.30) & 3.90(0.32) & 7.52(0.69) & 7.52(0.74) & 1.71(0.16) & 2.03(0.17) \\ 
        ~ & HCO+ & 1.02(0.05) & 1.22(0.05) & 2.27(0.12) & 2.27(0.11) & 0.52(0.03) & 0.62(0.03) \\ 
        G005.88-00.39 & HCN & 378.93(9.49) & 657.63(10.02) & 4212.03(105.49) & 4212.03(111.40) & … & … \\ 
        ~ & HCO+ & 412.87(11.80) & 875.24(14.18) & 4992.24(142.73) & 4992.24(171.51) & … & … \\ 
        G010.62-00.38 & HCN & 149.77(16.61) & 282.19(17.35) & 1185.04(131.41) & 1185.04(137.28) & … & … \\ 
        ~ & HCO+ & 61.37(9.88) & 161.06(11.23) & 472.16(76.03) & 472.16(86.36) & … & … \\ 
        G011.91-00.61 & HCN & 11.74(0.99) & 7.72(0.90) & 89.49(7.51) & 89.49(6.88) & … & … \\ 
        ~ & HCO+ & 2.41(0.68) & 1.57(0.68) & 13.85(3.93) & 13.85(3.93) & 3.12(0.88) & 2.03(0.88) \\ 
        G012.80-00.20 & HCN & 314.73(8.24) & 453.24(10.55) & 2129.79(55.77) & 2129.79(71.36) & 290.02(7.59) & 417.65(9.72) \\ 
        ~ & HCO+ & 84.00(3.15) & 130.45(3.86) & 491.80(18.47) & 491.80(22.63) & 47.35(1.78) & 73.54(2.18) \\ 
        G015.03-00.67 & HCN & 51.75(1.23) & 60.01(1.32) & 281.76(6.68) & 281.76(7.21) & … & … \\ 
        ~ & HCO+ & 24.79(1.03) & 33.04(1.03) & 141.28(5.88) & 141.28(5.88) & 7.87(0.33) & 10.49(0.33) \\ 
        G023.43-00.10 & HCN & 354.15(20.51) & 379.08(25.19) & 3473.39(201.17) & 3473.39(247.01) & 234.88(13.60) & 251.41(16.70) \\ 
        ~ & HCO+ & 128.02(18.73) & 107.94(17.57) & 1130.65(165.42) & 1130.65(155.19) & … & … \\ 
        G023.00-00.41 & HCN & 143.02(8.41) & 106.61(7.66) & 1112.47(65.44) & 1112.47(59.58) & 68.14(4.01) & 50.79(3.65) \\ 
        ~ & HCO+ & 149.02(8.06) & 49.10(7.63) & 1110.65(60.06) & 1110.65(56.85) & 96.21(5.20) & 31.70(4.93) \\ 
        G029.95-00.01 & HCN & 234.20(21.28) & 144.59(20.22) & 1462.28(132.86) & 1462.28(126.26) & … & … \\ 
        ~ & HCO+ & 122.90(22.14) & 83.23(22.14) & 756.04(136.17) & 756.04(136.17) & … & … \\ 
        G032.04+00.00 & HCN & 215.13(18.39) & 227.69(17.04) & 1565.33(133.80) & 1565.33(124.01) & … & … \\ 
        ~ & HCO+ & 49.58(7.43) & 63.78(6.99) & 280.49(42.01) & 280.49(39.54) & … & … \\ 
        G043.16+00.01 & HCN & 1219.82(179.30) & 2717.02(185.43) & 19829.91(2914.70) & 19829.91(3014.35) & … & … \\ 
        ~ & HCO+ & 1661.61(94.80) & 5483.82(127.09) & 28997.34(1654.39) & 28997.34(2217.98) & … & … \\ 
        G069.54-00.97 & HCN & 39.83(3.28) & 39.11(3.39) & 310.10(25.56) & 310.10(26.40) & 50.12(4.13) & 49.21(4.27) \\ 
        ~ & HCO+ & 23.89(1.48) & 23.44(1.44) & 177.23(10.99) & 177.23(10.68) & … & … \\ 
        G075.76+00.33 & HCN & 89.30(4.49) & 146.39(4.92) & 582.18(29.25) & 582.18(32.05) & 29.49(1.48) & 48.35(1.62) \\ 
        ~ & HCO+ & 60.40(3.09) & 101.50(3.44) & 389.08(19.90) & 389.08(22.16) & 19.71(1.01) & 33.12(1.12) \\ 
        G081.87+00.78 & HCN & 16.14(1.05) & 21.12(1.14) & 108.88(7.06) & 108.88(7.71) & 33.30(2.16) & 43.58(2.36) \\ 
        ~ & HCO+ & 13.72(0.76) & 16.02(0.76) & 92.56(5.13) & 92.56(5.13) & 28.31(1.57) & 33.06(1.57) \\ 
        G081.75+00.50 & HCN & 20.20(0.33) & 16.88(0.35) & 90.99(F1.47) & 90.99(1.57) & 7.63(0.12) & 6.37(0.13) \\ 
        ~ & HCO+ & 4.35(0.43) & 6.50(0.47) & 21.85(2.18) & 21.85(2.36) & … & … \\ 
        G109.87+02.11 & HCN & 4.03(0.23) & 5.73(0.23) & 26.64(1.49) & 26.64(1.49) & 14.21(2.03) & 15.21(0.60) \\ 
        ~ & HCO+ & 11.26(0.48) & 3.81(0.40) & 127.59(5.40) & 127.59(4.51) & 51.25(2.17) & 17.34(1.81) \\ 
        G028.86+00.00 & HCN & … & … & … & … & … & … \\ 
        ~ & HCO+ & 57.88(8.27) & 77.67(7.68) & 264.81(37.83) & 355.33(35.13) & 10.05(1.44) & 19.07(1.88) \\ 
        G043.79-00.12 & HCN & … & … & … & … & … & … \\ 
        ~ & HCO+ & 68.28(7.15) & 58.71(6.00) & 349.57(36.61) & 300.56(30.71) & … & … \\ 
        G031.28+00.00 & HCN & 22.49(3.67)  & 15.90(2.60) & 47.86(7.39) & 51.52(8.42) & 3.15(0.49) & 3.39(0.55) \\ 
        G031.58+00.00 & HCN & 222.14(13.70) & 157.08(9.69) & 1648.44(93.22) & 2369.25(152.33) & 141.68(8.74) & 100.19(6.18) \\ 
        G188.94+00.88 & HCN & … & 34.89(1.69) & … & 276.10(13.38) & … & 62.94(3.05) \\ 
        ~ & HCO+ & … & 17.72(0.57) & … & 136.32(4.35) & … & 30.97(0.99) \\ 
        G009.62+00.19 & HCN & … & 313.21(12.55) & … & 2387.45(95.64) & … & … \\ 
        ~ & HCO+ & … & 154.22(15.38) & … & 887.65(88.50) & … & … \\ 
        G011.49-03.60 & HCN & … & 6.92(0.52) & … & 46.80(3.51) & … & 14.89(1.12) \\ 
        ~ & HCO+ & … & 1.29(0.10) & … & 7.54(0.57) & … & 2.40(0.18) \\ 
                        \hline        
               \end{longtable}
               \footnotesize   {\textbf{Notes.}  Columns are: the source names/coordinates, line, the momentum of blue wing ($P_{Blue}$), the momentum of red wing ($P_{Red}$), the kinetic energy of blue wing ($E_k$ $_{Blue}$), the kinetic energy of red wing ($E_k$ $_{Red}$), the force of blue wing ($F_{Blue}$), the force of red wing ($F_{Red}$). }
 \end{threeparttable}
  \clearpage

\begin{figure}
\centering
\addtocounter{figure}{0}
\includegraphics[width=0.35\columnwidth]{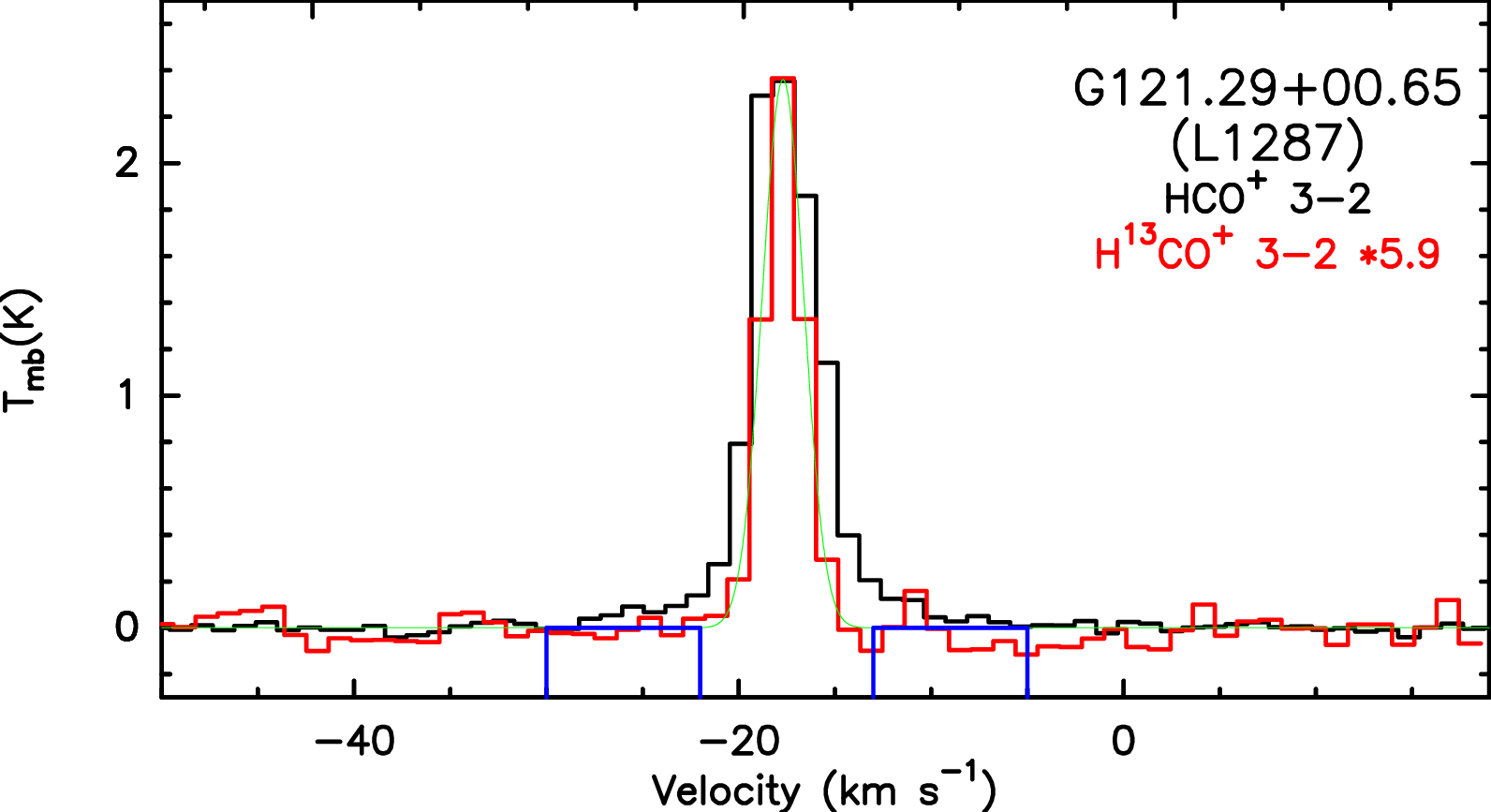} 
\includegraphics[width=0.35\columnwidth]{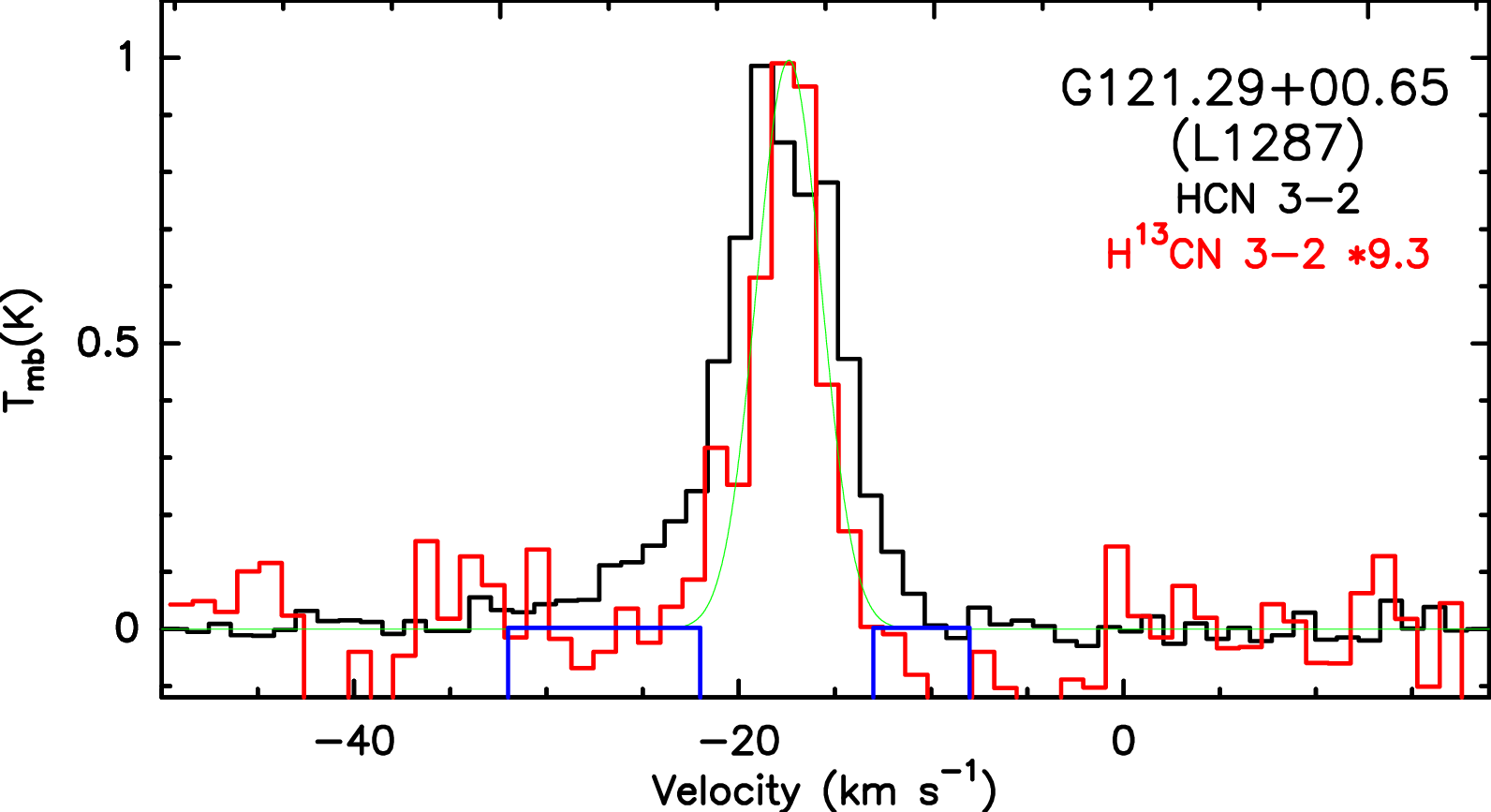} 
\includegraphics[width=0.35\columnwidth]{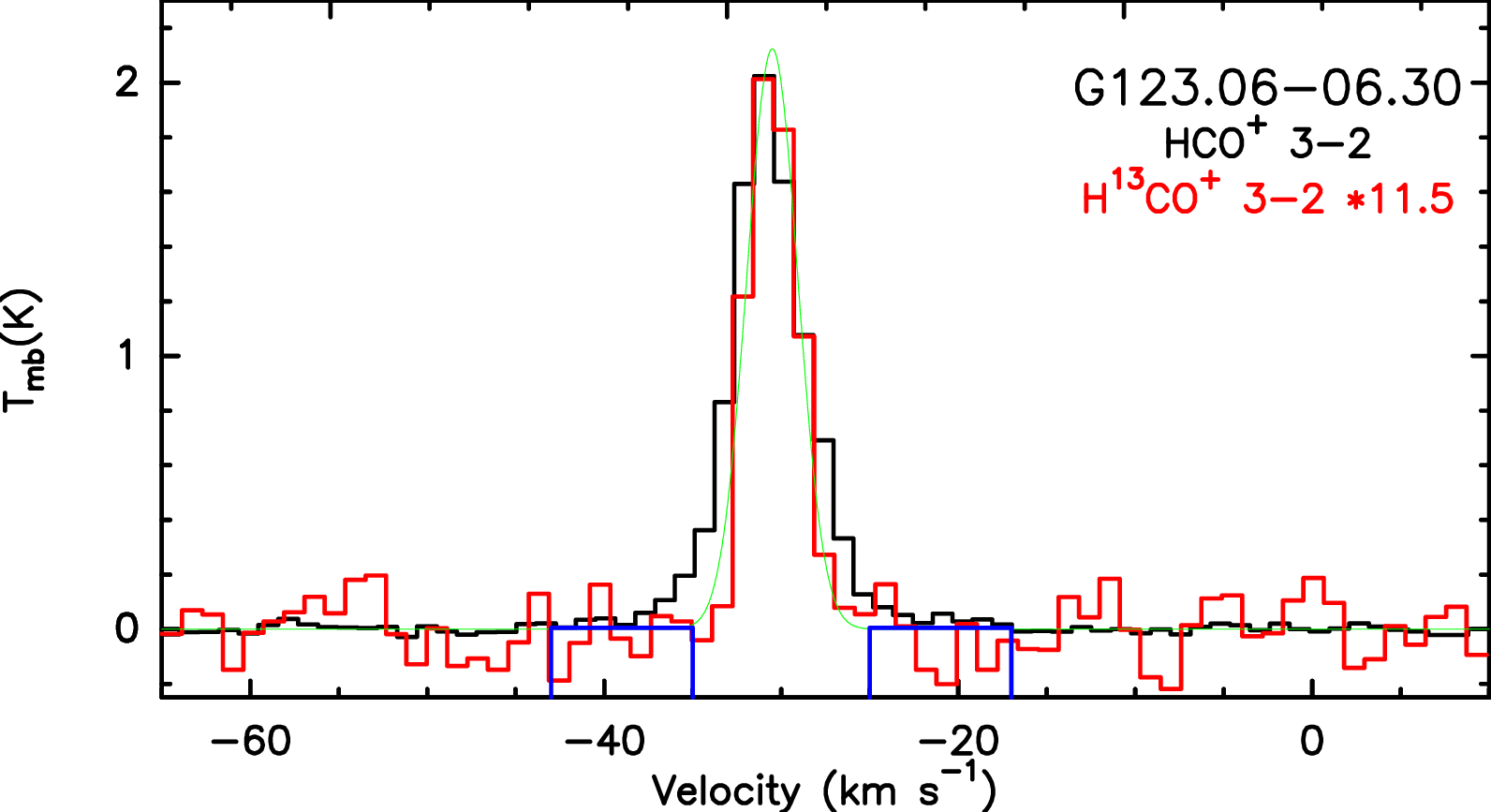} 
\includegraphics[width=0.35\columnwidth]{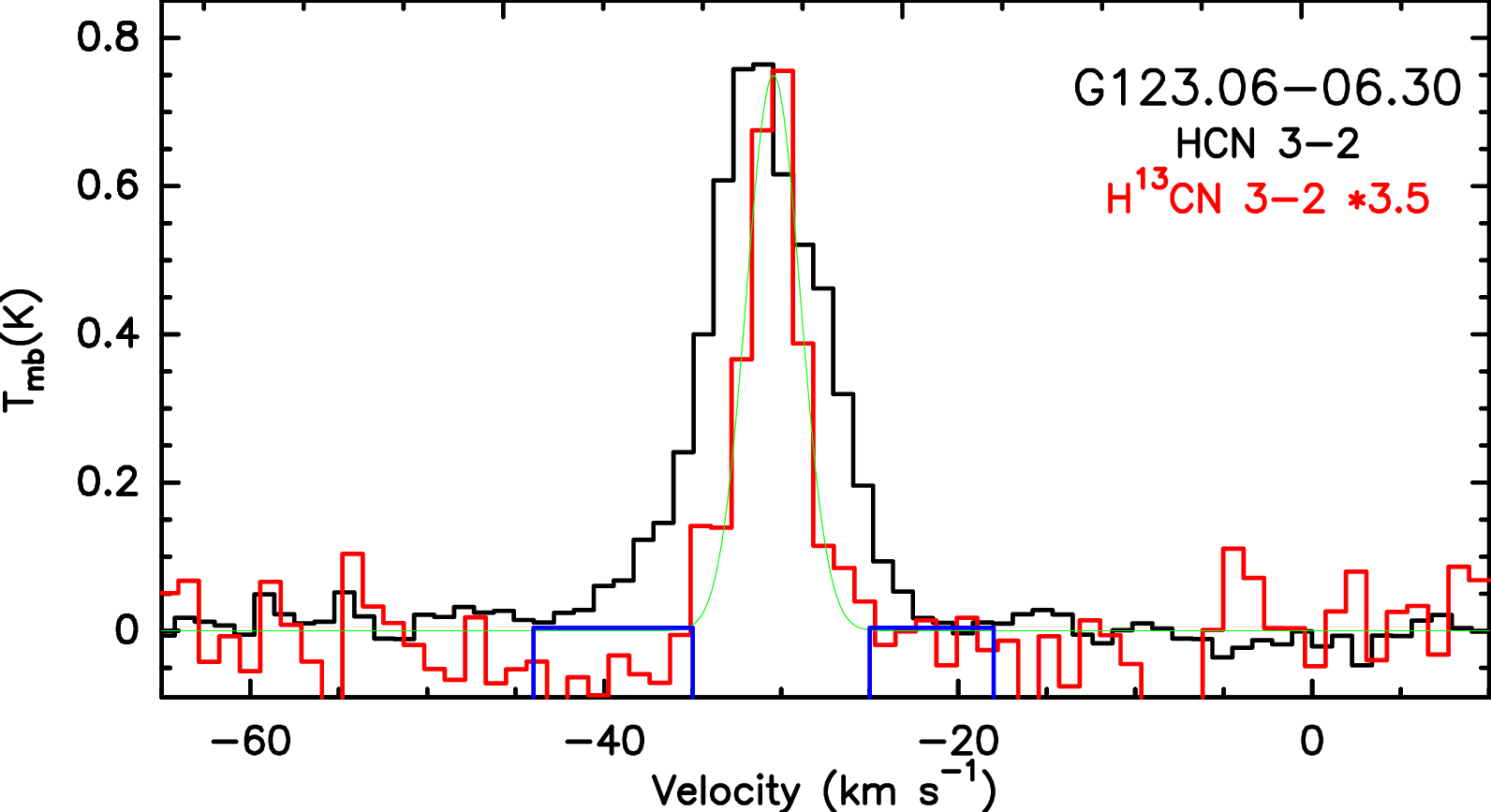} 
\includegraphics[width=0.35\columnwidth]{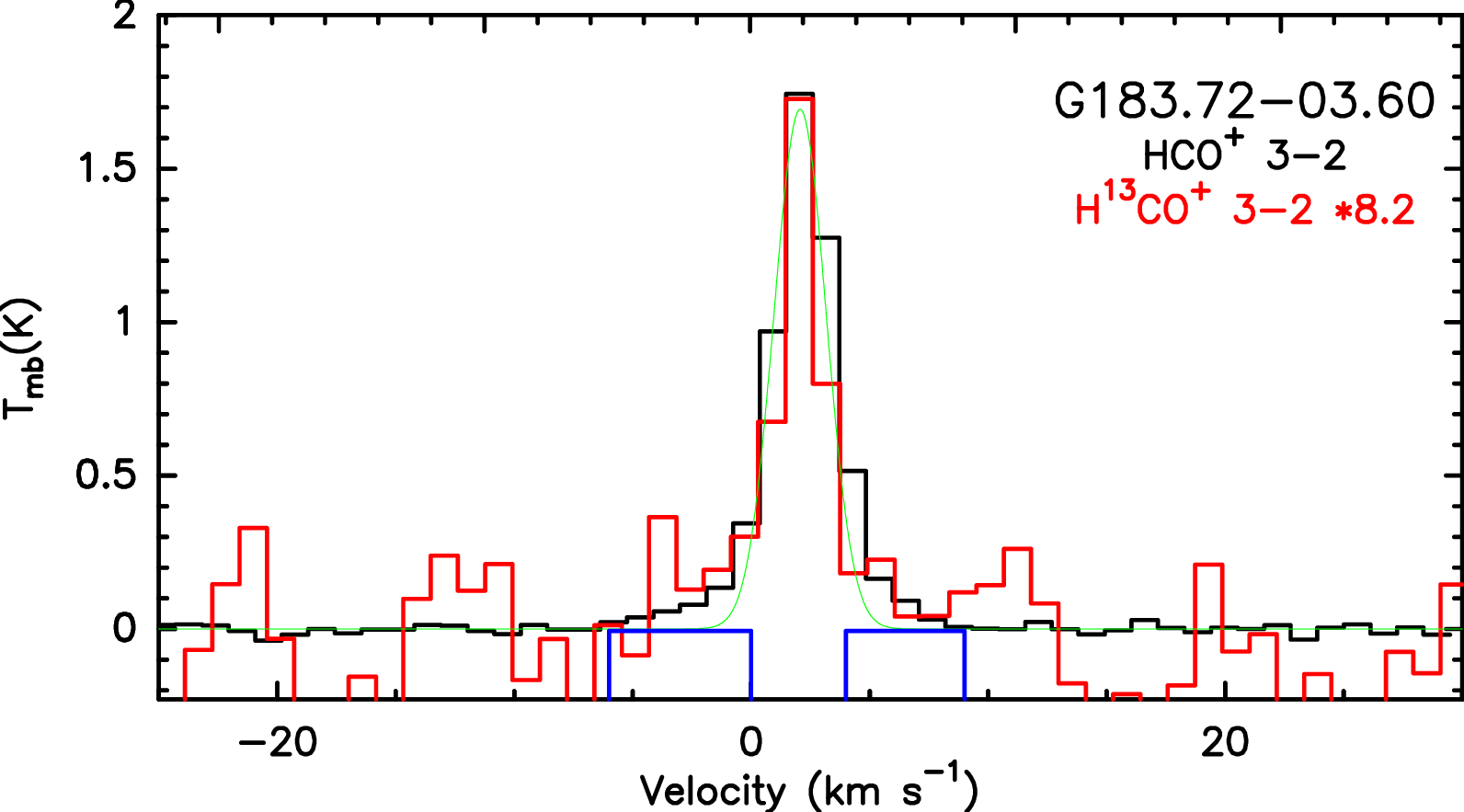} 
\includegraphics[width=0.35\columnwidth]{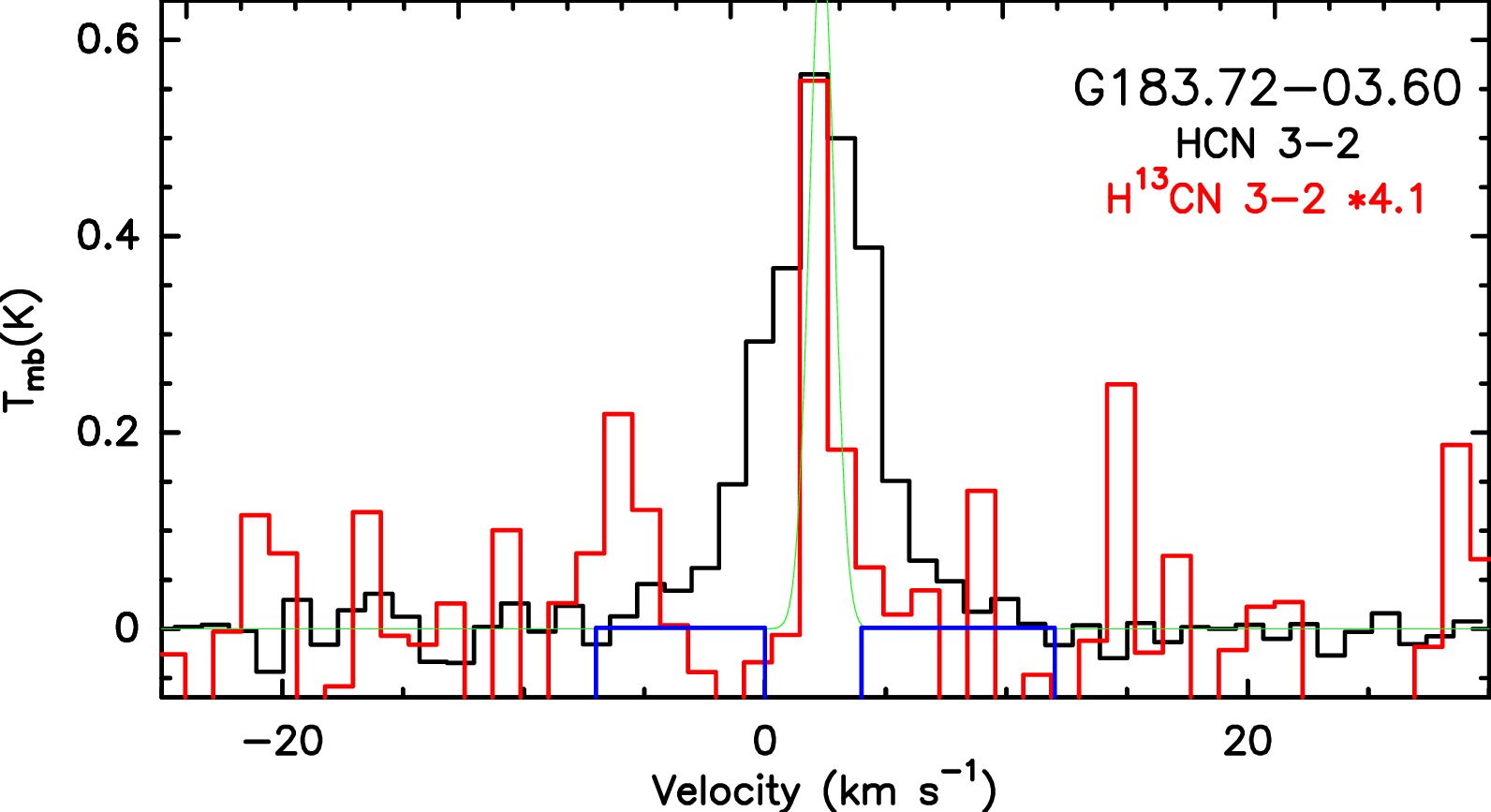}
\includegraphics[width=0.35\columnwidth]{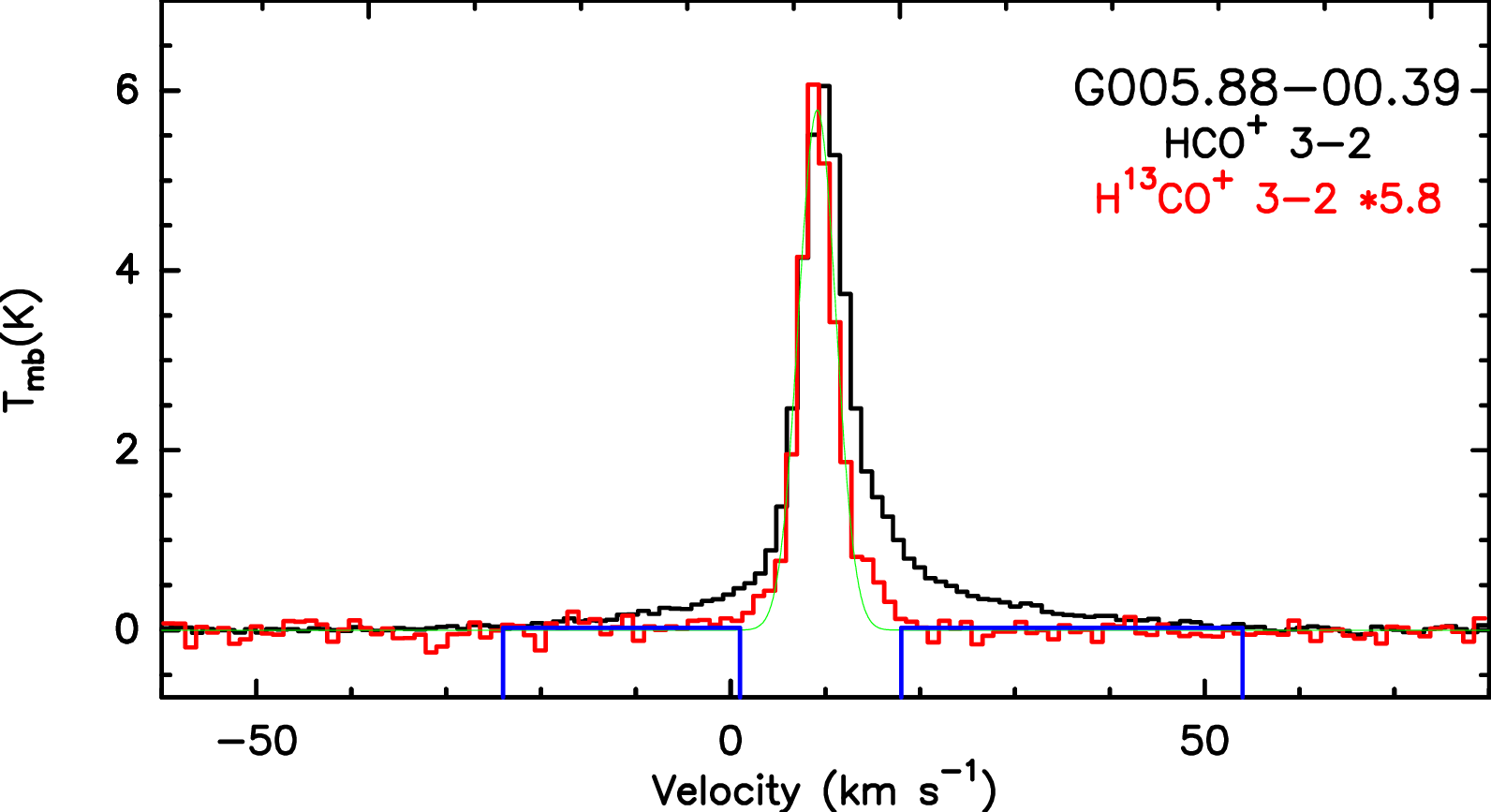} 
\includegraphics[width=0.35\columnwidth]{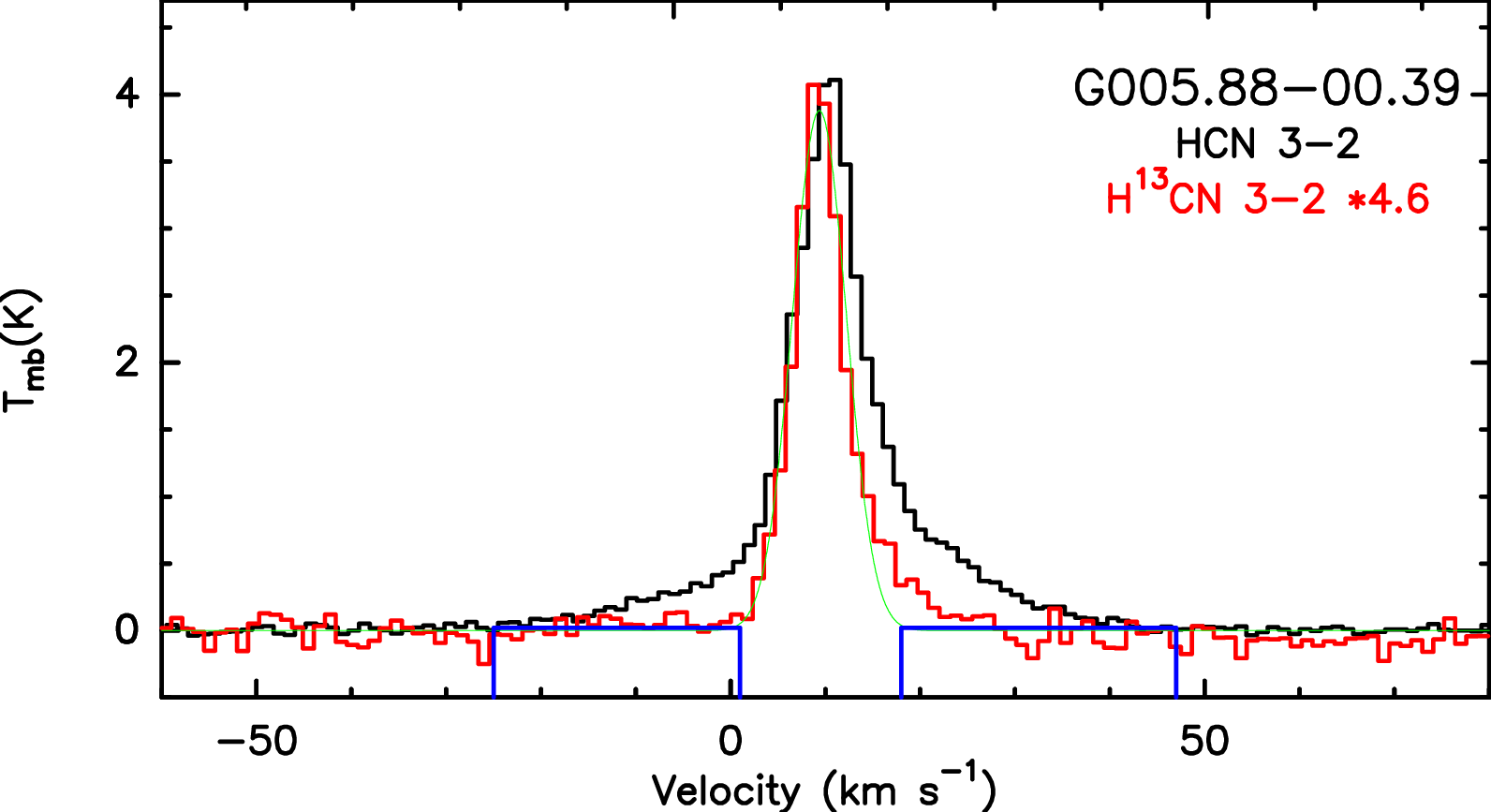} 
\includegraphics[width=0.35\columnwidth]{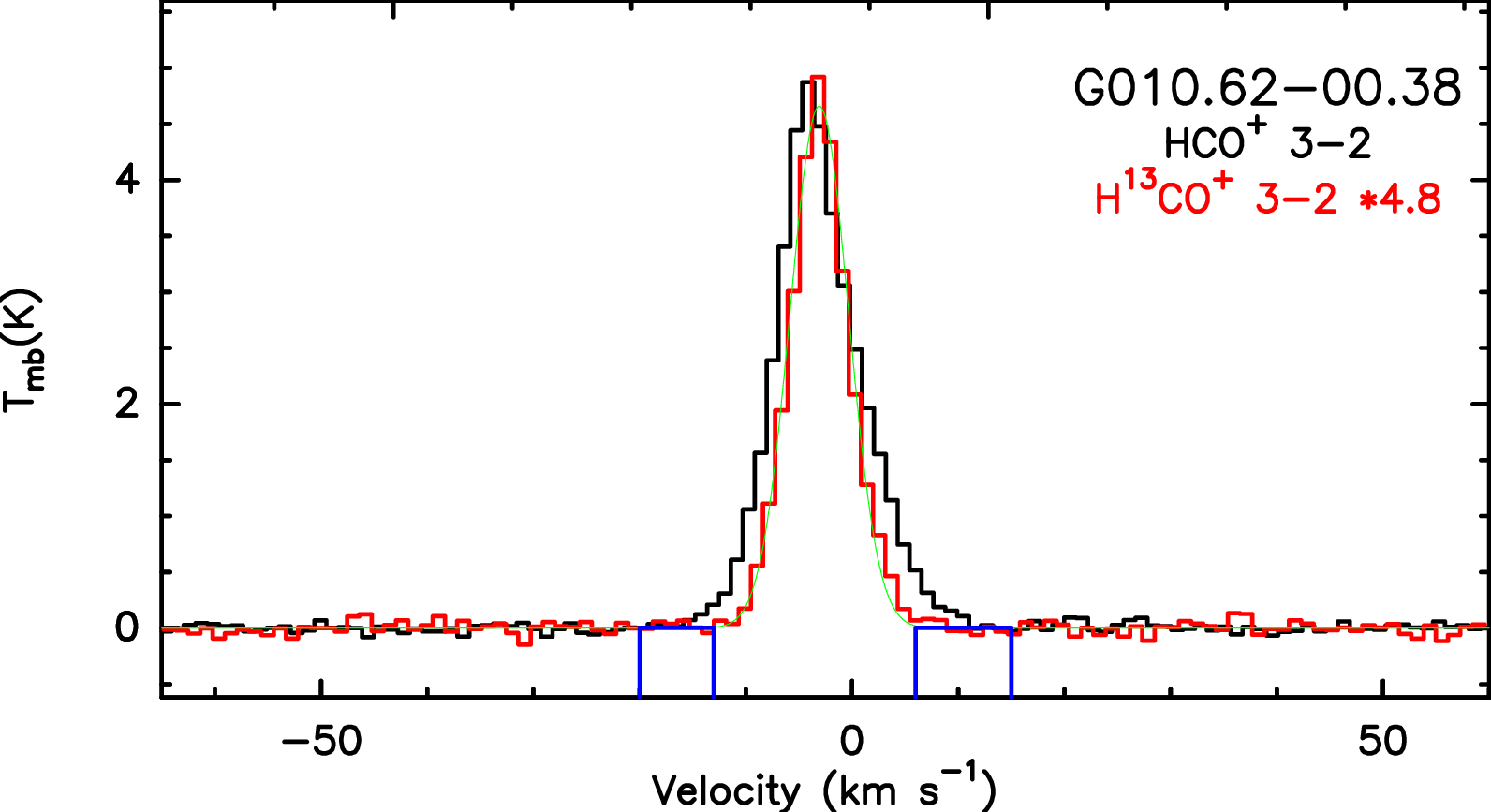} 
\includegraphics[width=0.35\columnwidth]{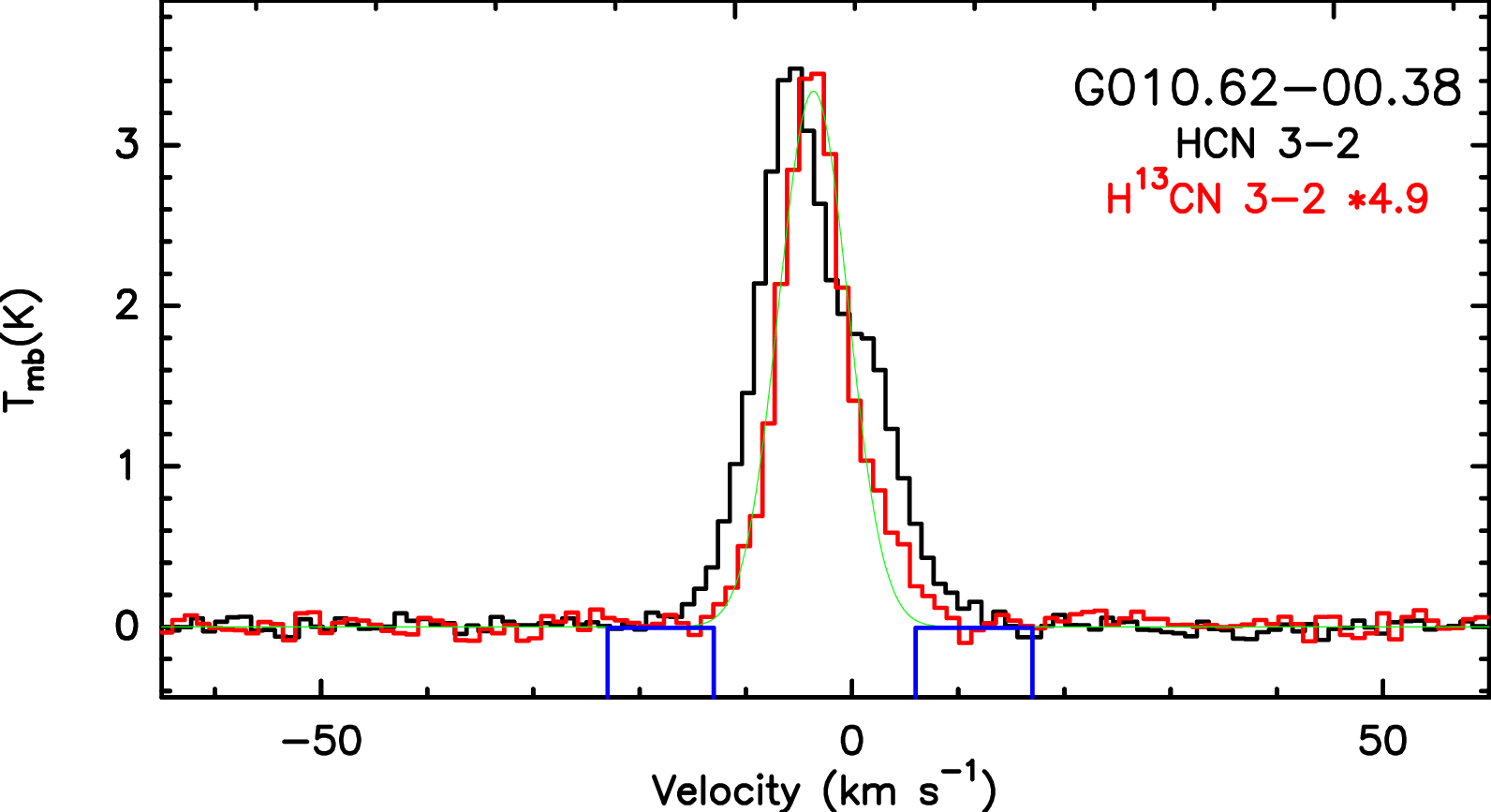} 
\includegraphics[width=0.35\columnwidth]{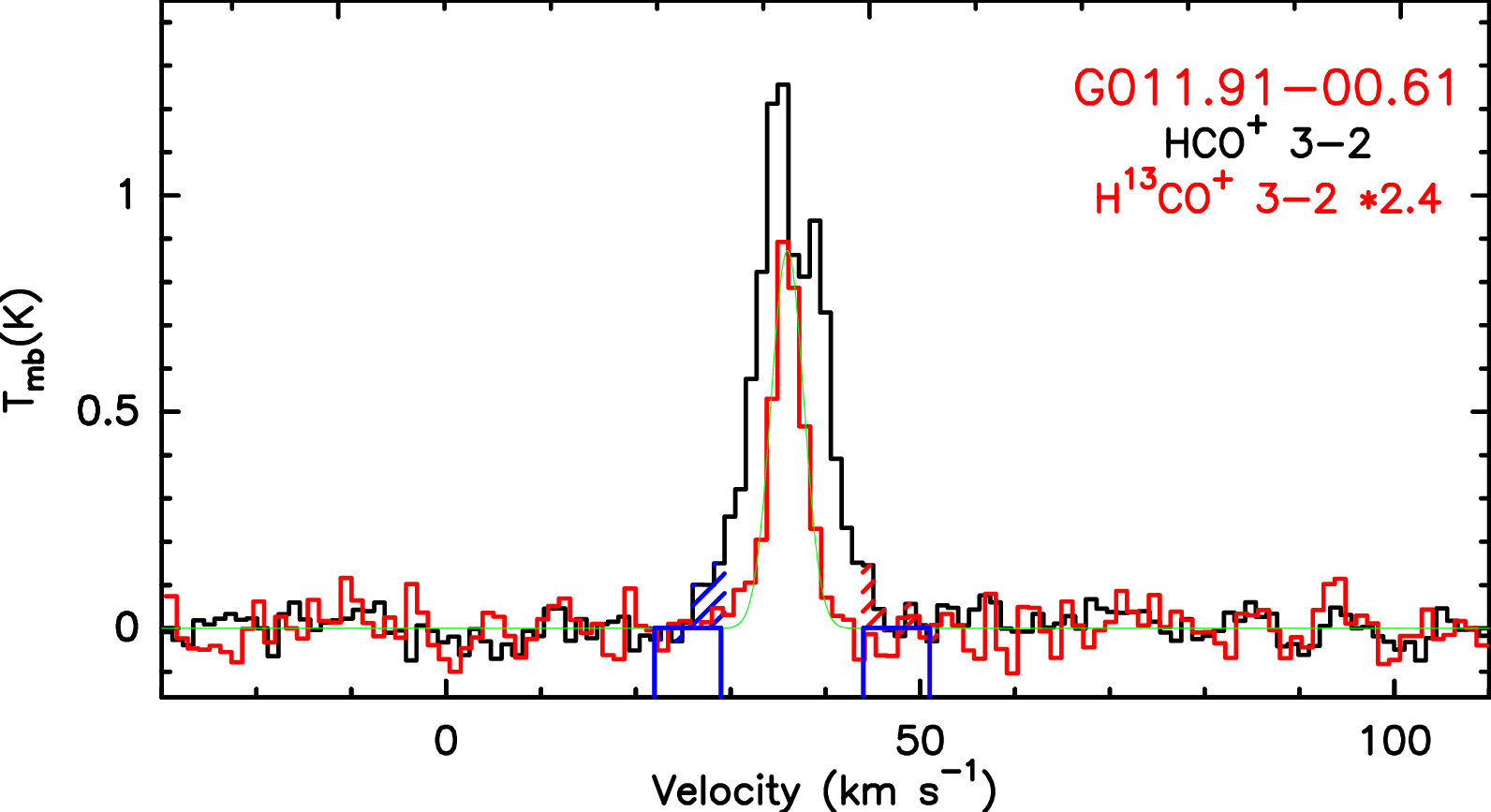} 
\includegraphics[width=0.35\columnwidth]{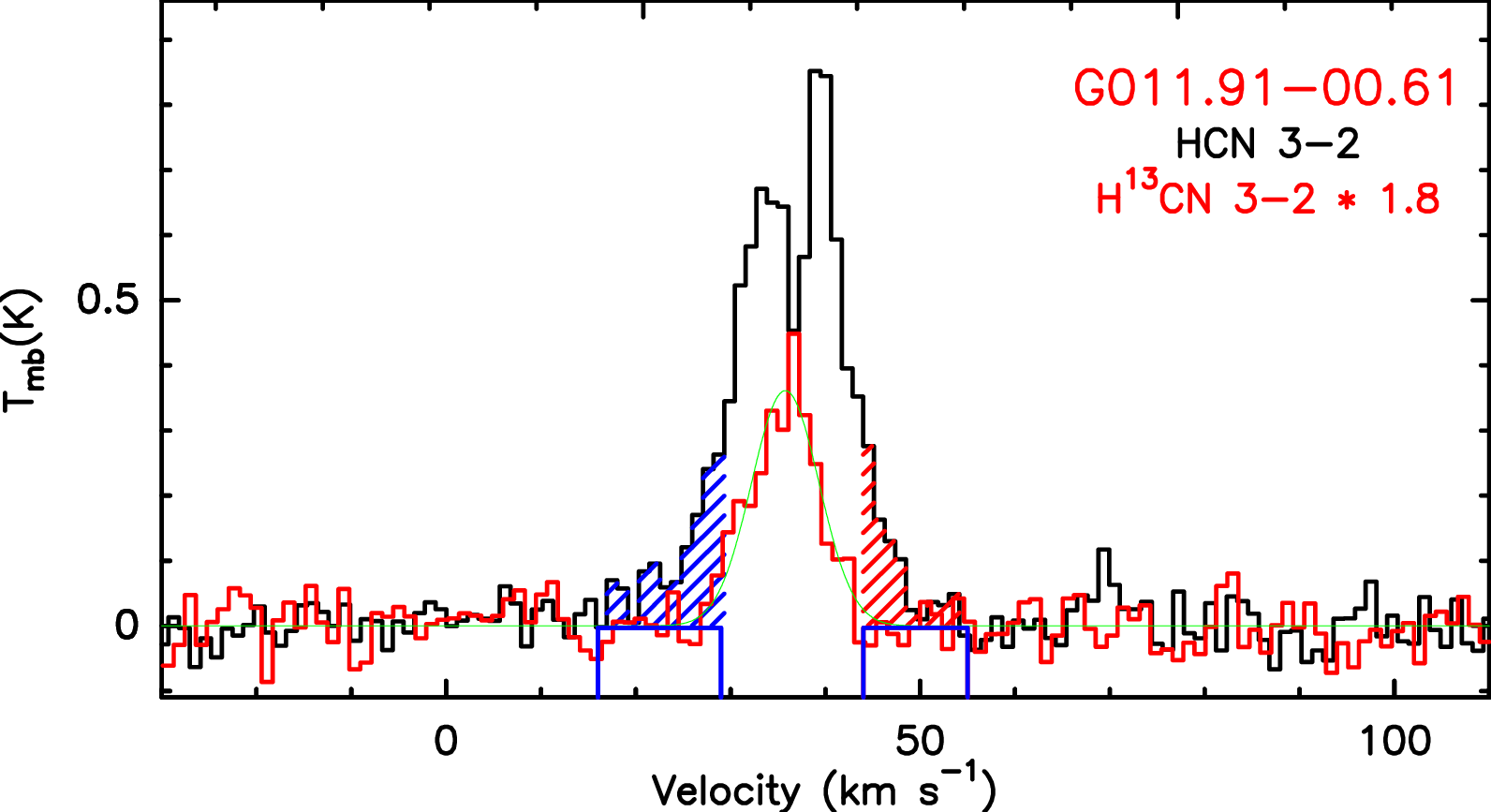} 

\addtocounter{figure}{0}
  \caption{The averaged spectral lines of HCN and HCO$^+$ 3-2. The source names are shown in the panels. Each panel presents the spectra of HCO$^+$ or HCN(black) and H$^{13}$CO$^+$ or H$^{13}$CN(red) 3-2 after averaging all the spectral lines within the signal area. The blue and red line wing emissions are labeled with blue windows. The green line is the Gaussian fitting of the spectral line of H$^{13}$CO$^+$ or H$^{13}$CN 3-2.}\label{fig1}
\end{figure}

\begin{figure}
\centering
\addtocounter{figure}{-1}
\includegraphics[width=0.35\columnwidth]{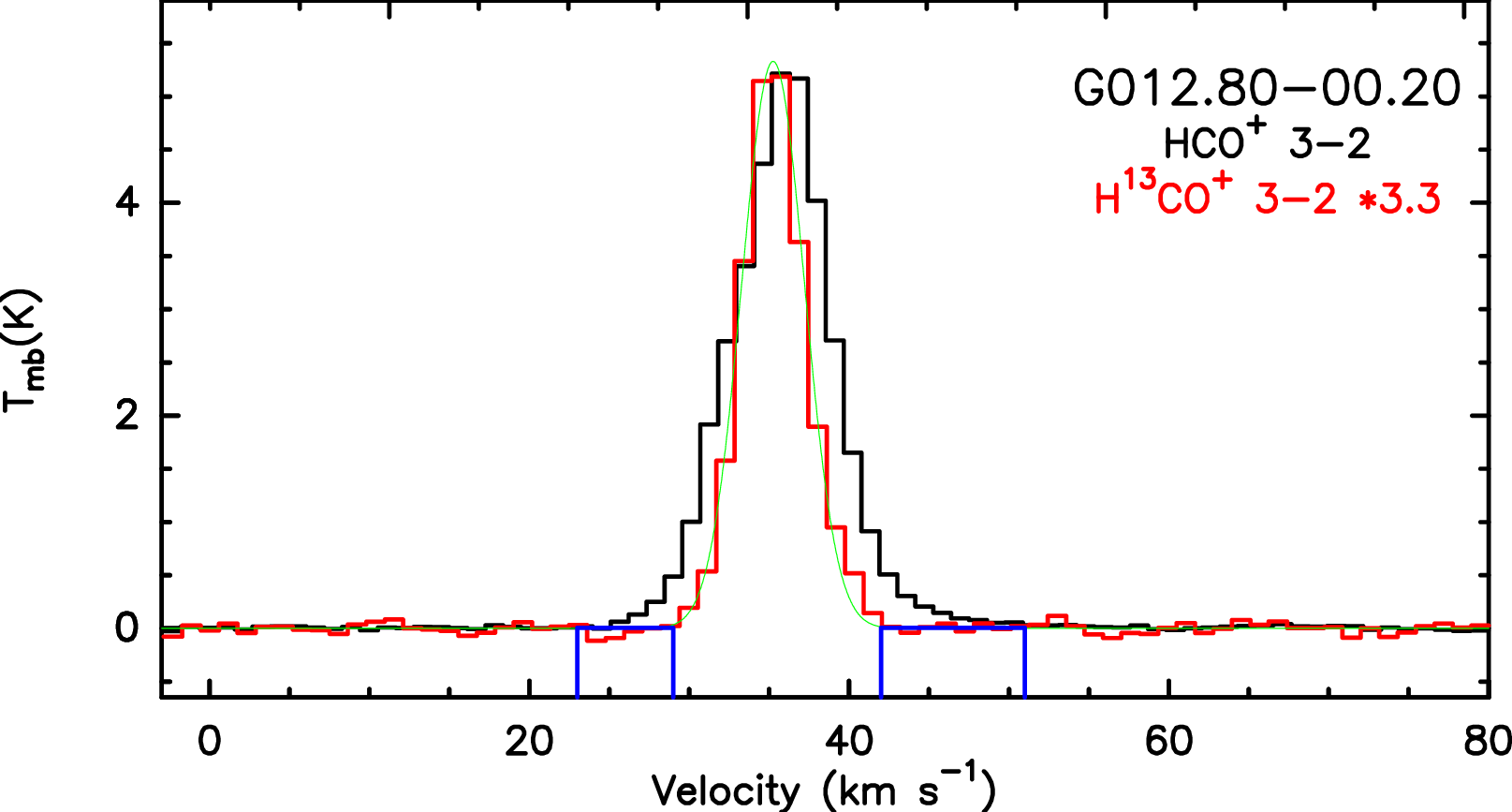} 
\includegraphics[width=0.35\columnwidth]{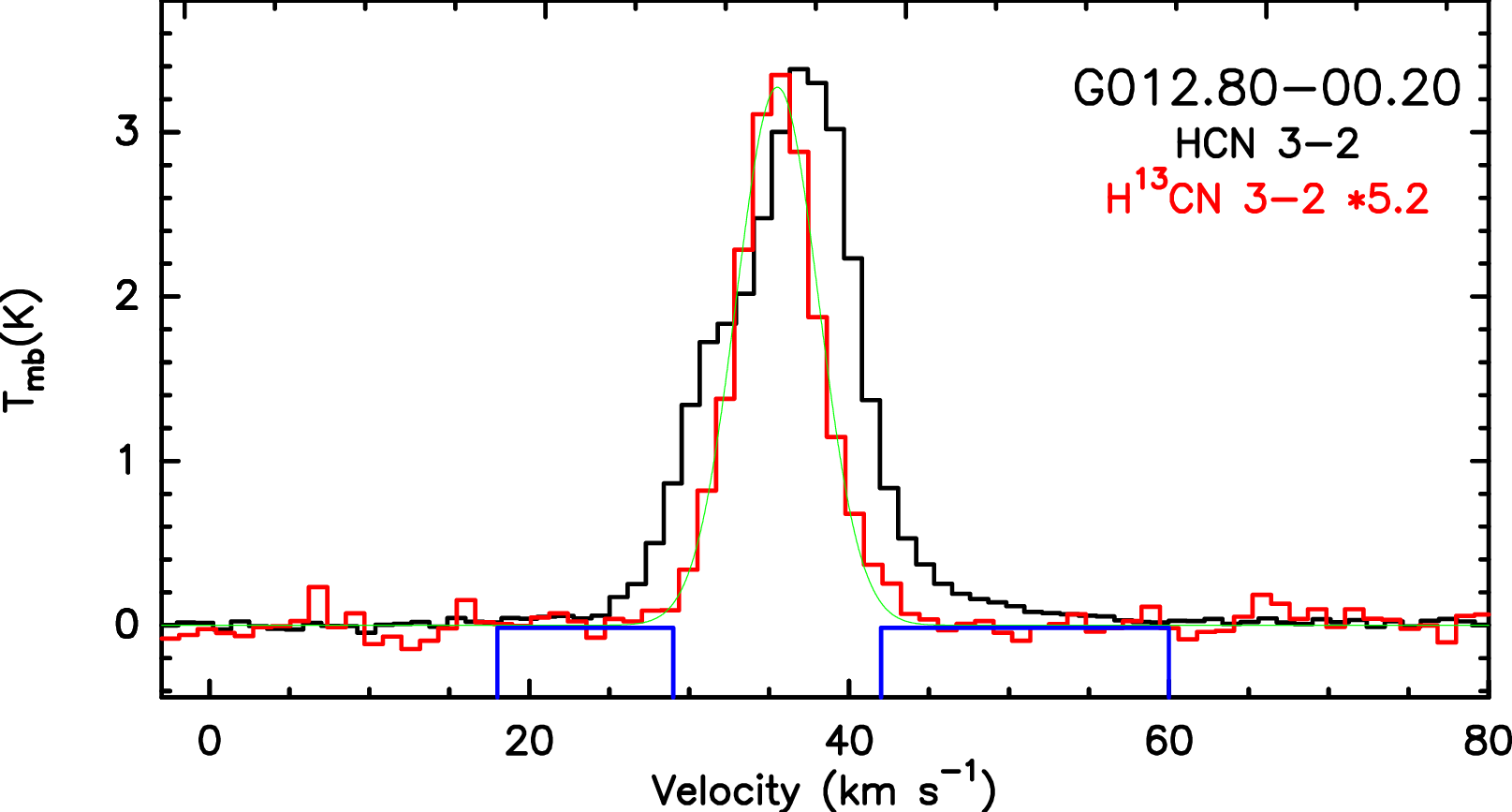} 
\includegraphics[width=0.35\columnwidth]{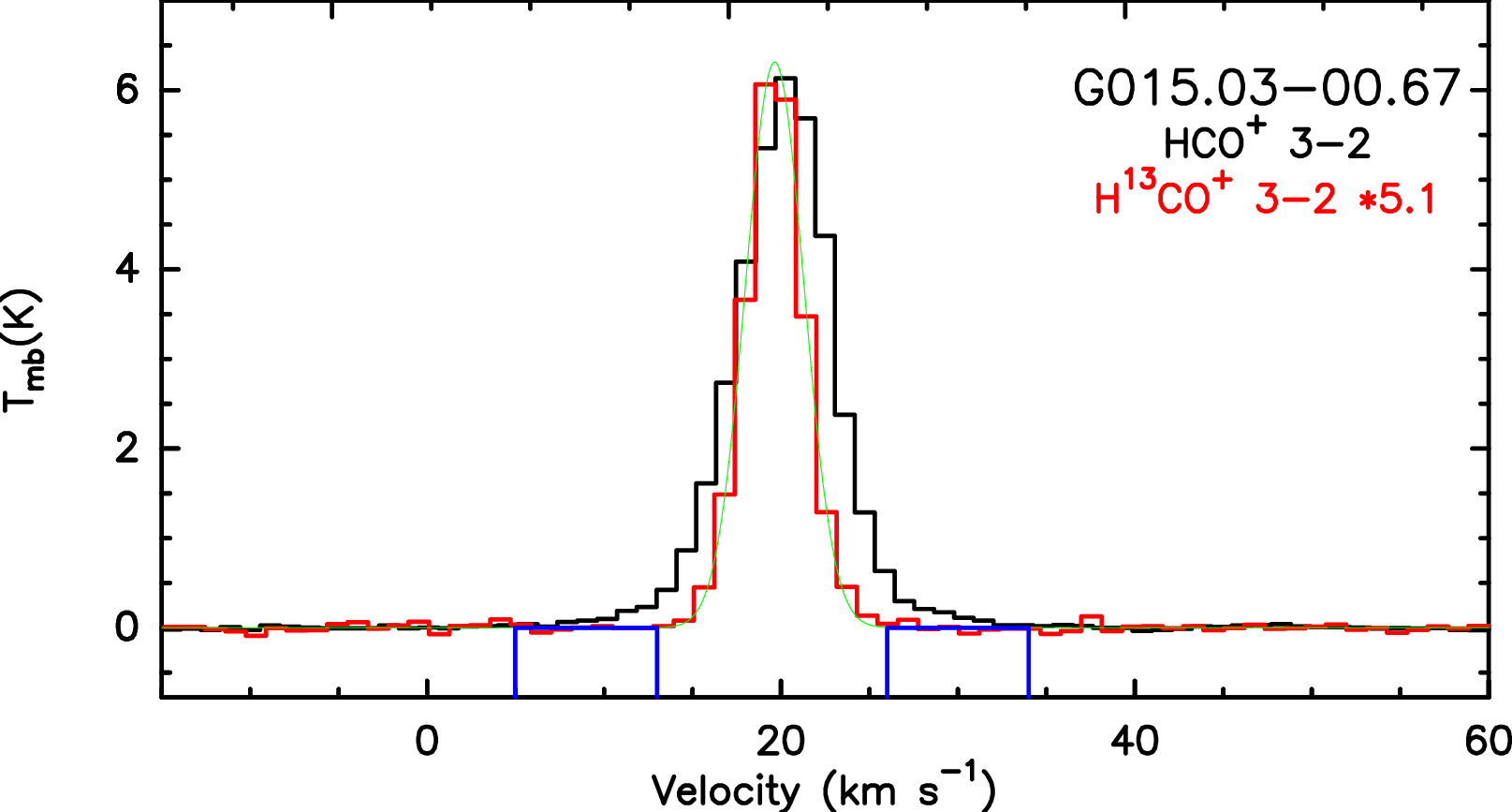} 
\includegraphics[width=0.35\columnwidth]{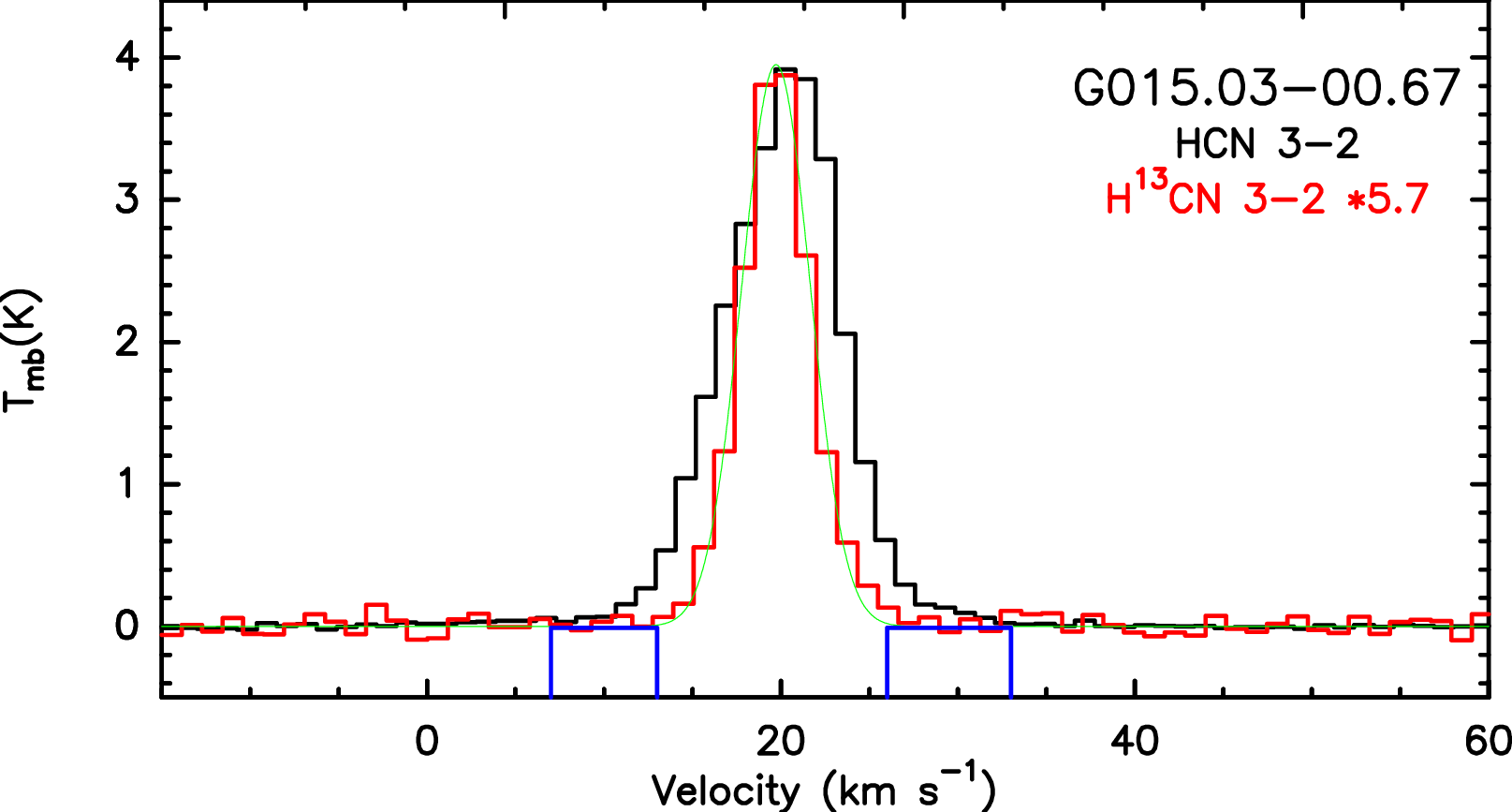} 
\includegraphics[width=0.35\columnwidth]{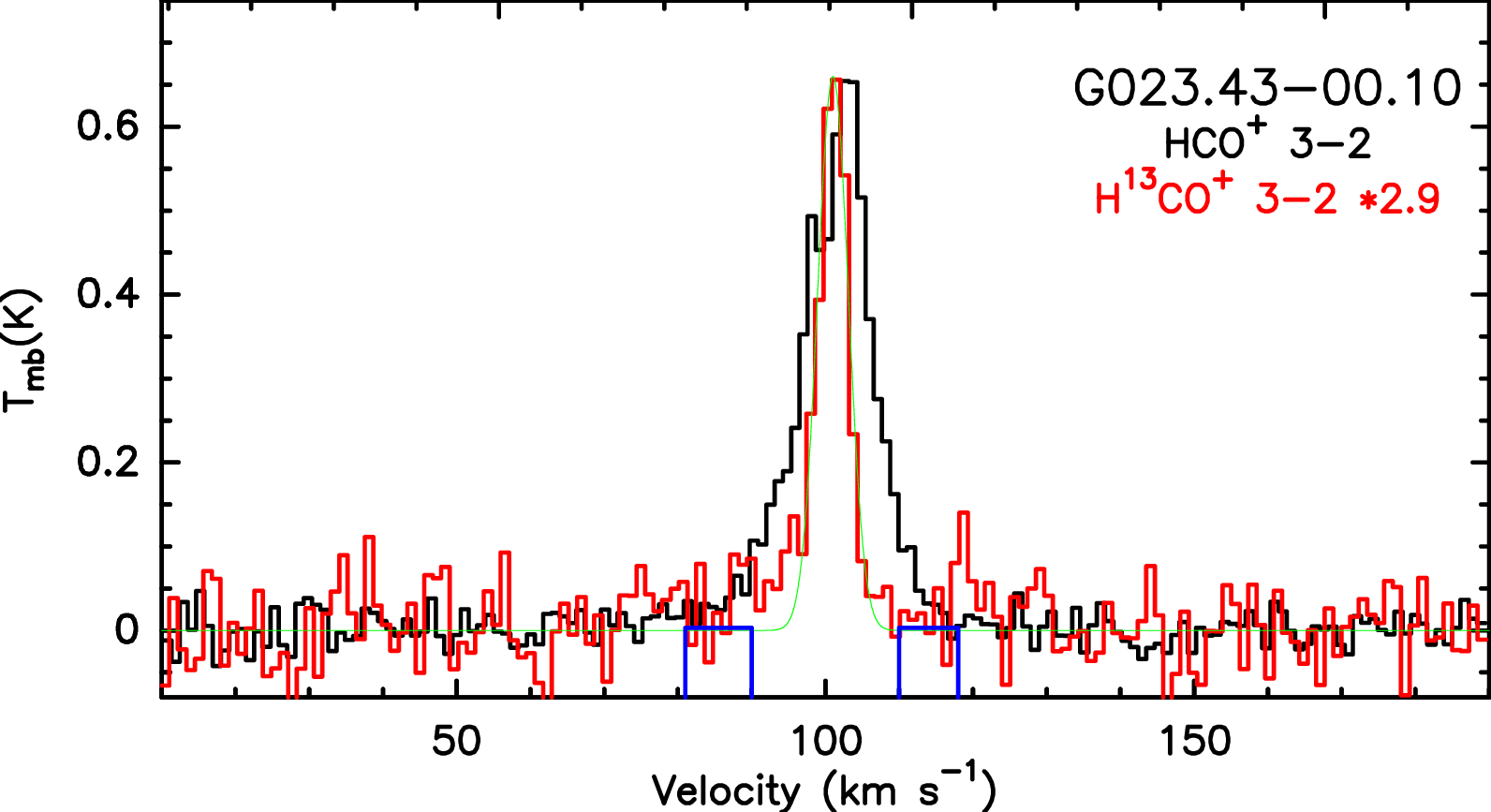} 
\includegraphics[width=0.35\columnwidth]{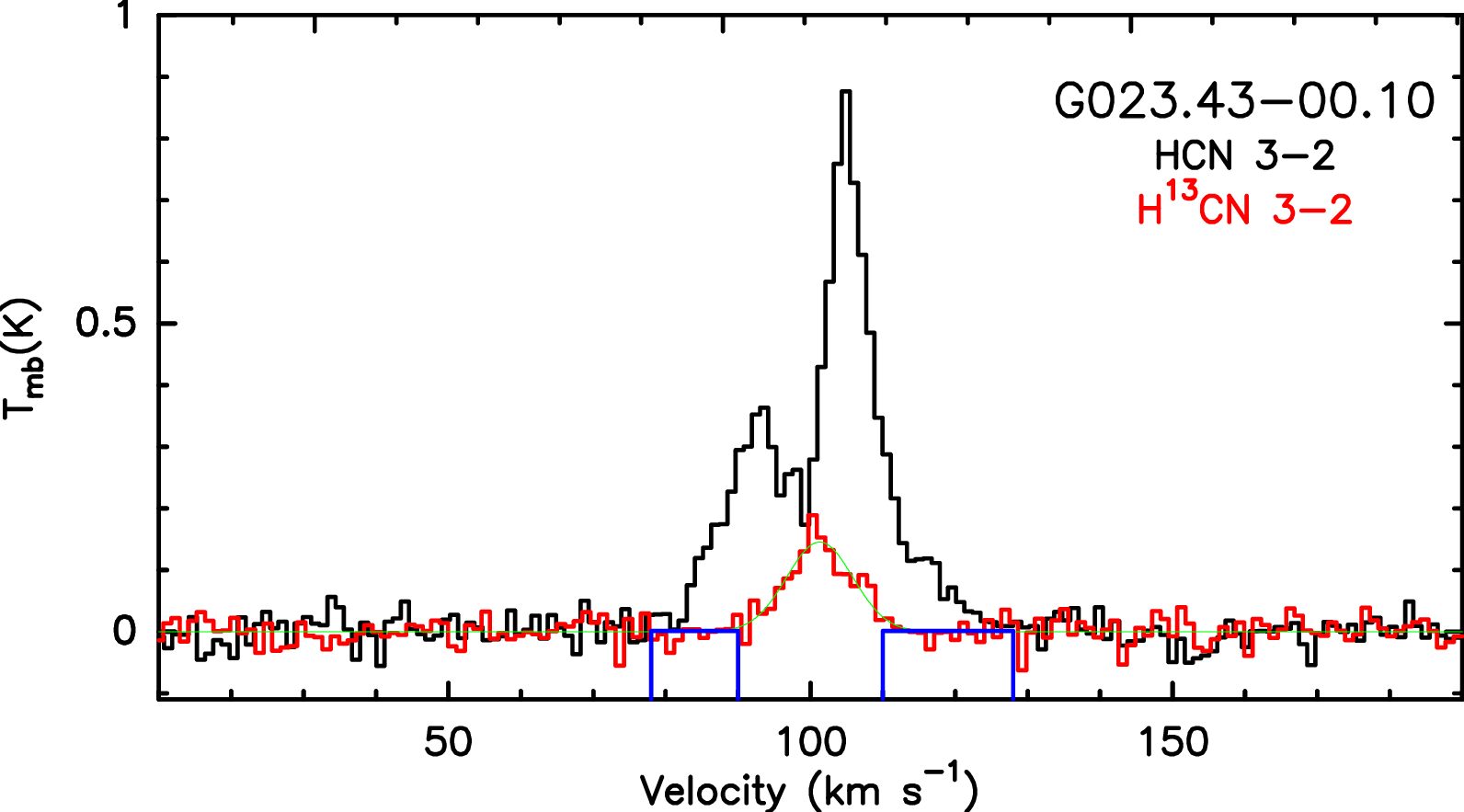} 
\includegraphics[width=0.35\columnwidth]{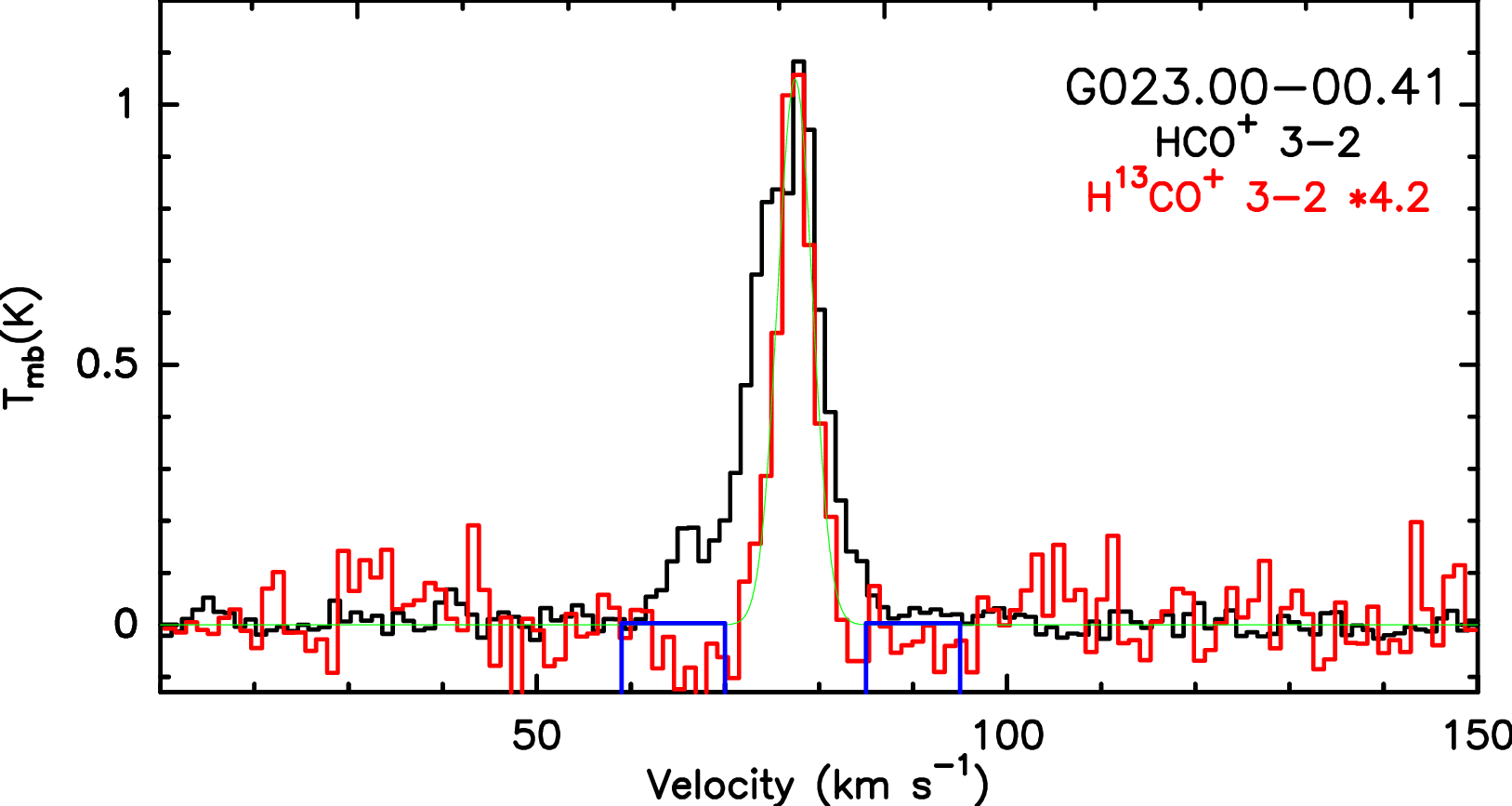} 
\includegraphics[width=0.35\columnwidth]{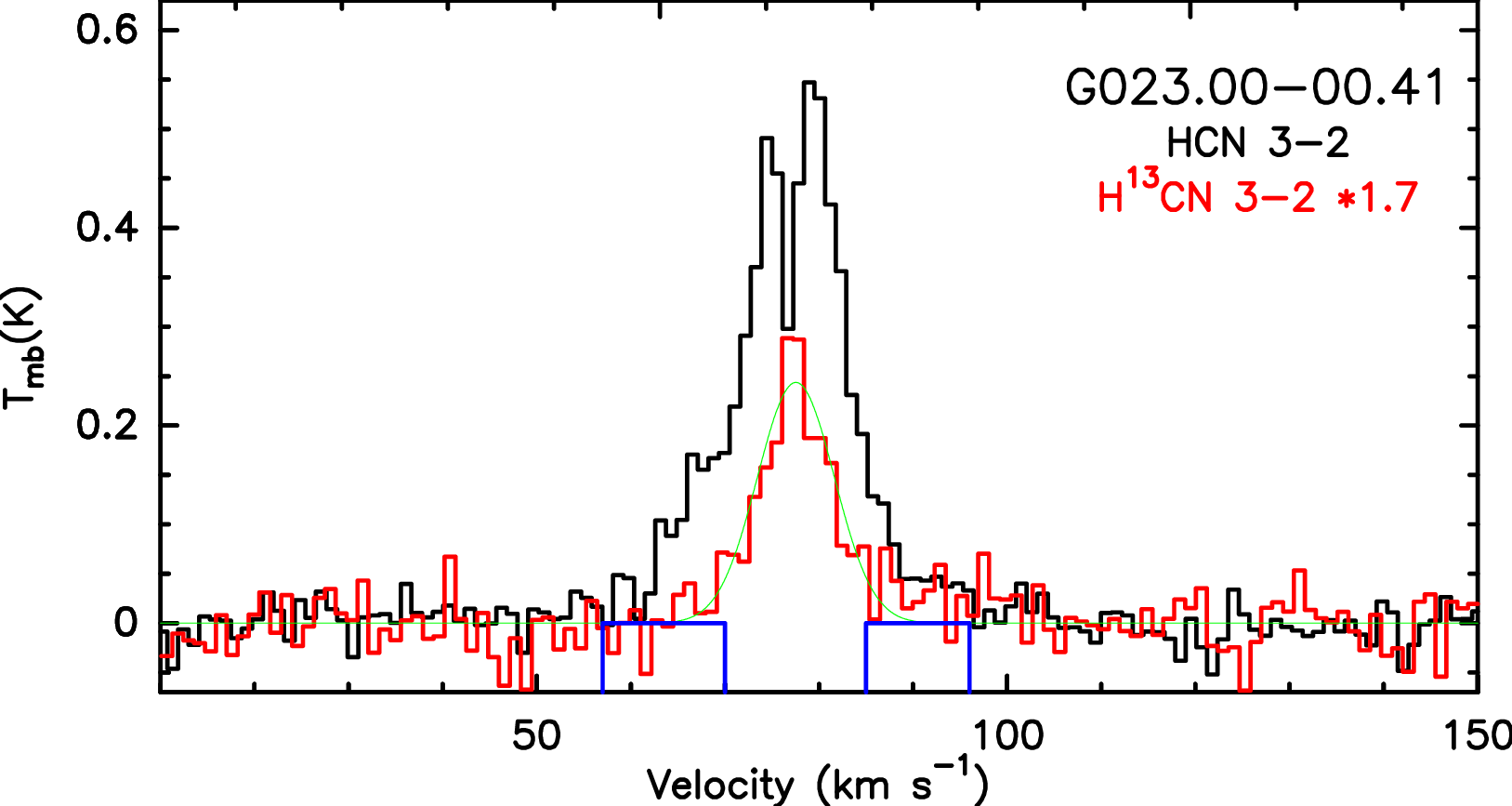} 
\includegraphics[width=0.35\columnwidth]{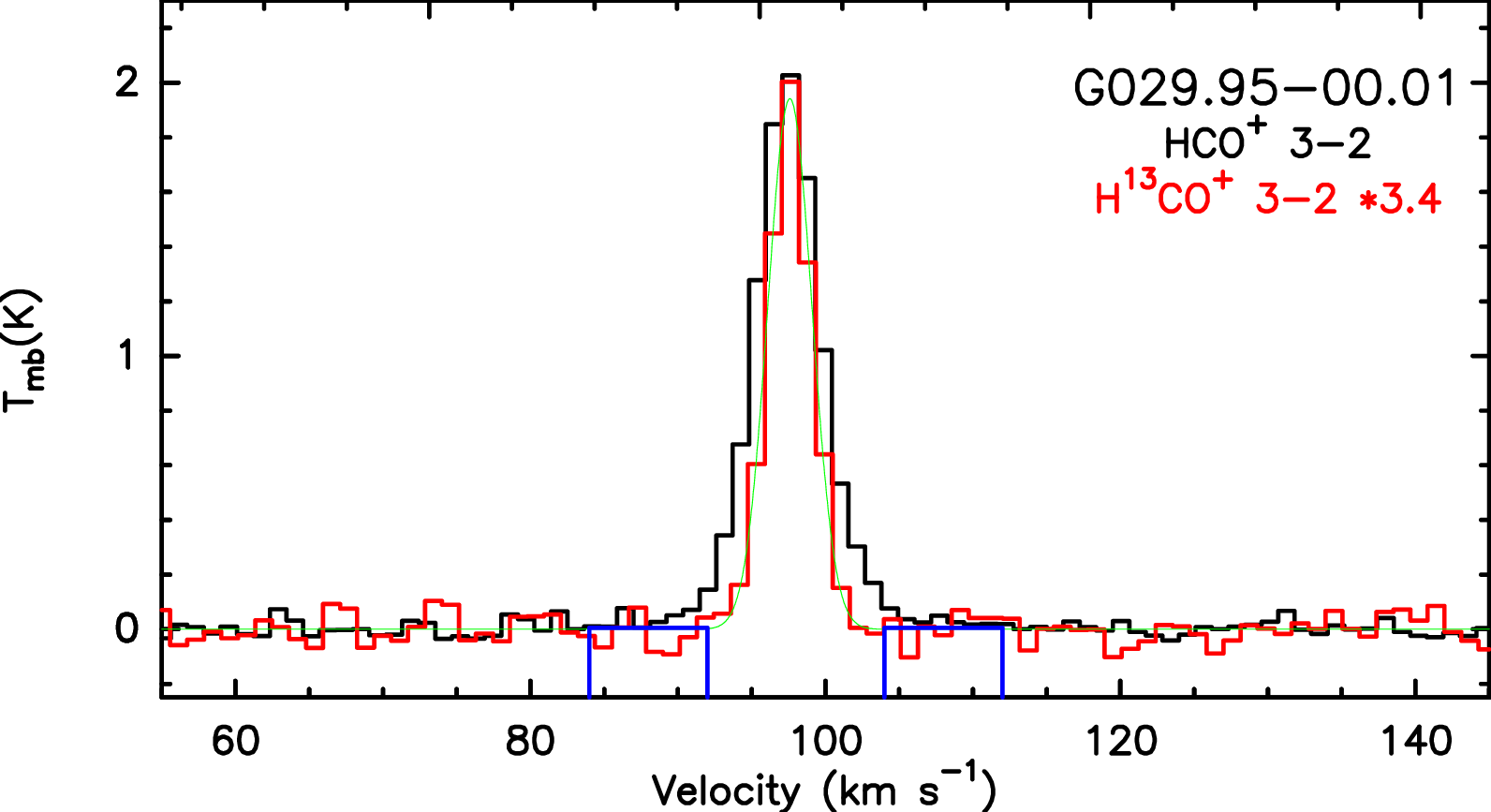} 
\includegraphics[width=0.35\columnwidth]{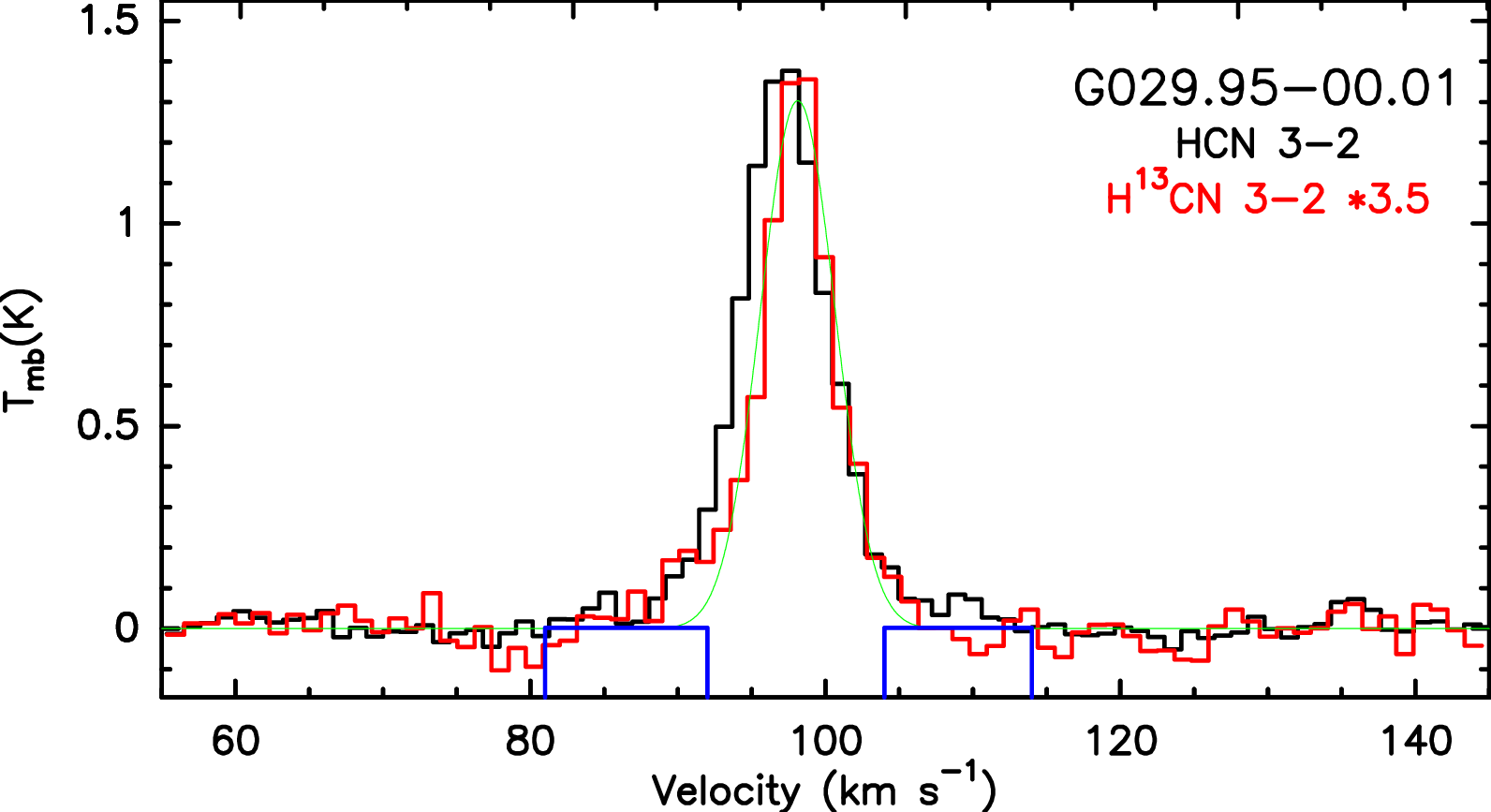} 
\includegraphics[width=0.35\columnwidth]{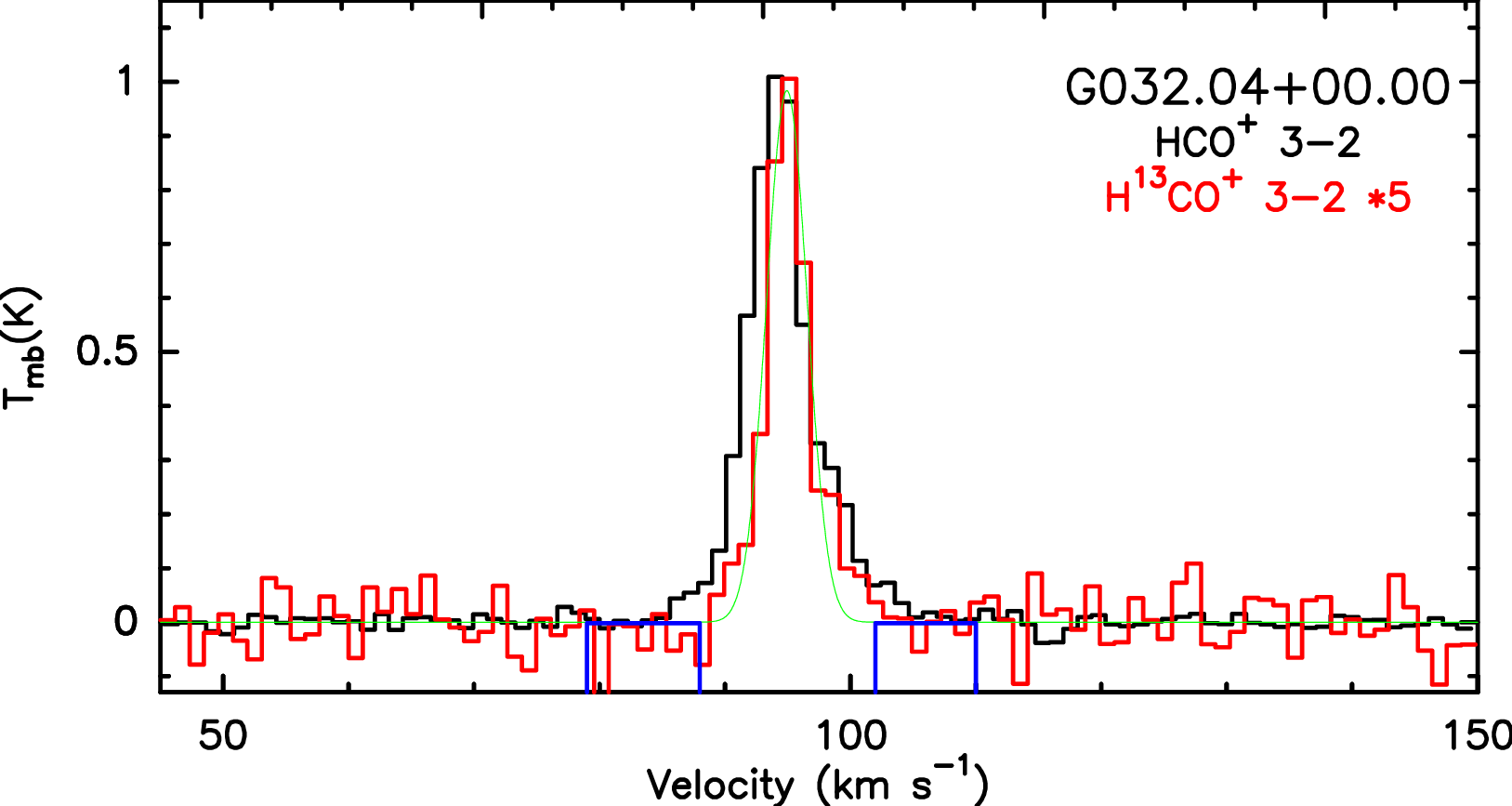} 
\includegraphics[width=0.35\columnwidth]{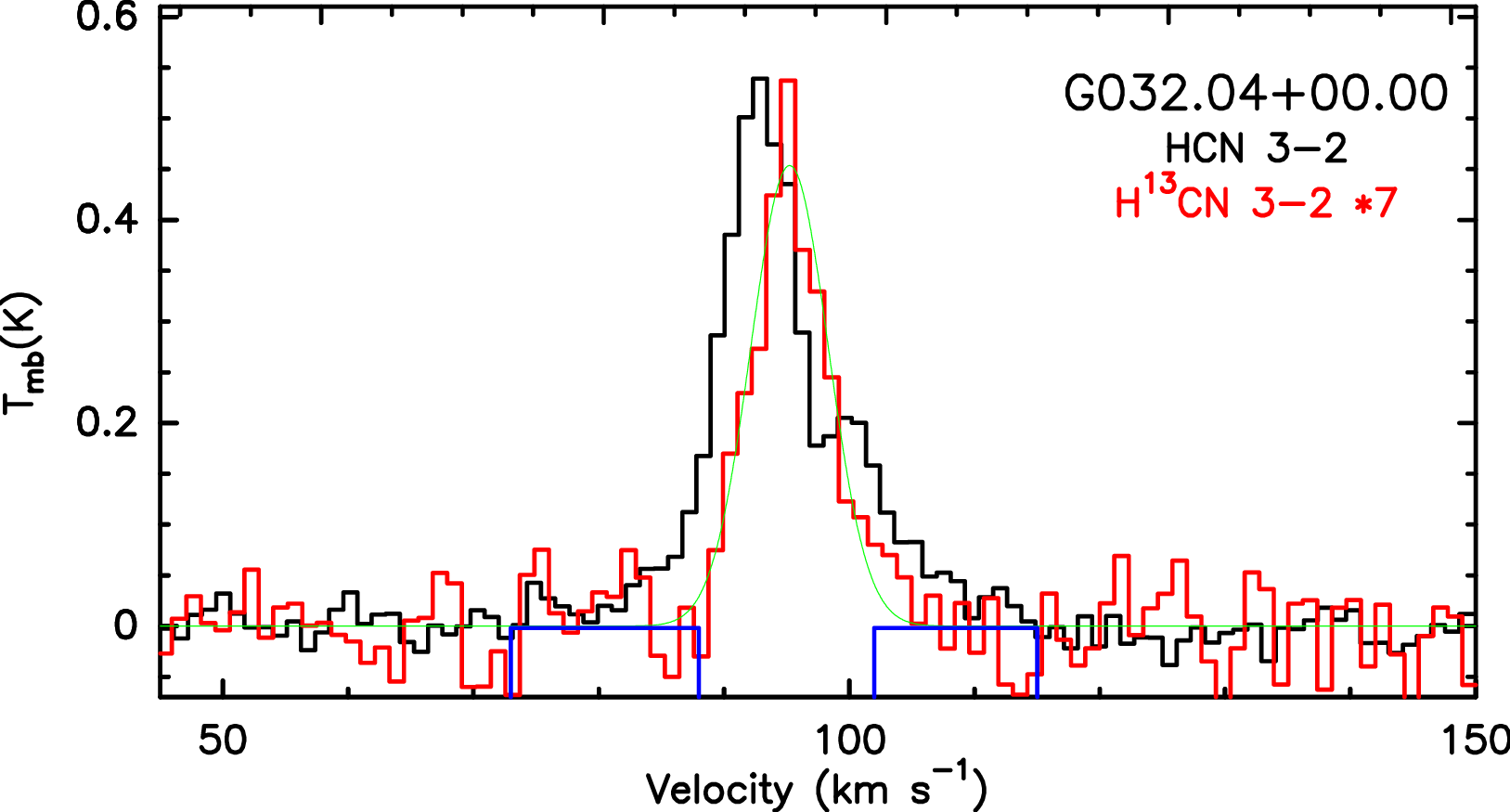} 
\includegraphics[width=0.35\columnwidth]{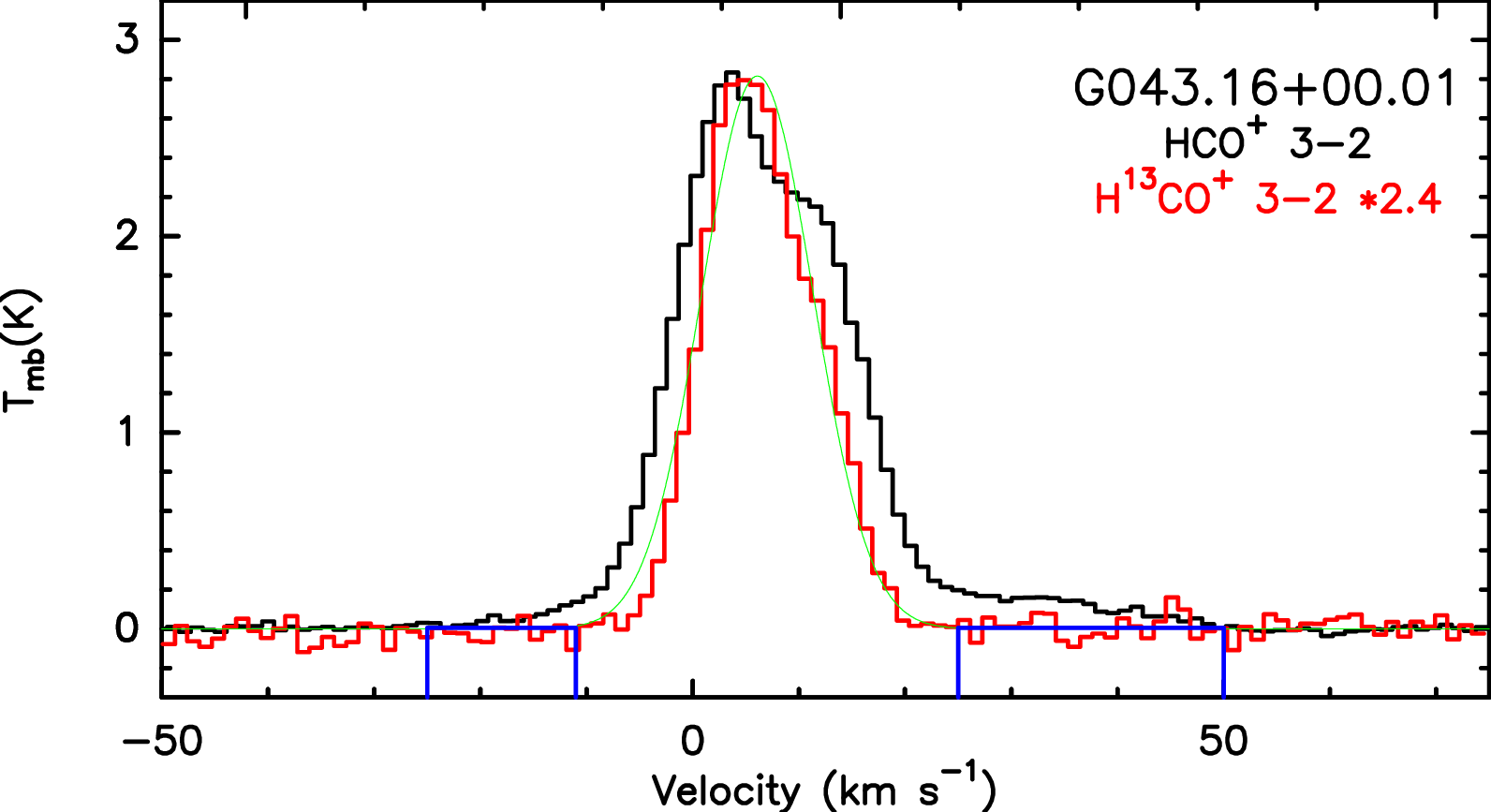} 
\includegraphics[width=0.35\columnwidth]{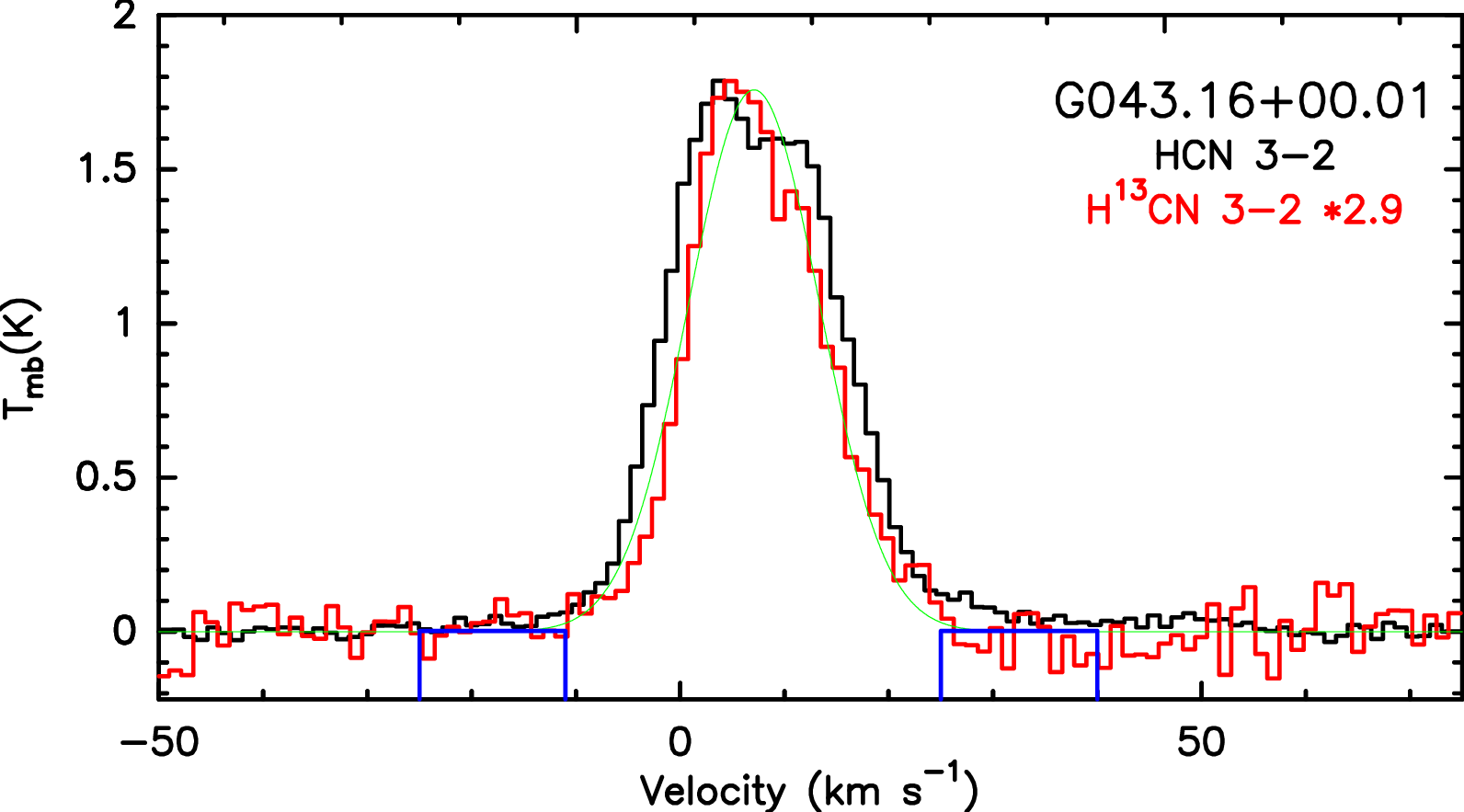} 
\addtocounter{figure}{0}
\caption{Continued.}
\label{fig1}
\end{figure}

\begin{figure}
\centering
\addtocounter{figure}{-1}
\centering
\includegraphics[width=0.35\columnwidth]{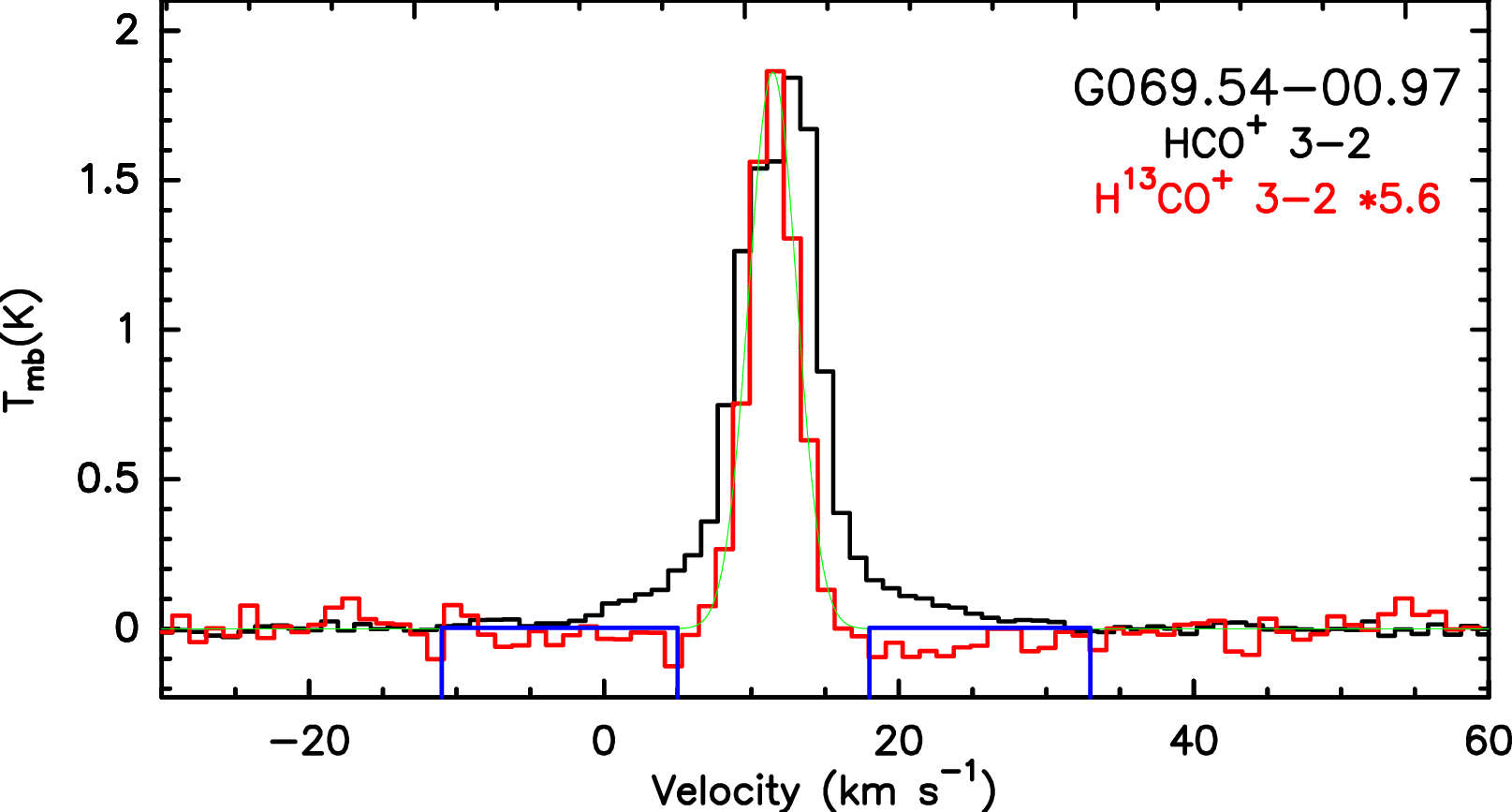} 
\includegraphics[width=0.35\columnwidth]{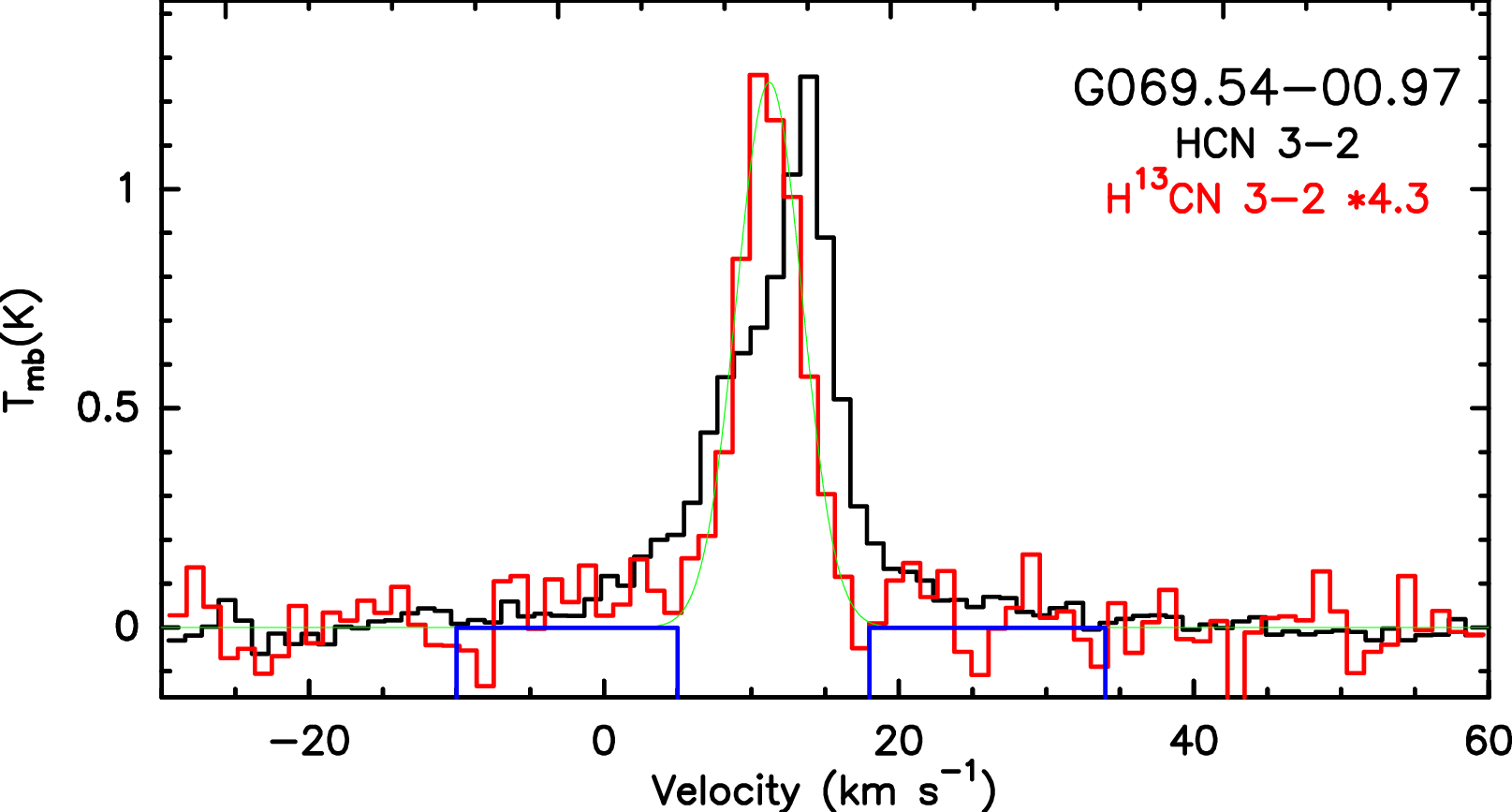}  
\includegraphics[width=0.35\columnwidth]{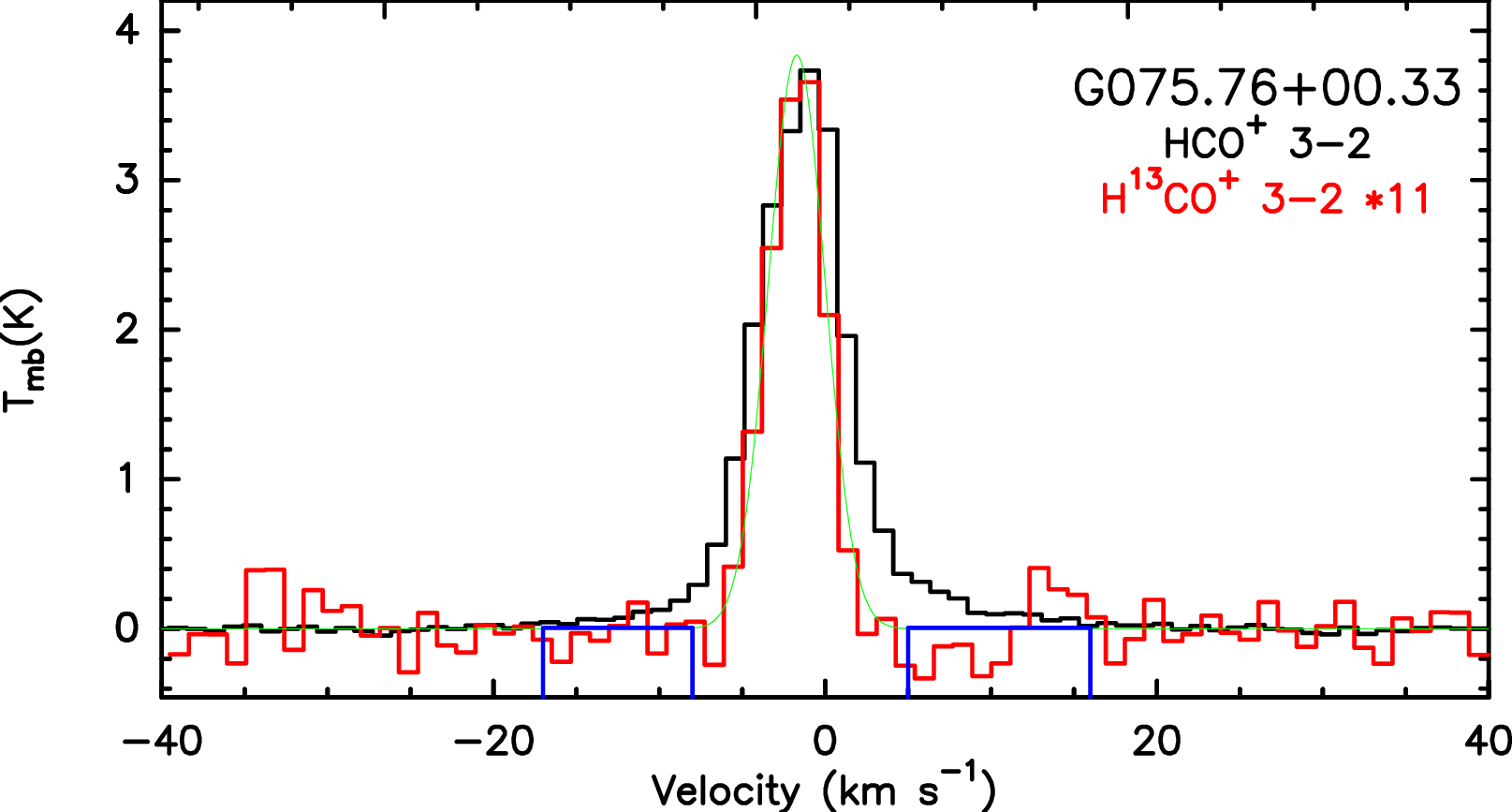} 
\includegraphics[width=0.35\columnwidth]{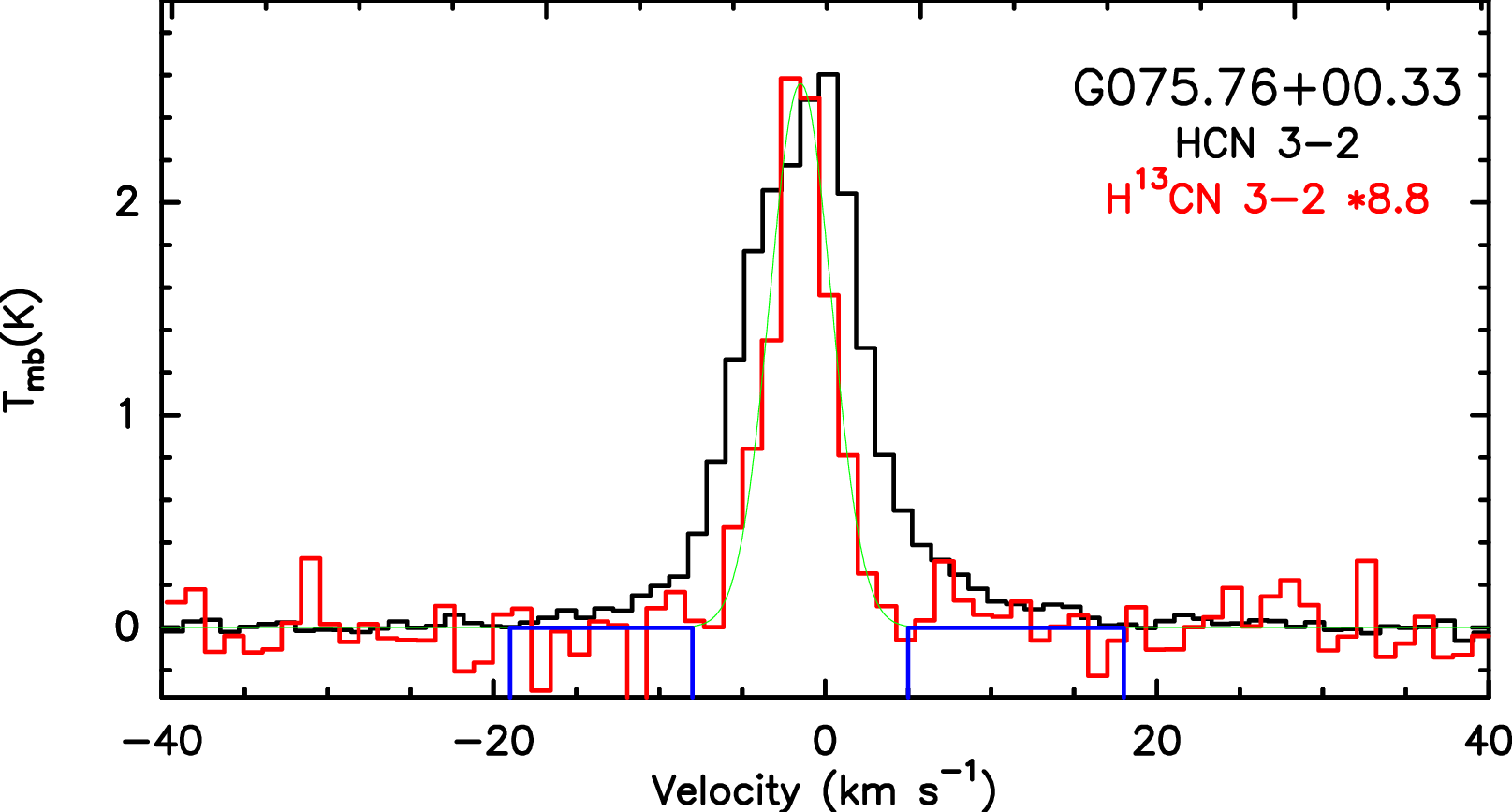} 
\includegraphics[width=0.35\columnwidth]{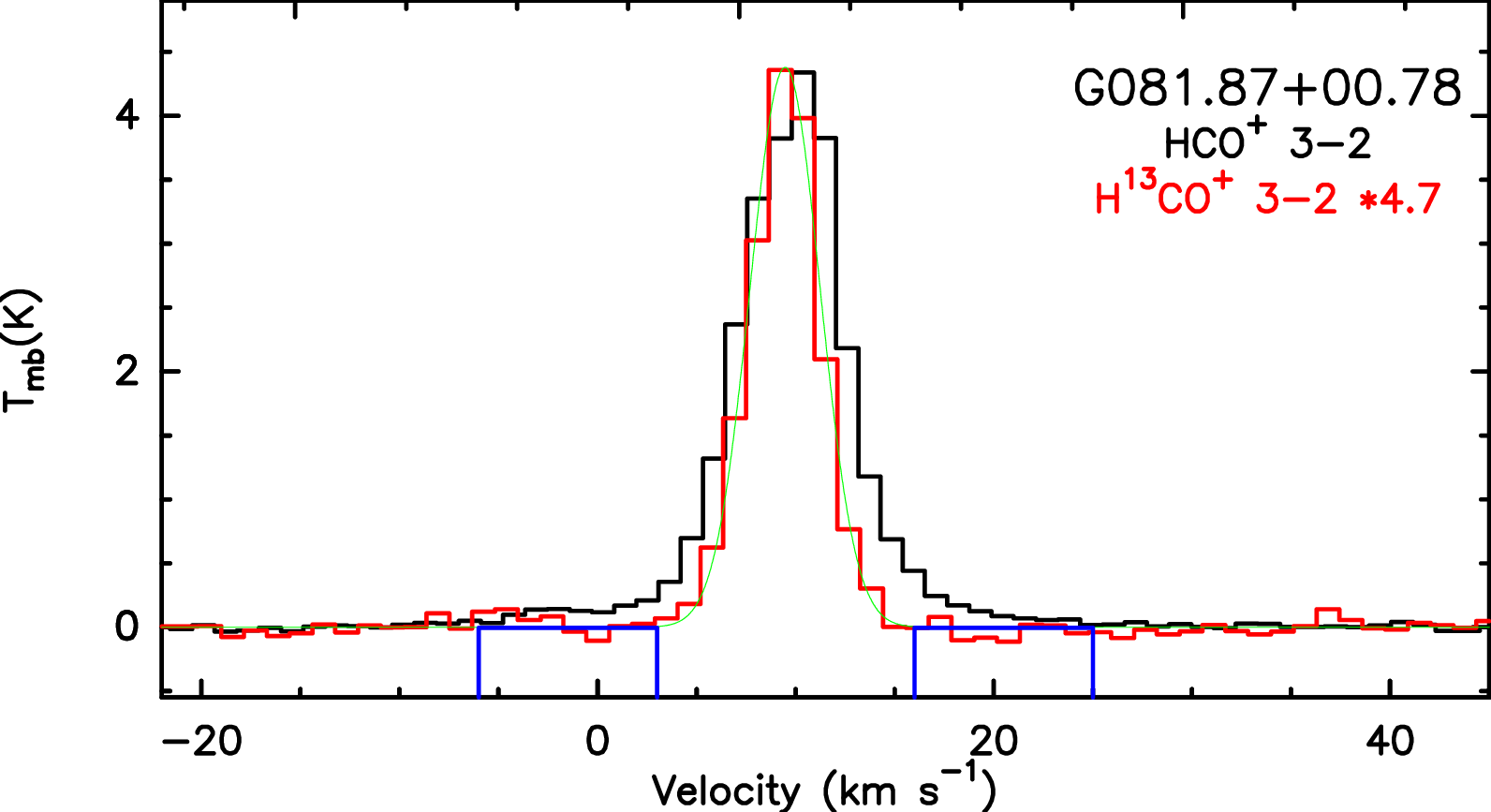} 
\includegraphics[width=0.35\columnwidth]{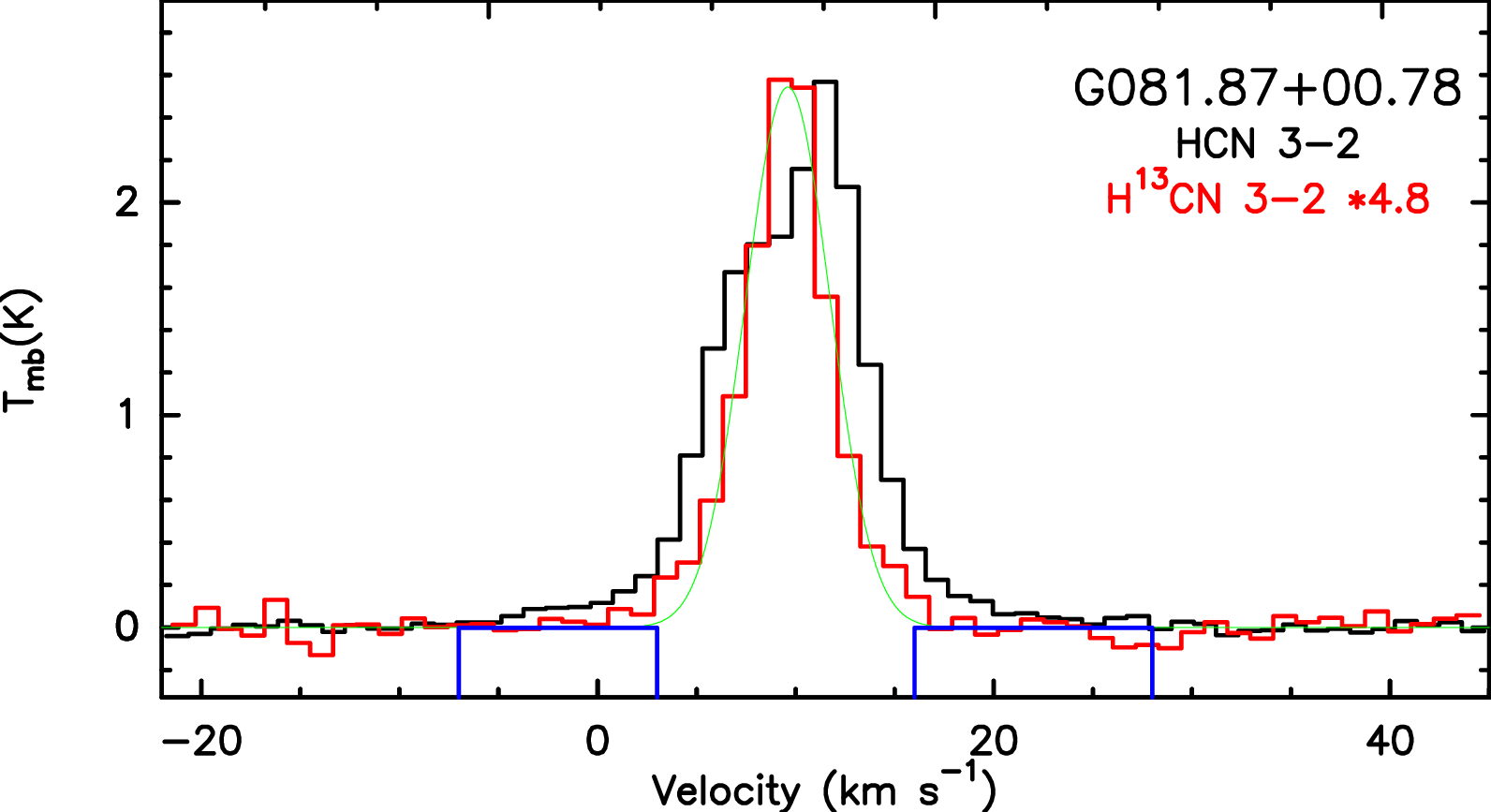} 
\includegraphics[width=0.35\columnwidth]{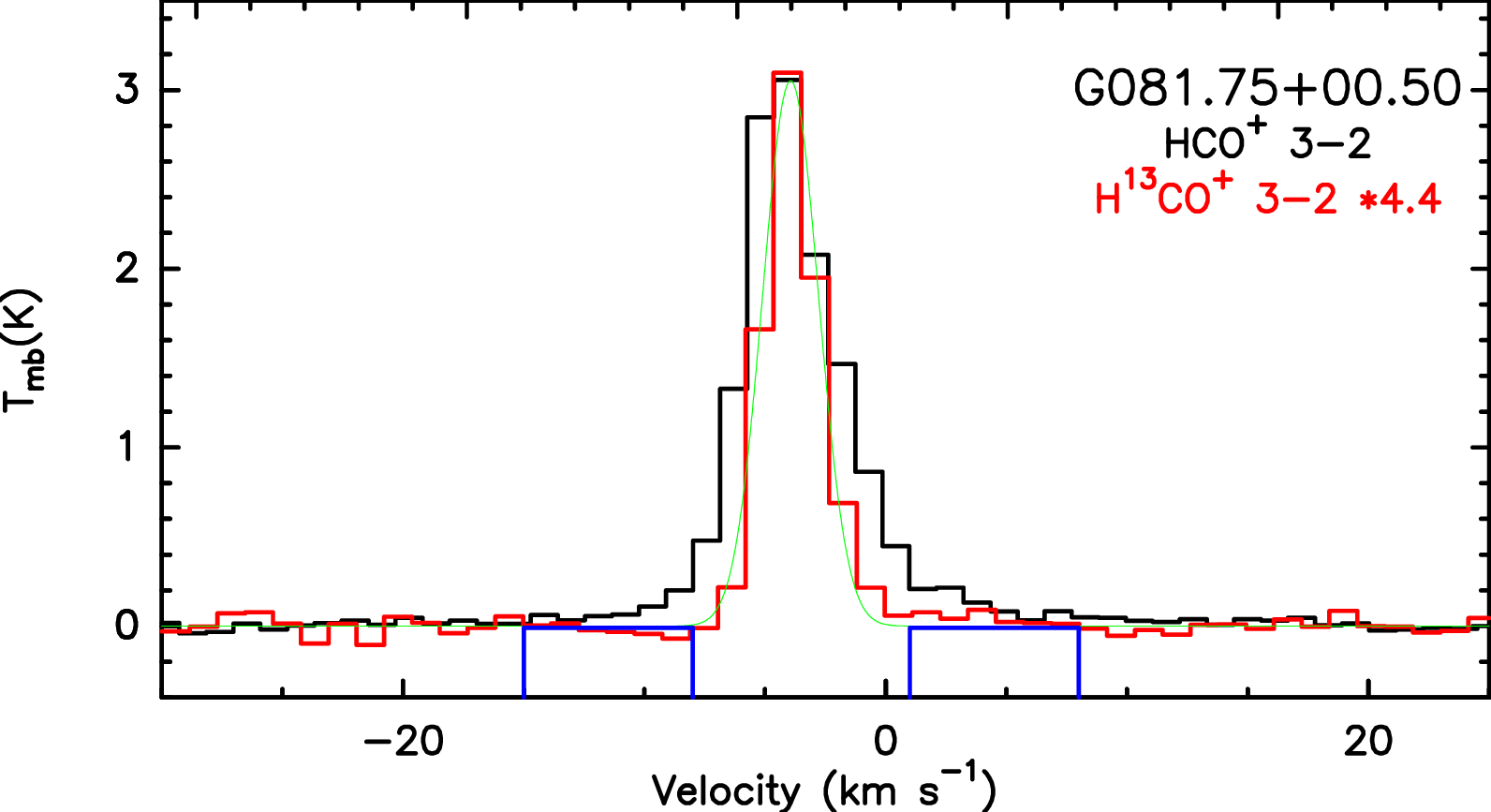} 
\includegraphics[width=0.35\columnwidth]{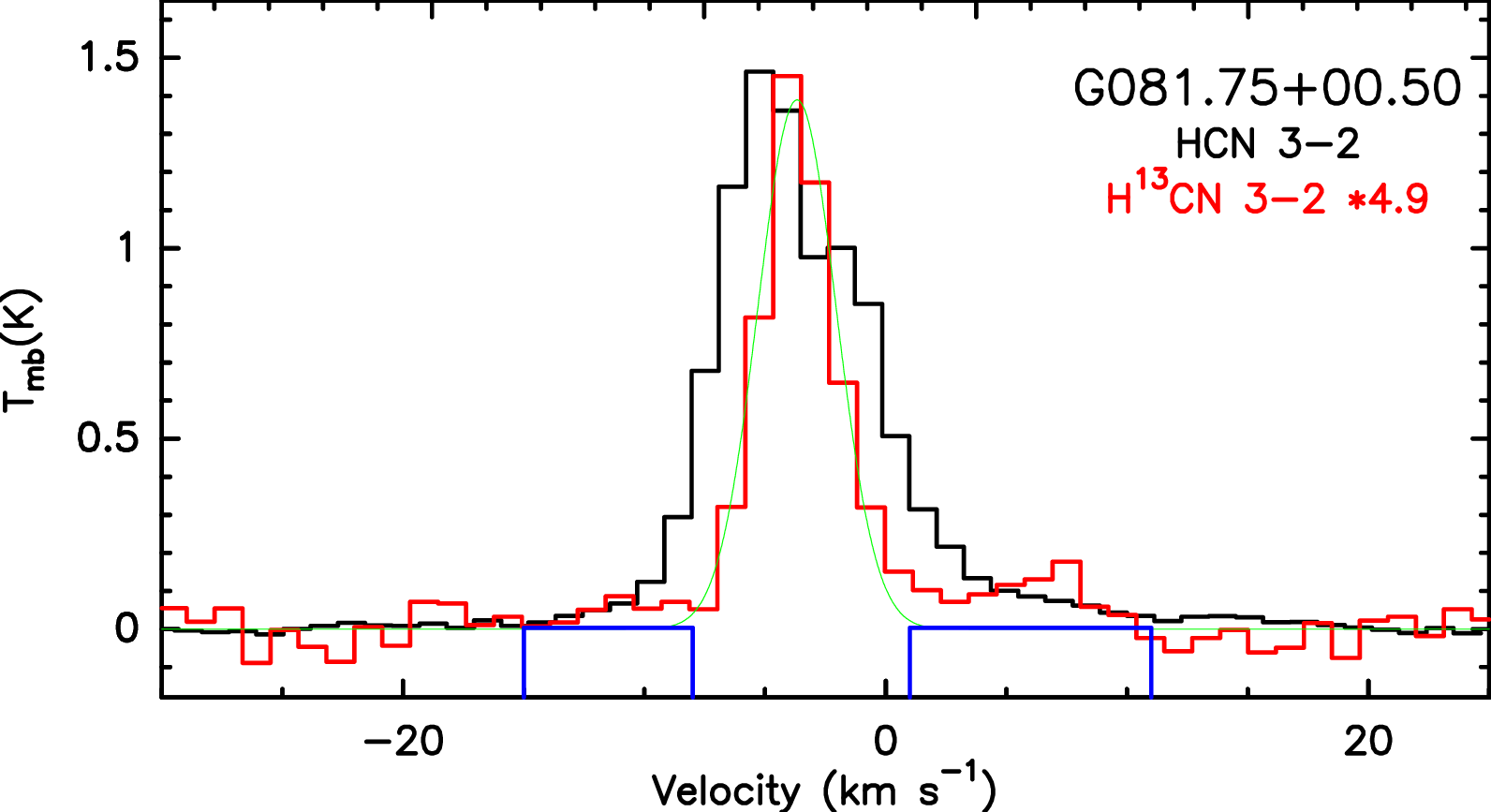} 
\includegraphics[width=0.35\columnwidth]{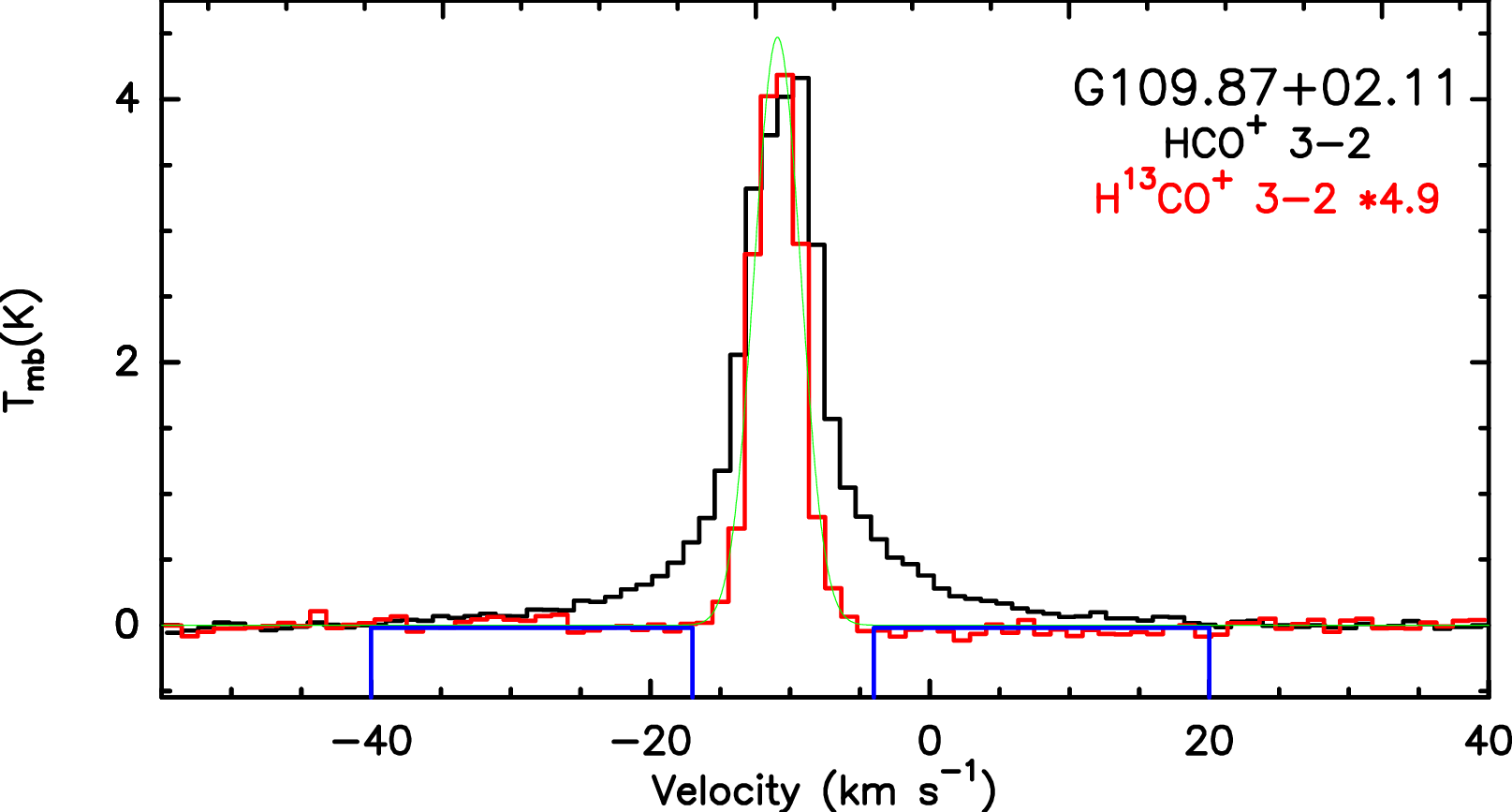} 
\includegraphics[width=0.35\columnwidth]{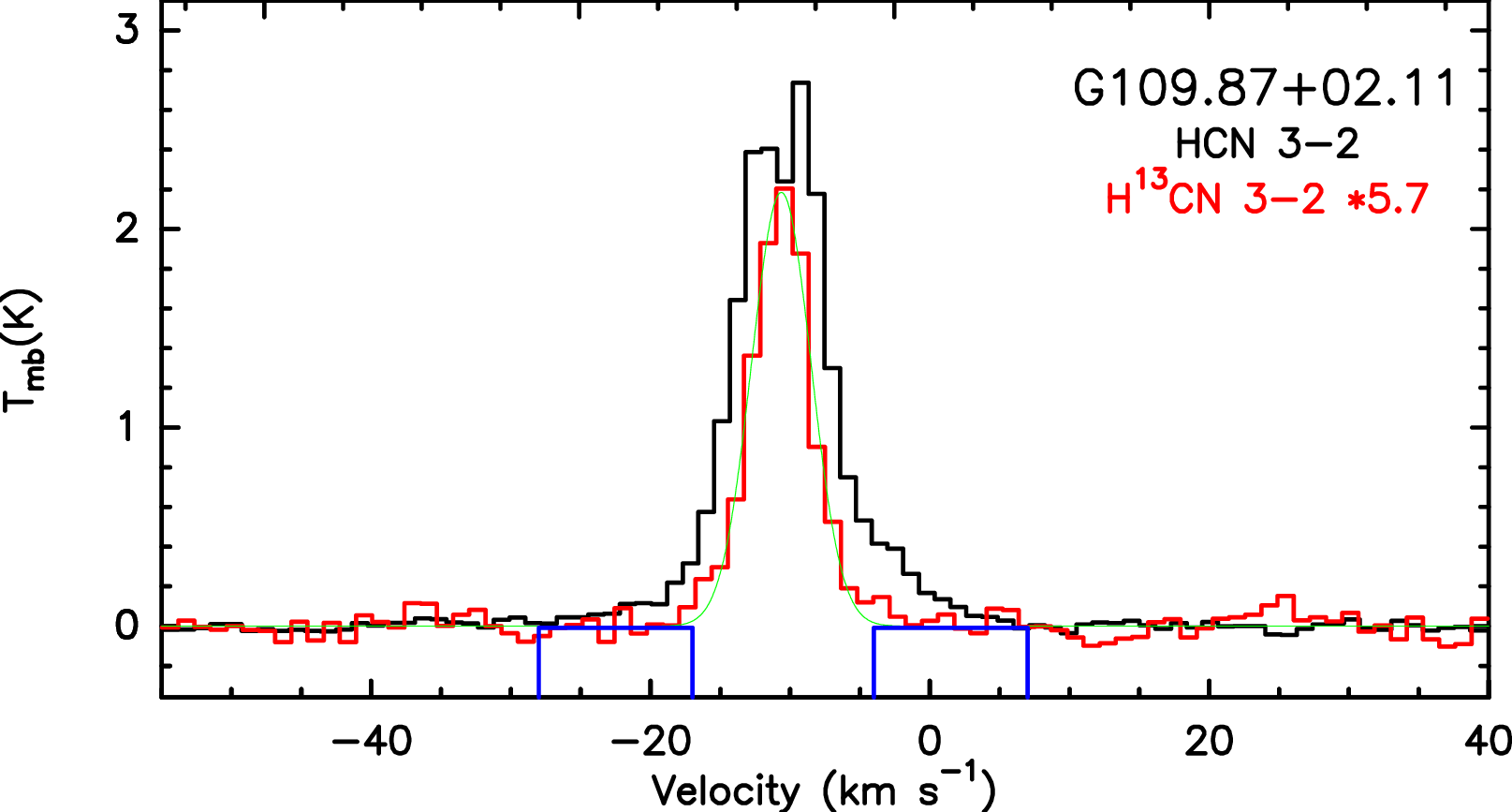} 
\includegraphics[width=0.35\columnwidth]{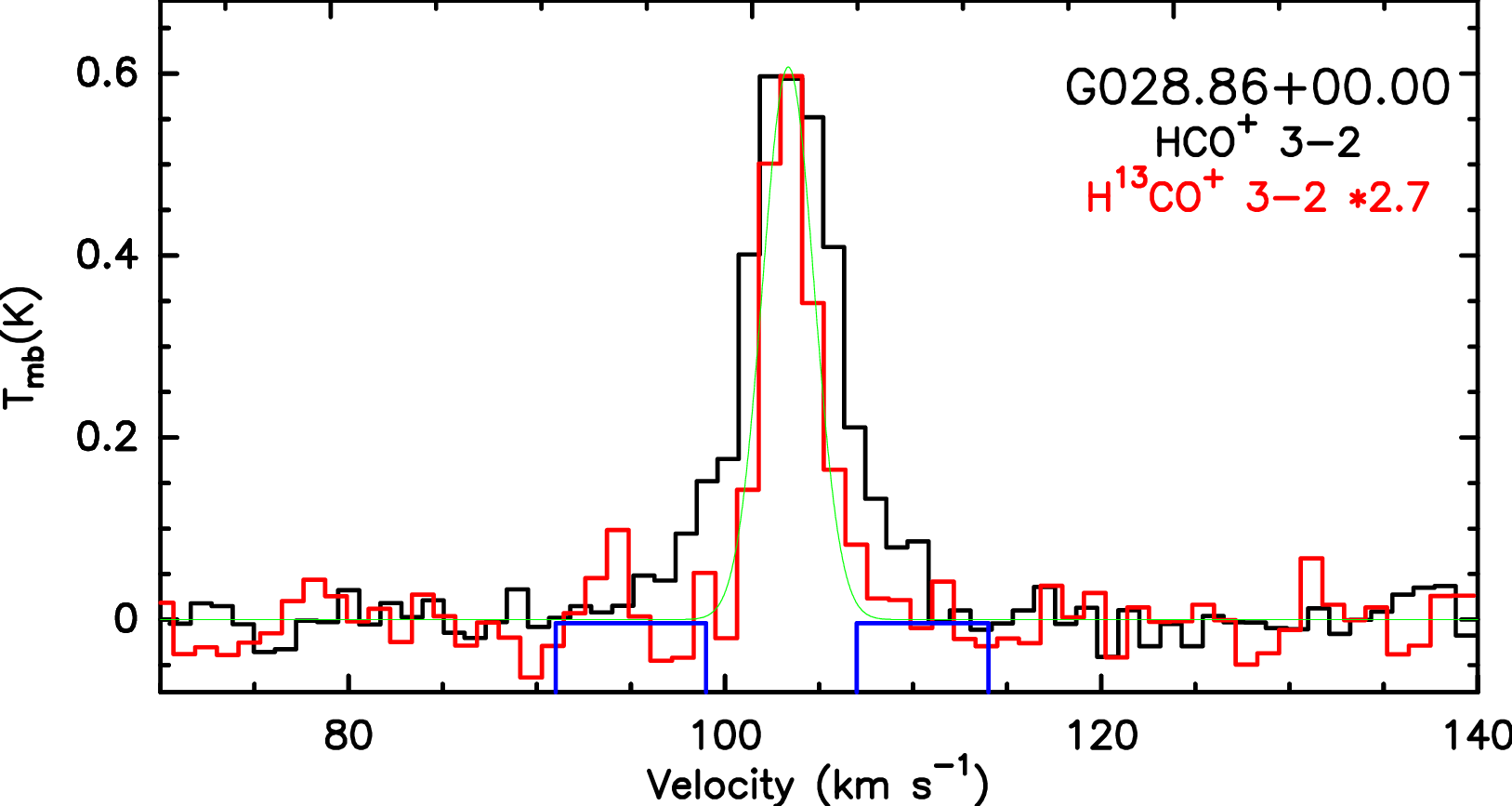}
\includegraphics[width=0.35\columnwidth]{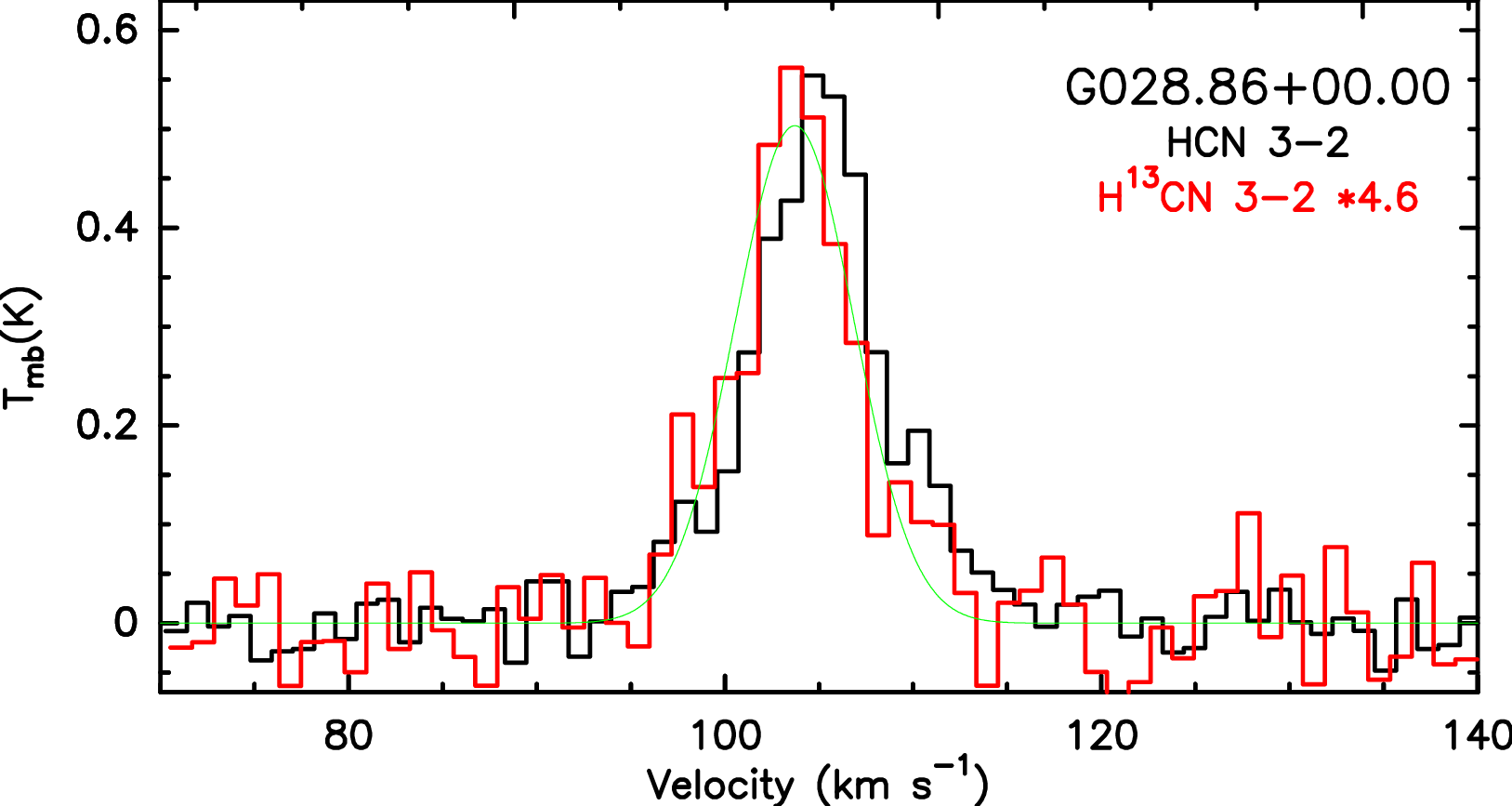} 
\addtocounter{figure}{0}
\caption{Continued.}
\label{fig1}
\end{figure}

\begin{figure}
\centering
\addtocounter{figure}{-1}
\includegraphics[width=0.35\columnwidth]{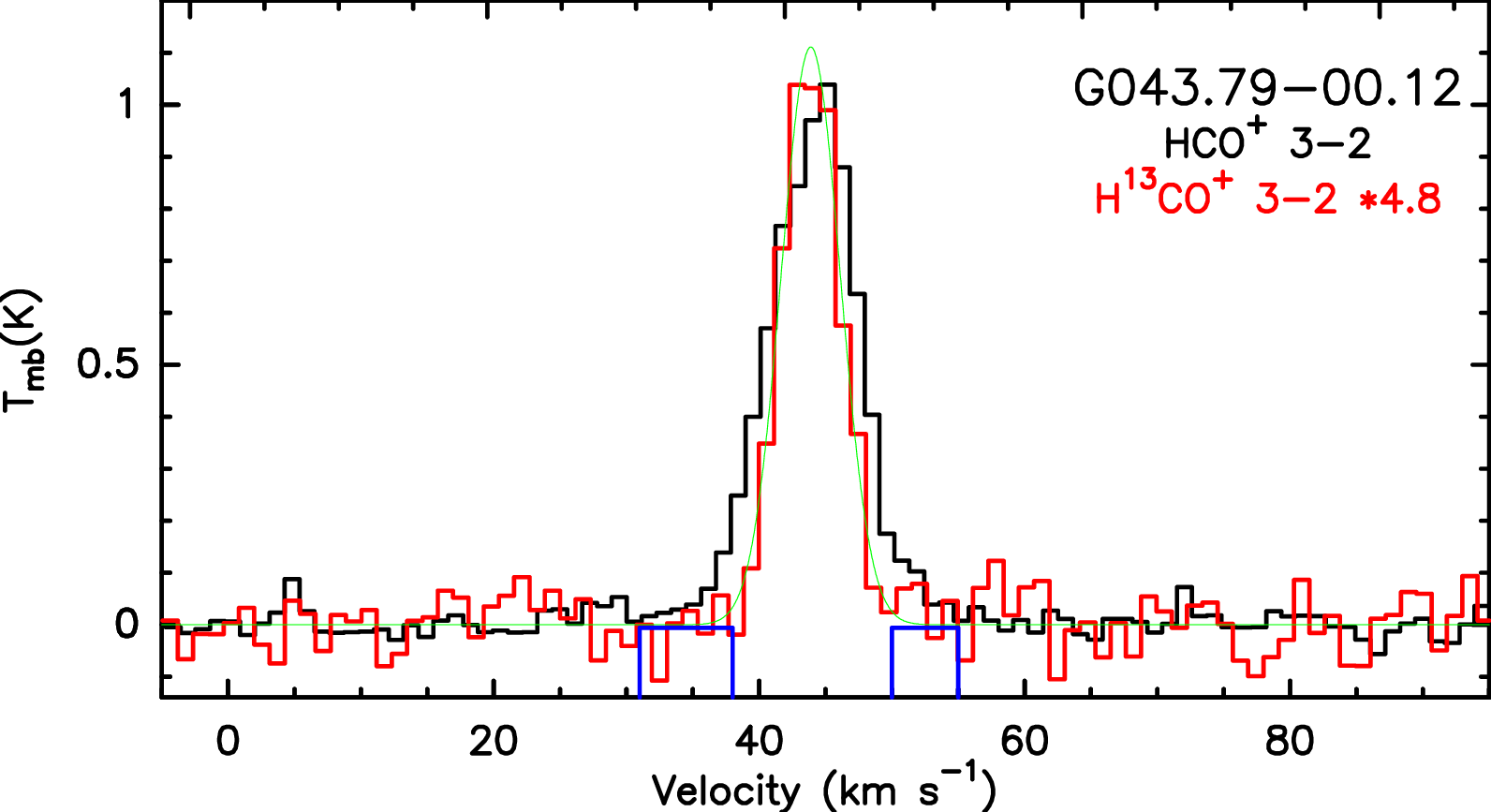}  
\includegraphics[width=0.35\columnwidth]{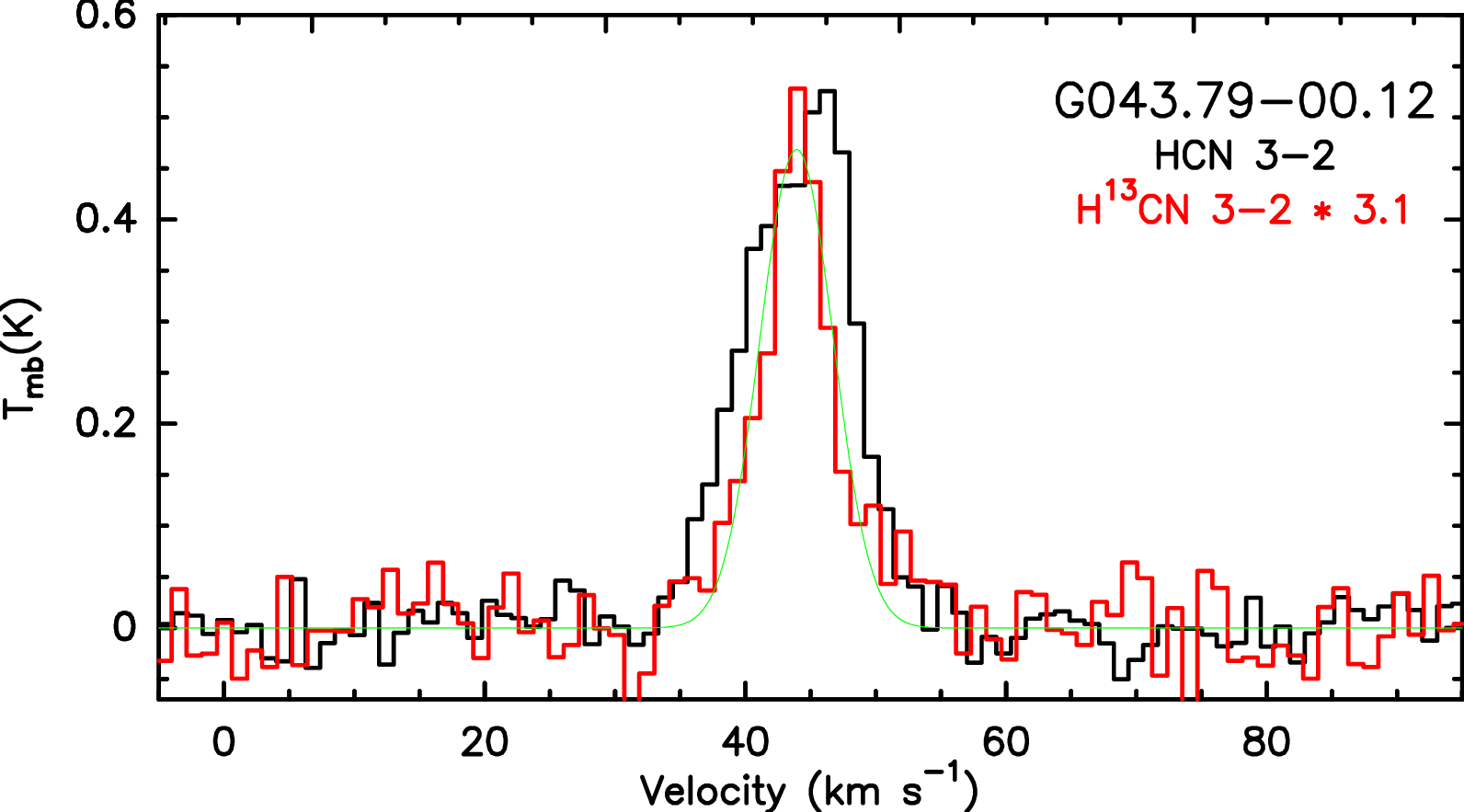} 
\includegraphics[width=0.35\columnwidth]{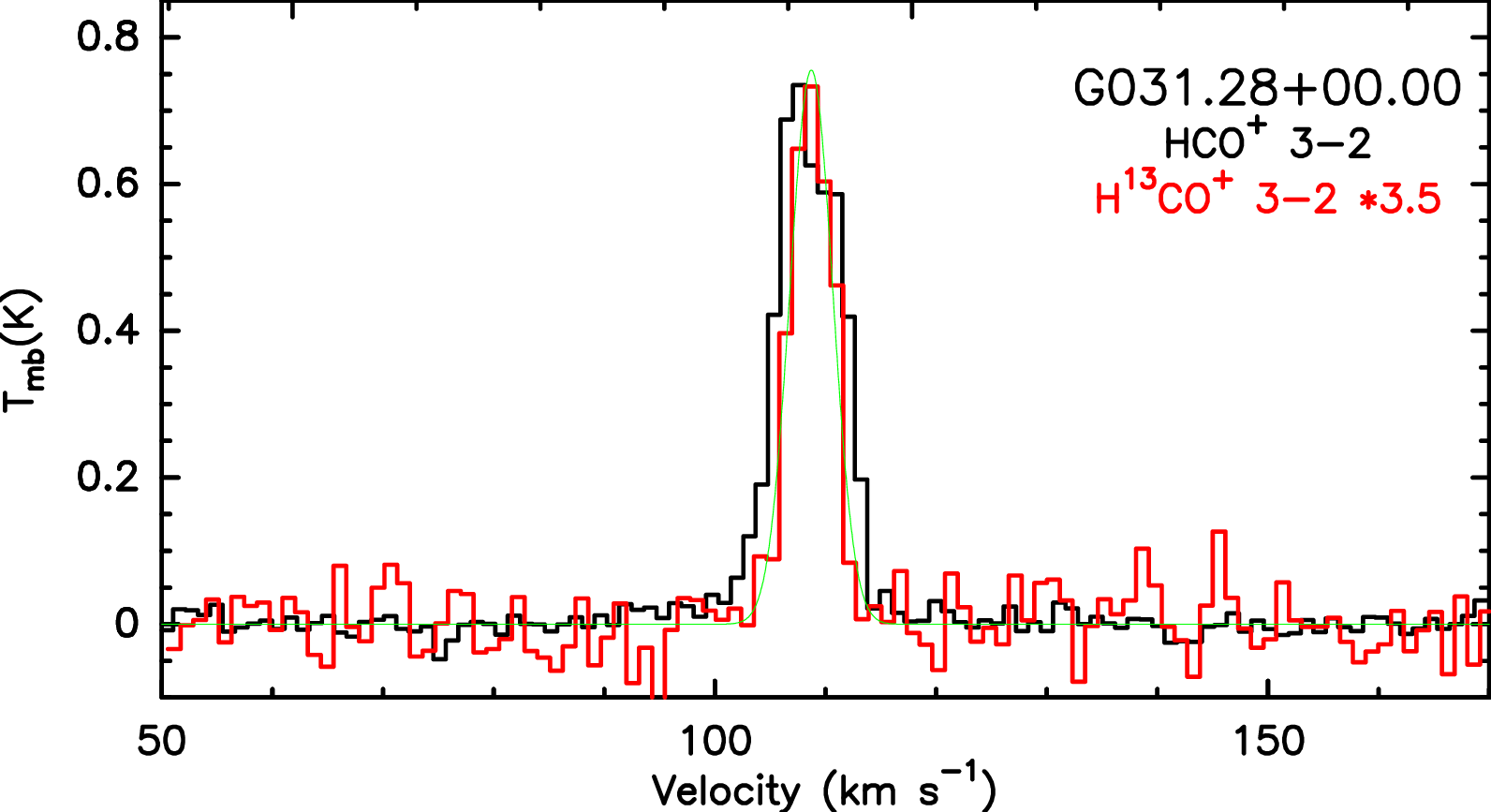} 
\includegraphics[width=0.35\columnwidth]{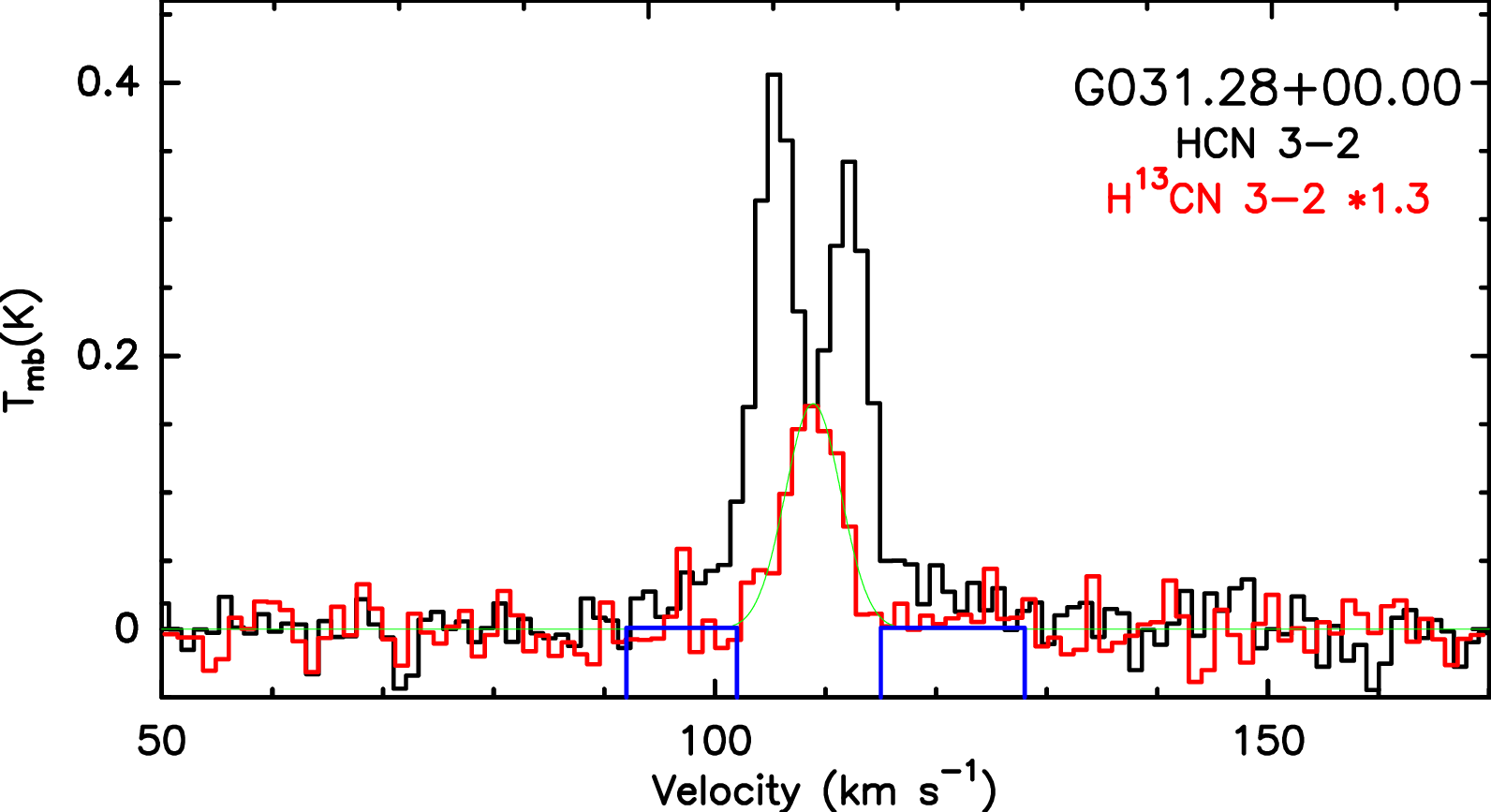} 
\includegraphics[width=0.35\columnwidth]{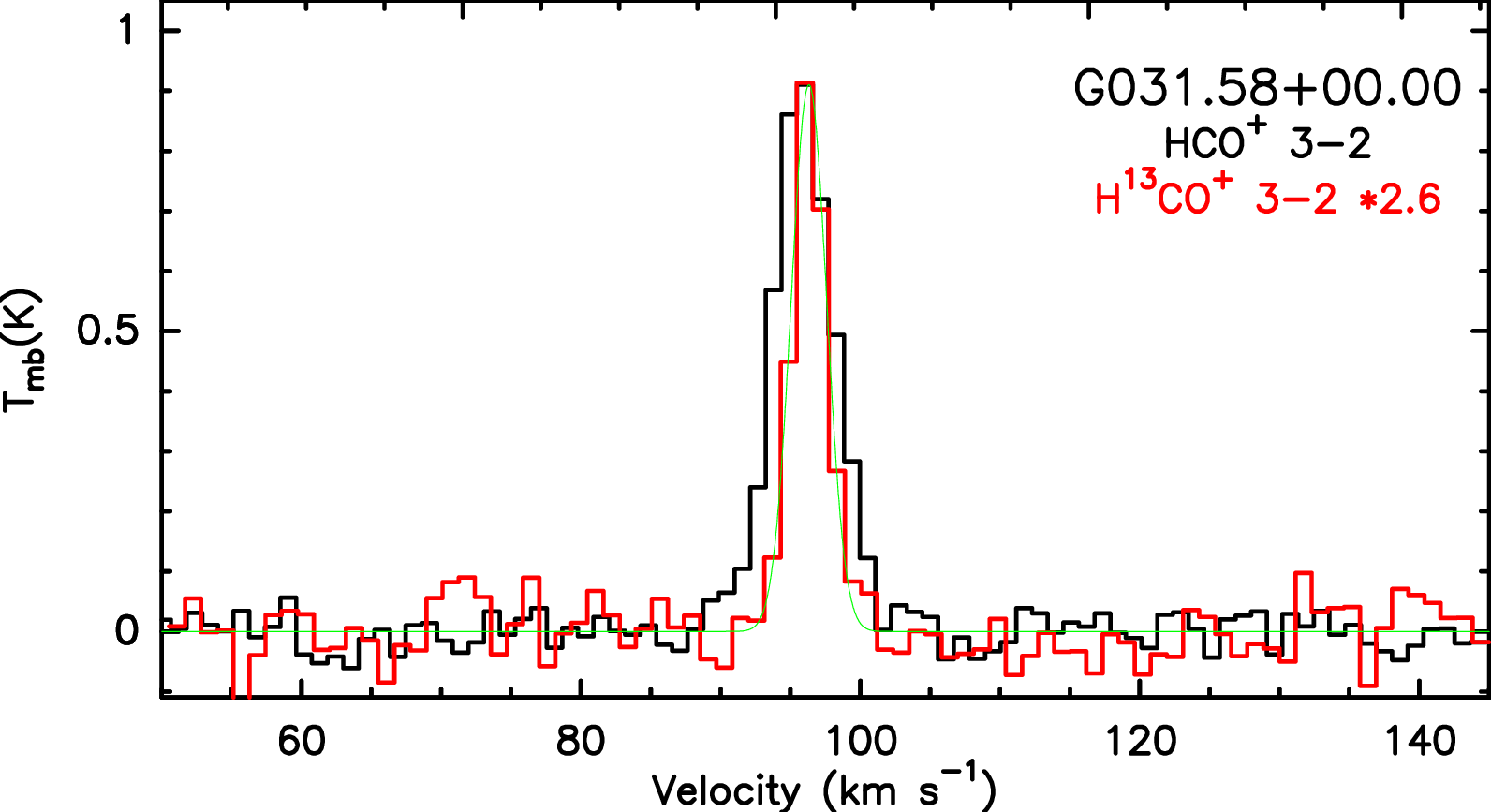} 
\includegraphics[width=0.35\columnwidth]{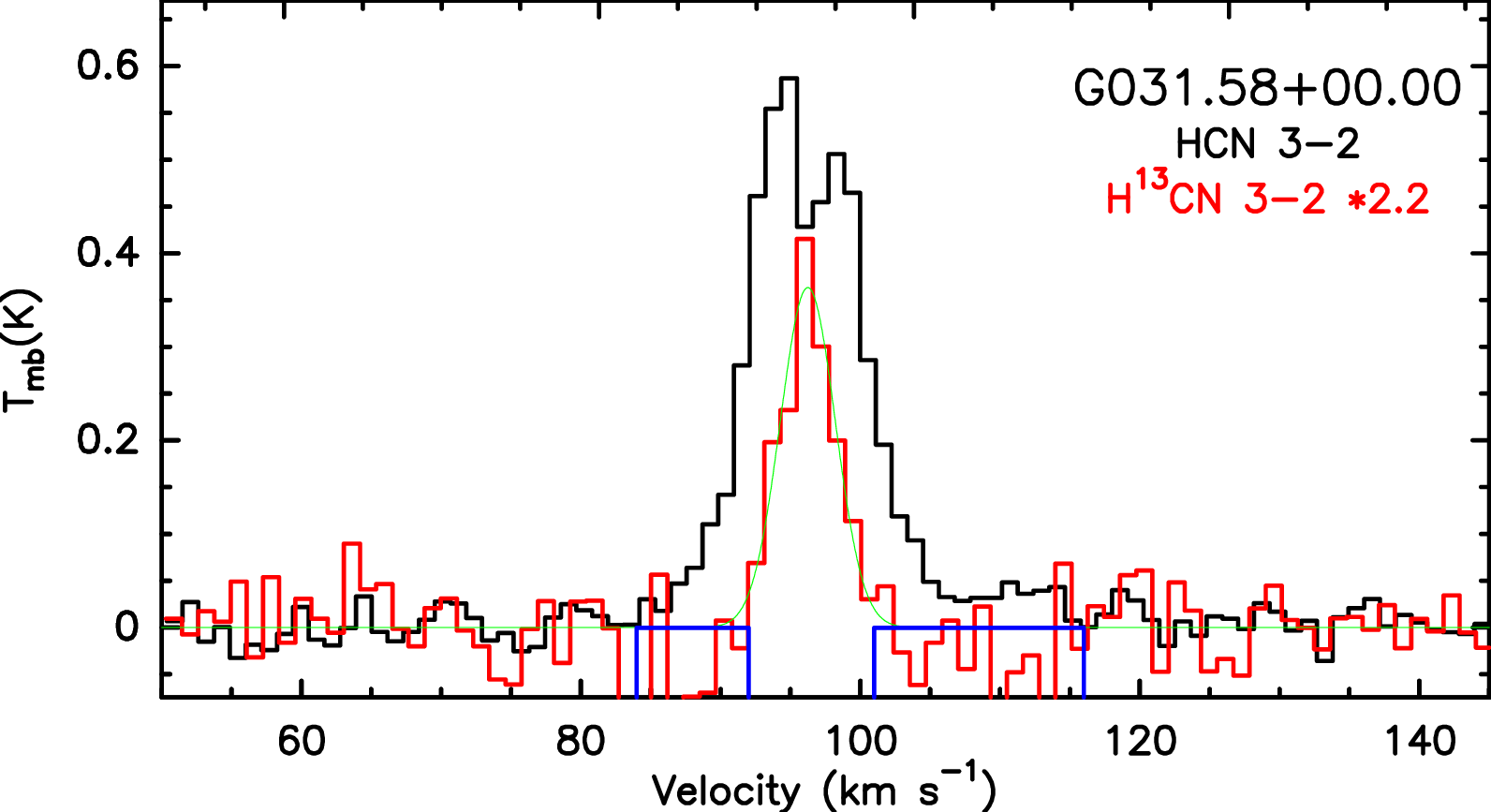} 
\includegraphics[width=0.35\columnwidth]{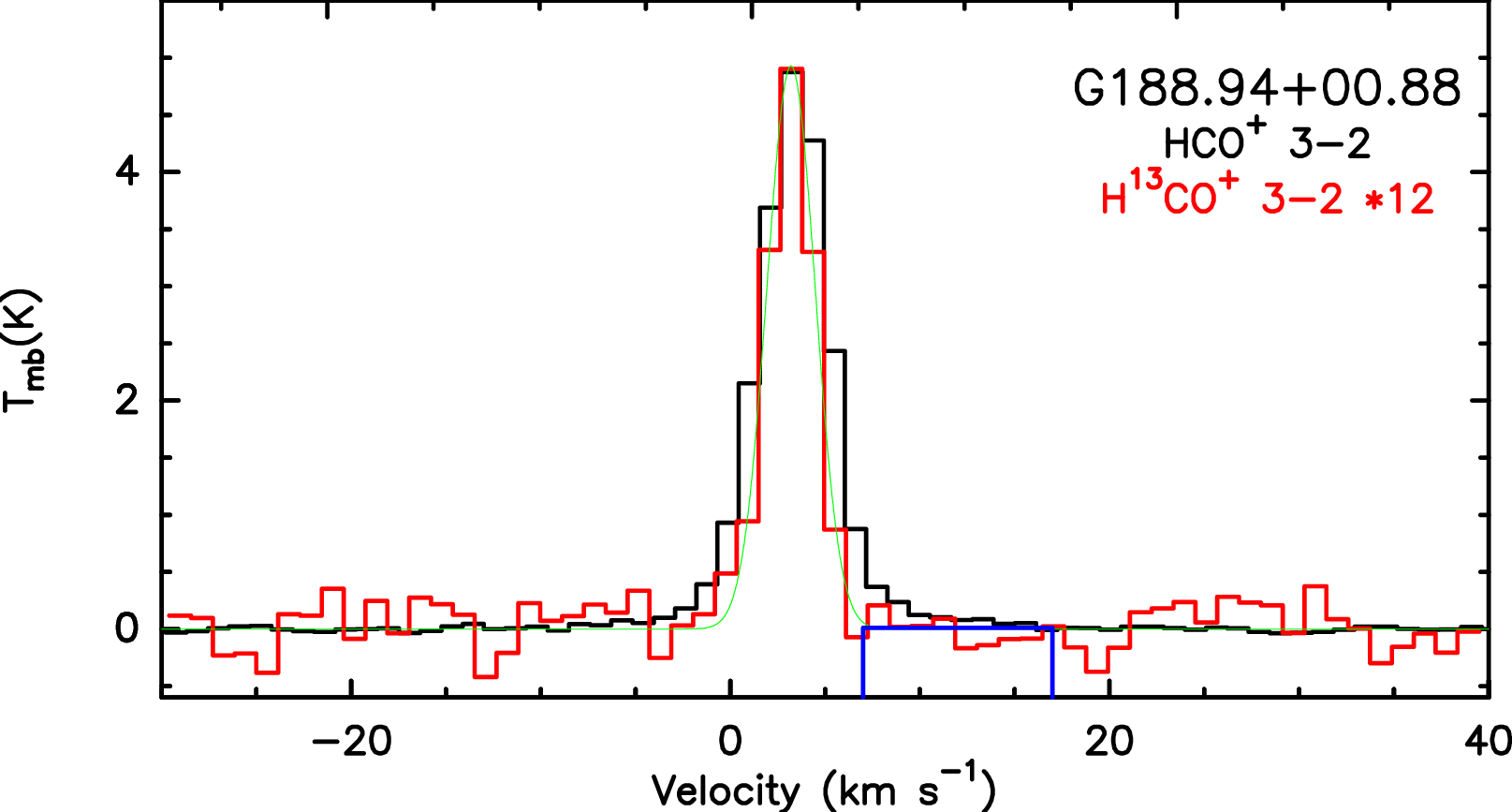} 
\includegraphics[width=0.35\columnwidth]{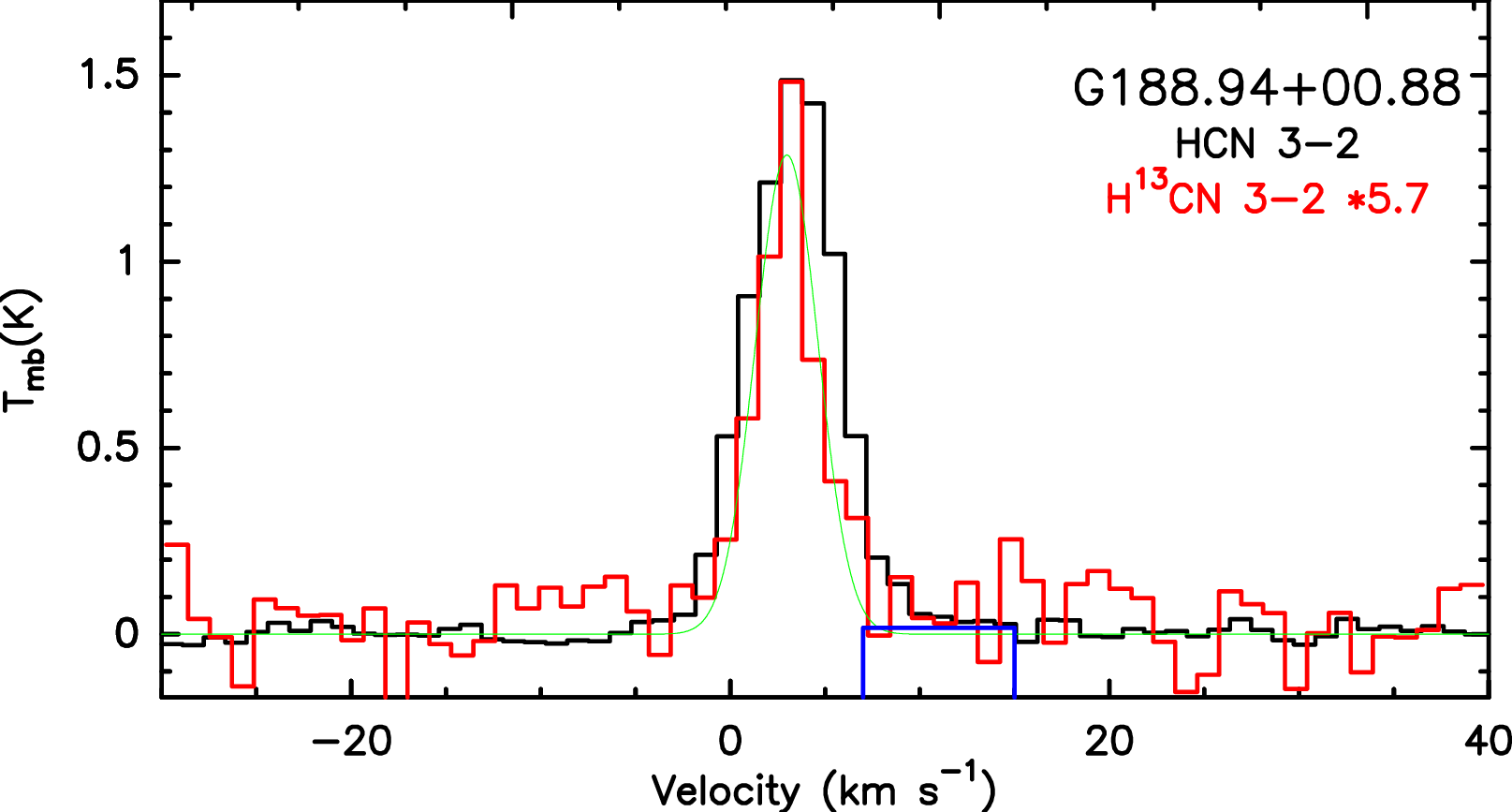} 
\includegraphics[width=0.35\columnwidth]{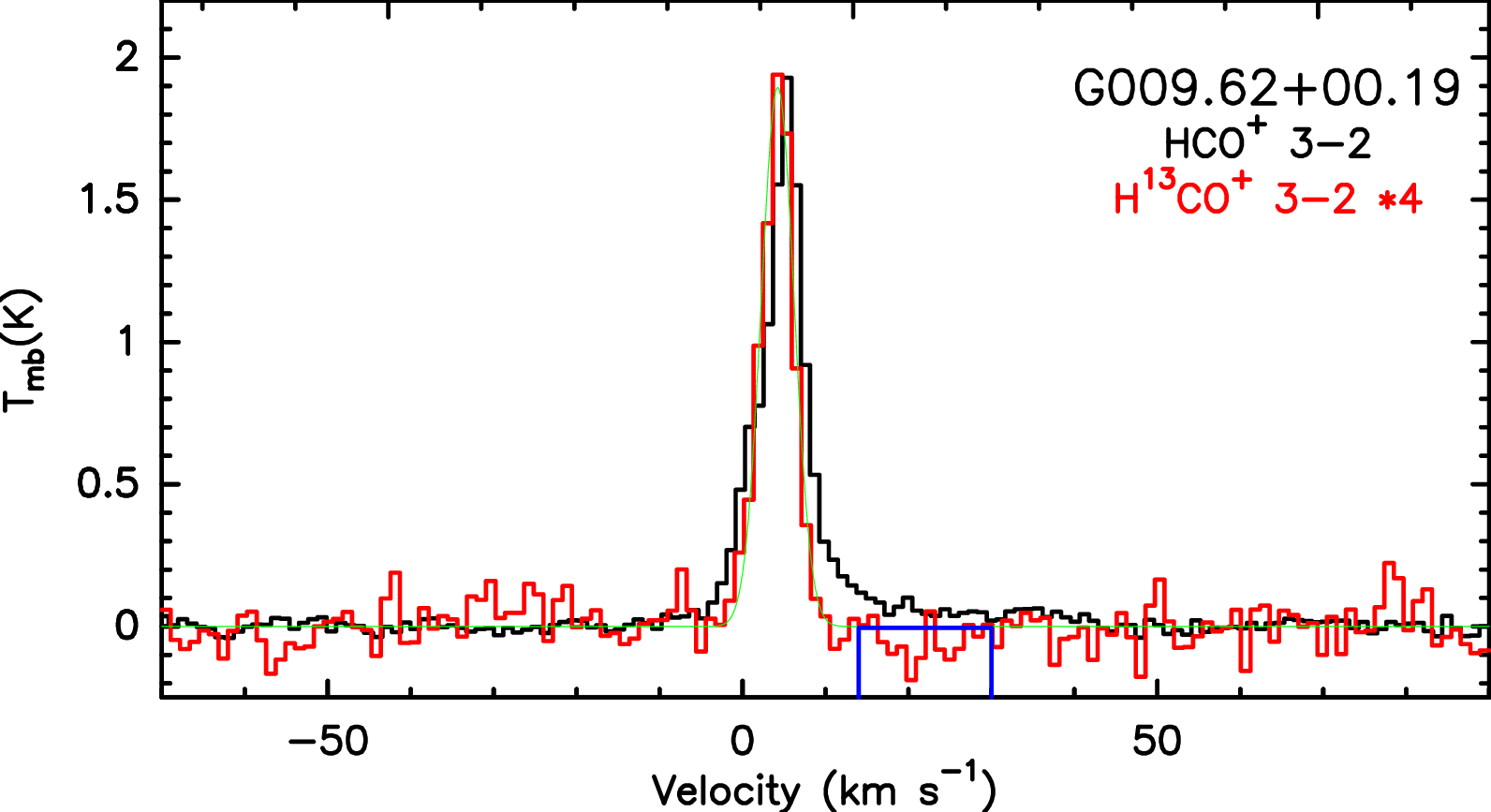} 
\includegraphics[width=0.35\columnwidth]{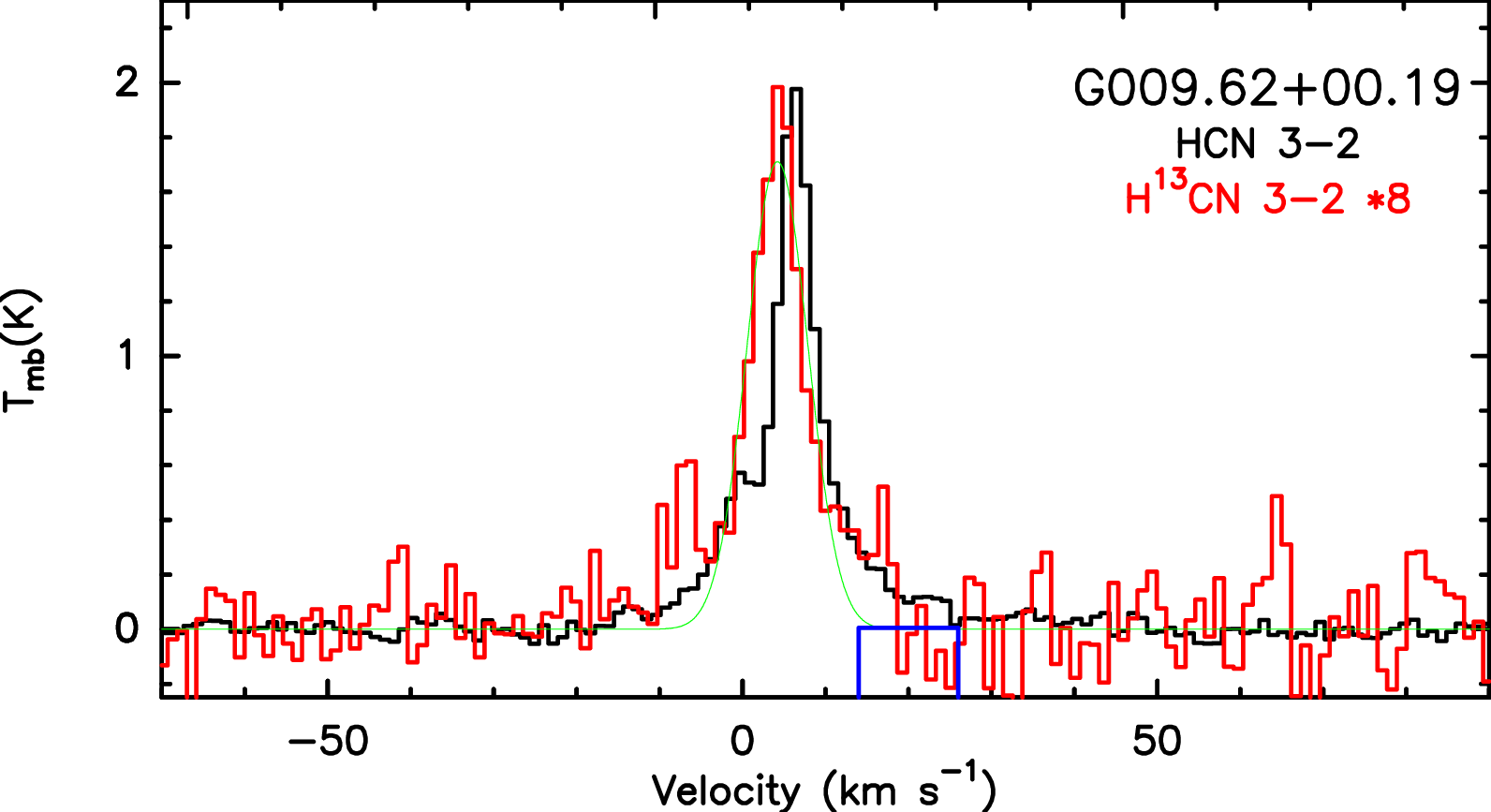}  
\includegraphics[width=0.35\columnwidth]{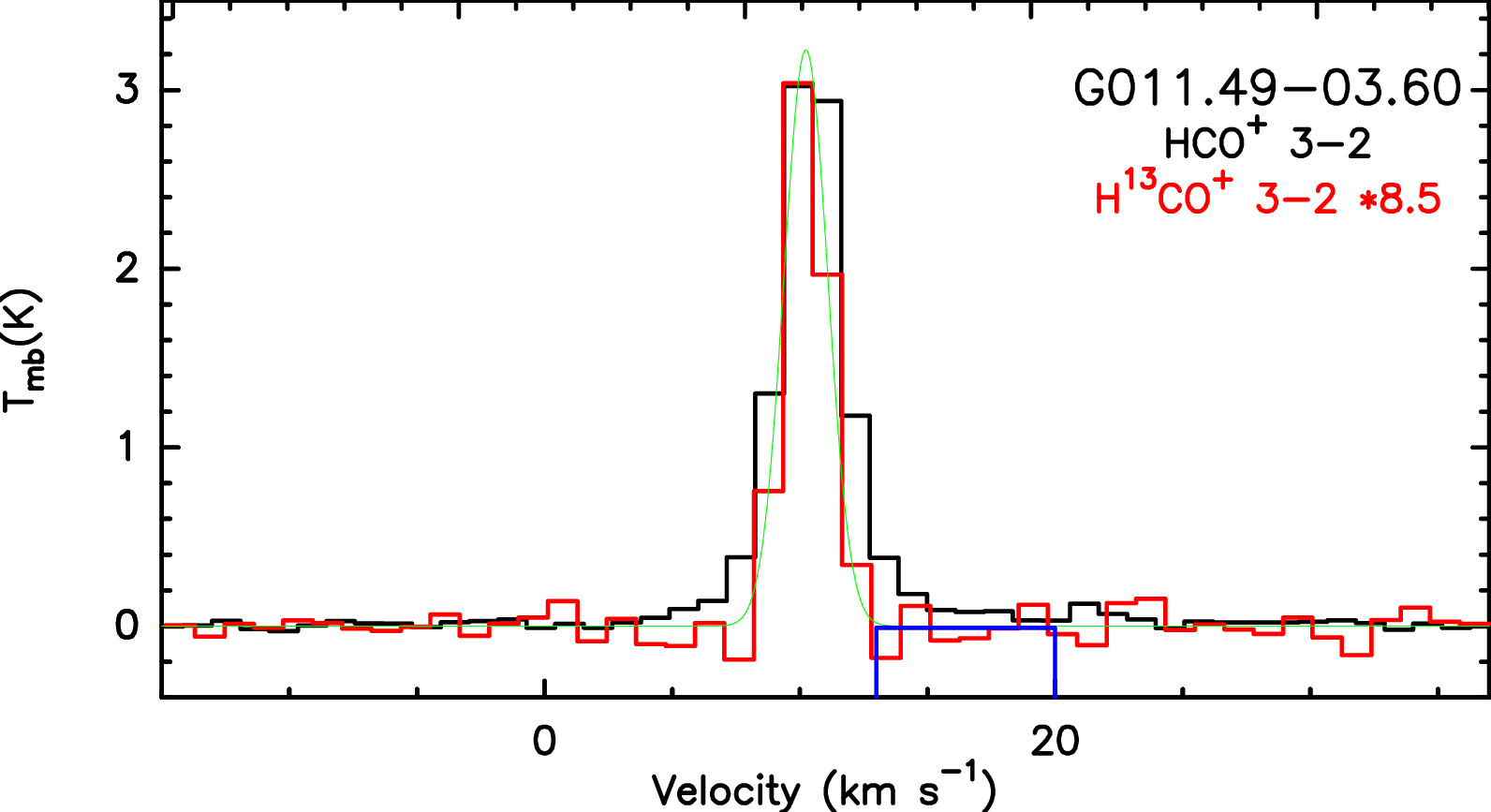} 
\includegraphics[width=0.35\columnwidth]{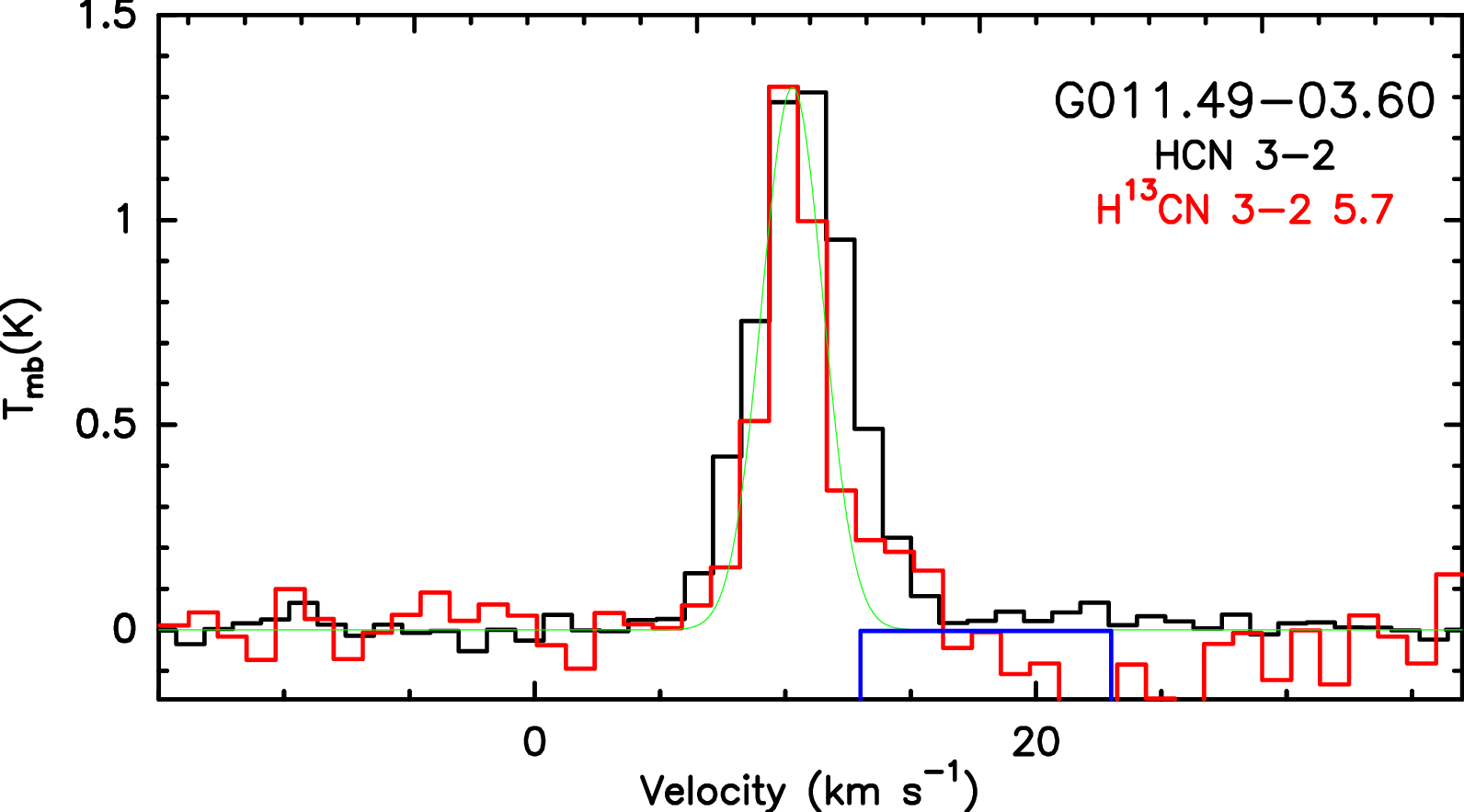} 
\addtocounter{figure}{0}
\caption{Continued.}
\label{fig1}
\end{figure}

\begin{figure*}
\addtocounter{figure}{-1}
\includegraphics[width=0.35\columnwidth]{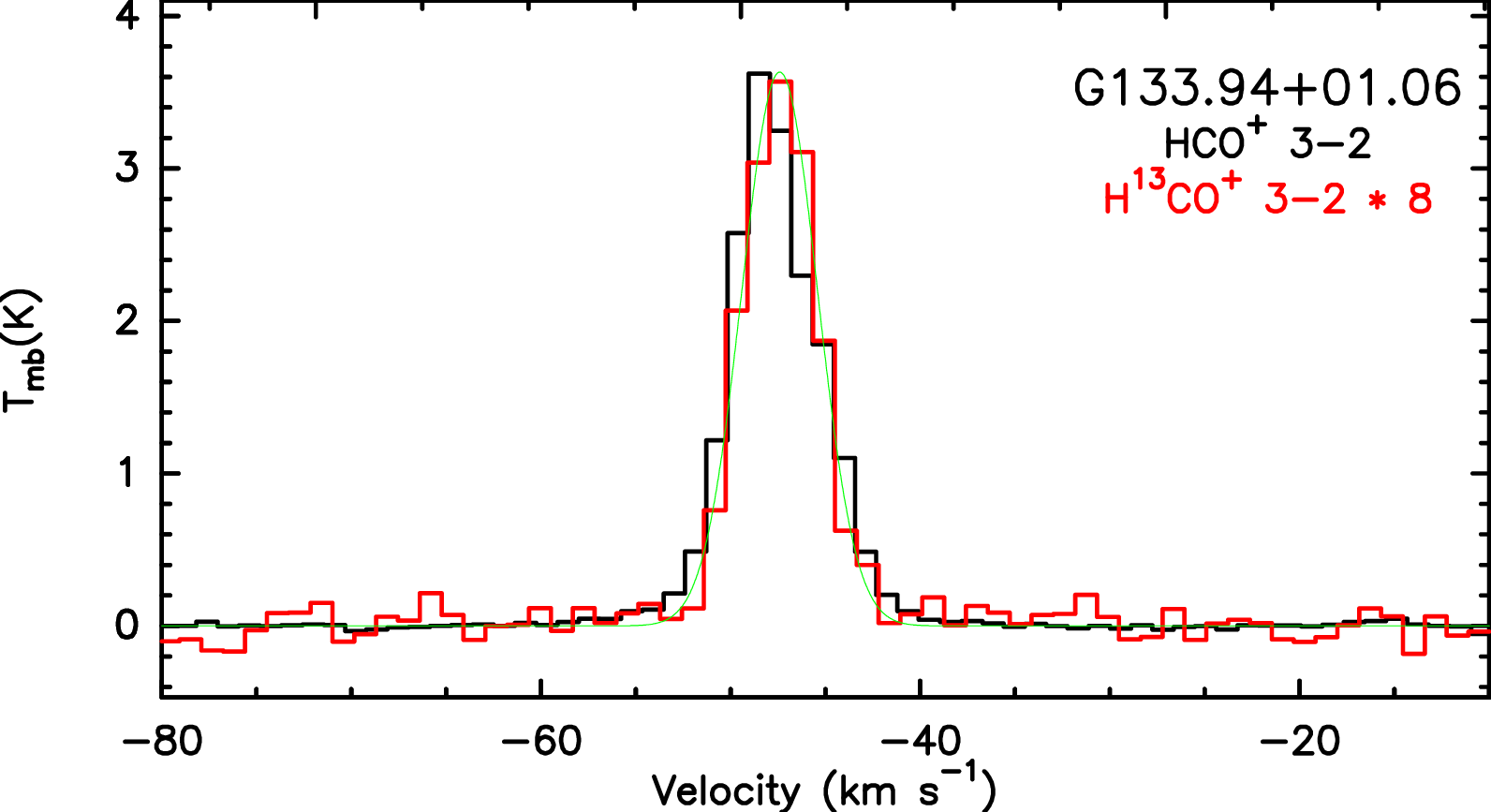} 
\includegraphics[width=0.35\columnwidth]{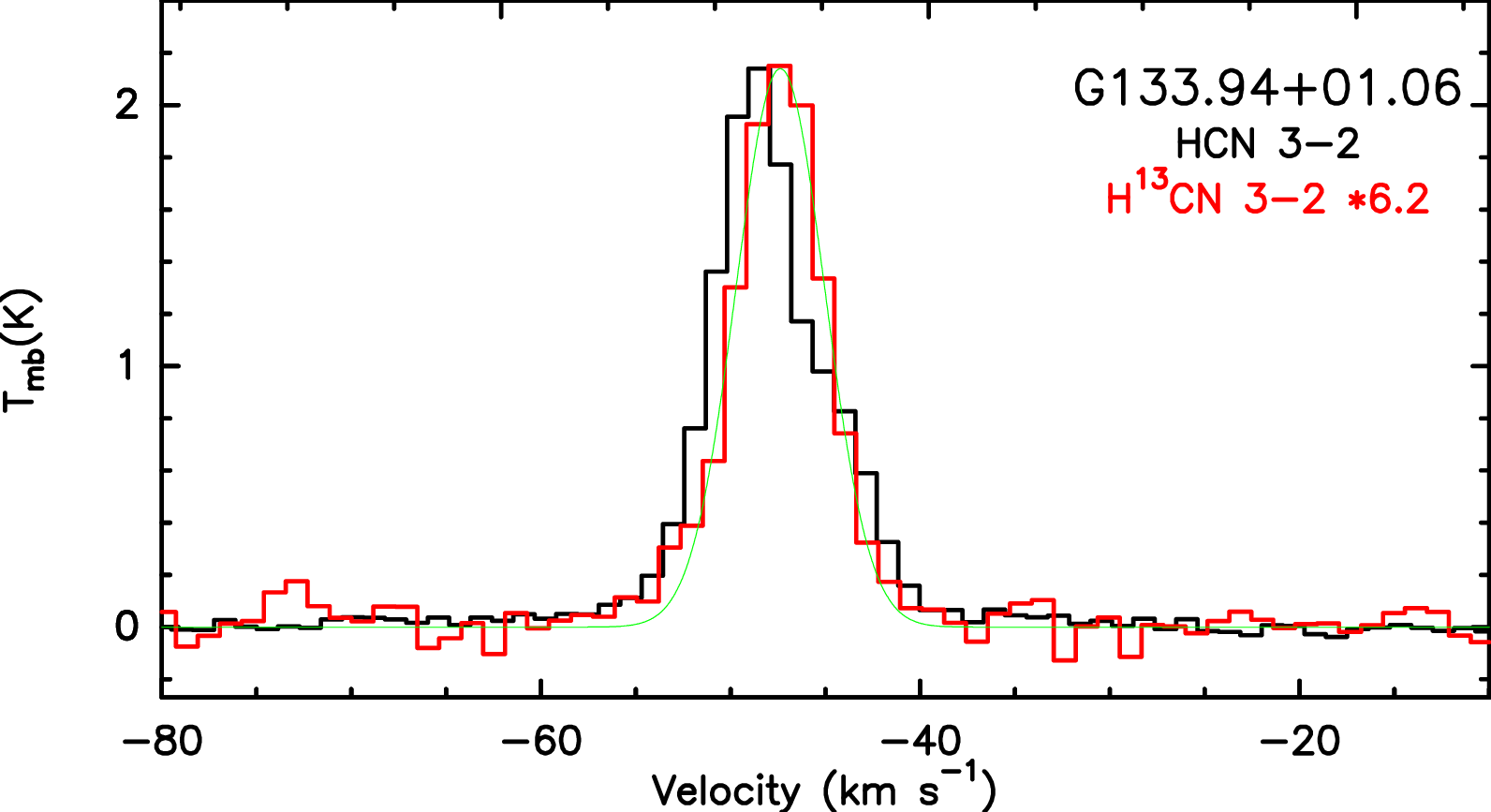} 
\includegraphics[width=0.35\columnwidth]{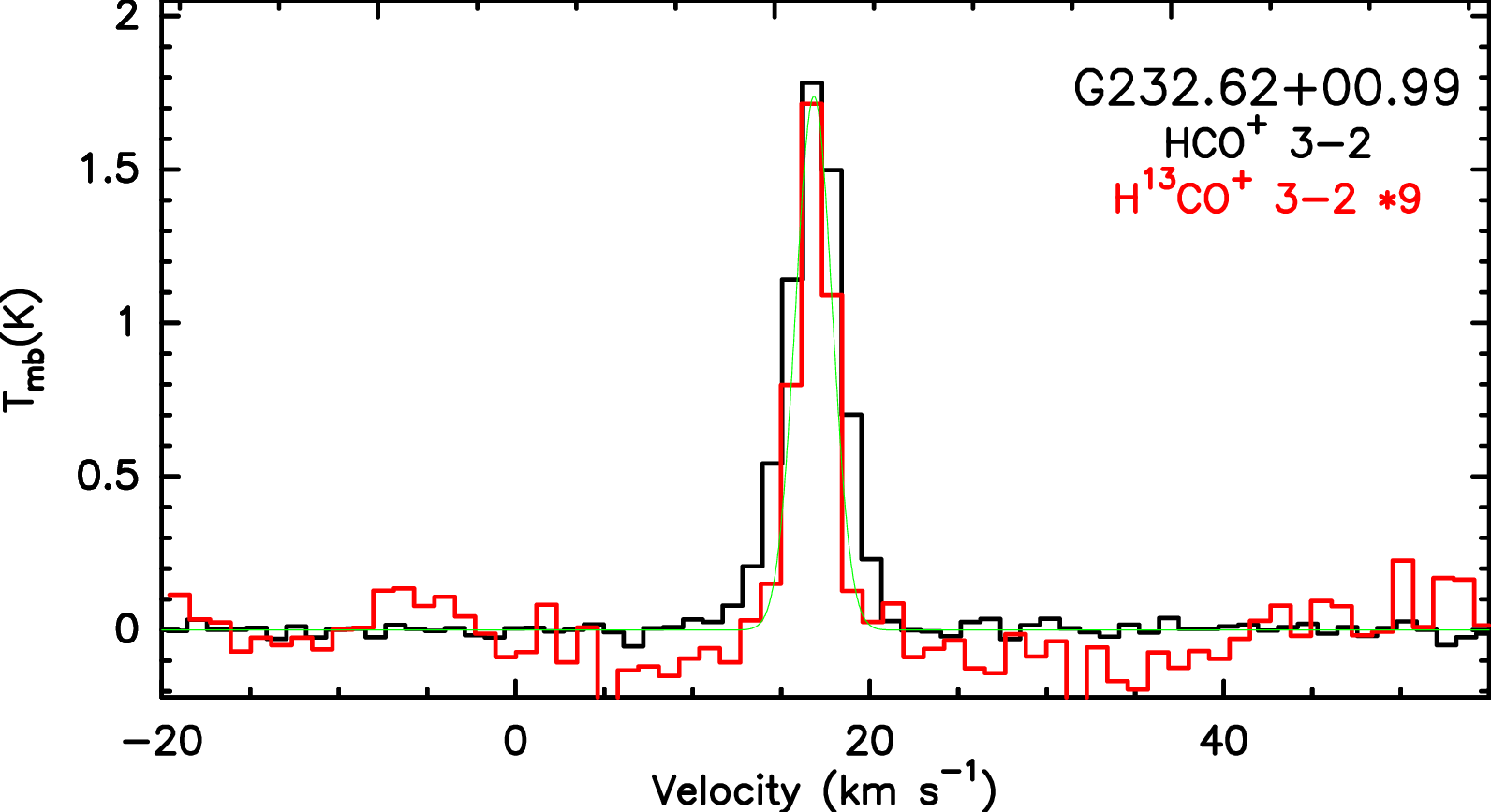} 
\includegraphics[width=0.35\columnwidth]{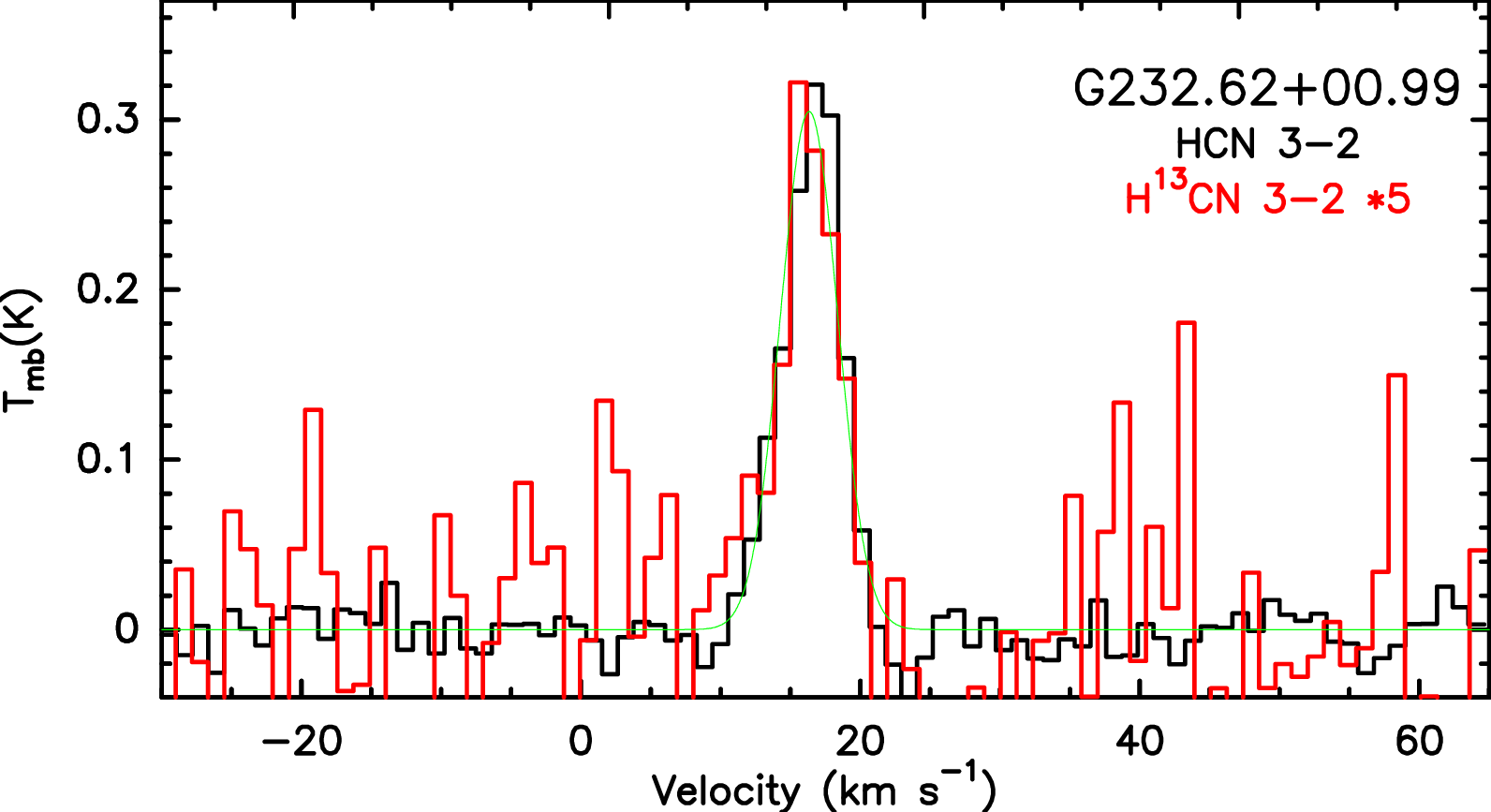} 
\includegraphics[width=0.35\columnwidth]{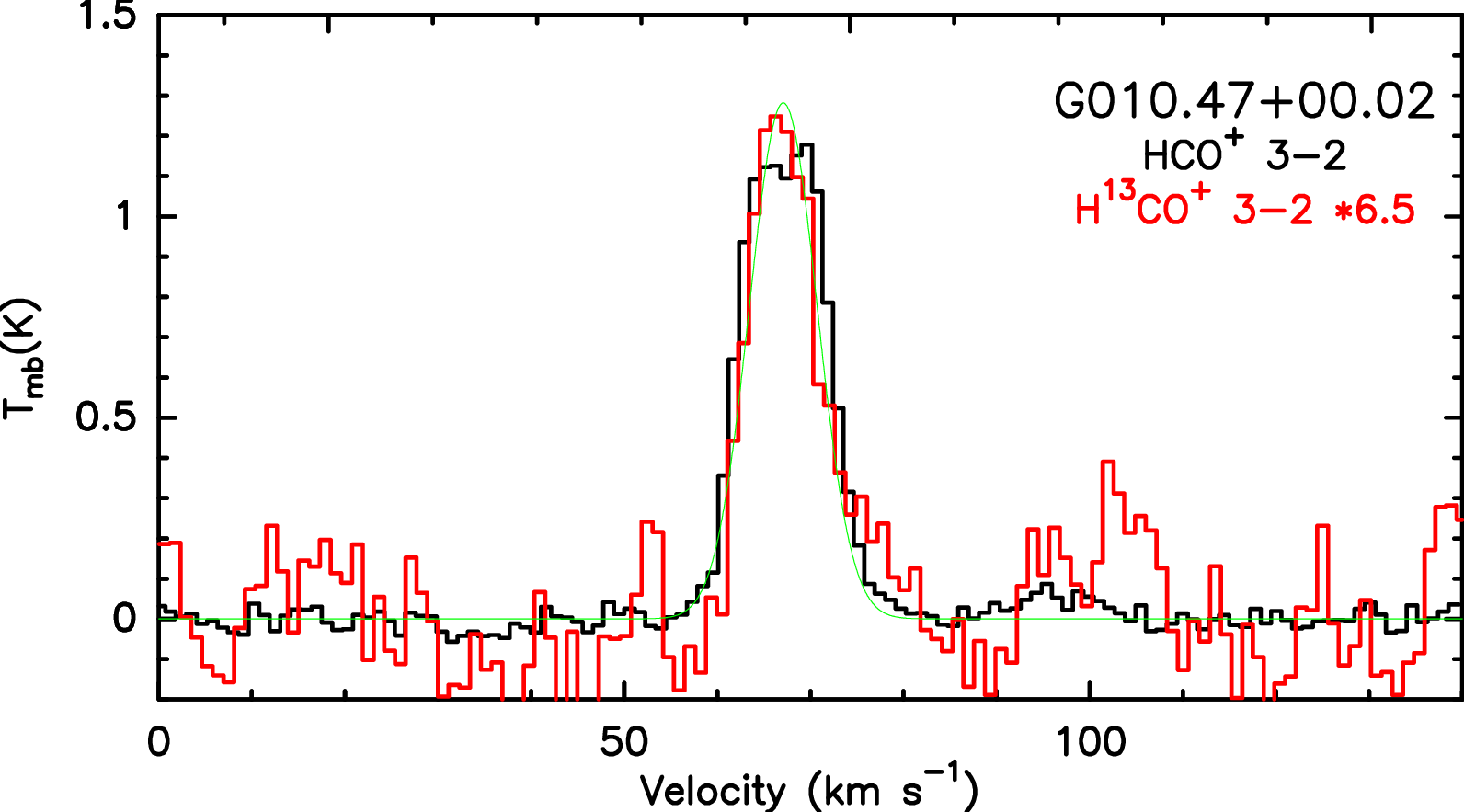} 
\includegraphics[width=0.35\columnwidth]{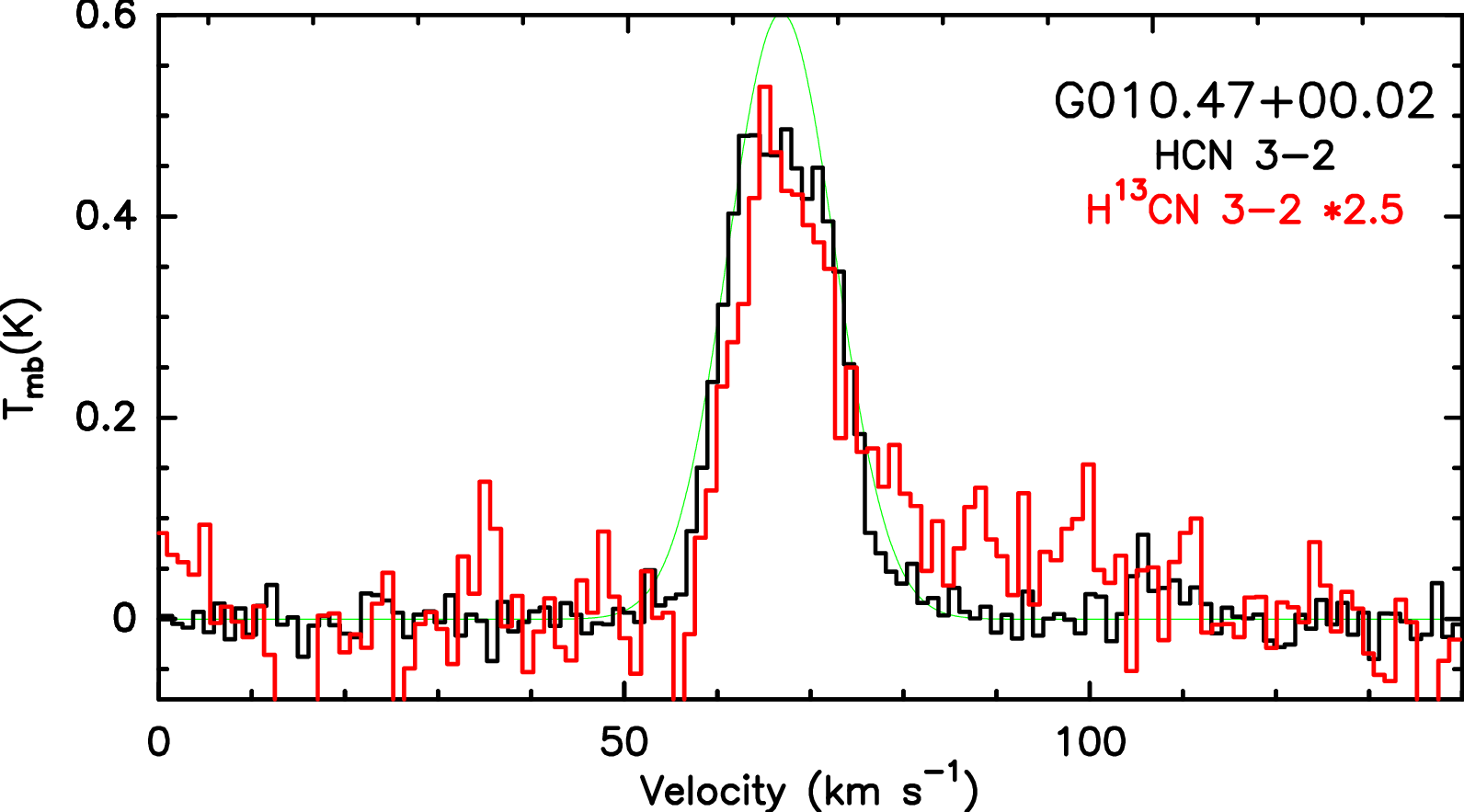} 
\includegraphics[width=0.35\columnwidth]{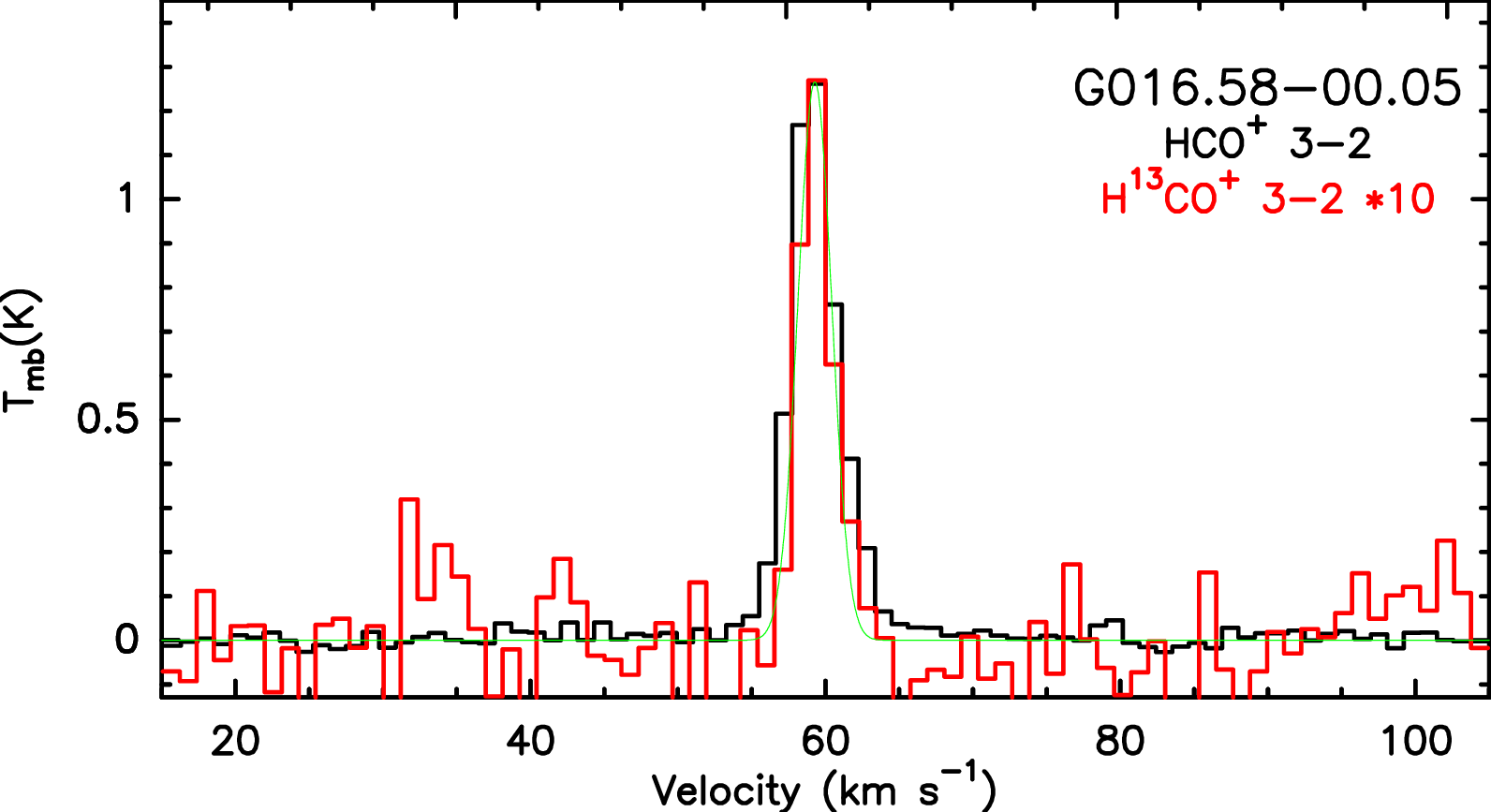} 
\includegraphics[width=0.35\columnwidth]{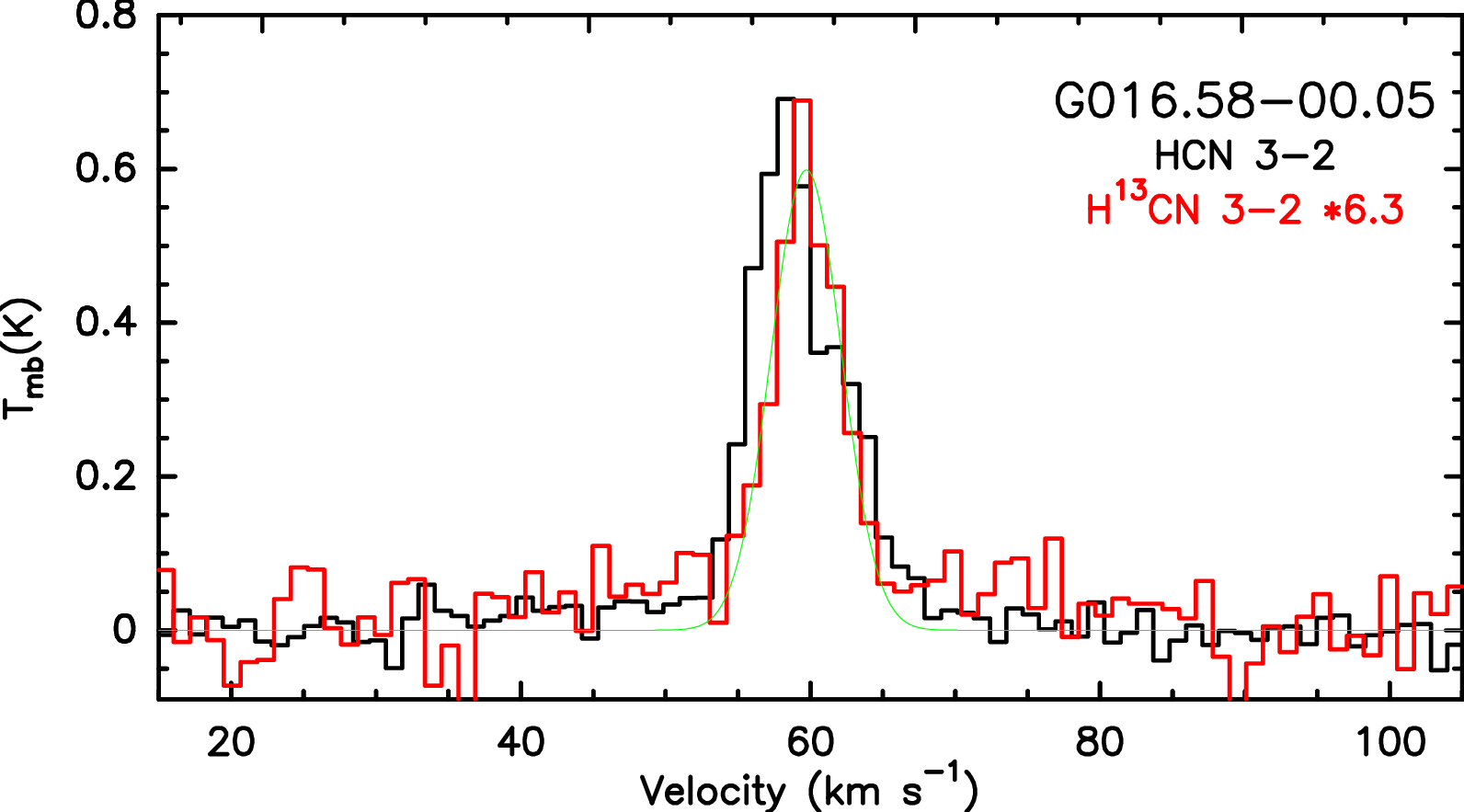} 
\includegraphics[width=0.35\columnwidth]{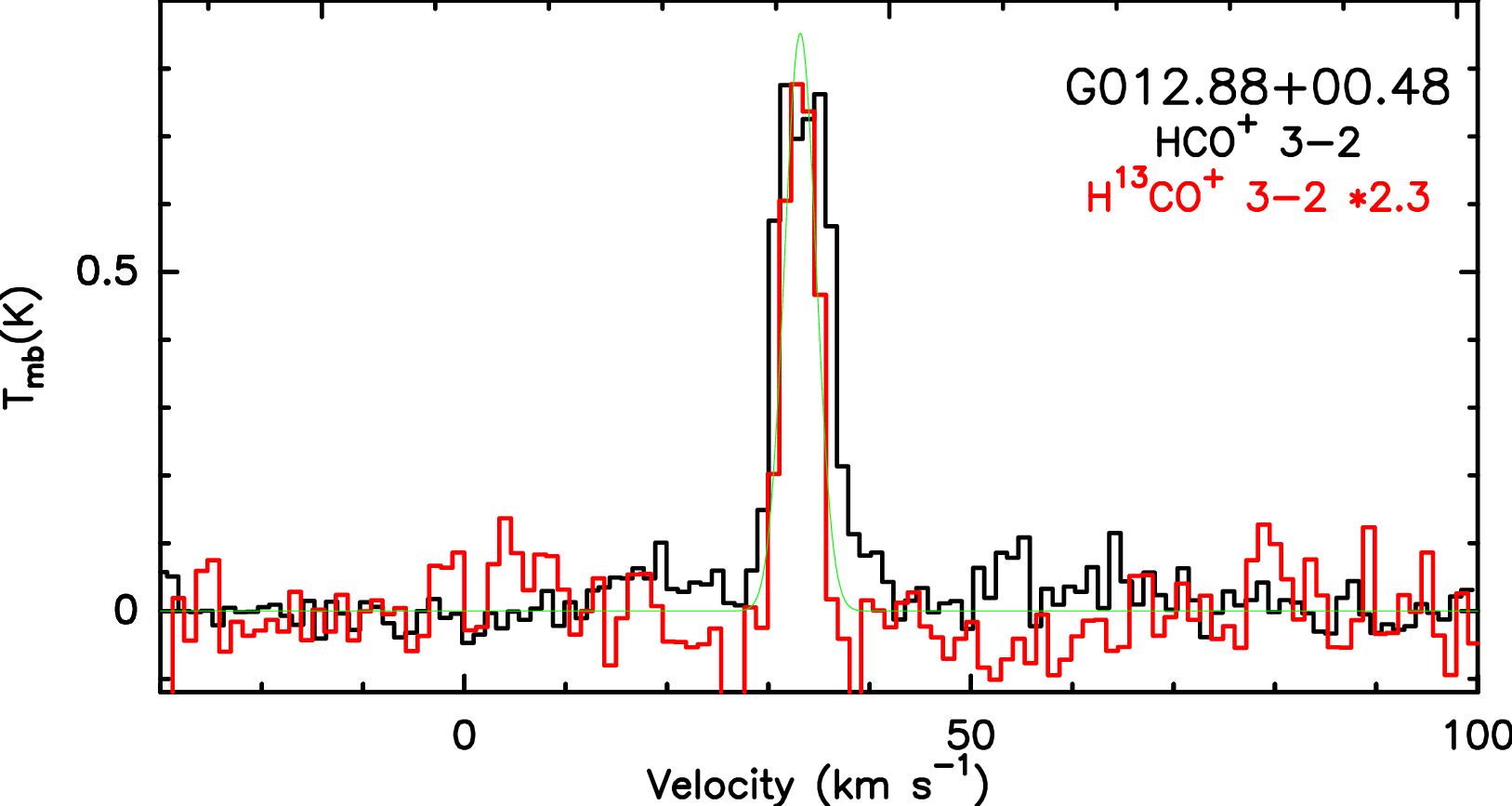} 
\includegraphics[width=0.35\columnwidth]{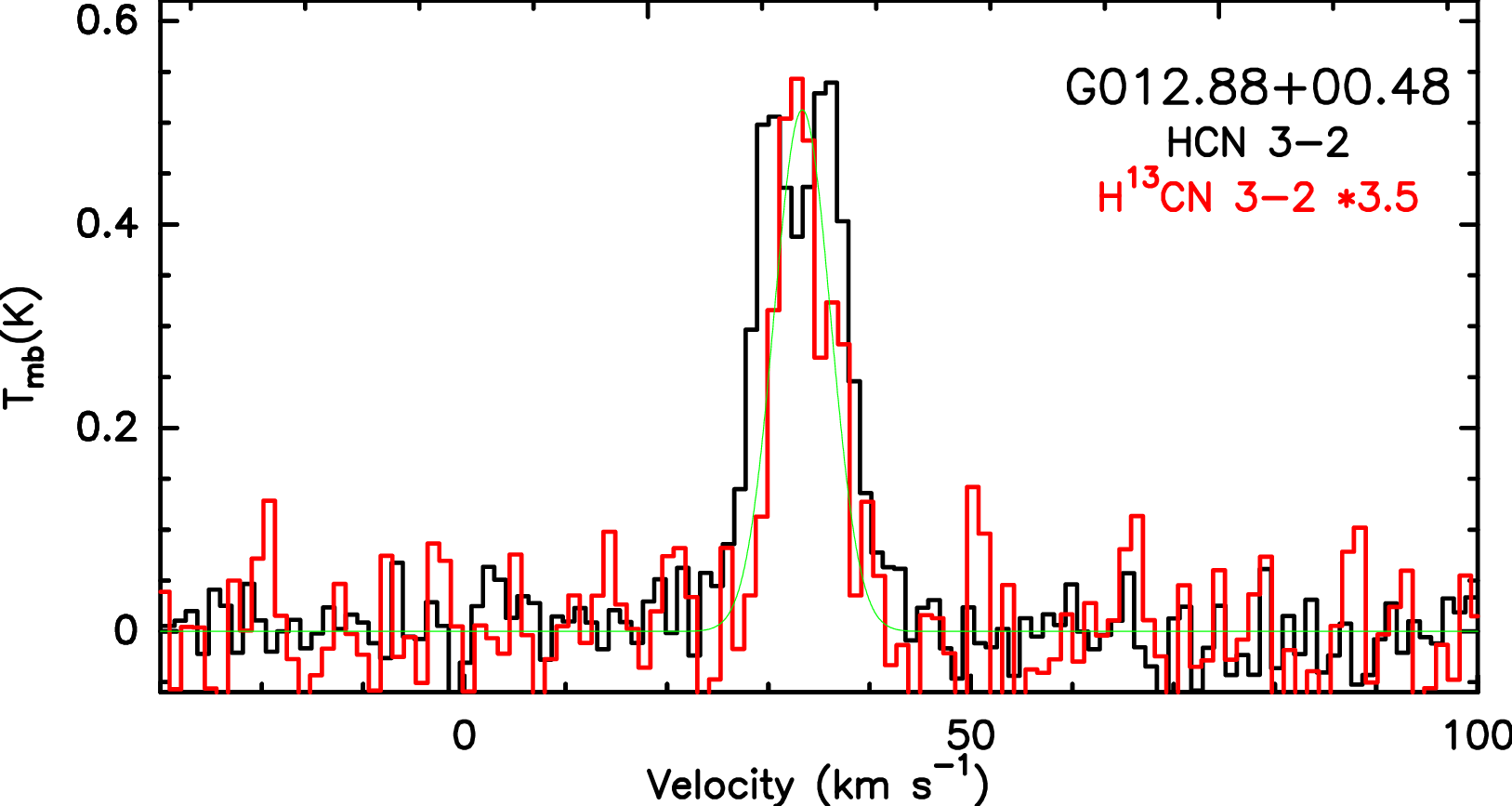} 
\includegraphics[width=0.35\columnwidth]{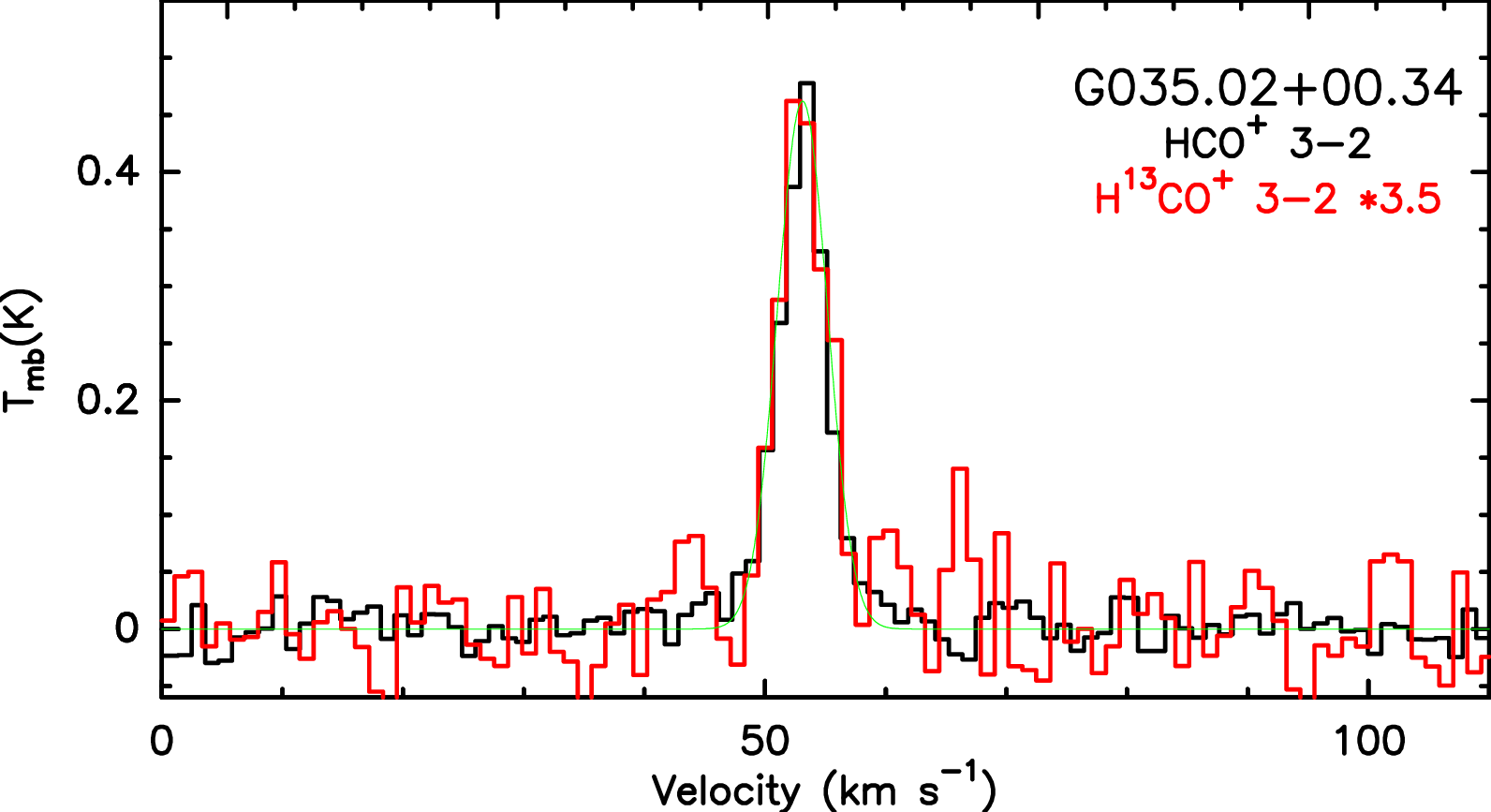} 
\includegraphics[width=0.35\columnwidth]{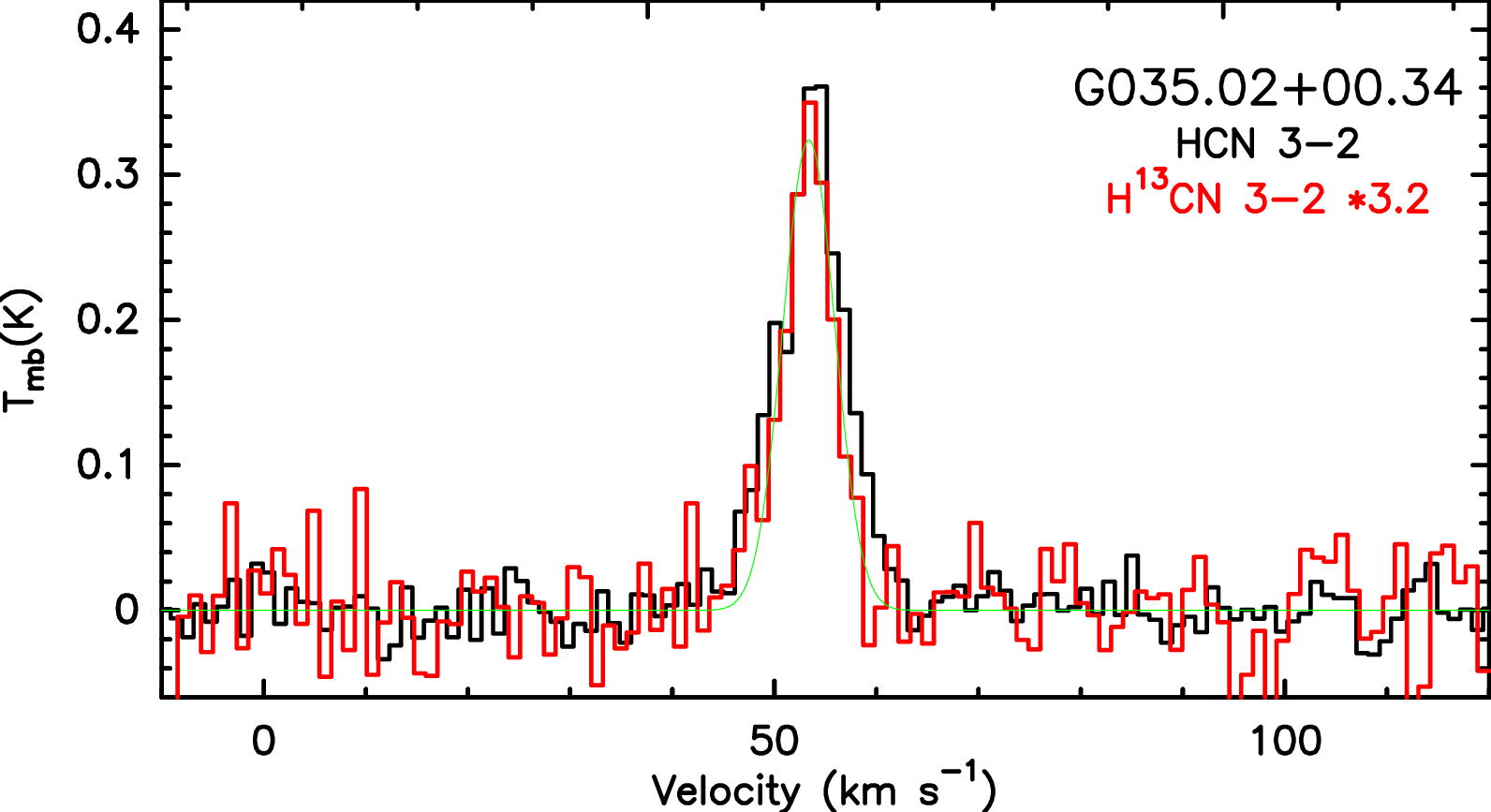} 
\includegraphics[width=0.35\columnwidth]{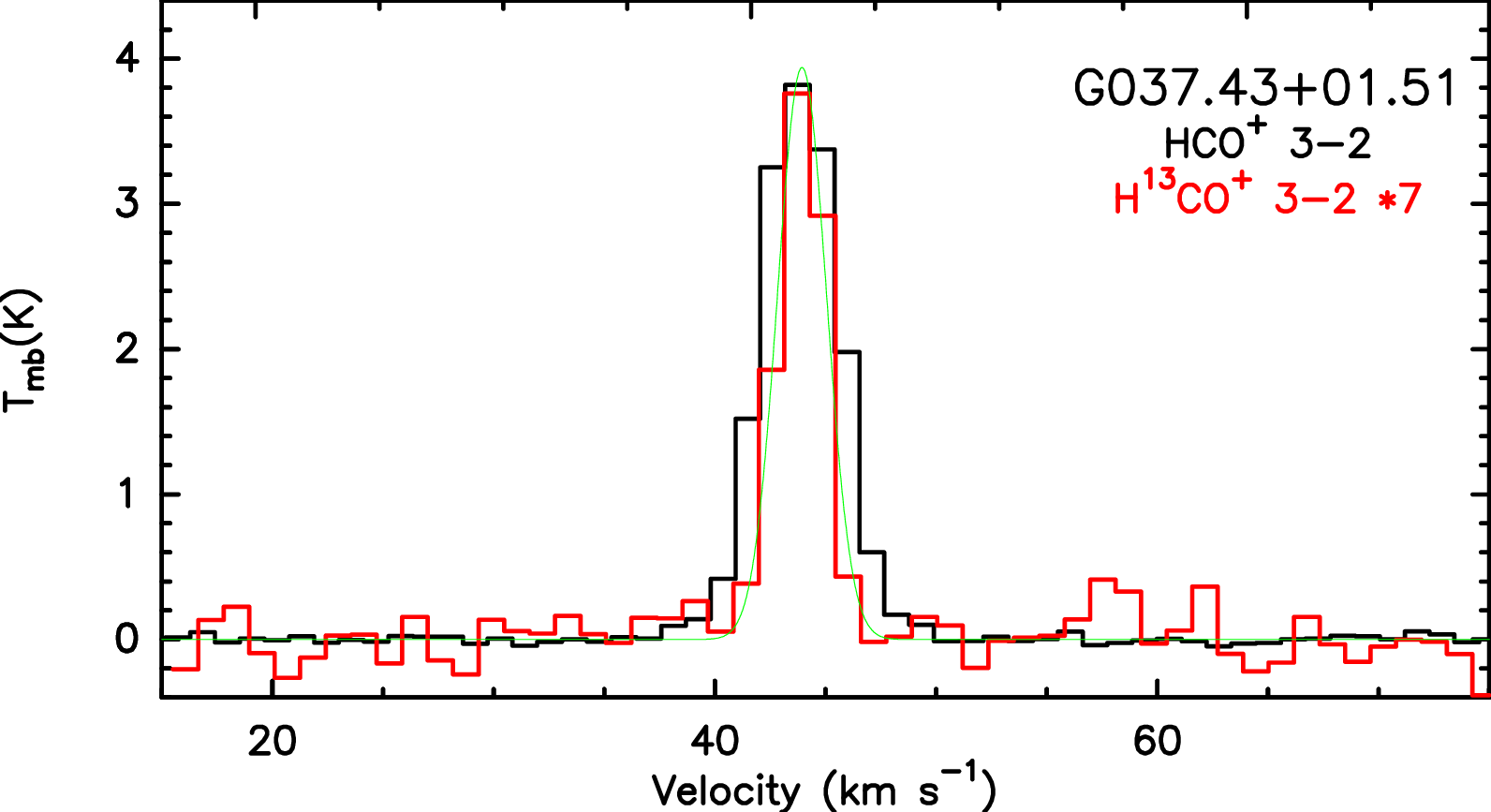} 
\includegraphics[width=0.35\columnwidth]{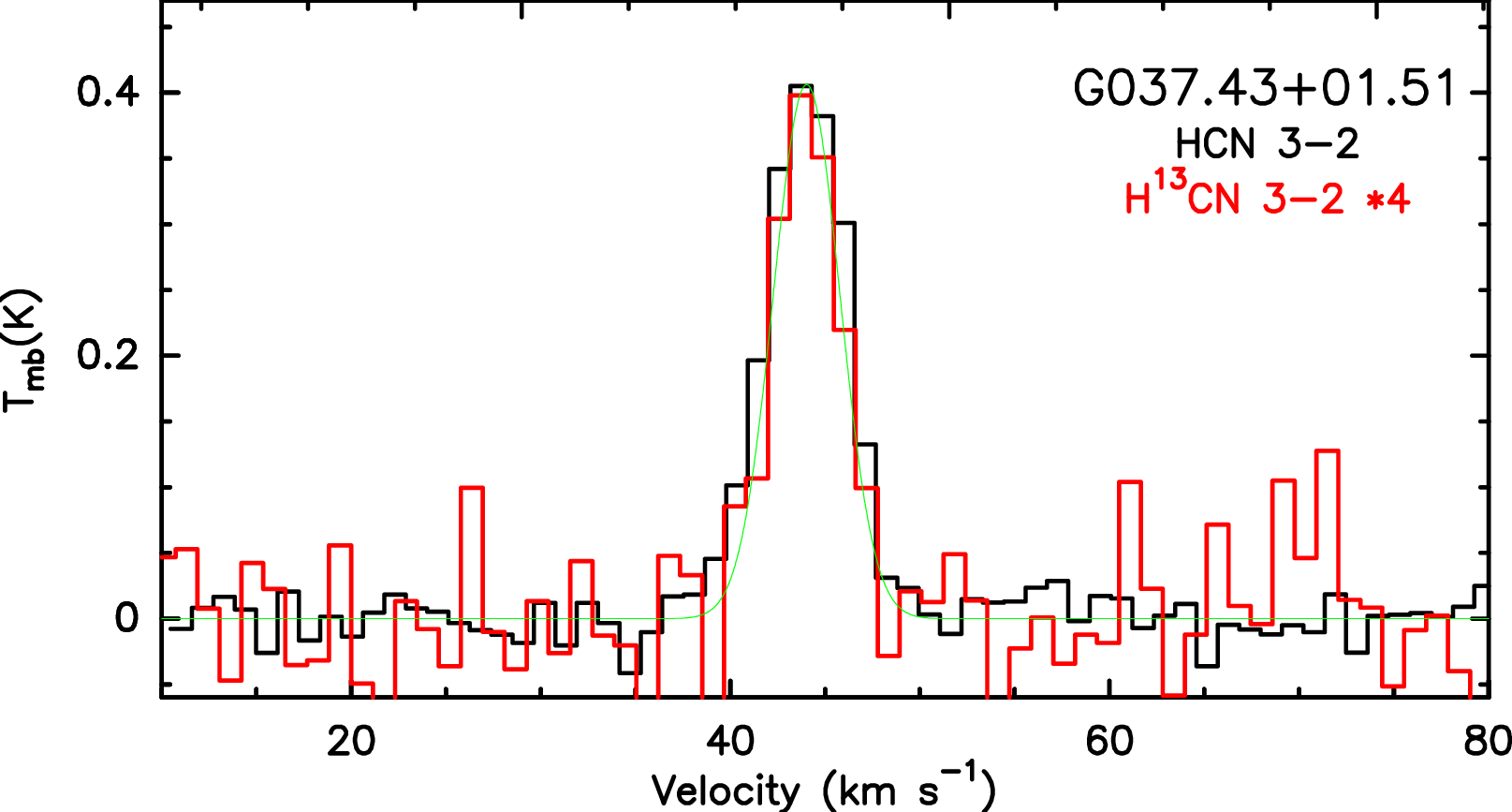}
\includegraphics[width=0.35\columnwidth]{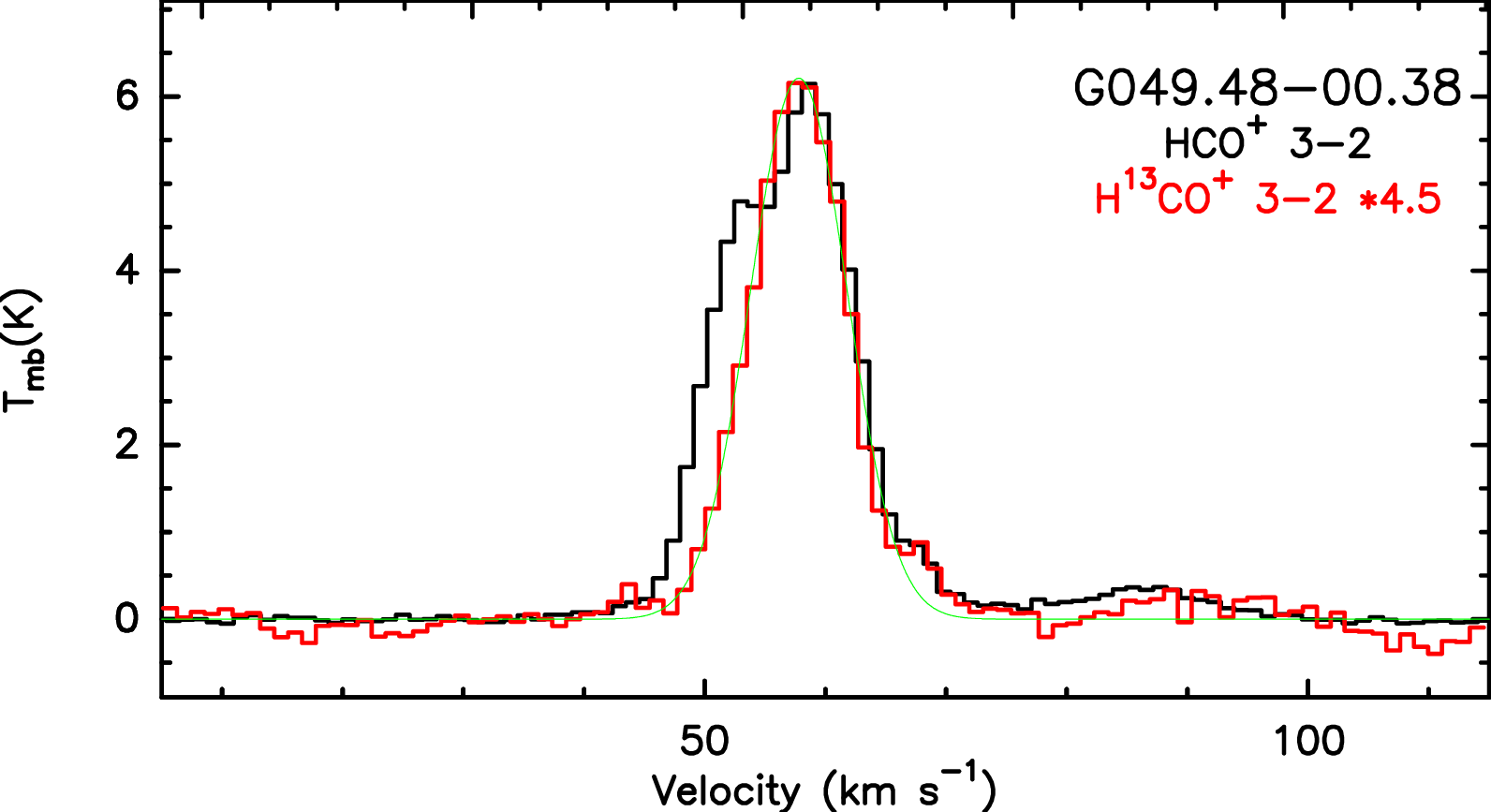} 
\includegraphics[width=0.35\columnwidth]{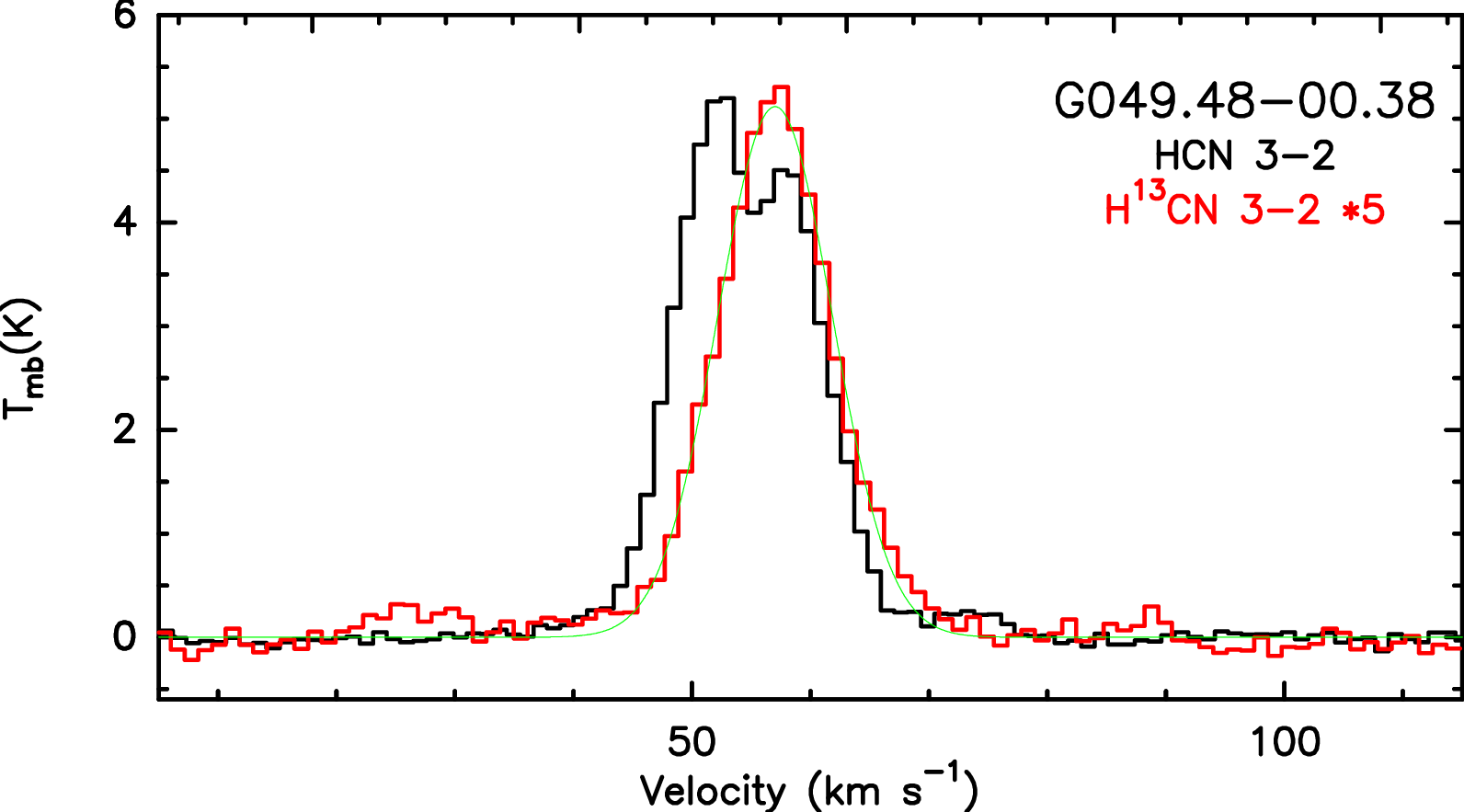}
\addtocounter{figure}{0}
\caption{Continued.}
\label{fig1}
\end{figure*}

\begin{figure*}[htbp]
\includegraphics[width=0.35\columnwidth]{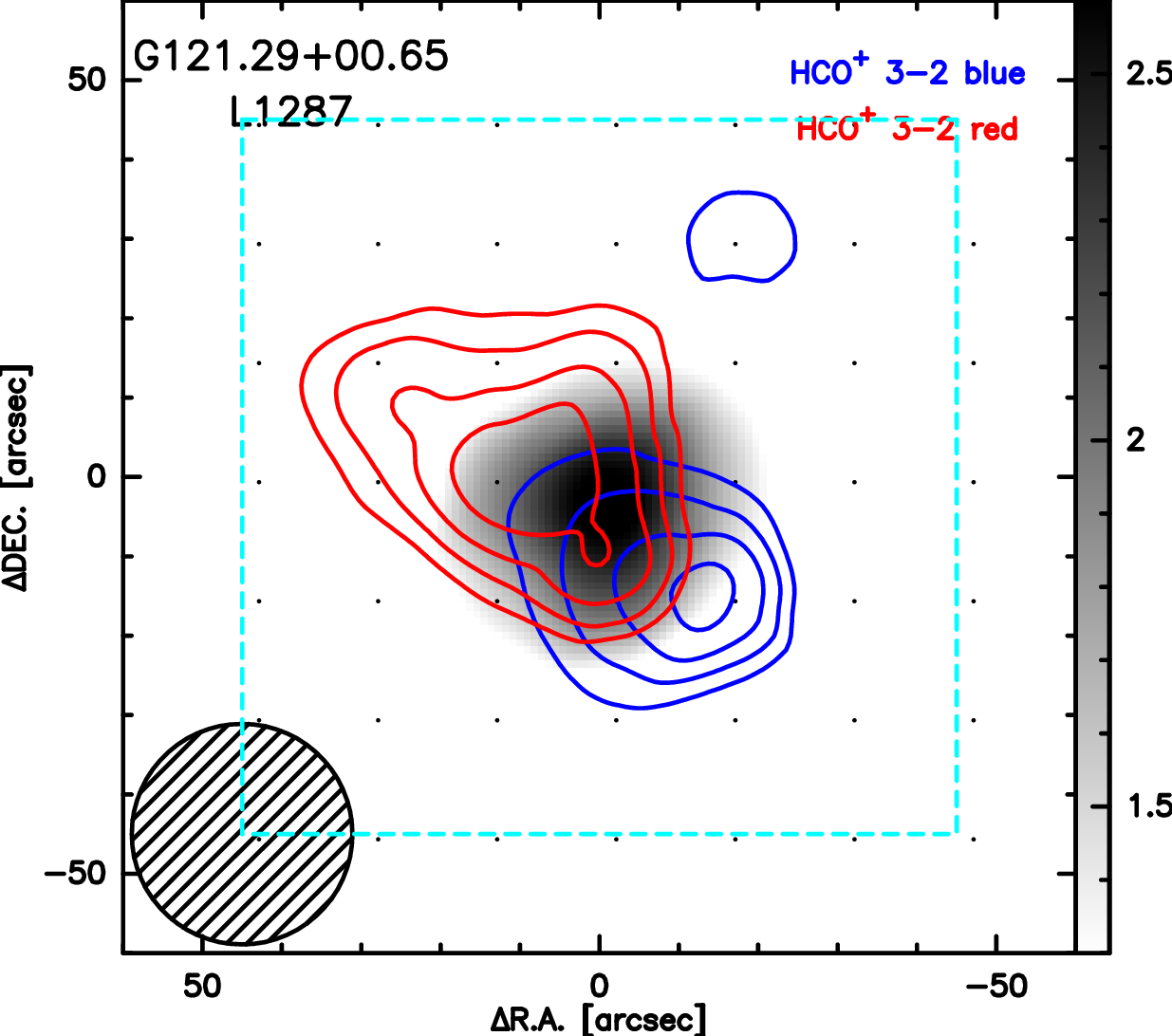} 
\includegraphics[width=0.35\columnwidth]{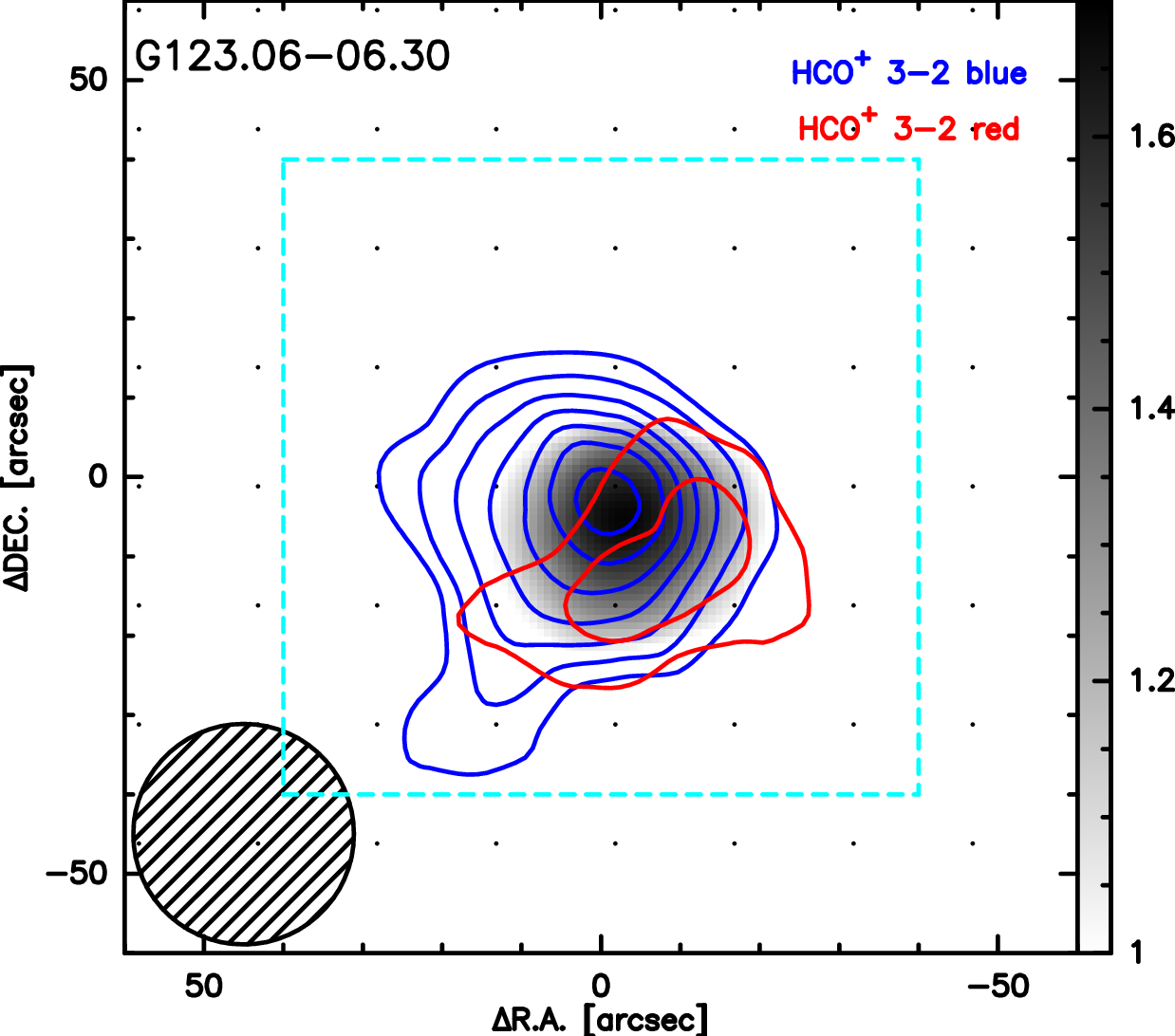} 
\includegraphics[width=0.35\columnwidth]{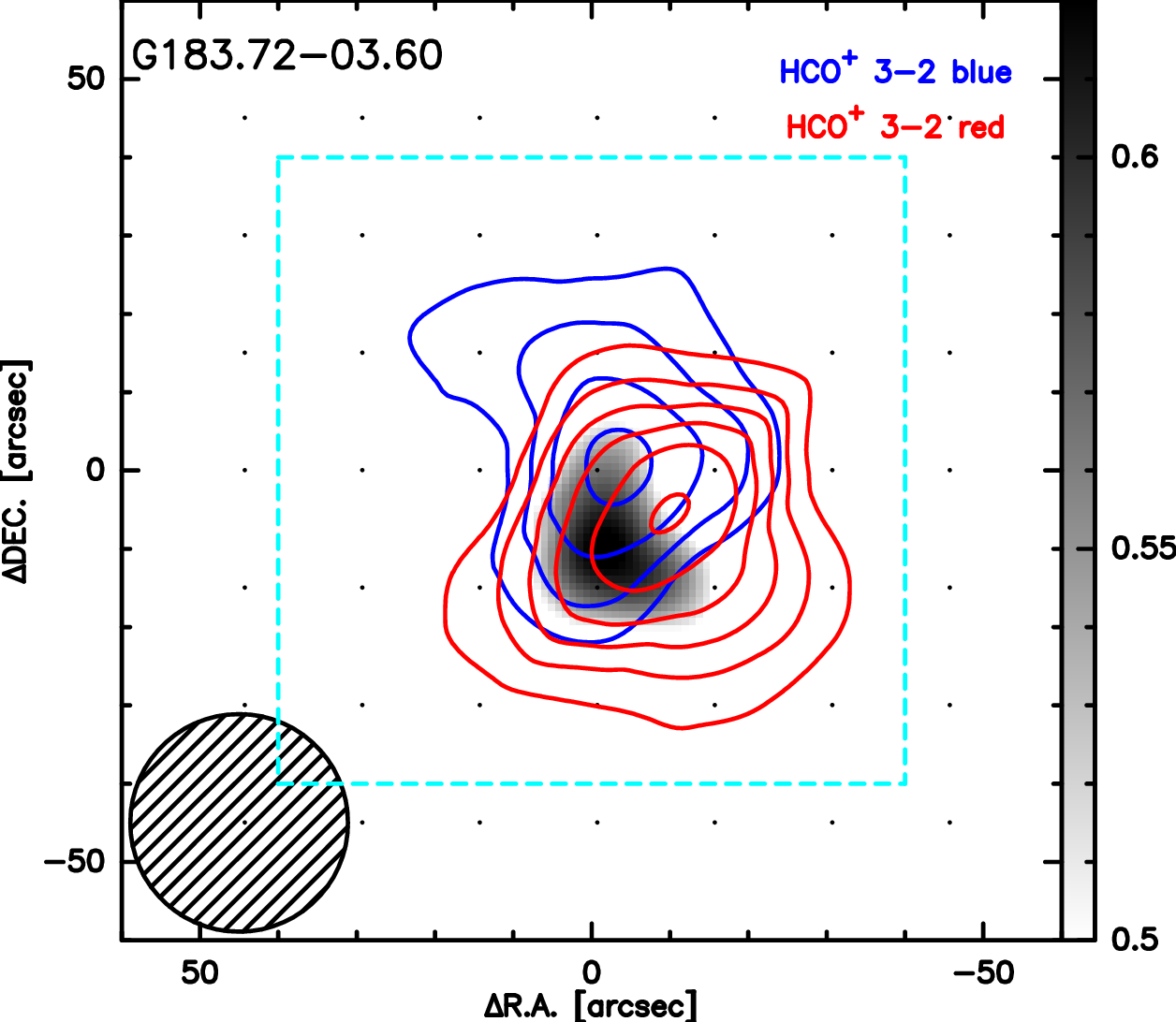} 
\includegraphics[width=0.35\columnwidth]{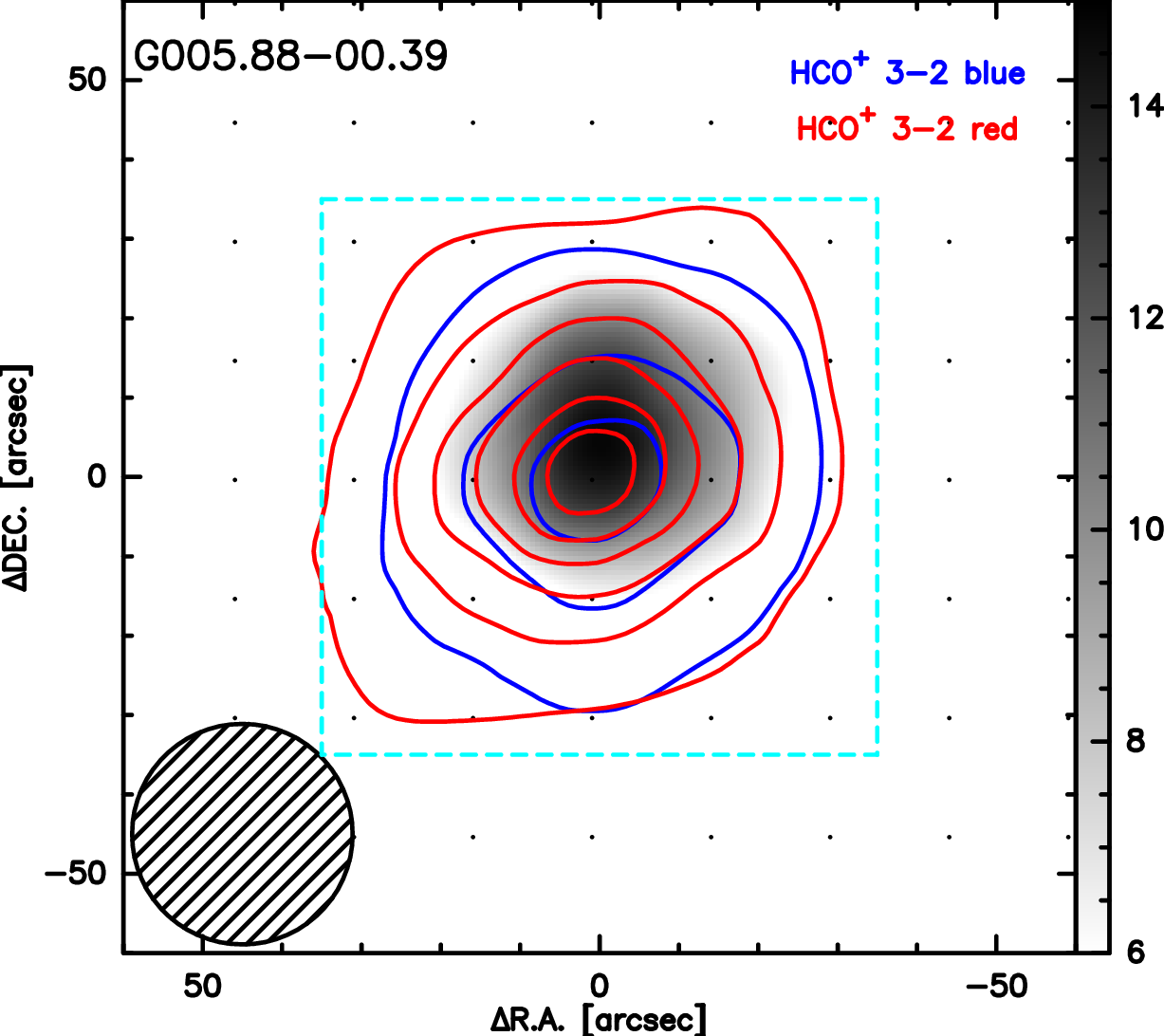} 
\includegraphics[width=0.35\columnwidth]{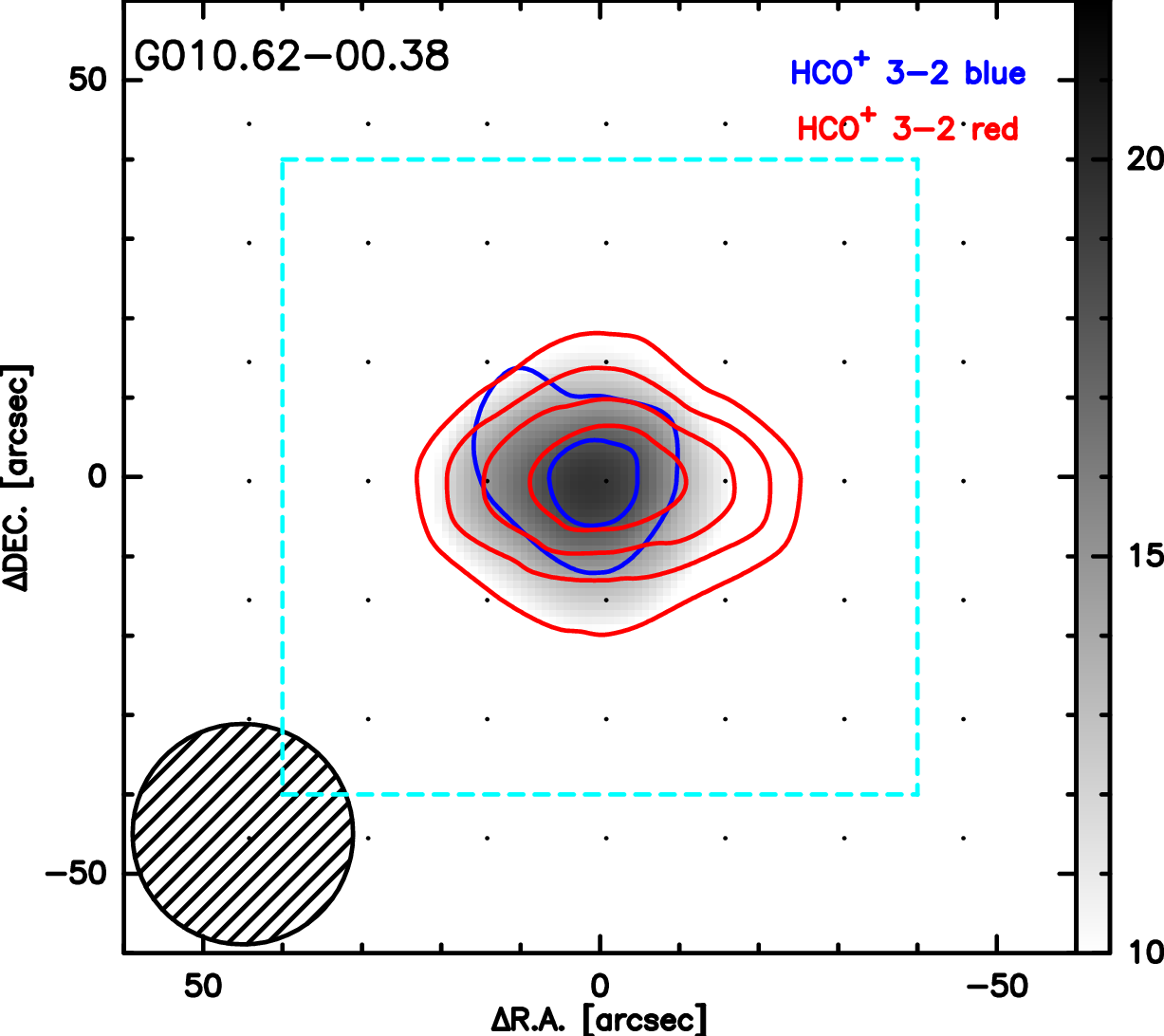} 
\includegraphics[width=0.35\columnwidth]{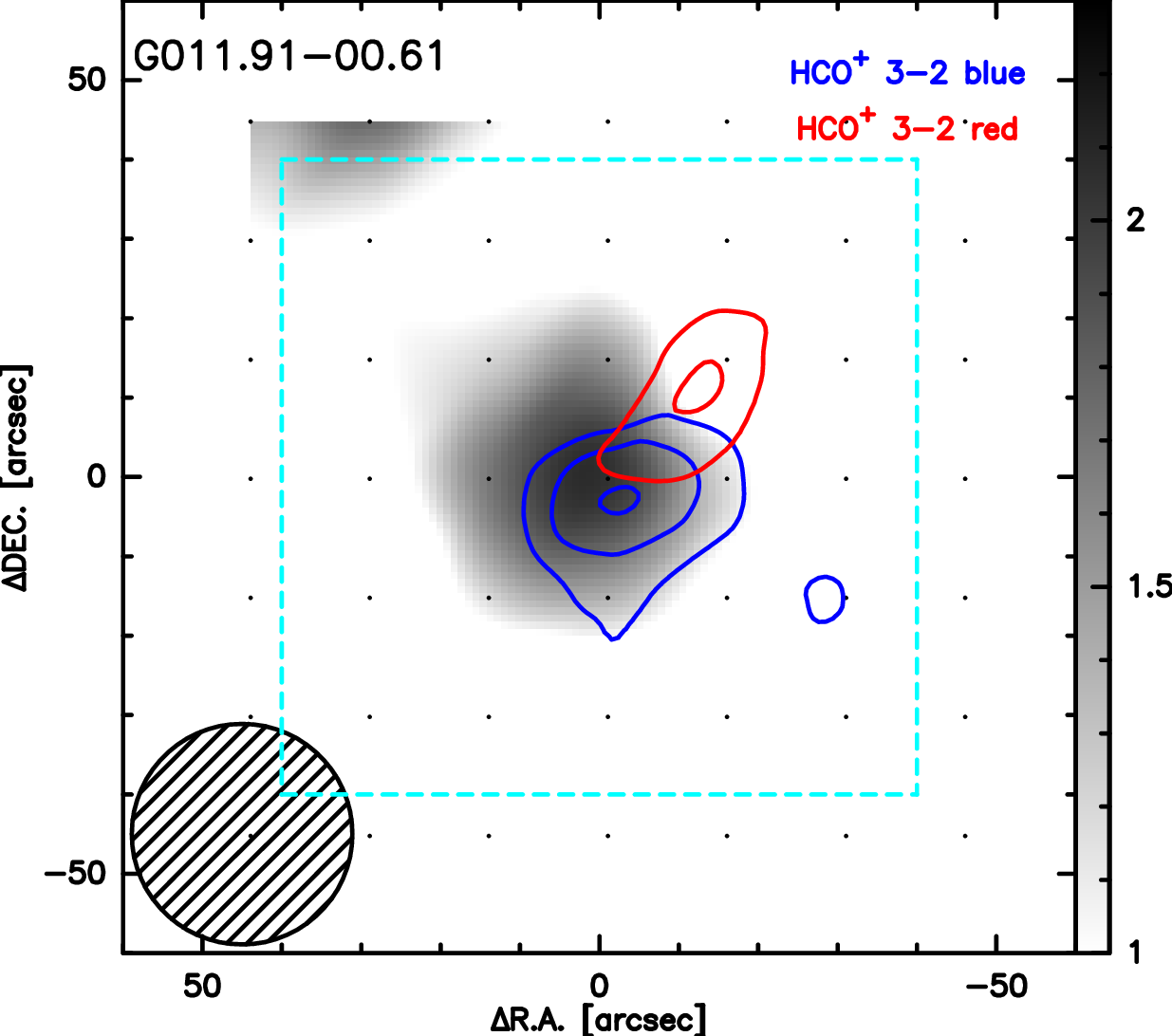} 
\includegraphics[width=0.35\columnwidth]{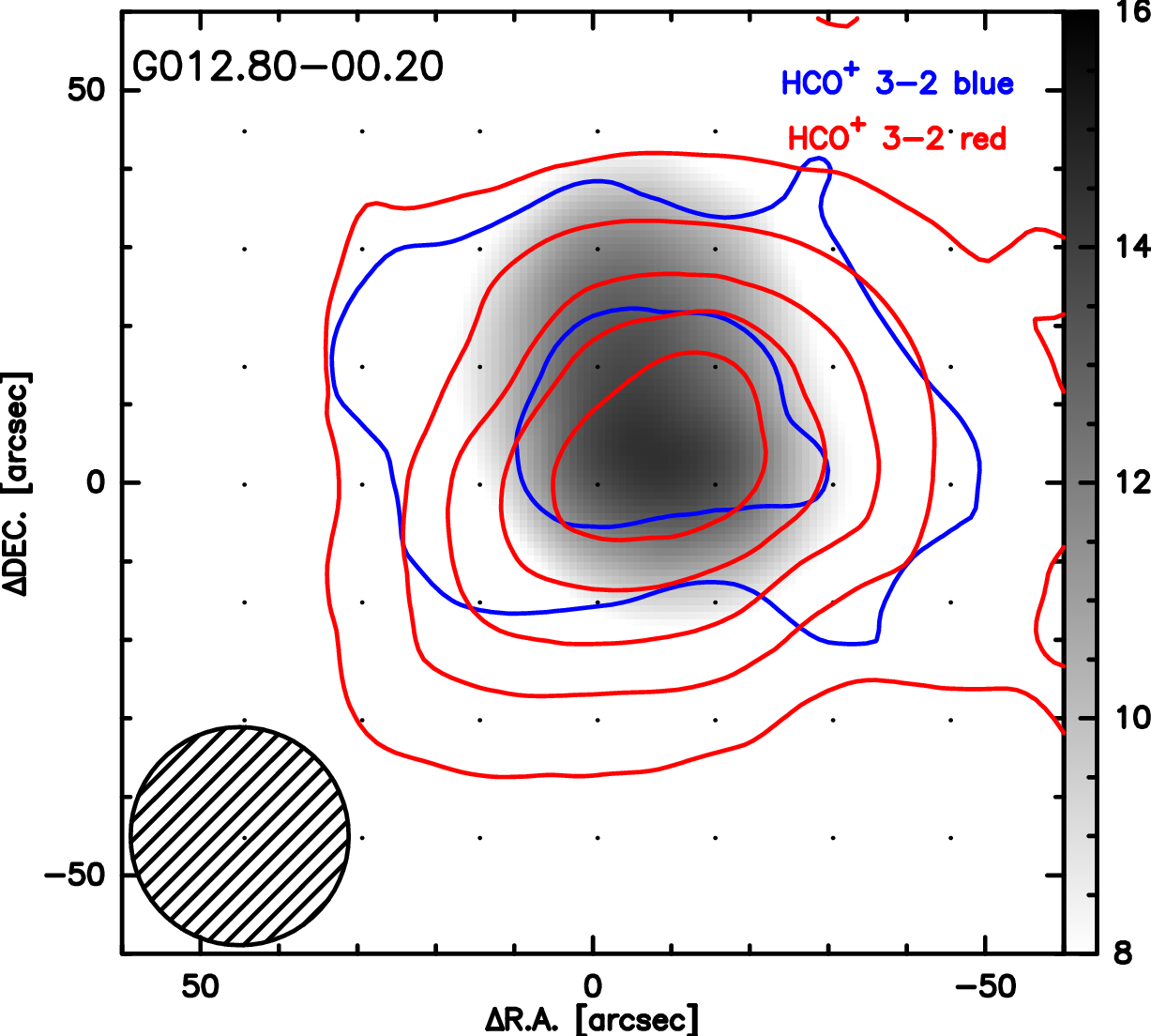} 
\includegraphics[width=0.35\columnwidth]{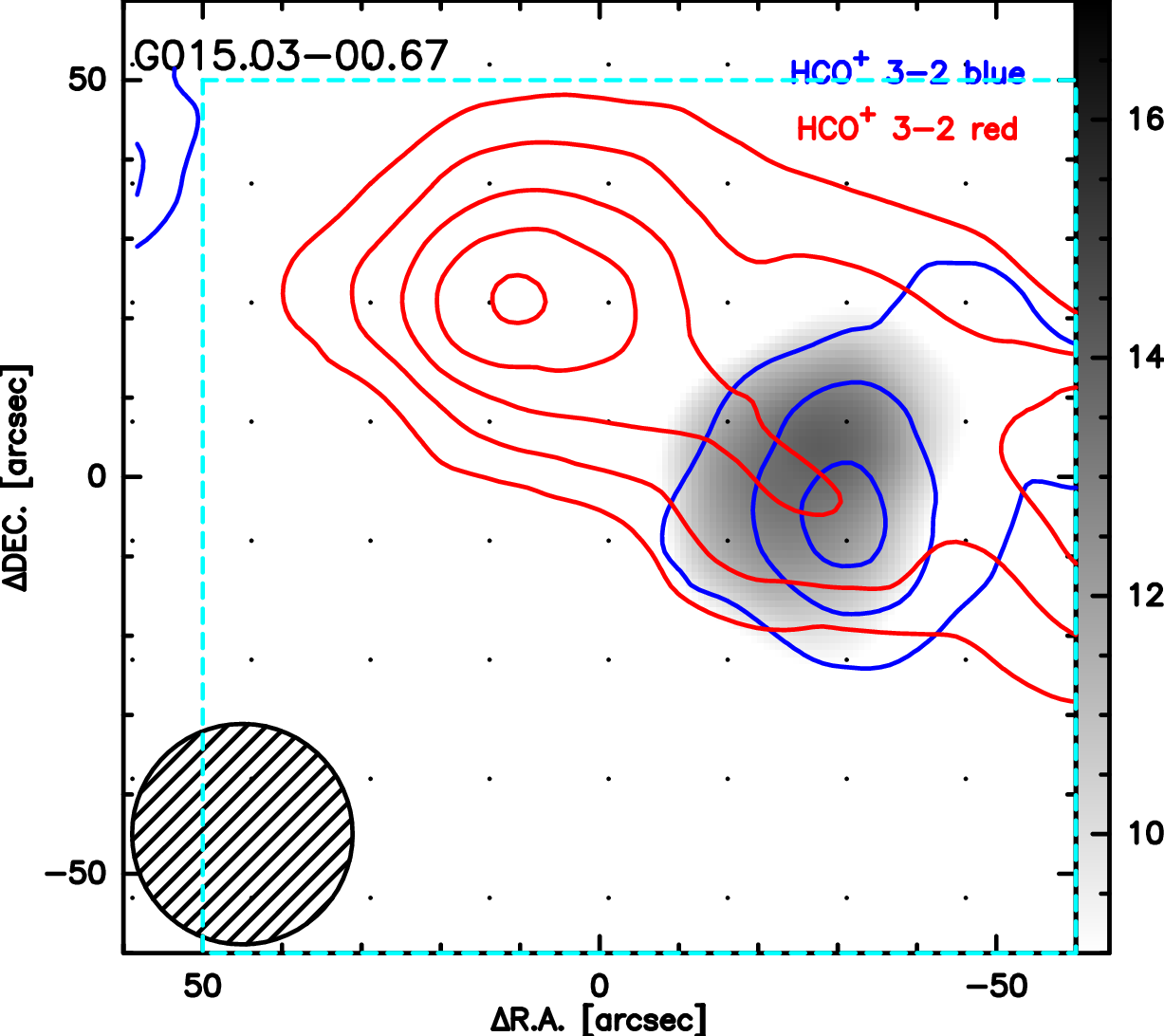} 
\includegraphics[width=0.35\columnwidth]{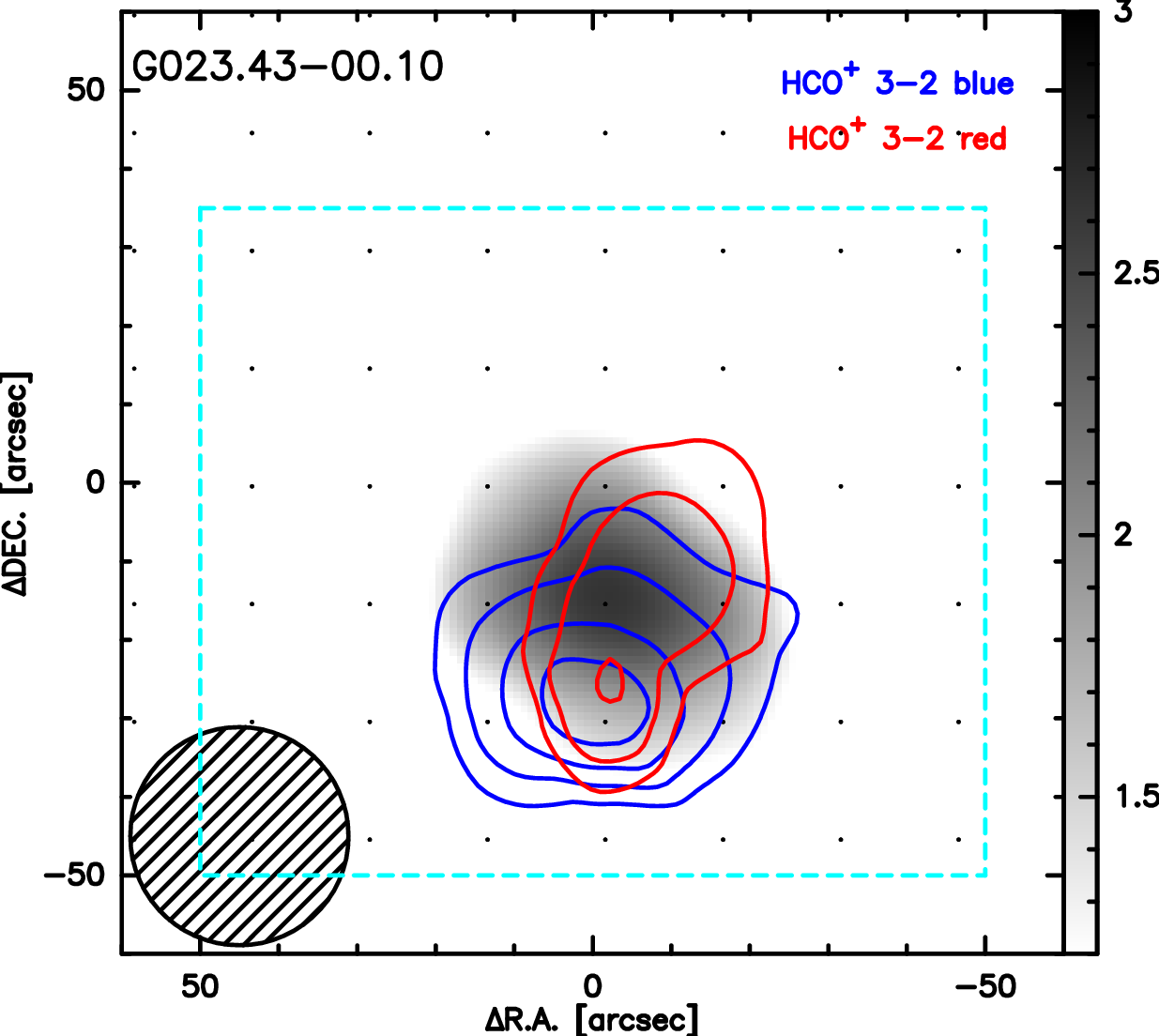} 
\includegraphics[width=0.35\columnwidth]{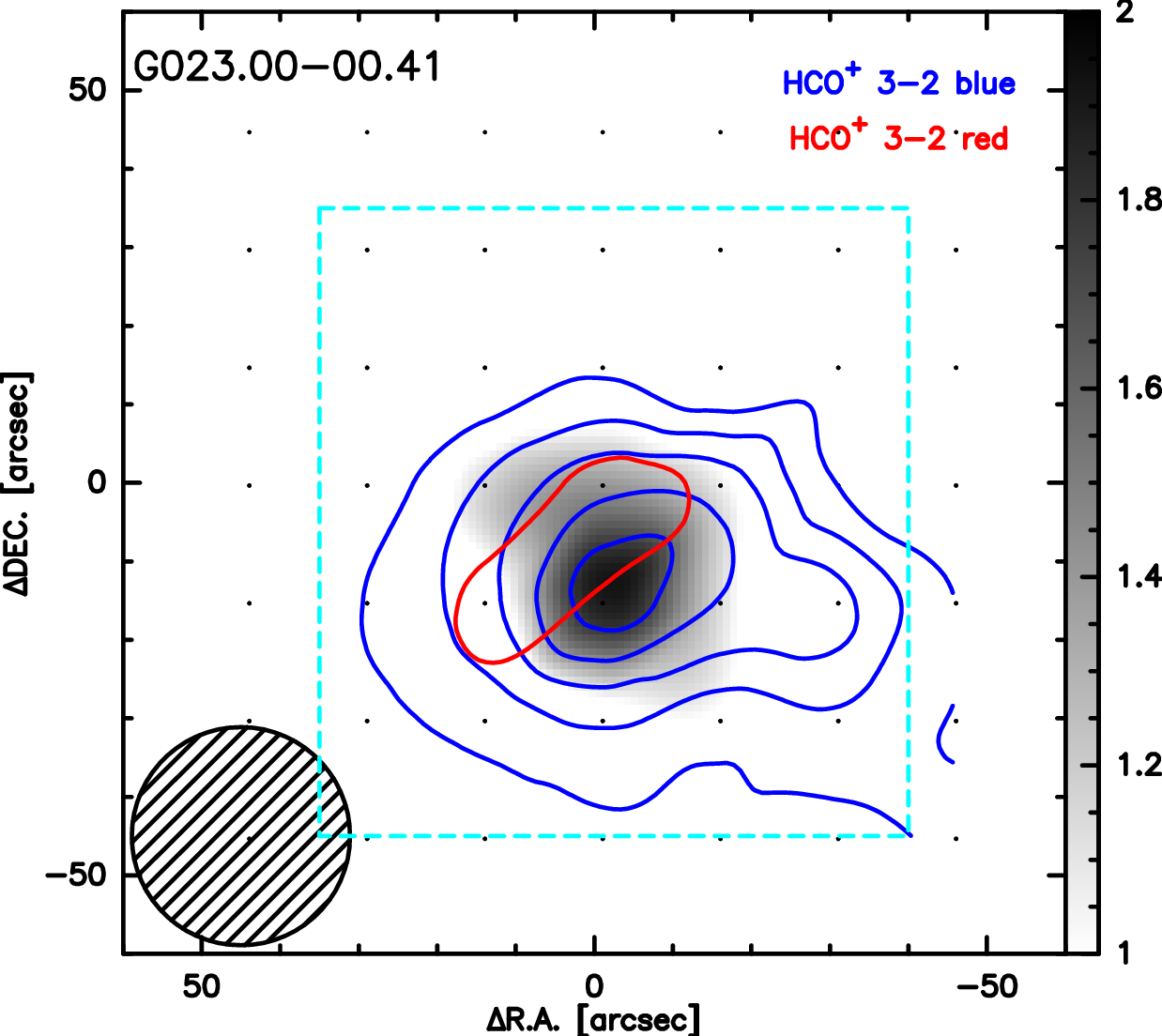} 
\includegraphics[width=0.35\columnwidth]{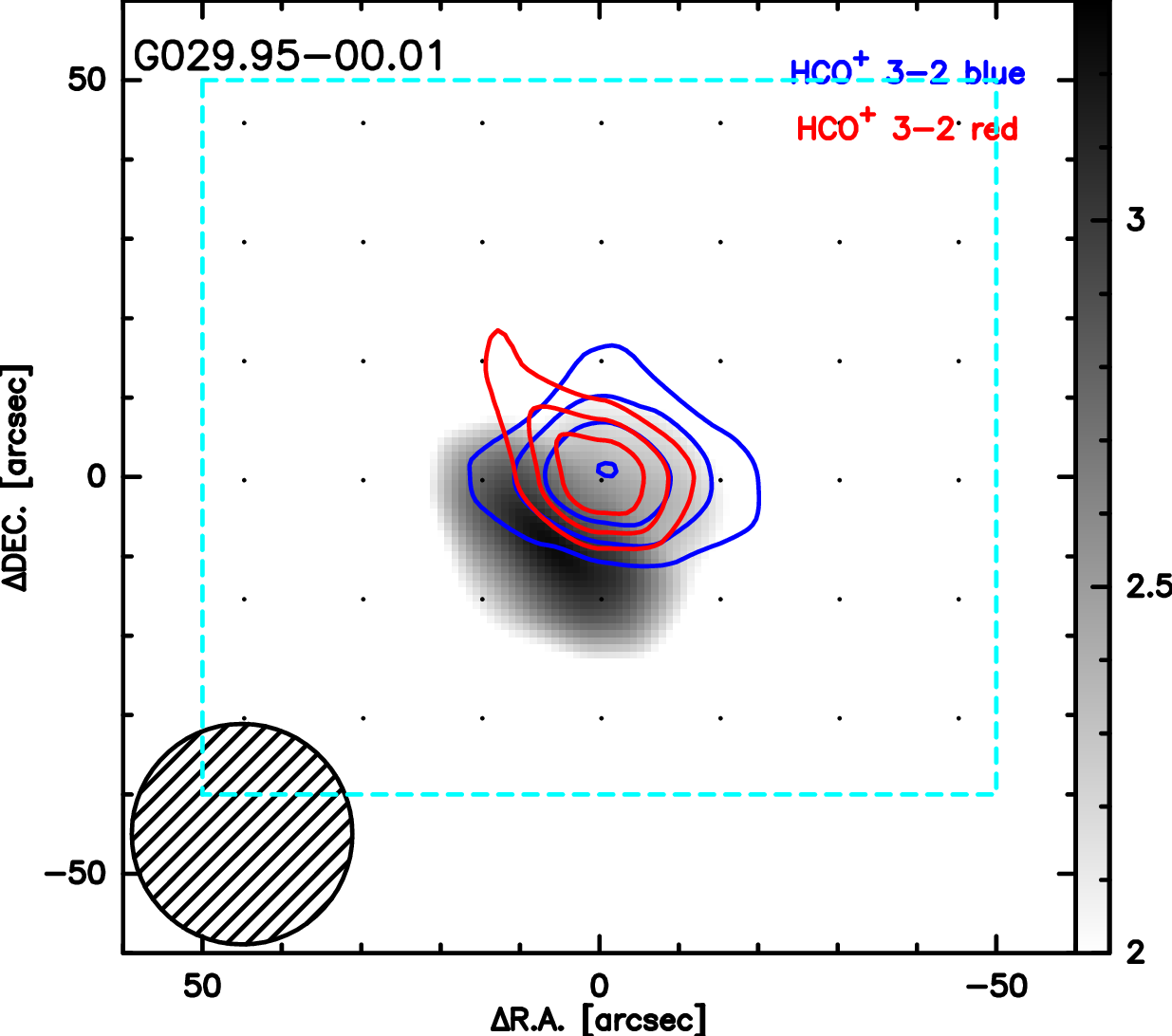} 
\includegraphics[width=0.35\columnwidth]{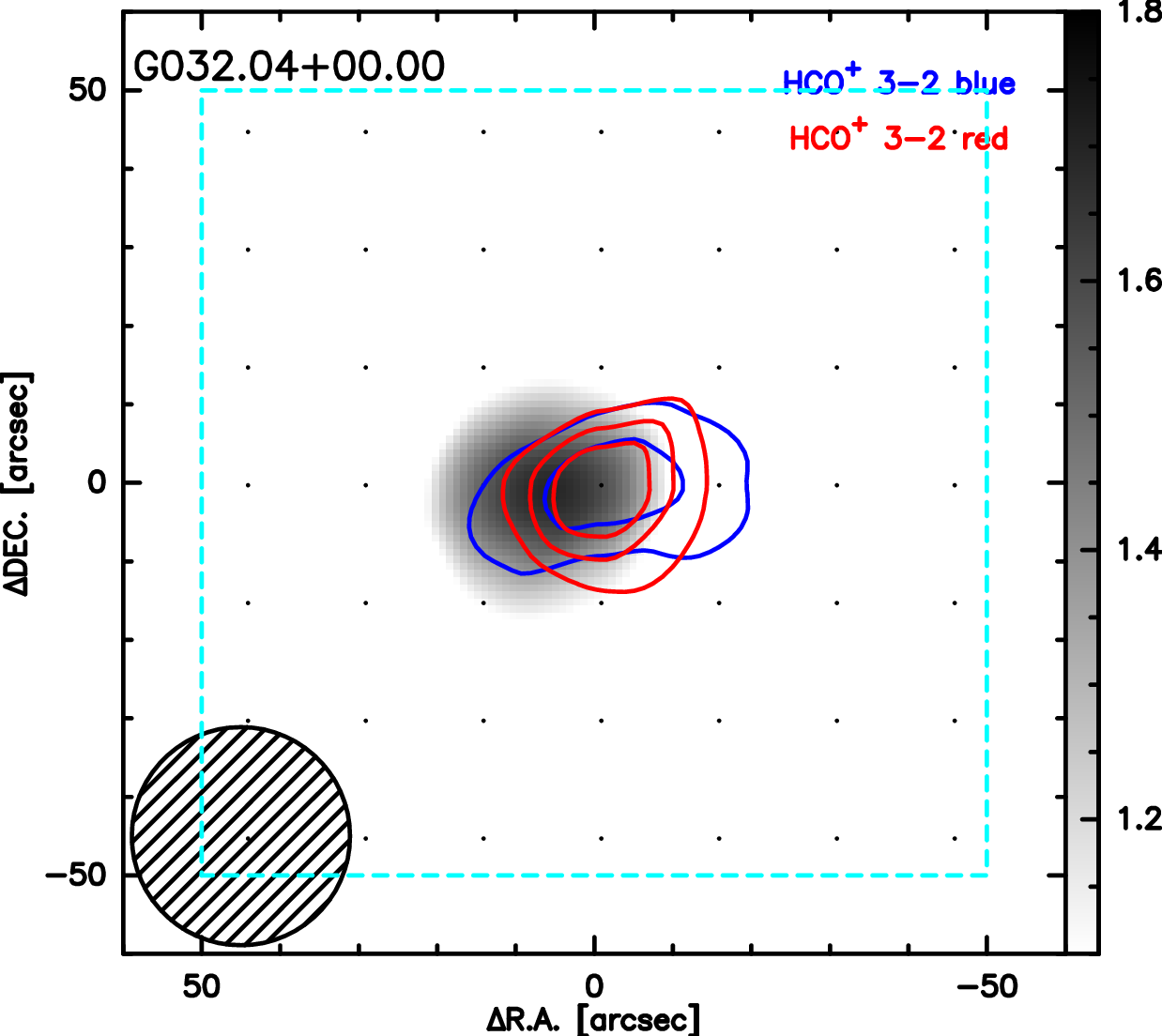} 

 \renewcommand{\thefigure}{2}
\caption{The line wing spatial distributions of HCO$^+$ 3-2. The source names are shown in the panels. Each panel shows the velocity integrated intensity distribution of blue and/or red wings traced by HCO$^+$ 3-2.  The greyscales illustrate the positions of cloud cores traced by H$^{13}$CO$^+$ 3-2. The cyan dotted line boxes indicate signal coverage. The detailed parameters for maps of this and other sources are presented in Table \ref{tab4}.}\label{fig2}
\end{figure*}

\begin{figure*}
\addtocounter{figure}{-1}
\includegraphics[width=0.35\columnwidth]{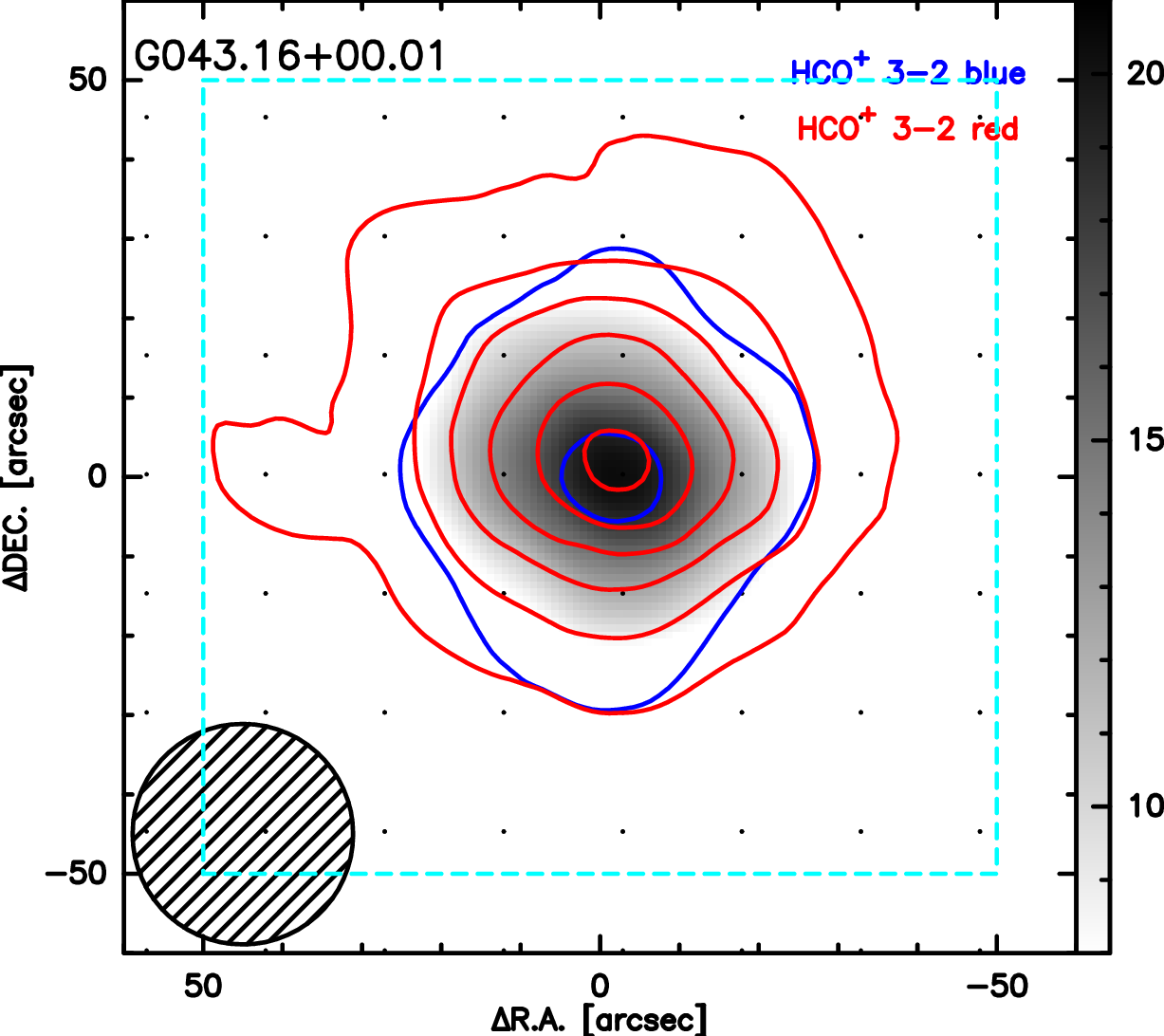}
\includegraphics[width=0.35\columnwidth]{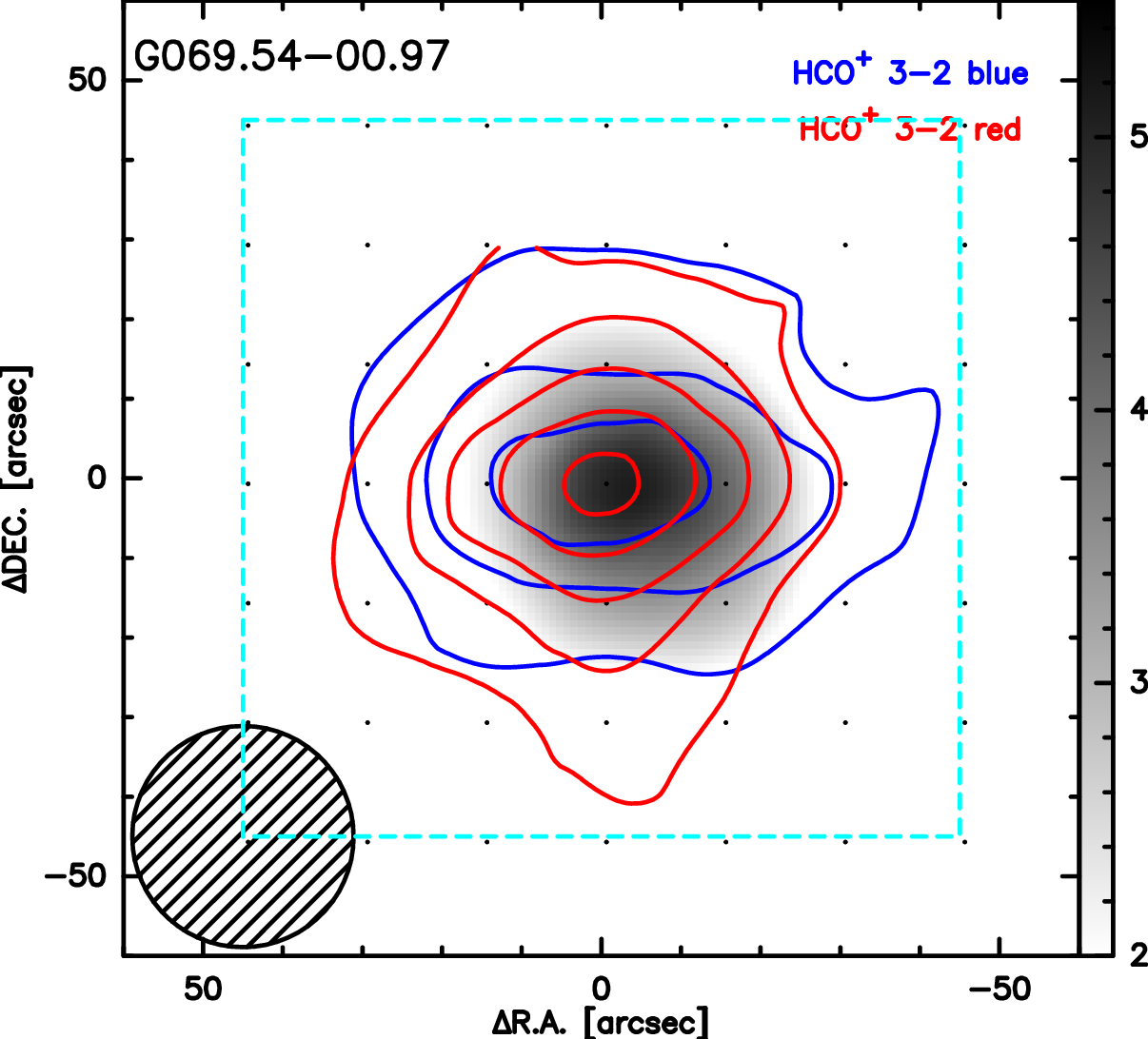} 
\includegraphics[width=0.35\columnwidth]{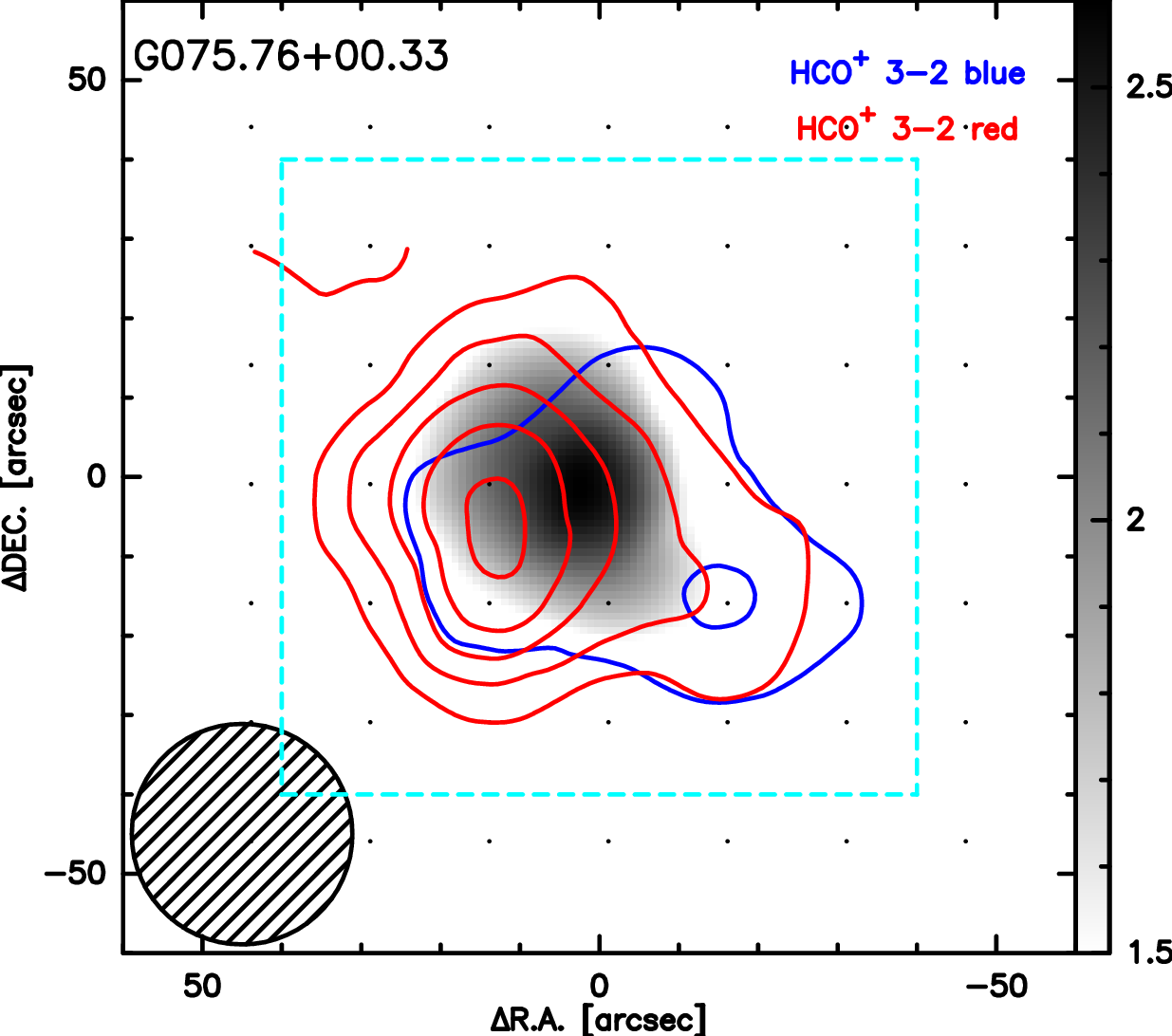} 
\includegraphics[width=0.35\columnwidth]{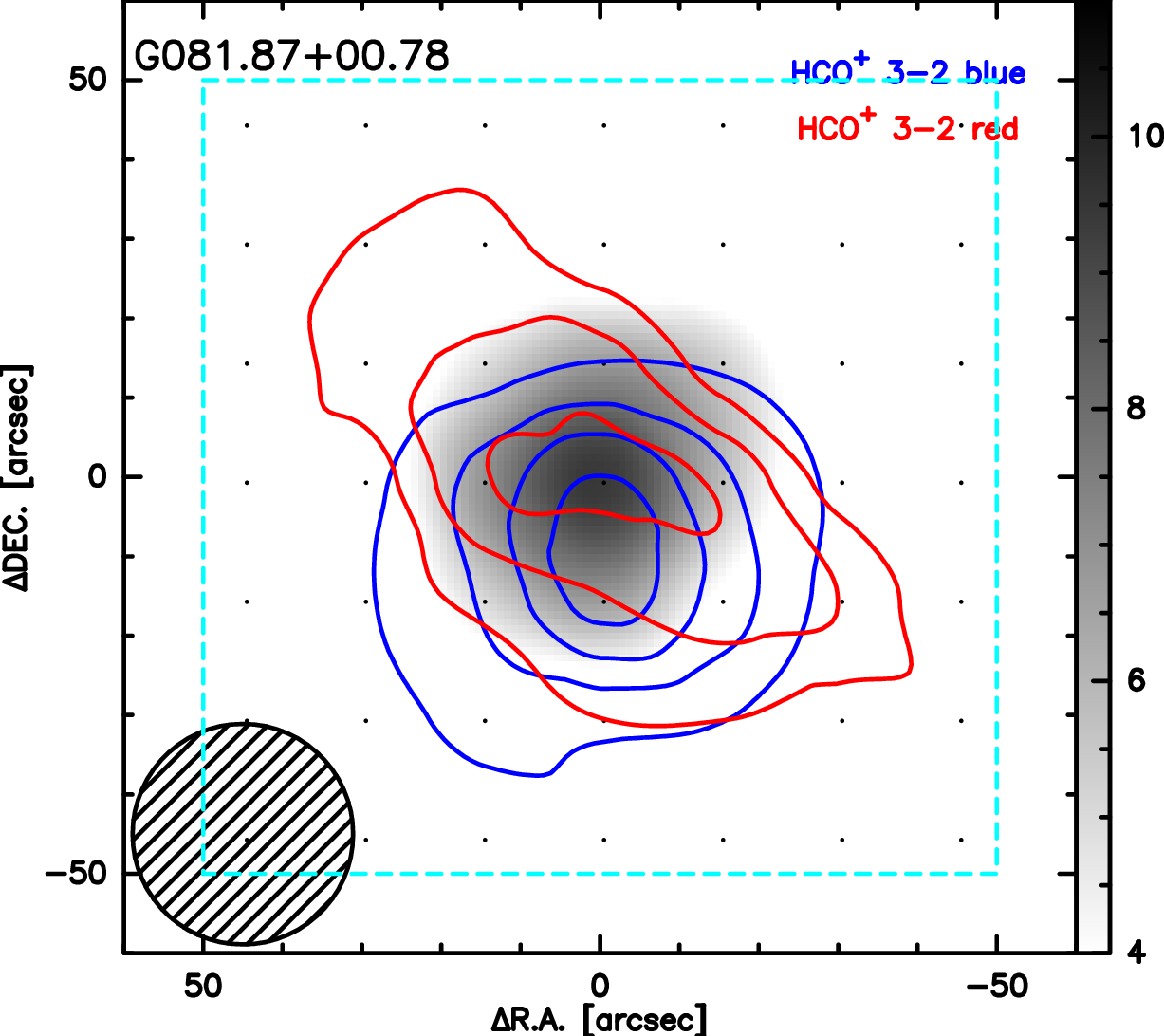} 
\includegraphics[width=0.35\columnwidth]{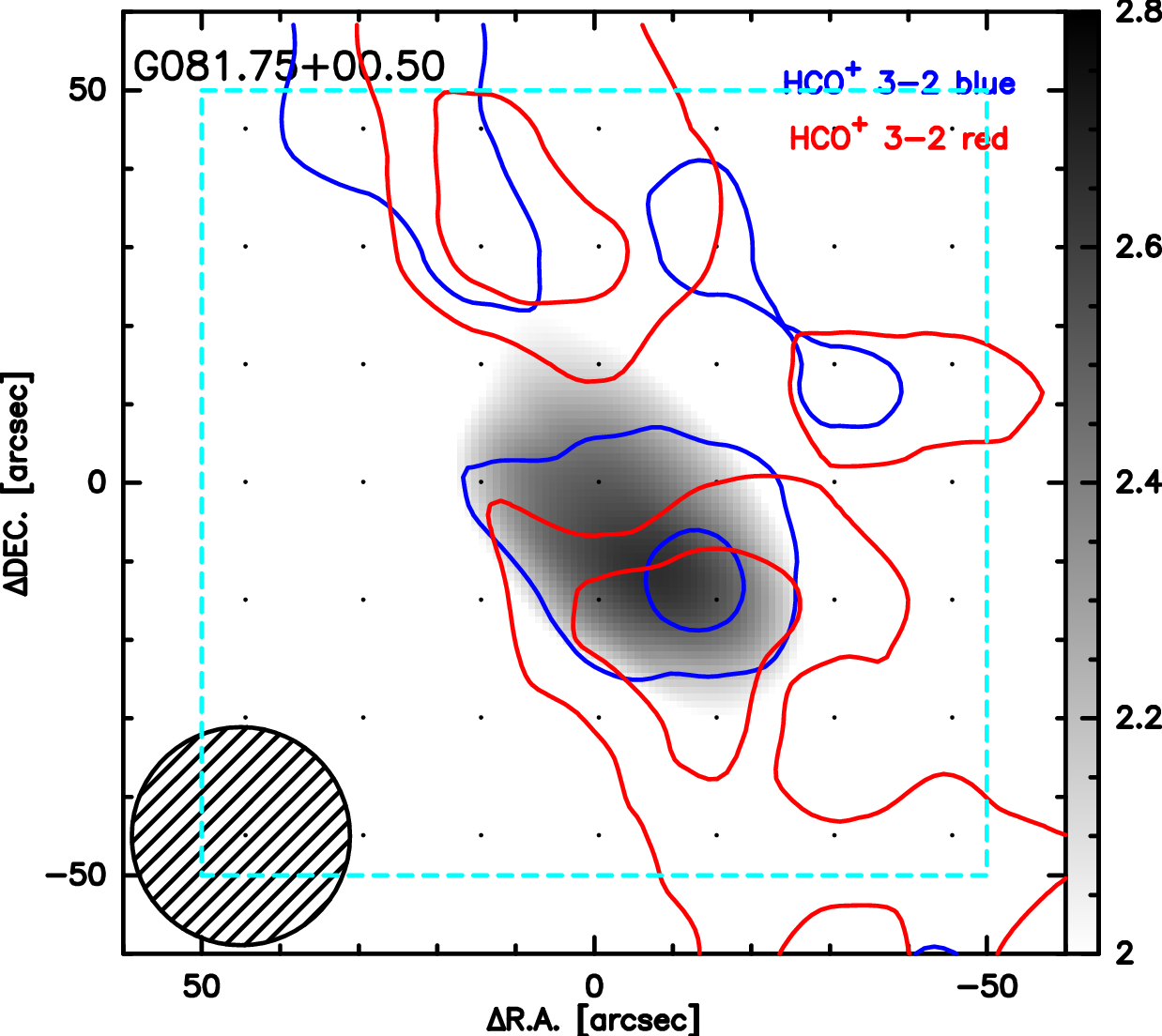} 
\includegraphics[width=0.35\columnwidth]{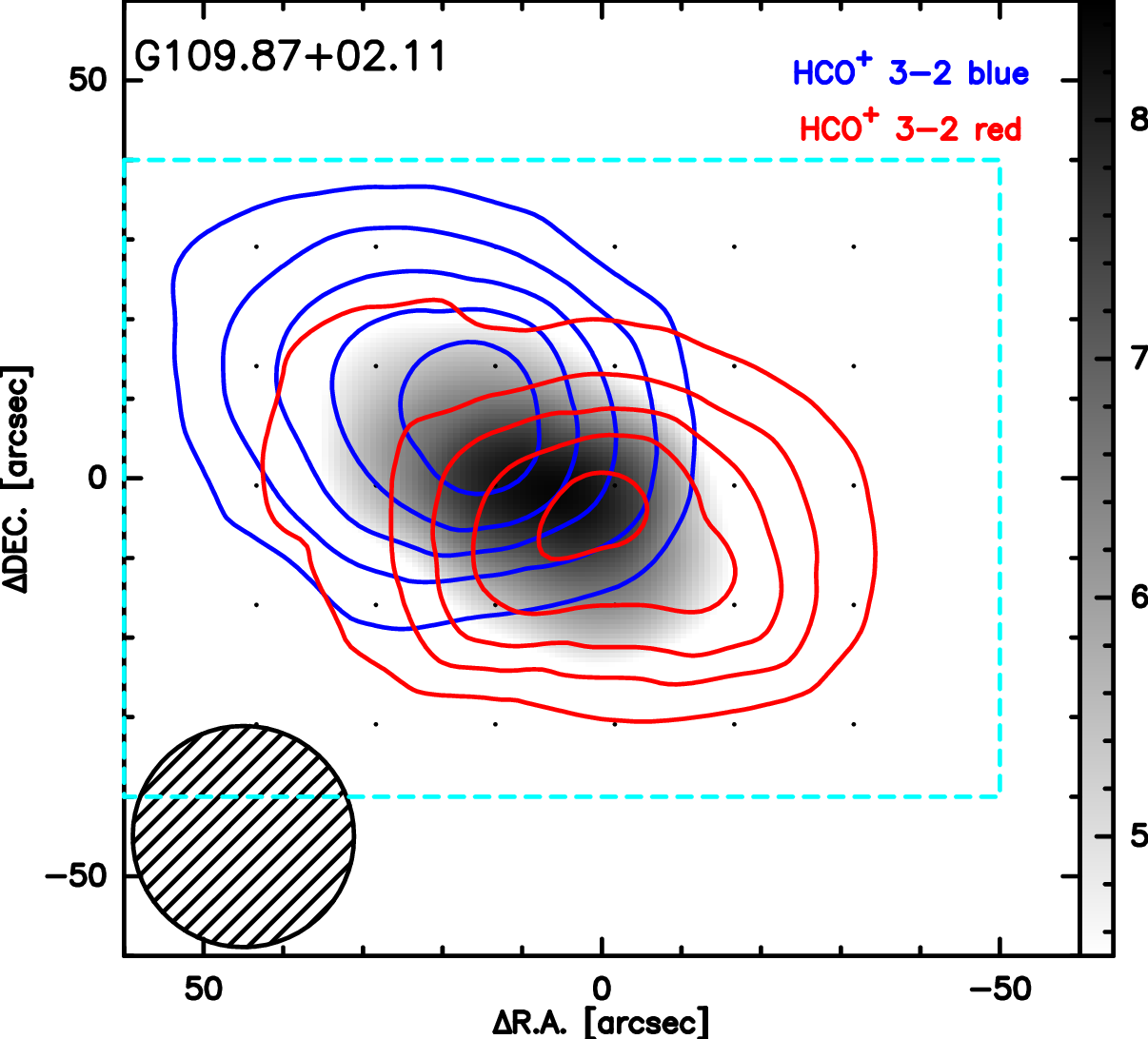} 
\includegraphics[width=0.35\columnwidth]{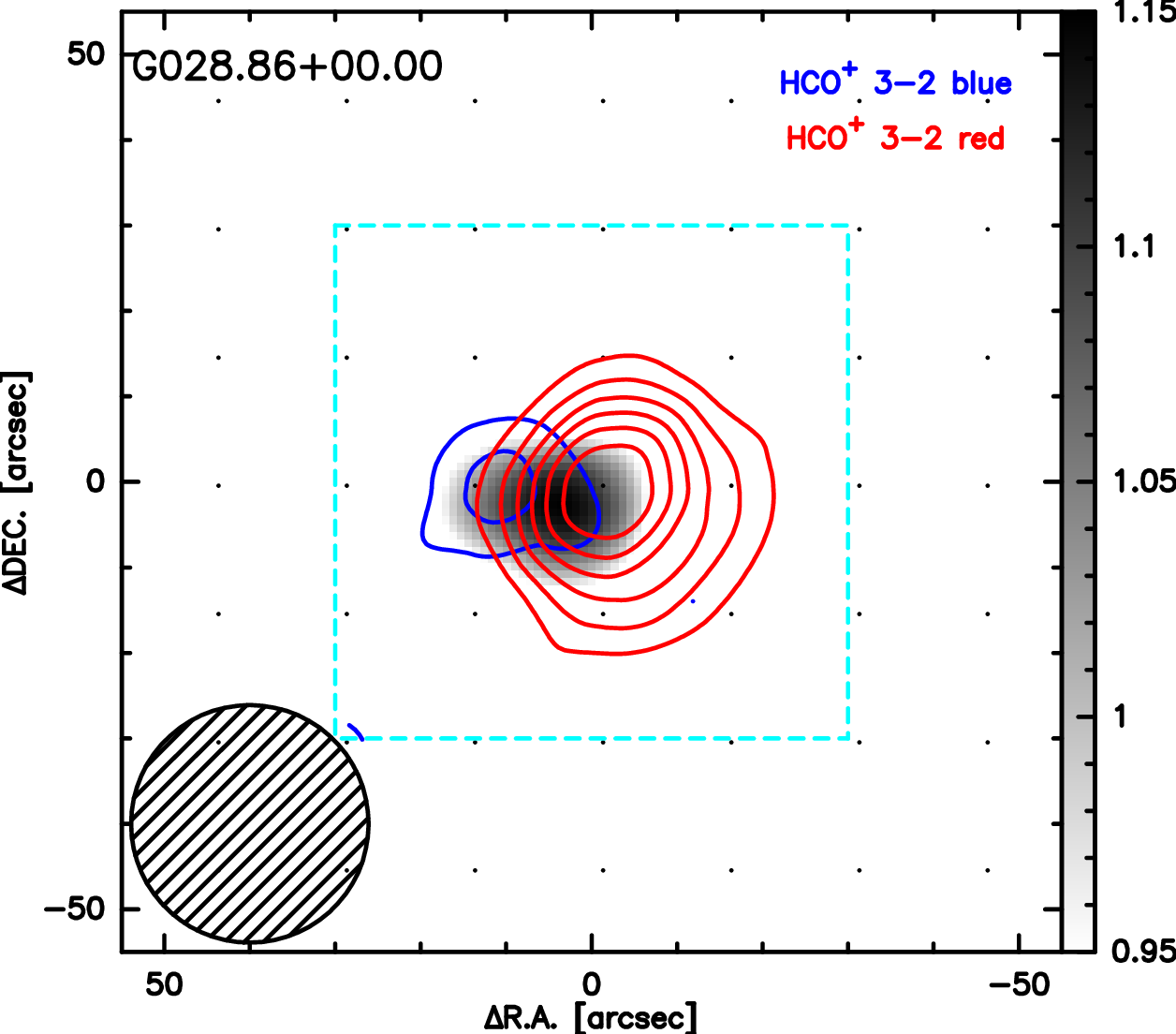} 
\includegraphics[width=0.35\columnwidth]{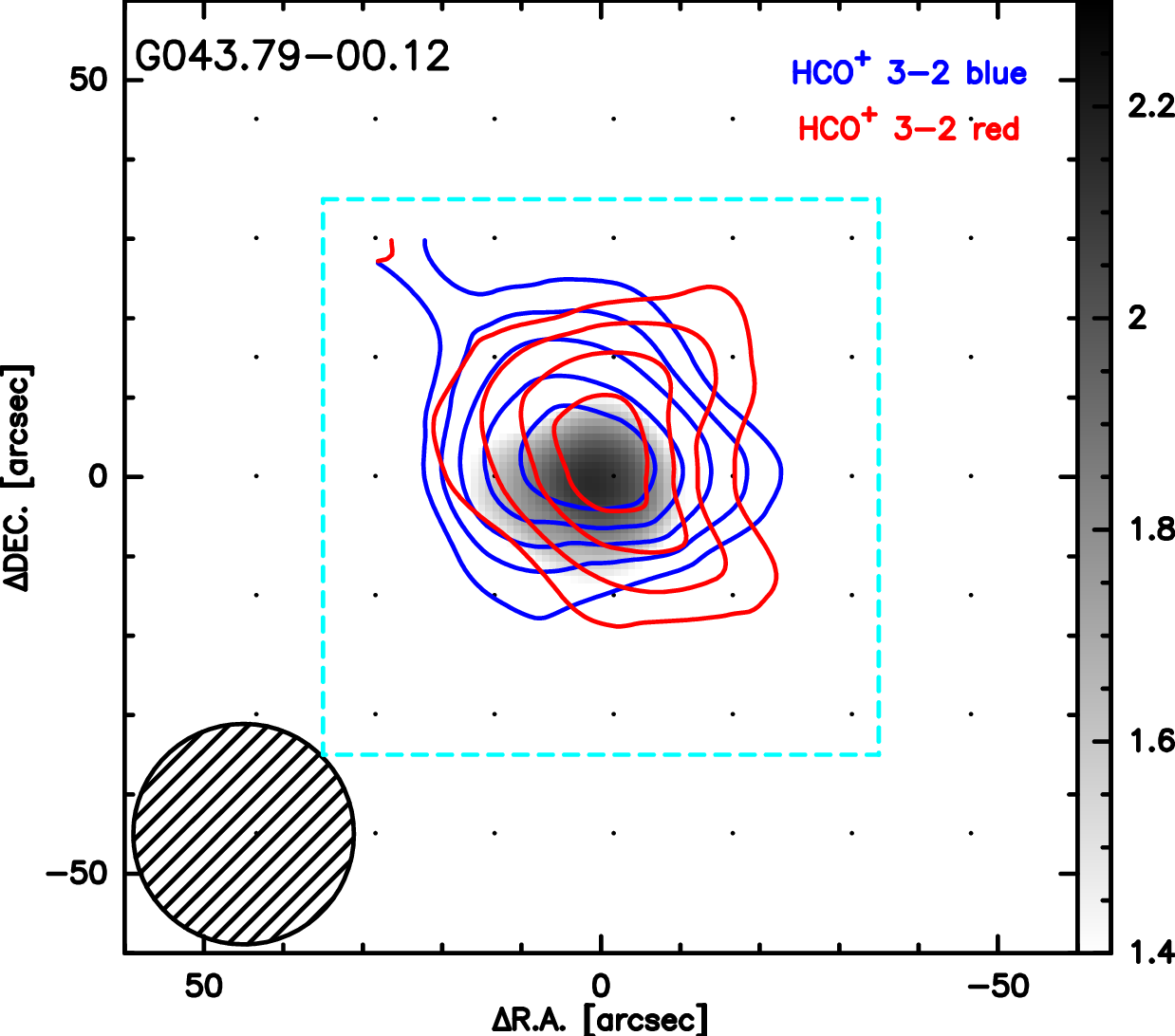} 
\includegraphics[width=0.35\columnwidth]{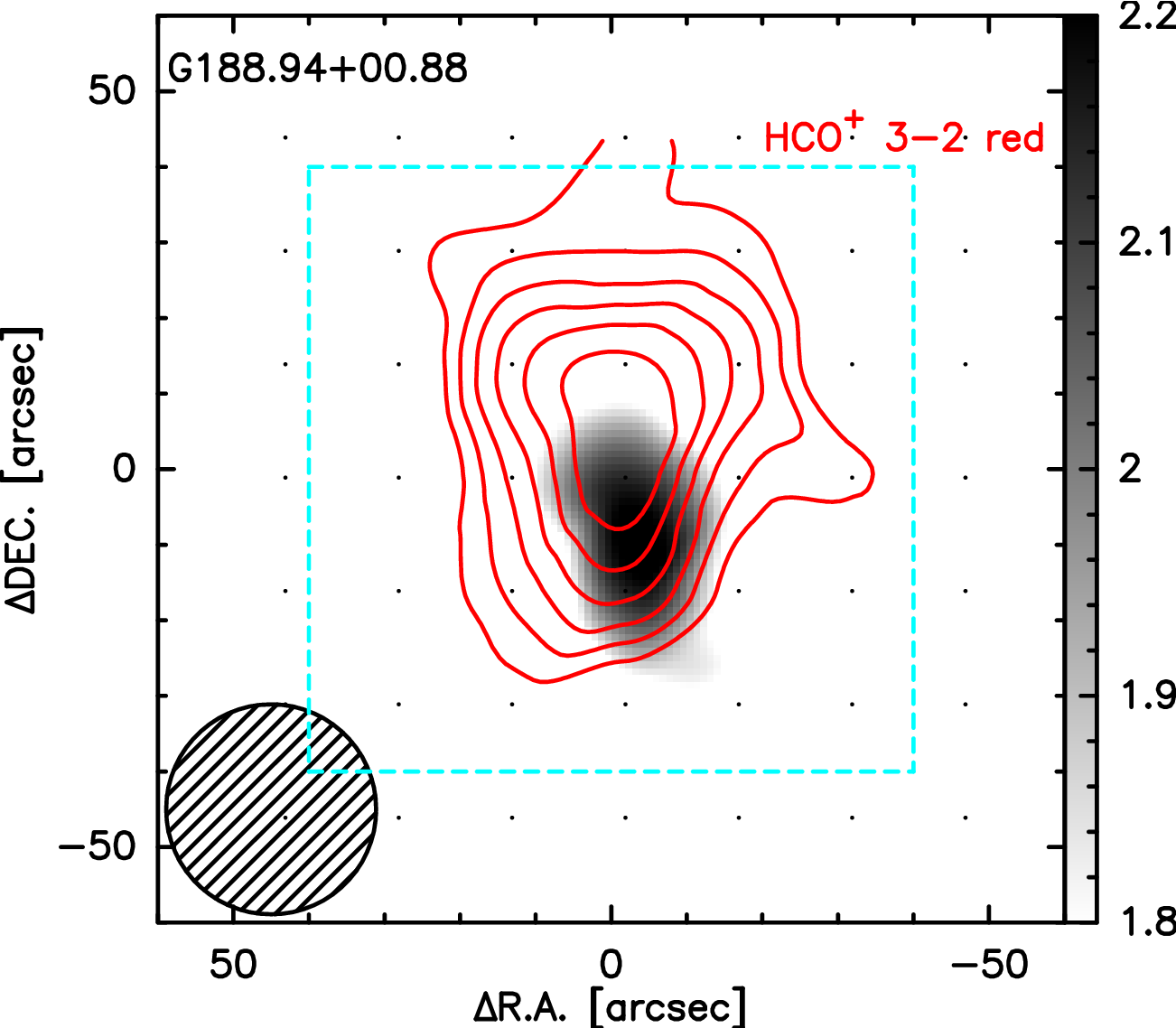} 
\includegraphics[width=0.35\columnwidth]{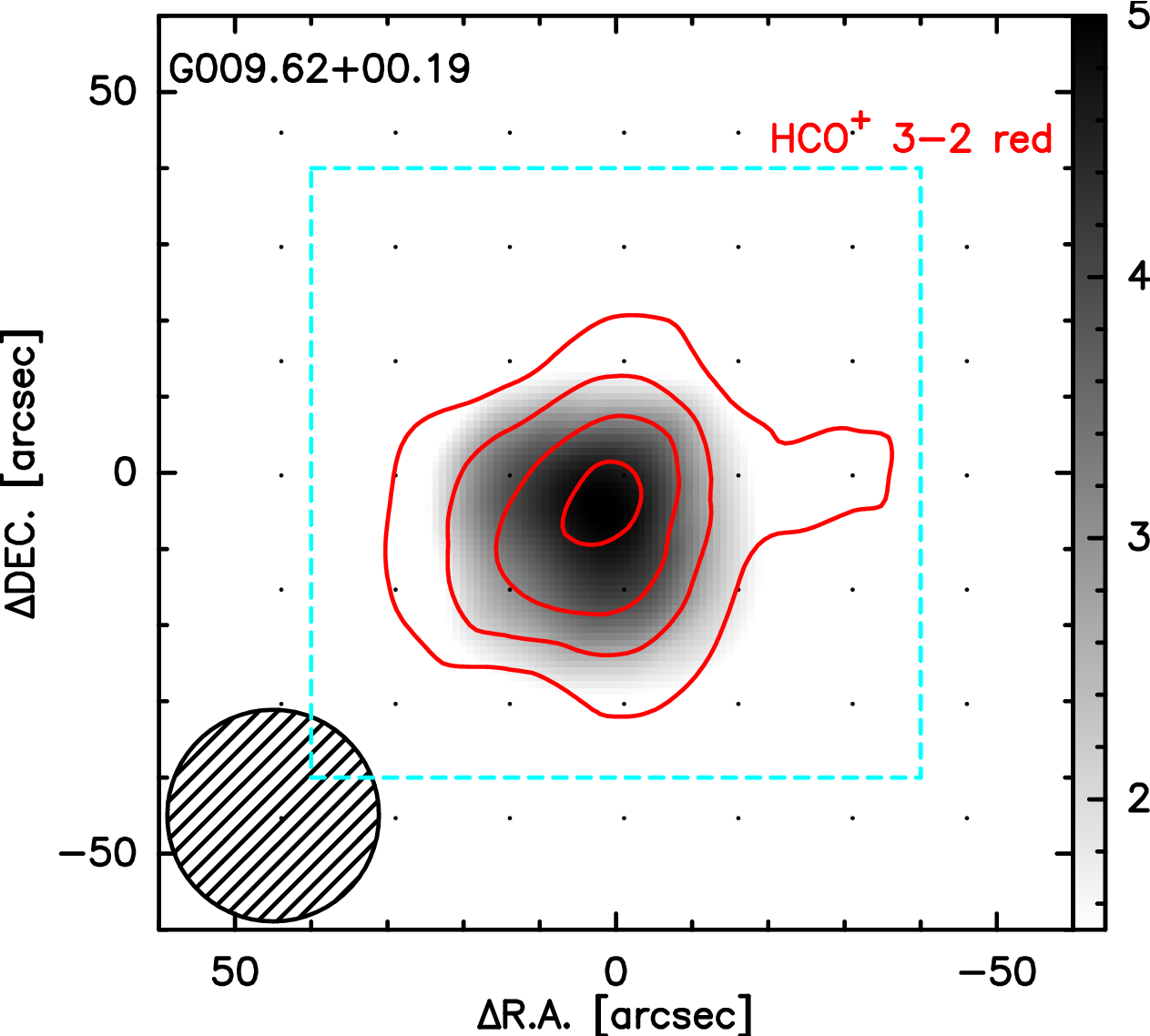} 
\includegraphics[width=0.35\columnwidth]{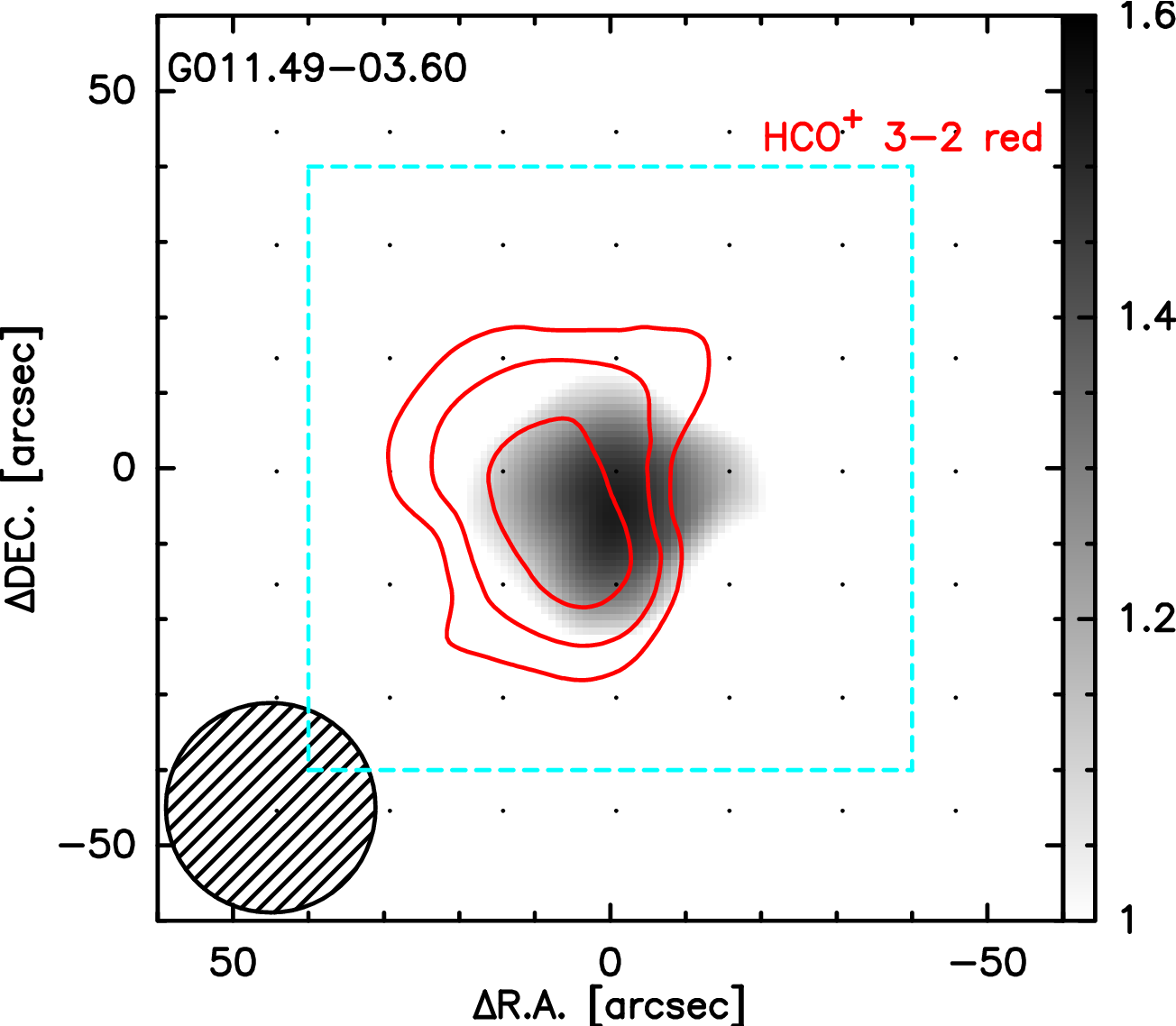} 
 \renewcommand{\thefigure}{2}
\addtocounter{figure}{-1}
\caption{Continued.}
\label{fig2}
\end{figure*}

\begin{figure}[htbp]
\includegraphics[width=0.35\columnwidth]{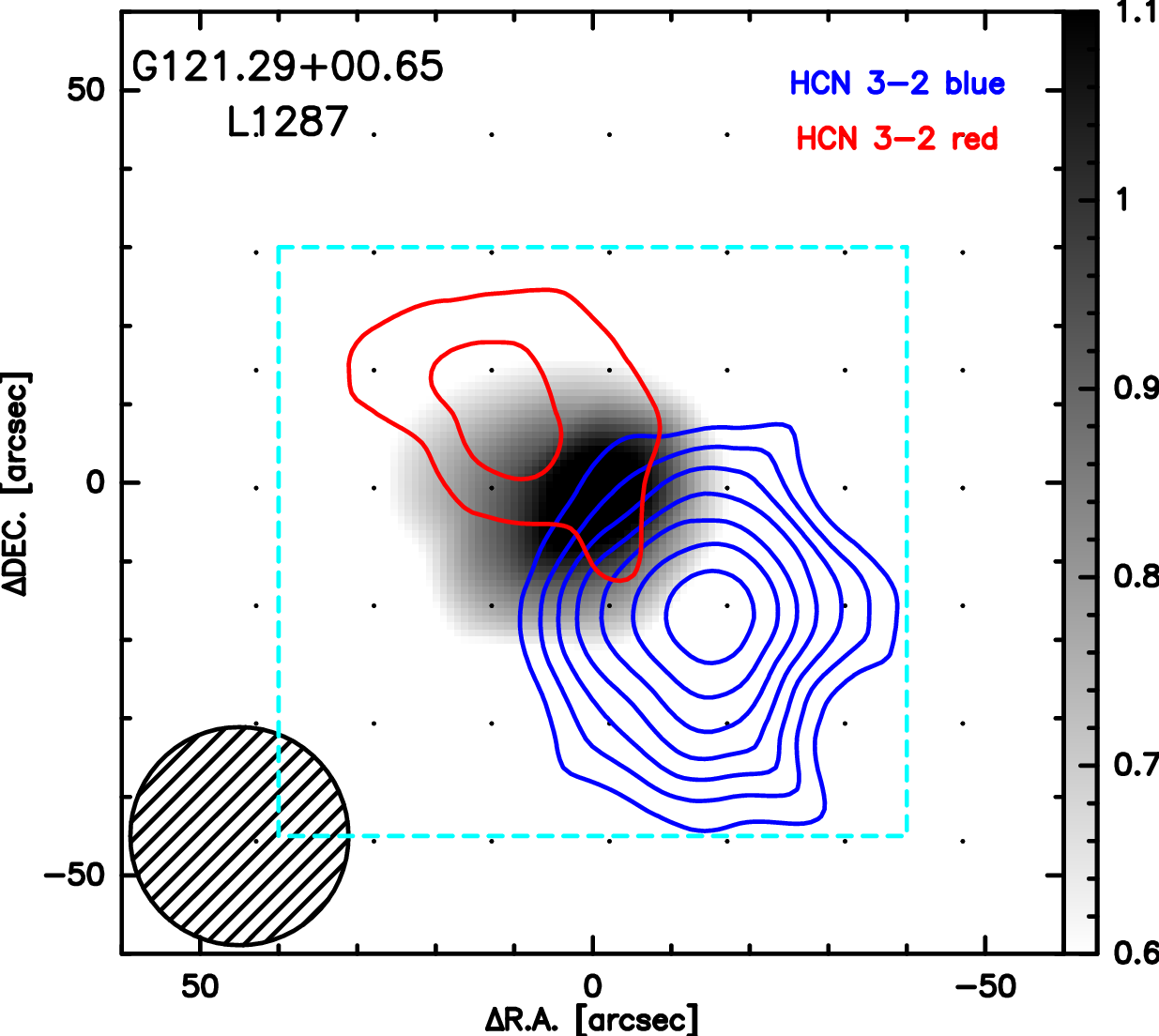} 
\includegraphics[width=0.35\columnwidth]{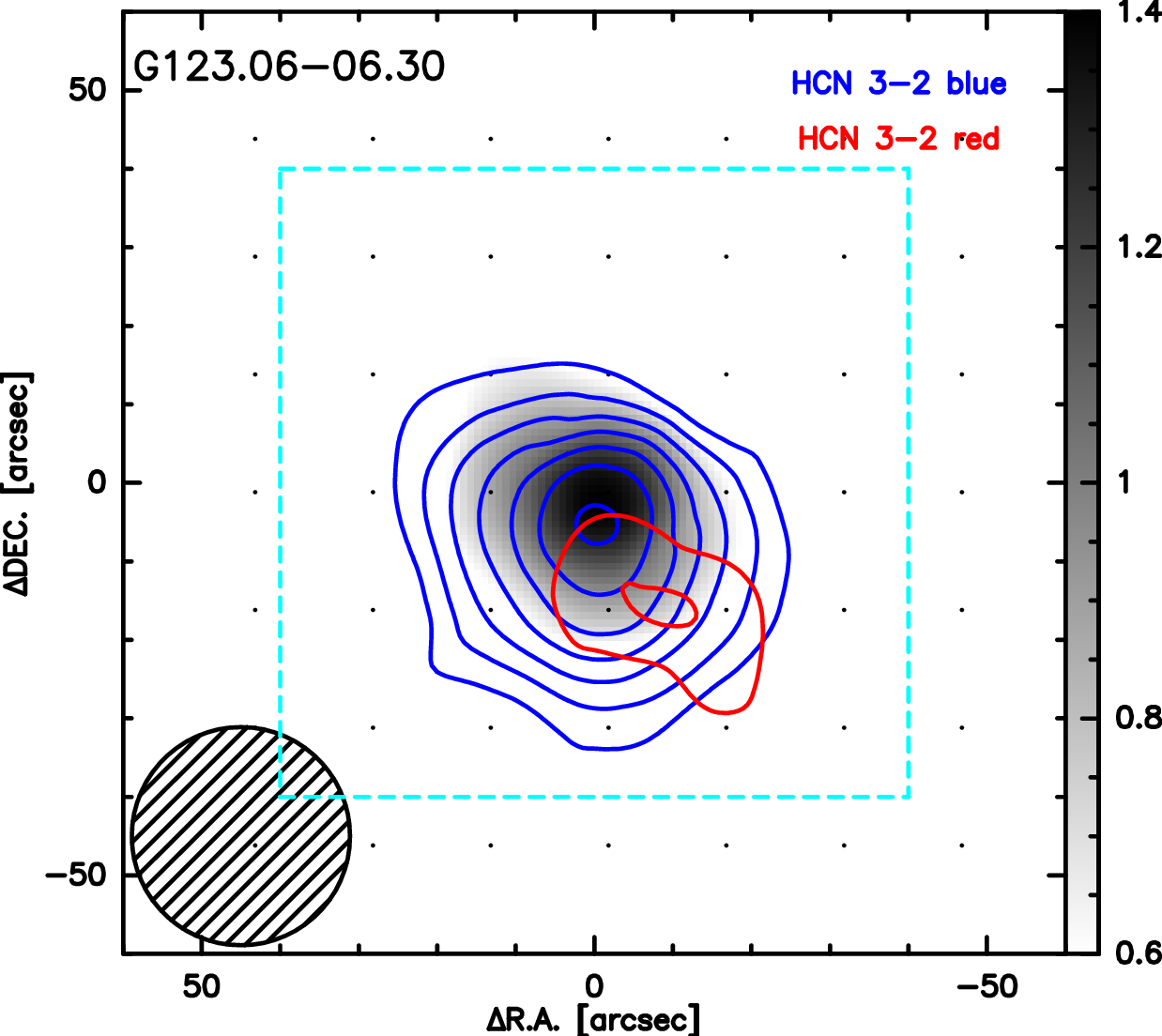} 
\includegraphics[width=0.35\columnwidth]{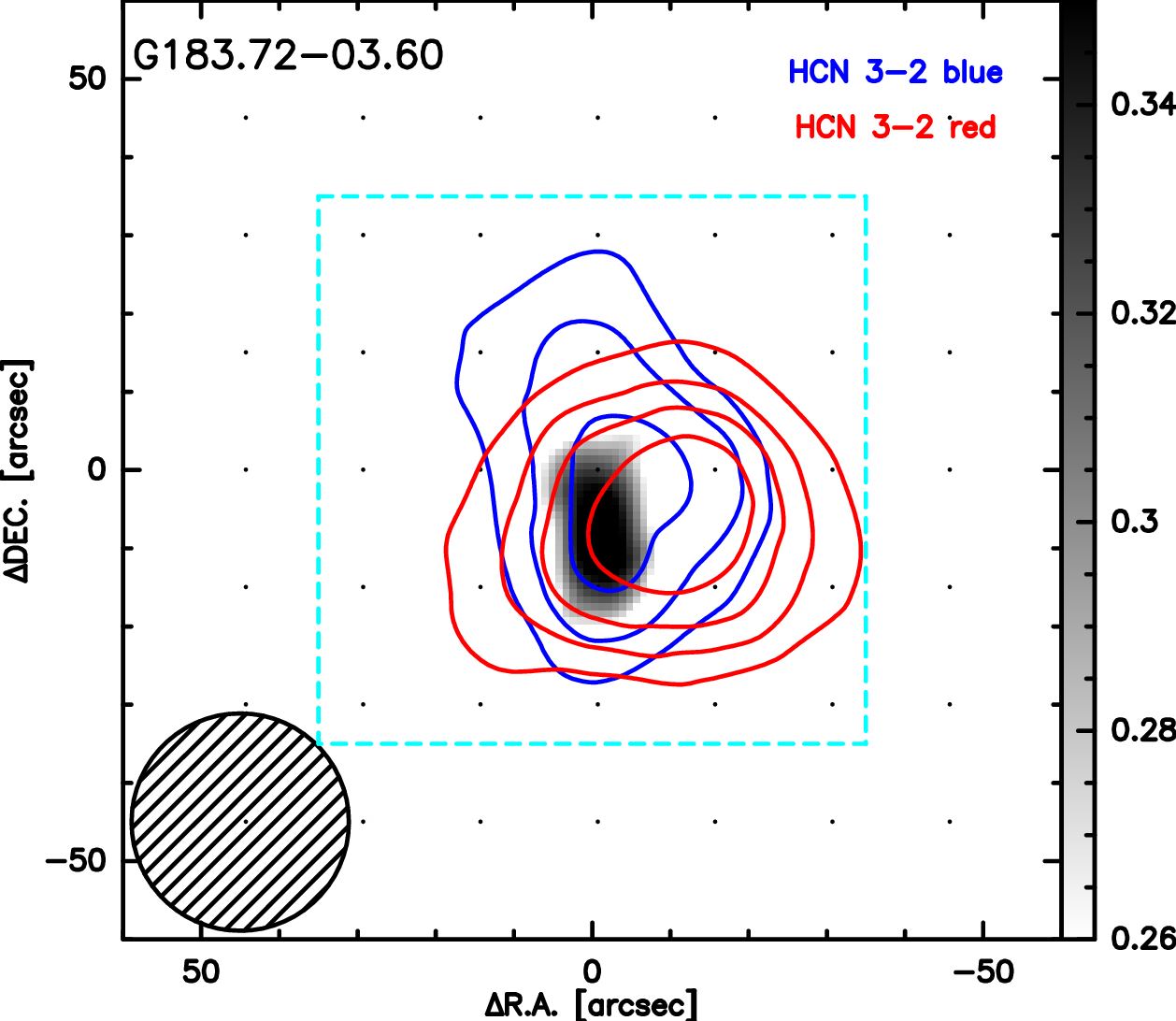} 
\includegraphics[width=0.35\columnwidth]{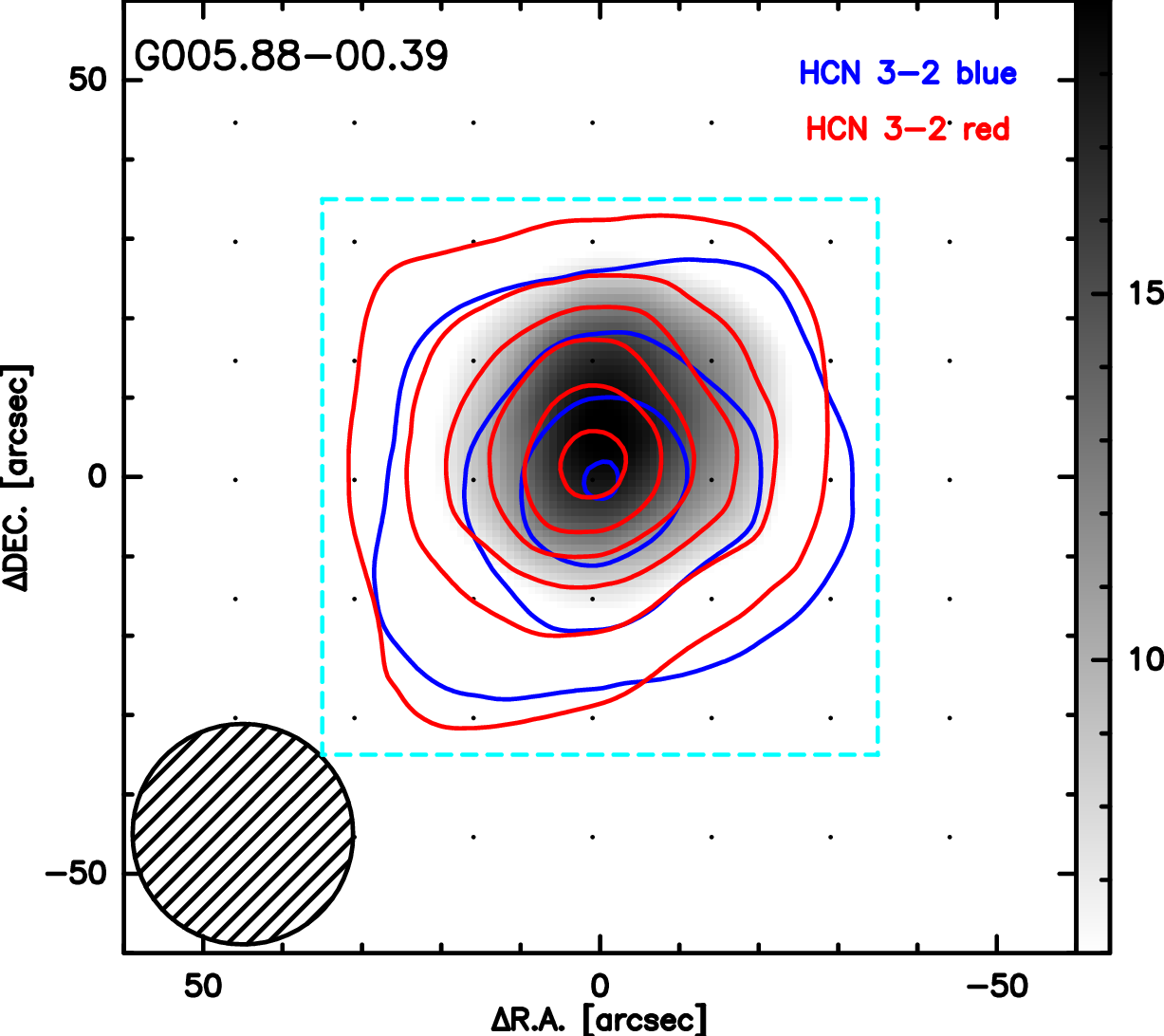} 
\includegraphics[width=0.35\columnwidth]{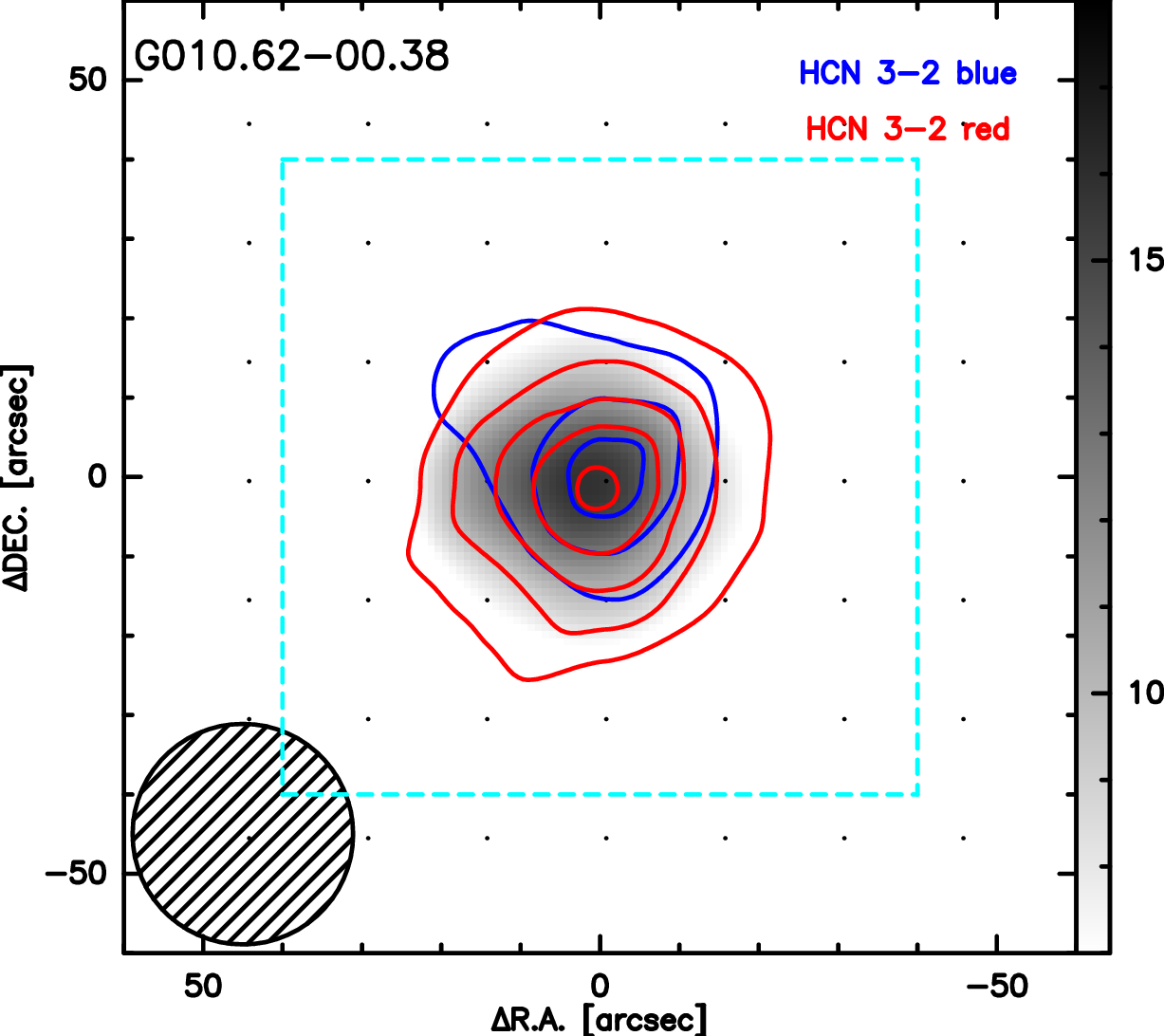} 
\includegraphics[width=0.35\columnwidth]{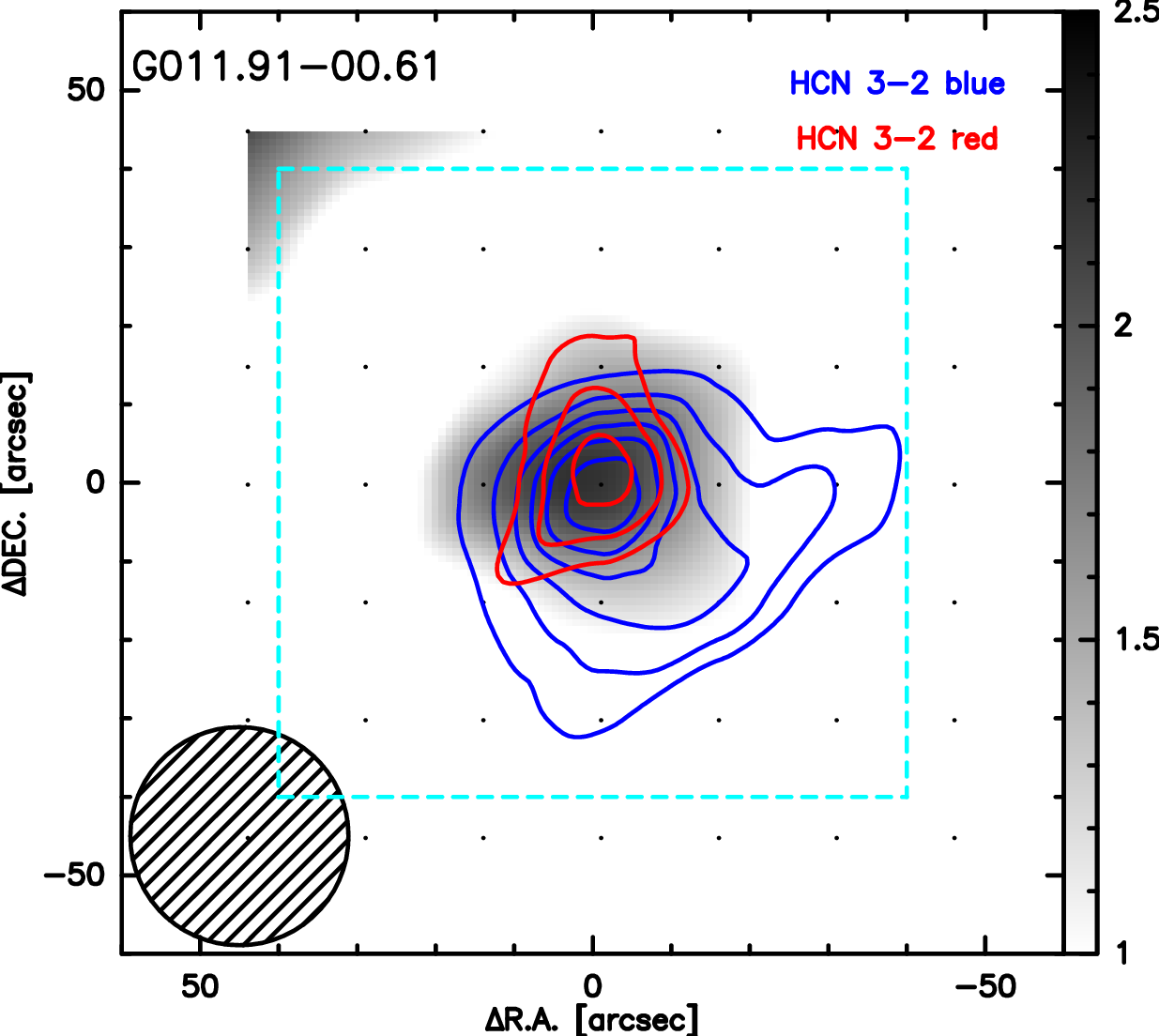} 
\includegraphics[width=0.35\columnwidth]{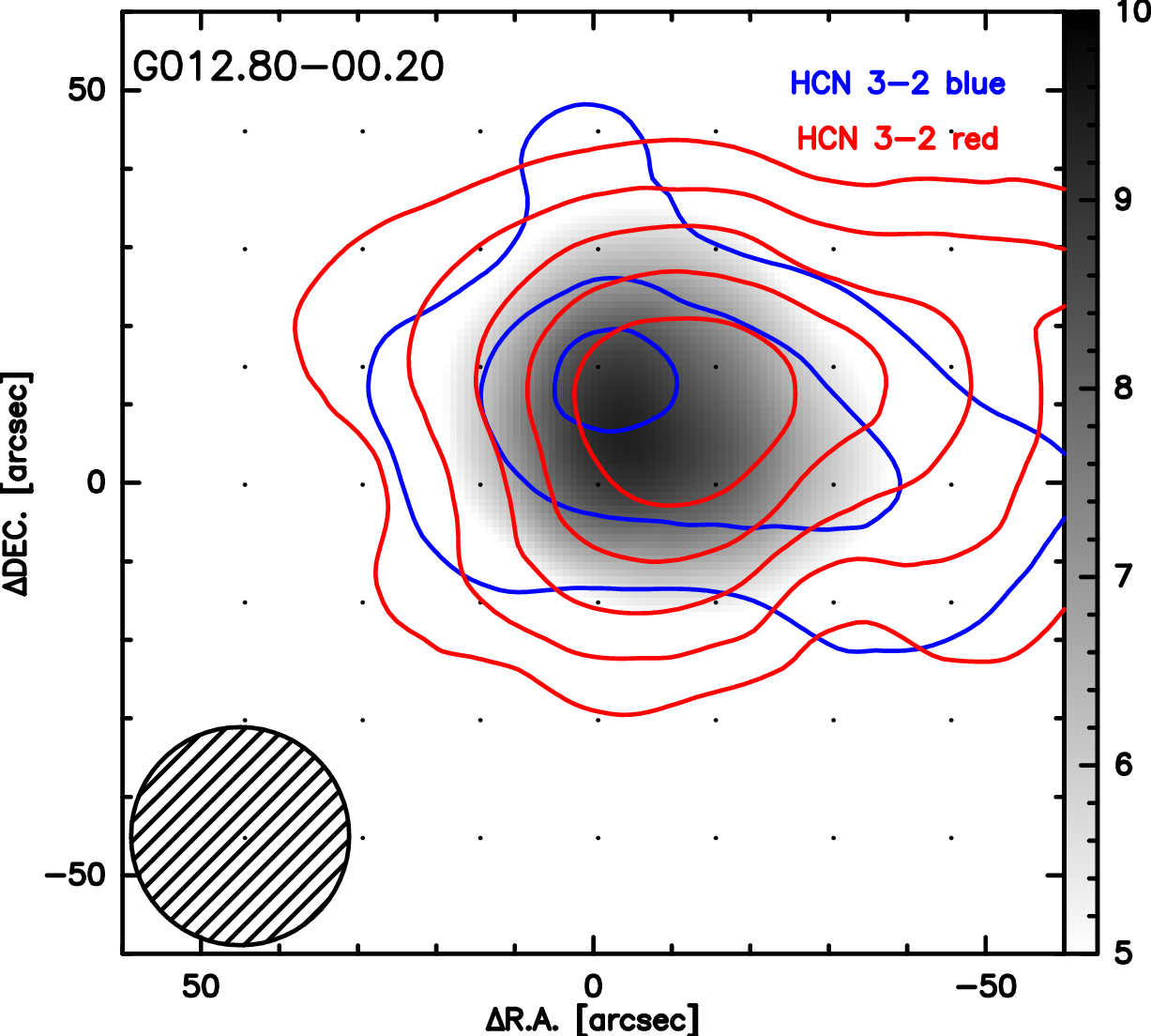} 
\includegraphics[width=0.35\columnwidth]{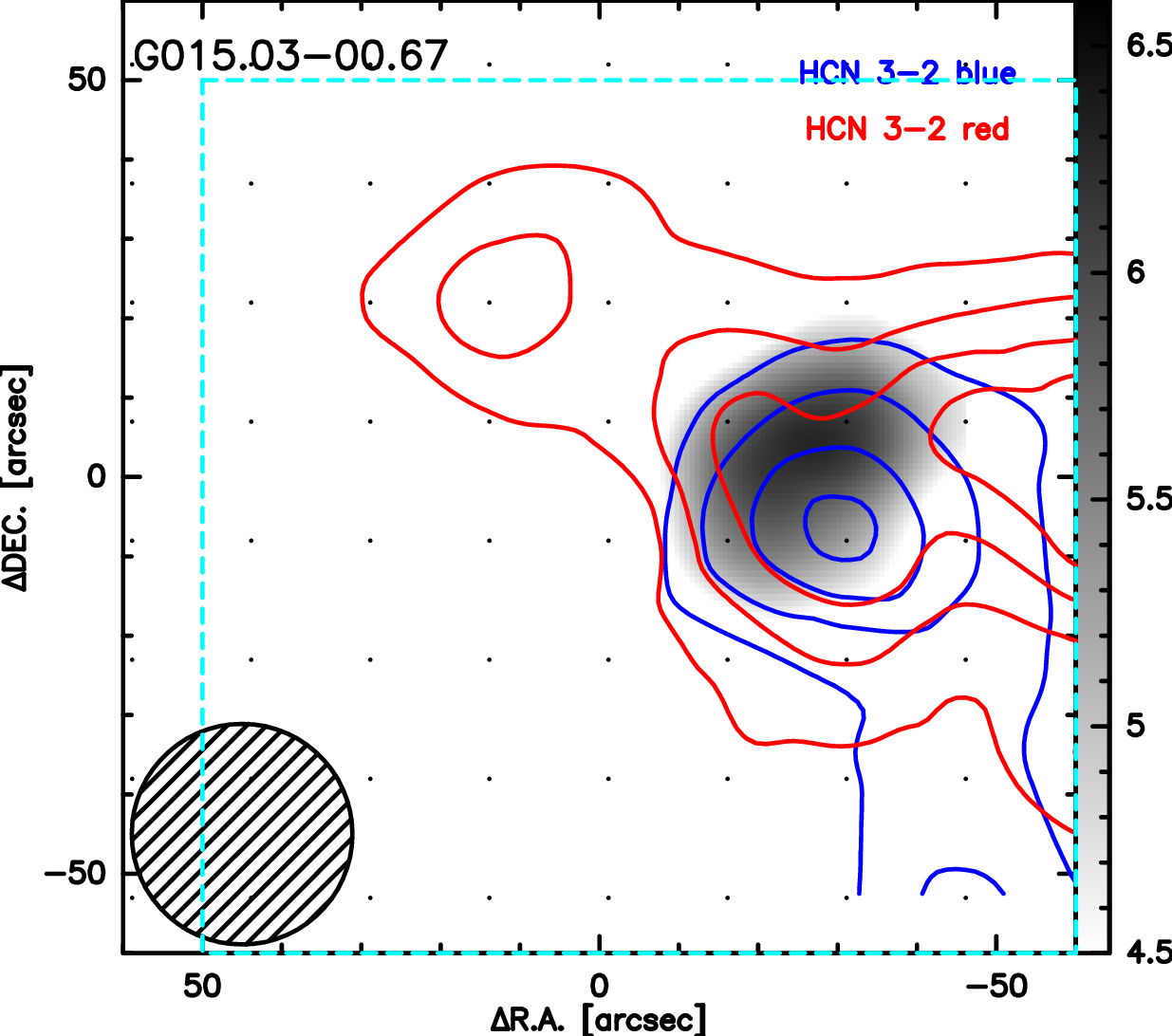} 
\includegraphics[width=0.35\columnwidth]{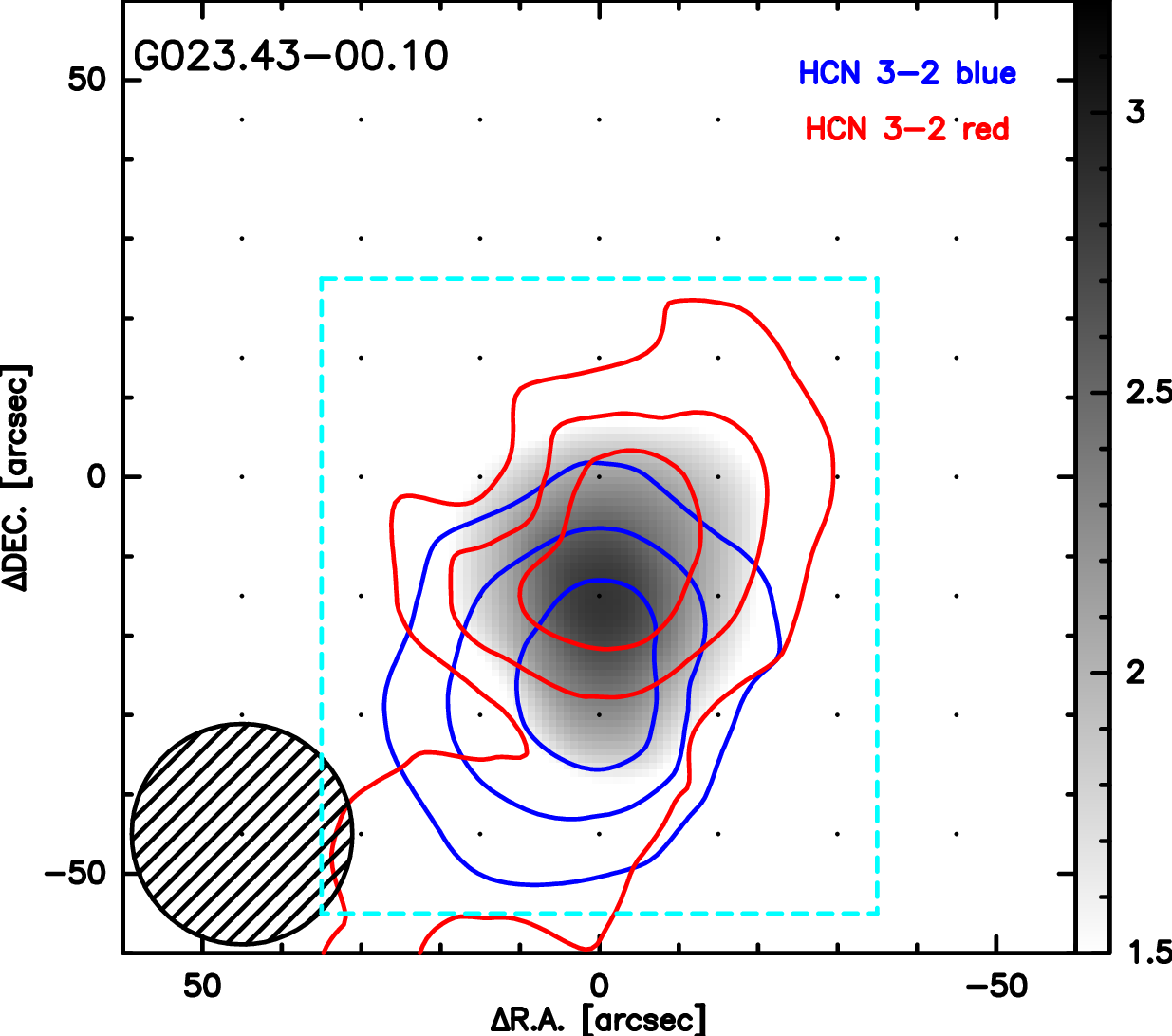}
\includegraphics[width=0.35\columnwidth]{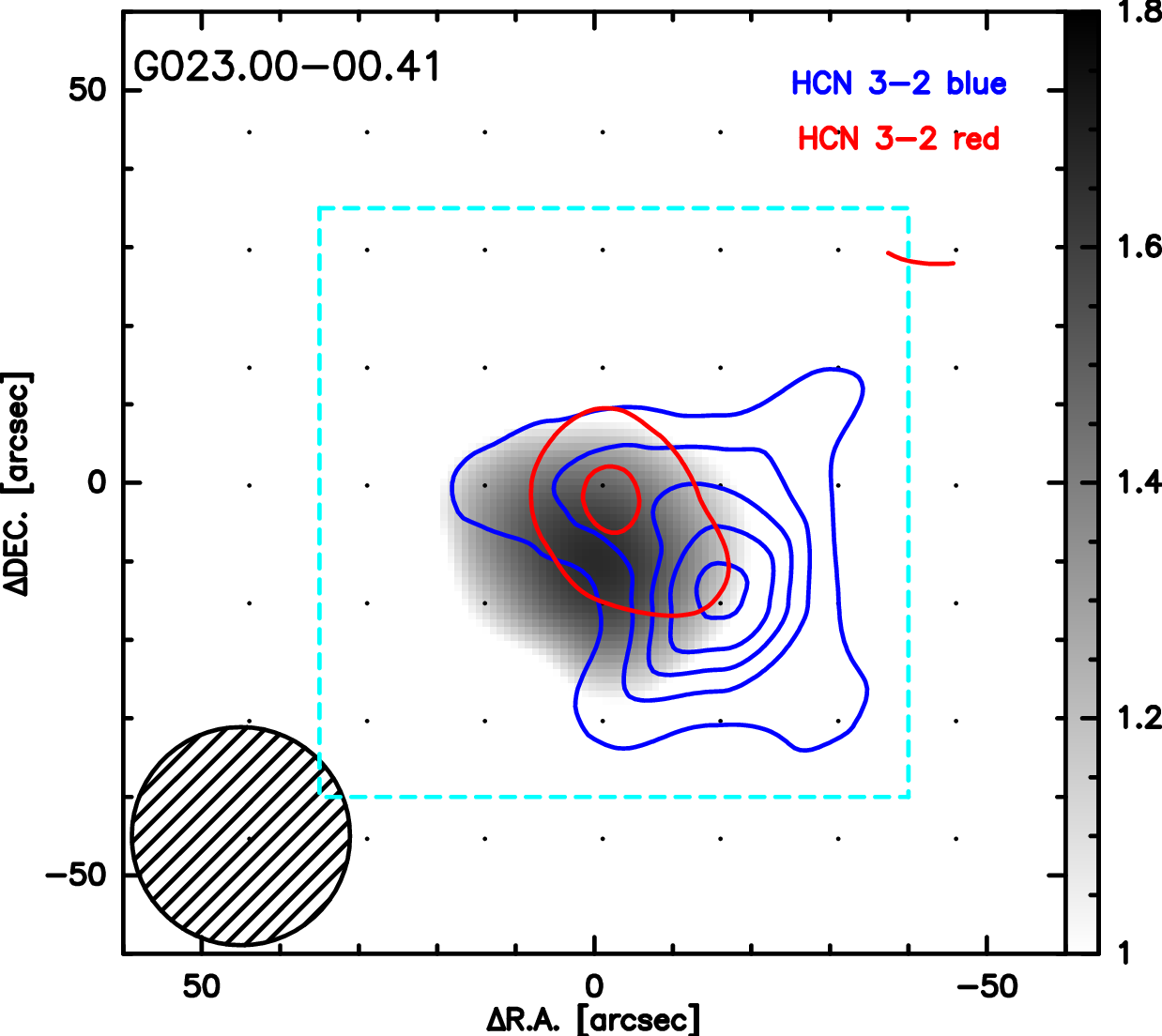} 
\includegraphics[width=0.35\columnwidth]{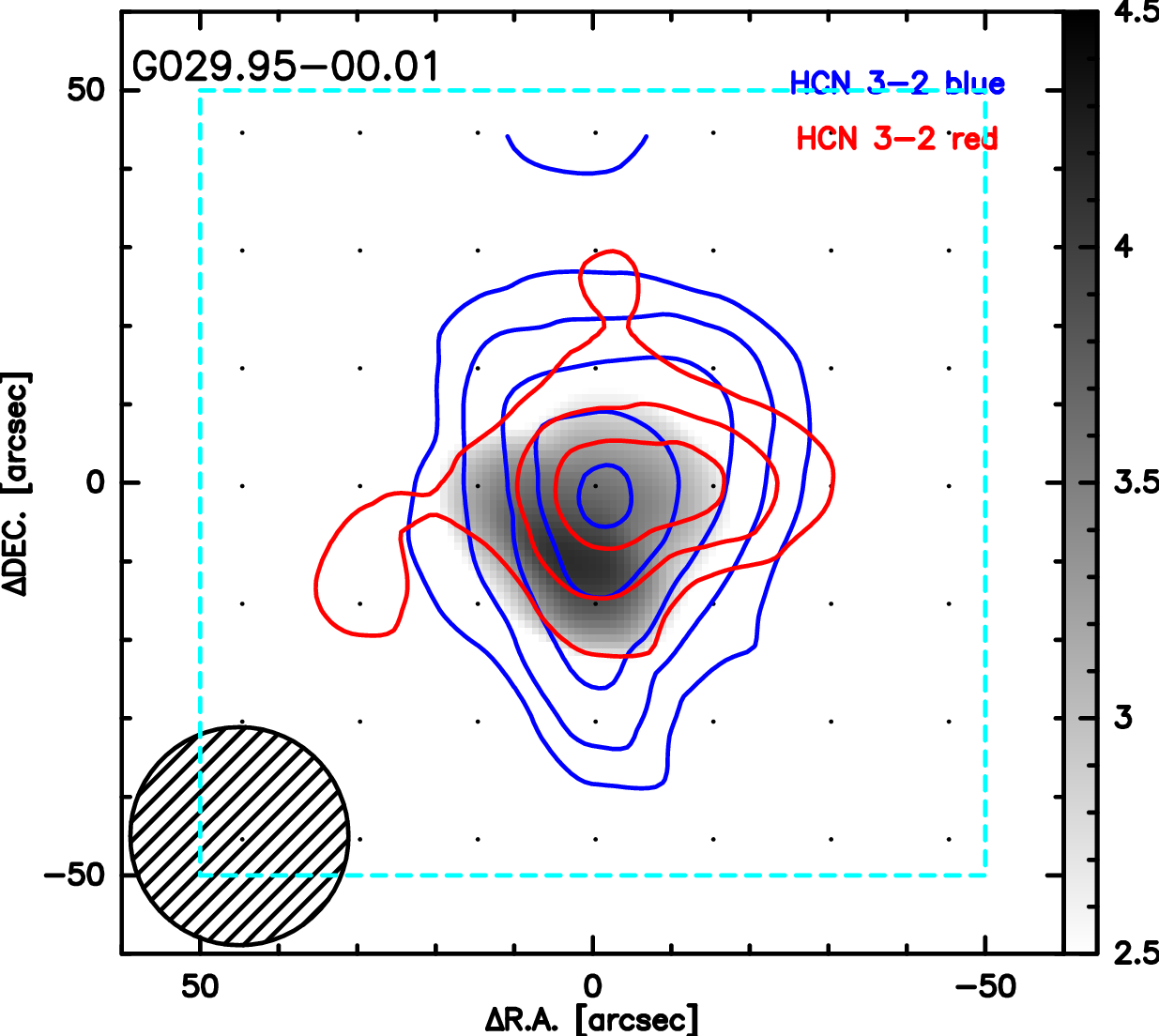} 
\includegraphics[width=0.35\columnwidth]{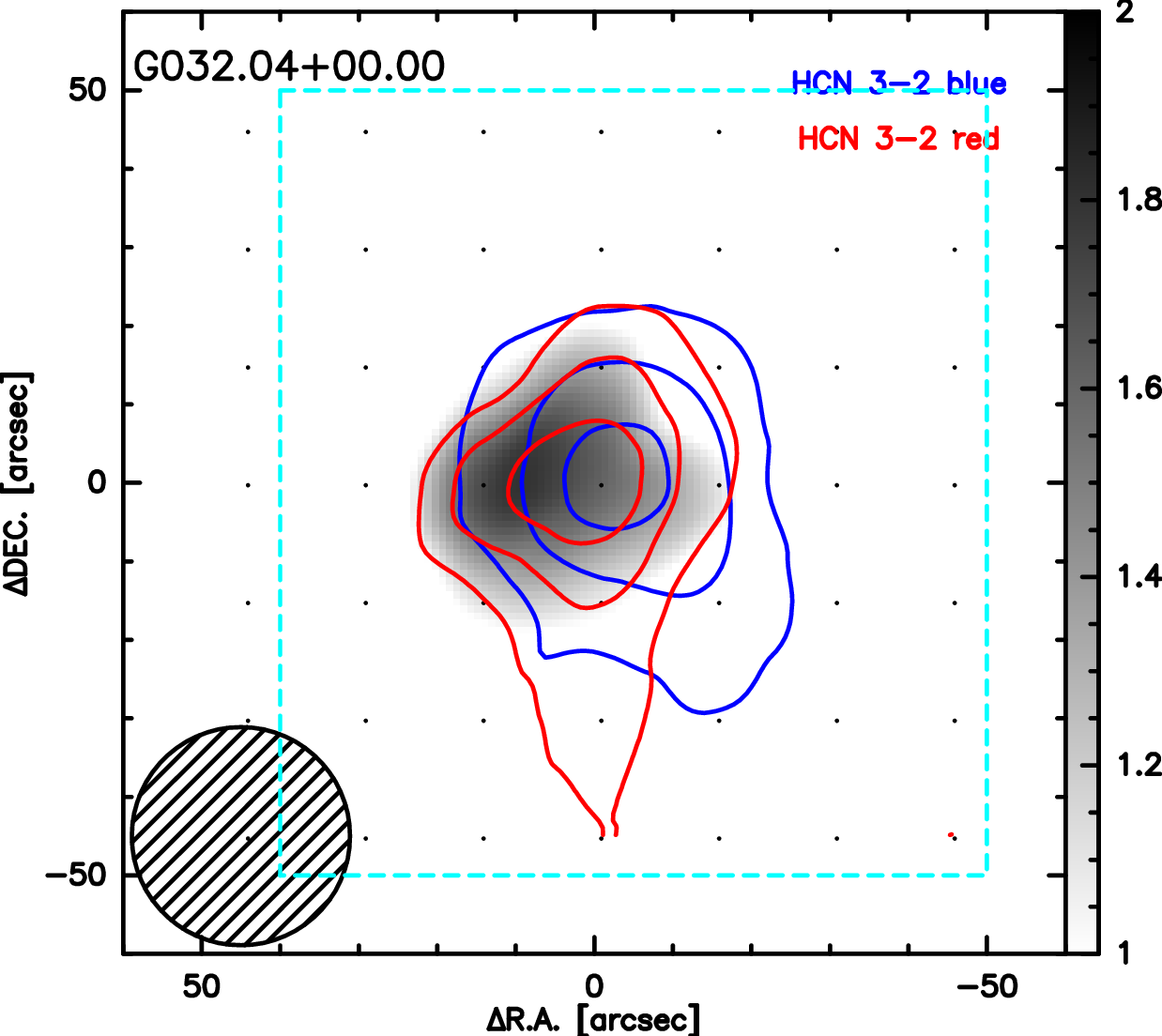} 

 \renewcommand{\thefigure}{3}
\caption{The line wing spatial distributions of HCN 3-2. The source names are shown in the panels. Each panel shows the velocity integrated intensity distribution of blue and/or red wings traced by HCN 3-2. The greyscales illustrate the positions of cloud cores traced by H$^{13}$CN 3-2. The cyan dotted line boxes indicate signal coverage. The detailed parameters for maps of this and other sources are presented in Table \ref{tab4}.  }\label{fig3}
\end{figure}

\begin{figure*}
\addtocounter{figure}{-1}
\includegraphics[width=0.35\columnwidth]{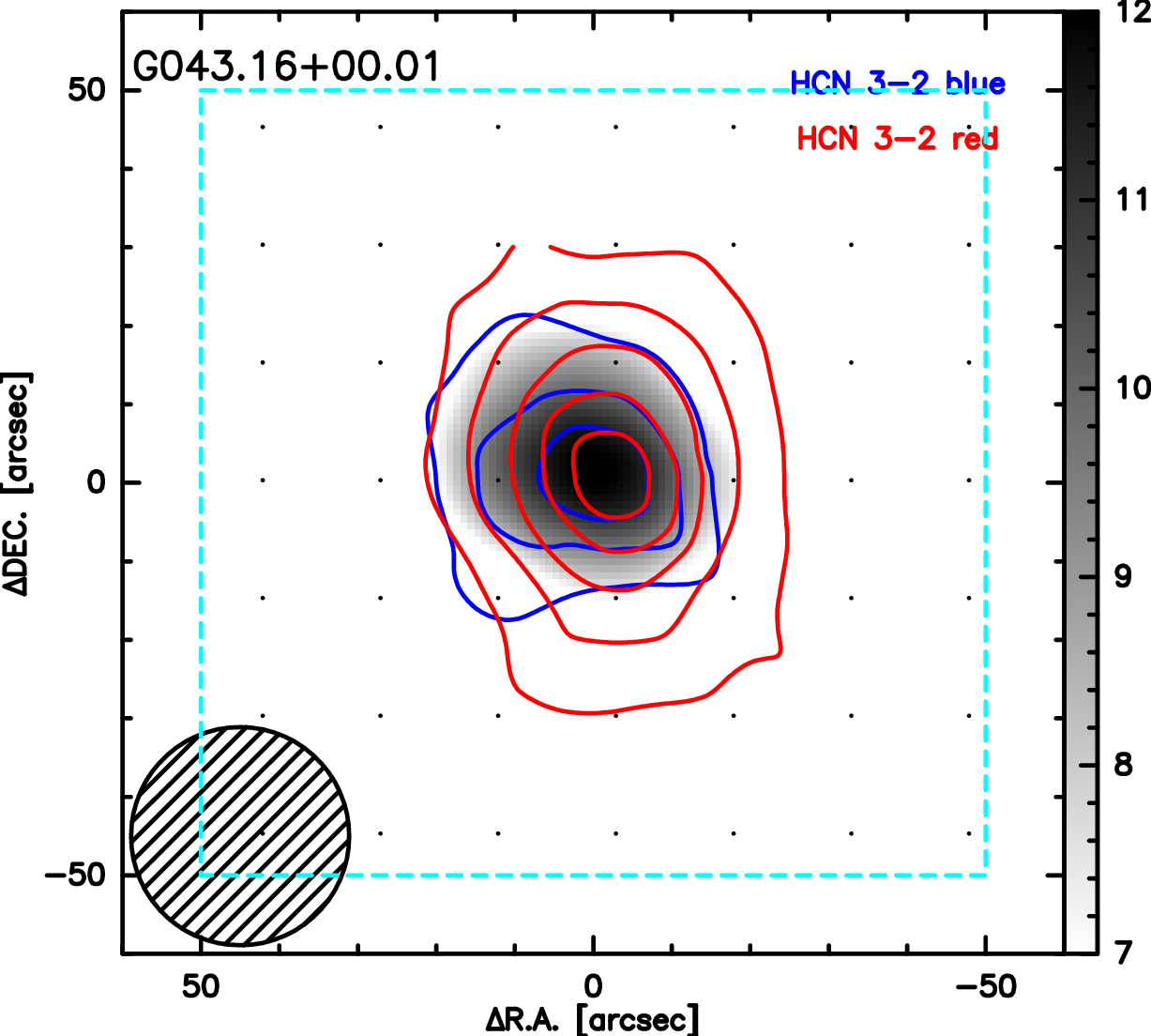}
\includegraphics[width=0.35\columnwidth]{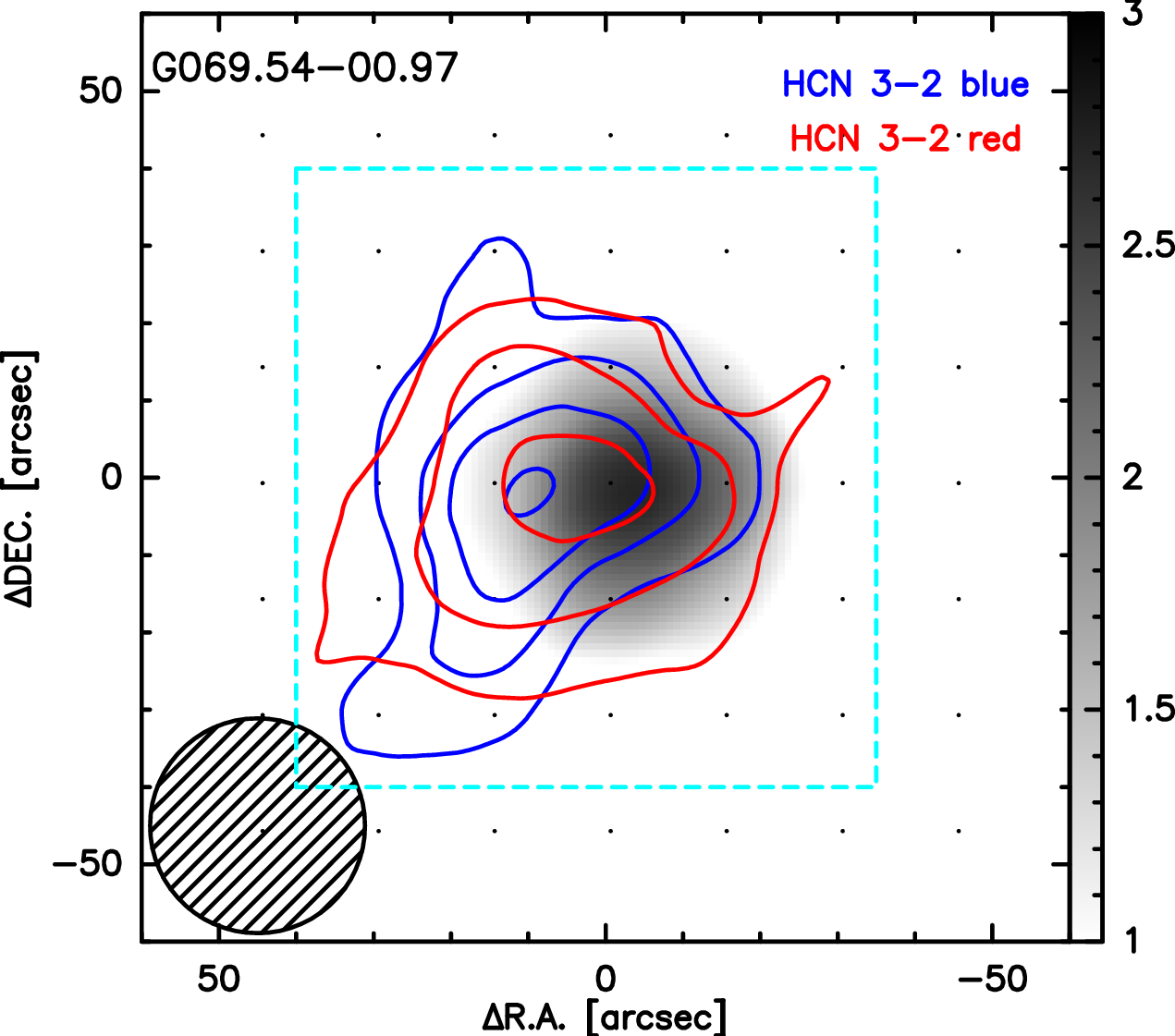} 
\includegraphics[width=0.35\columnwidth]{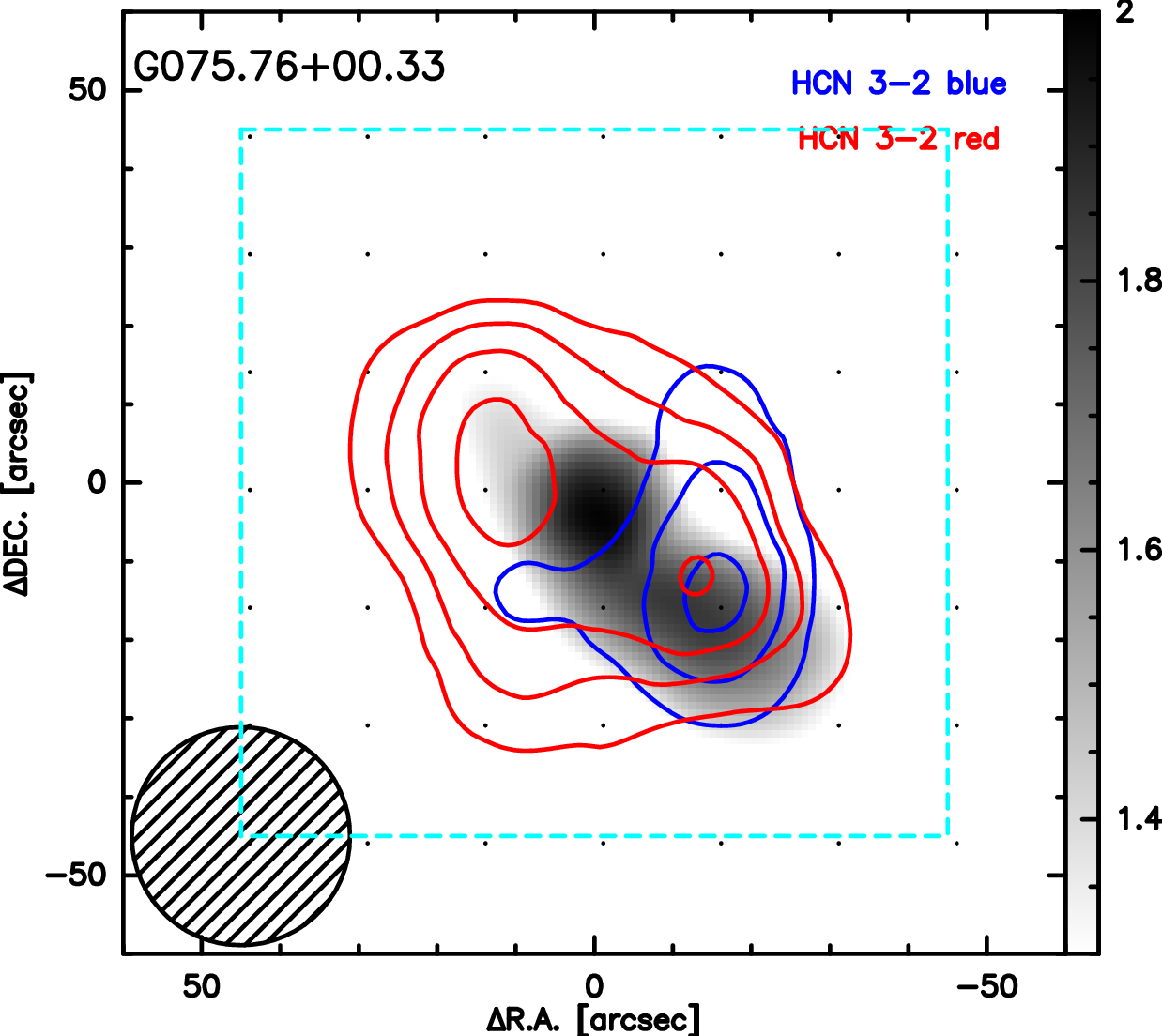} 
\includegraphics[width=0.35\columnwidth]{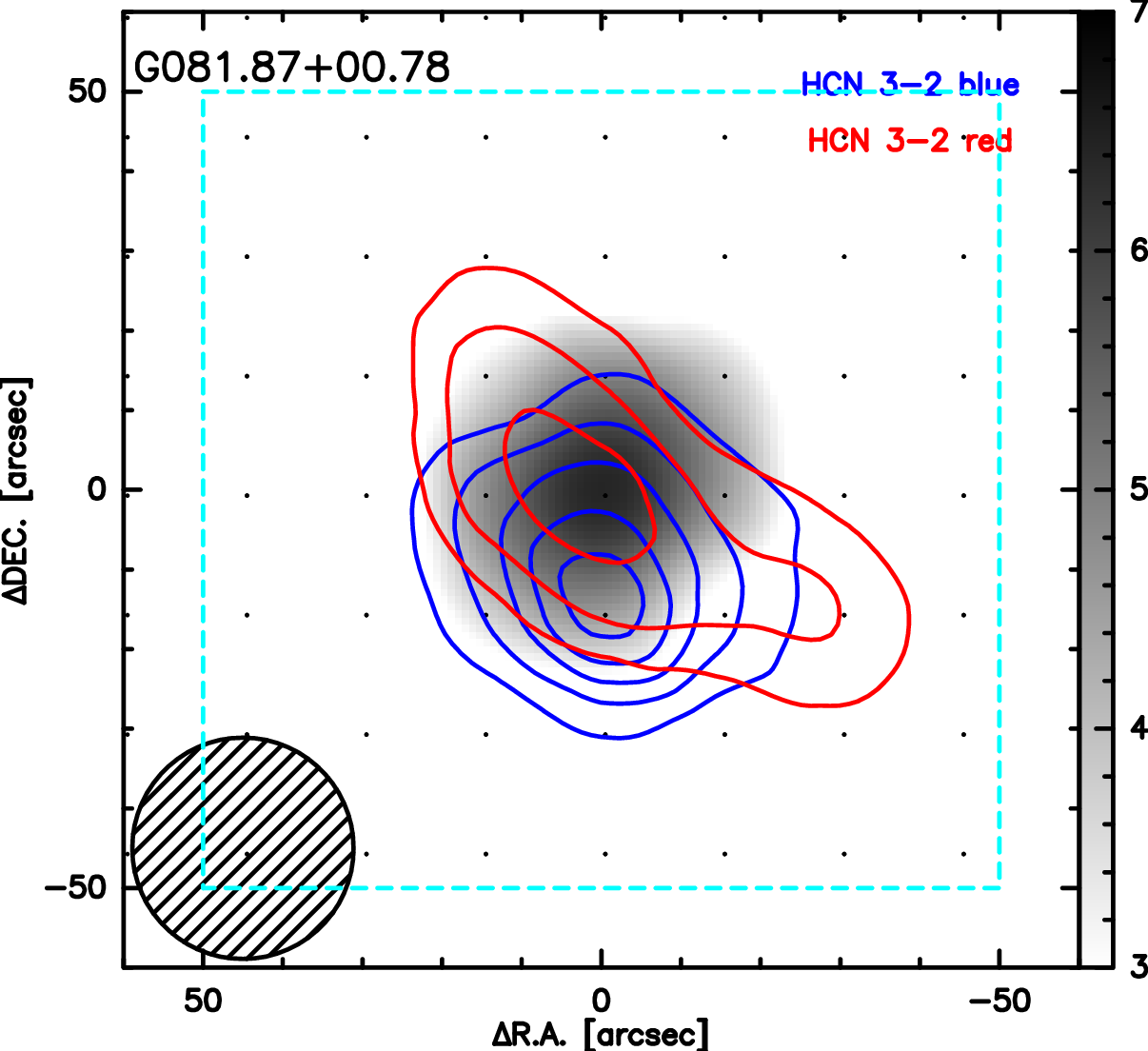} 
\includegraphics[width=0.35\columnwidth]{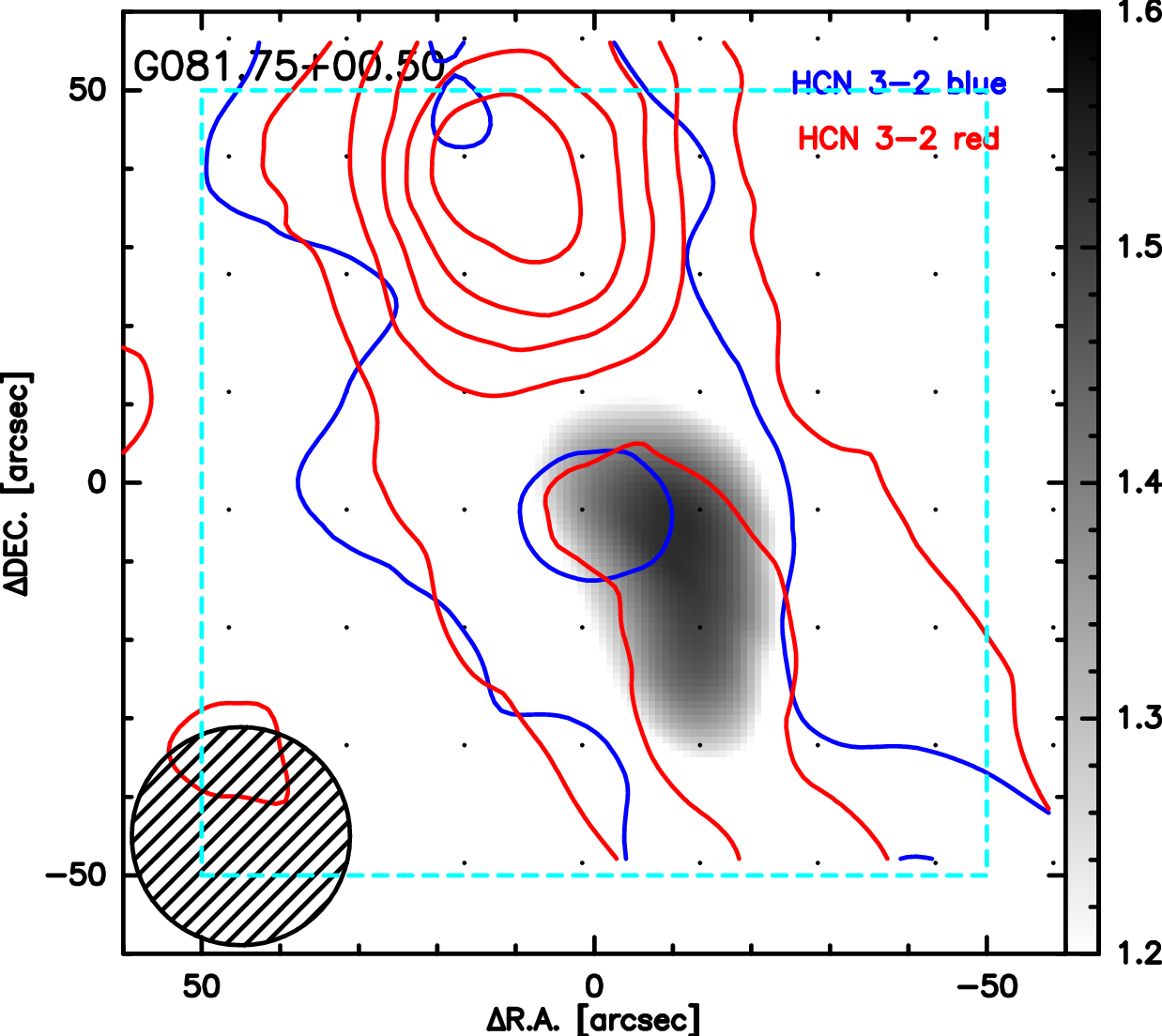} 
\includegraphics[width=0.35\columnwidth]{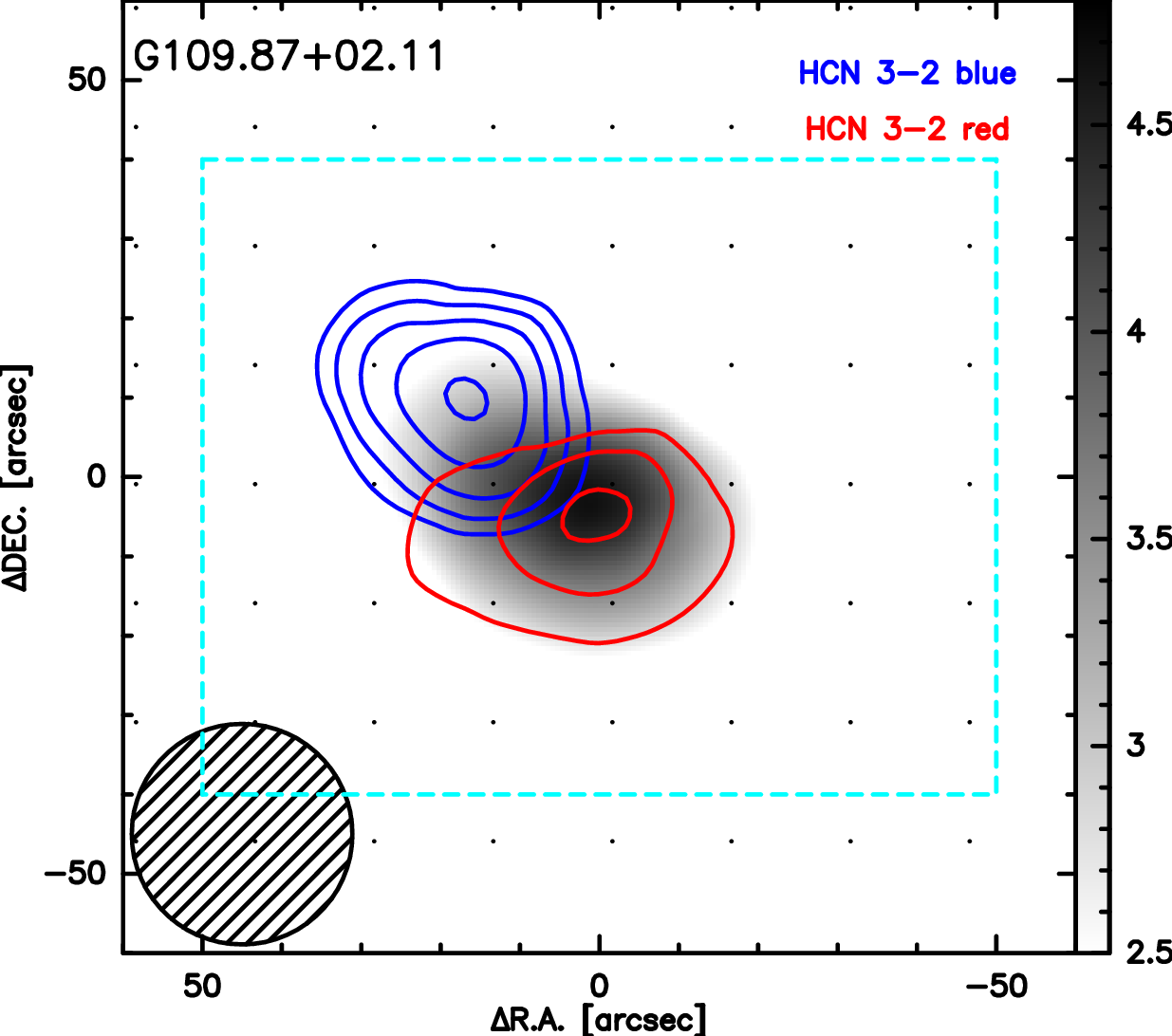}
\includegraphics[width=0.35\columnwidth]{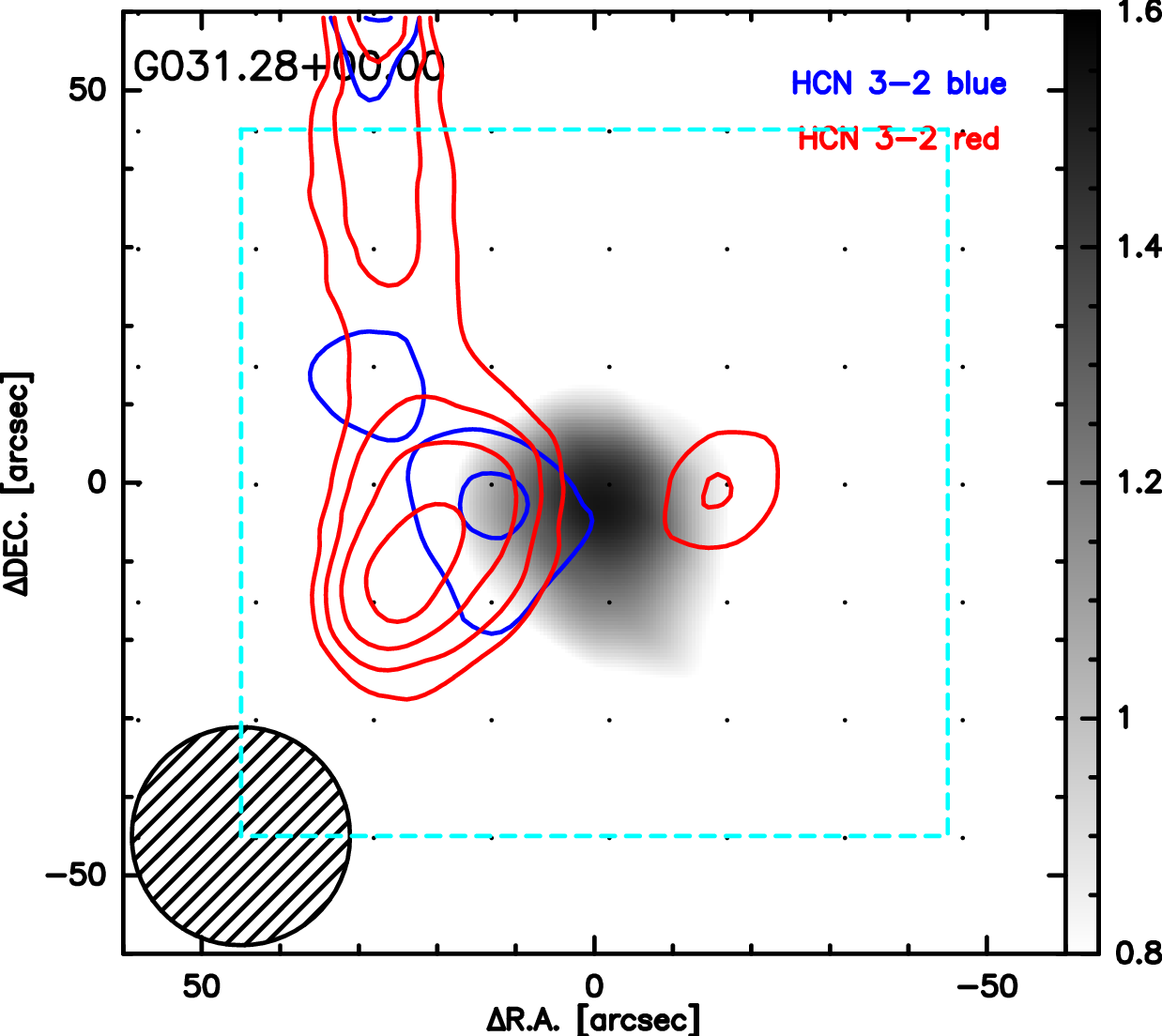} 
\includegraphics[width=0.35\columnwidth]{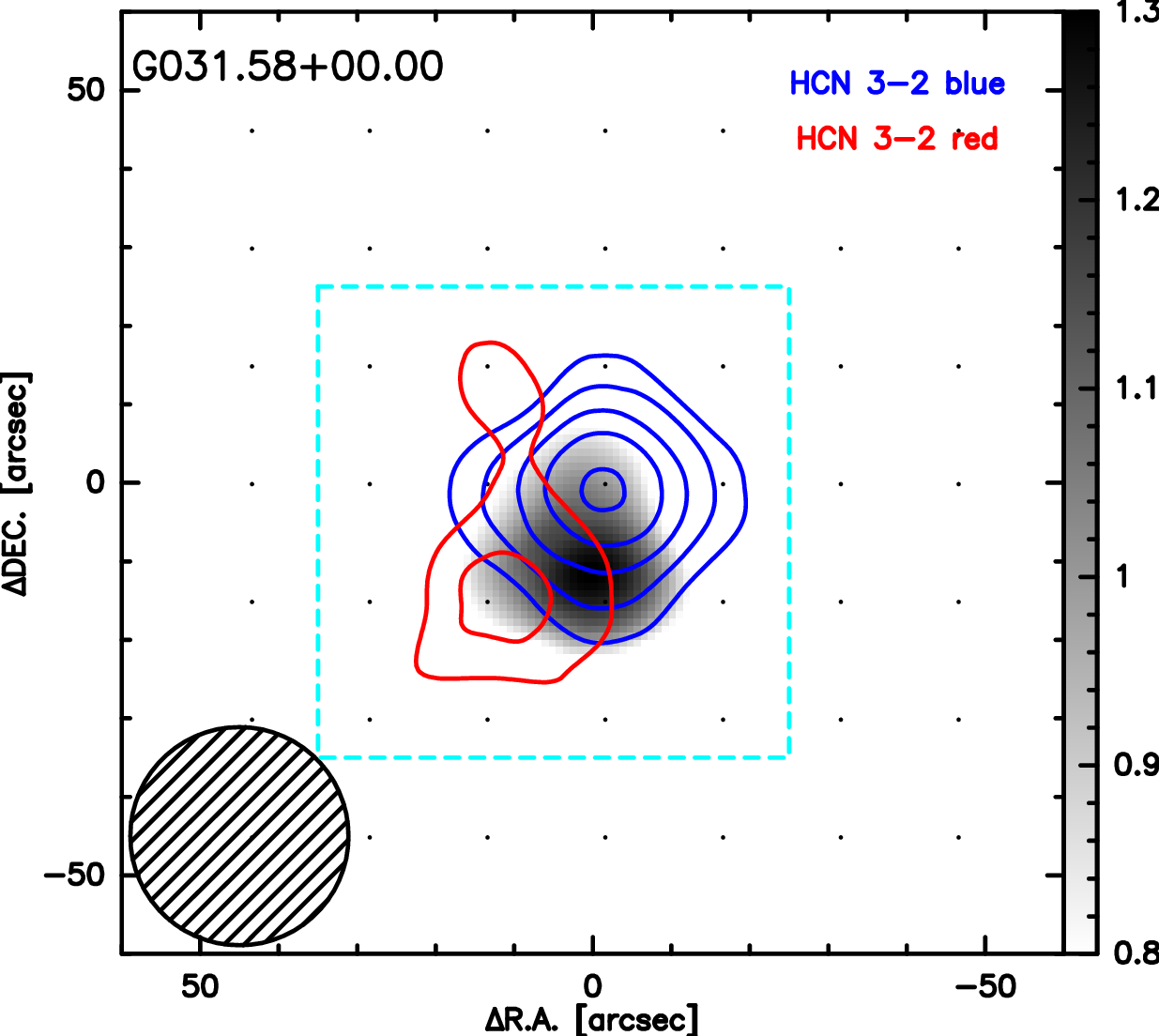} 
\includegraphics[width=0.35\columnwidth]{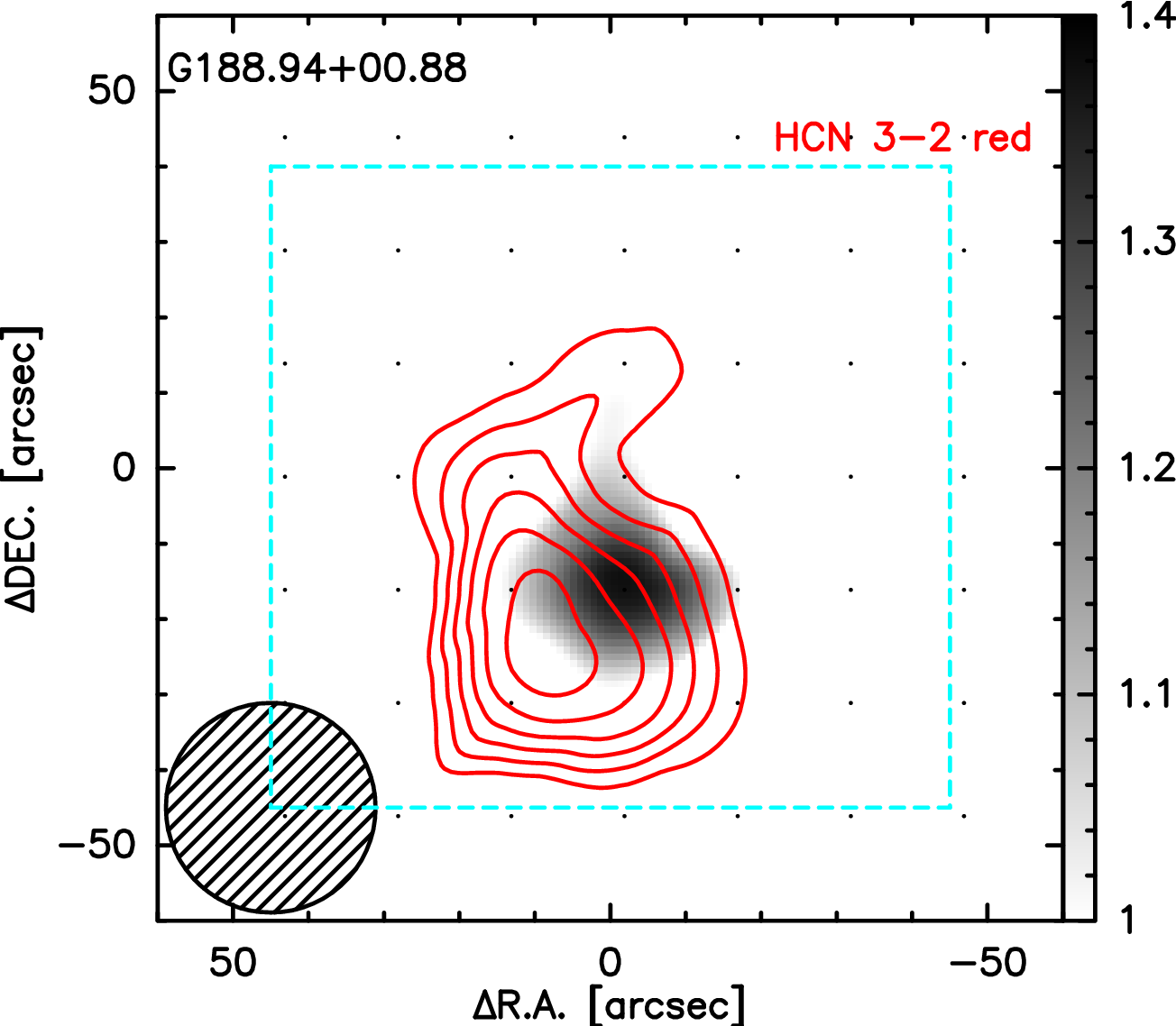} 
\includegraphics[width=0.35\columnwidth]{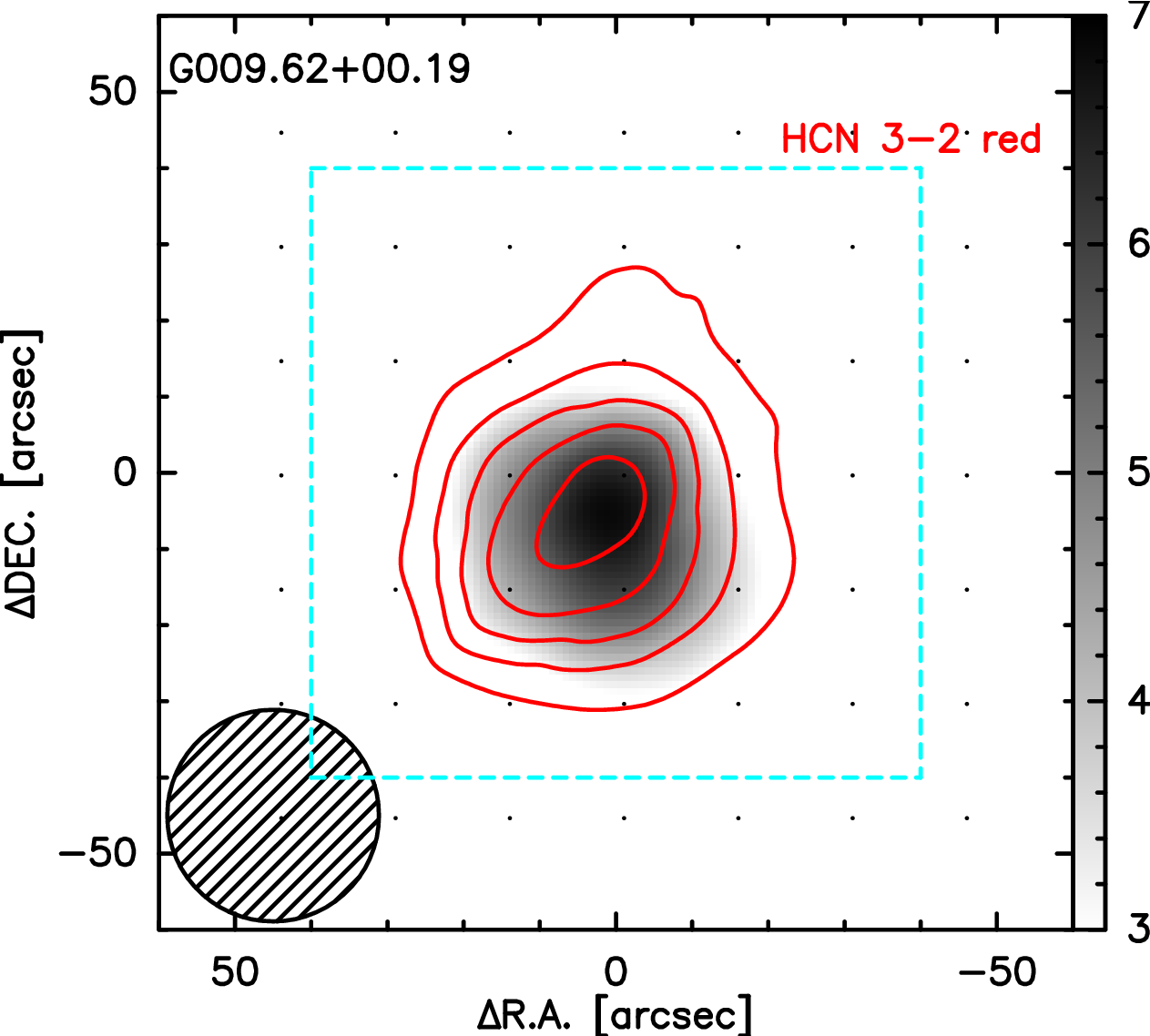} 
\includegraphics[width=0.35\columnwidth]{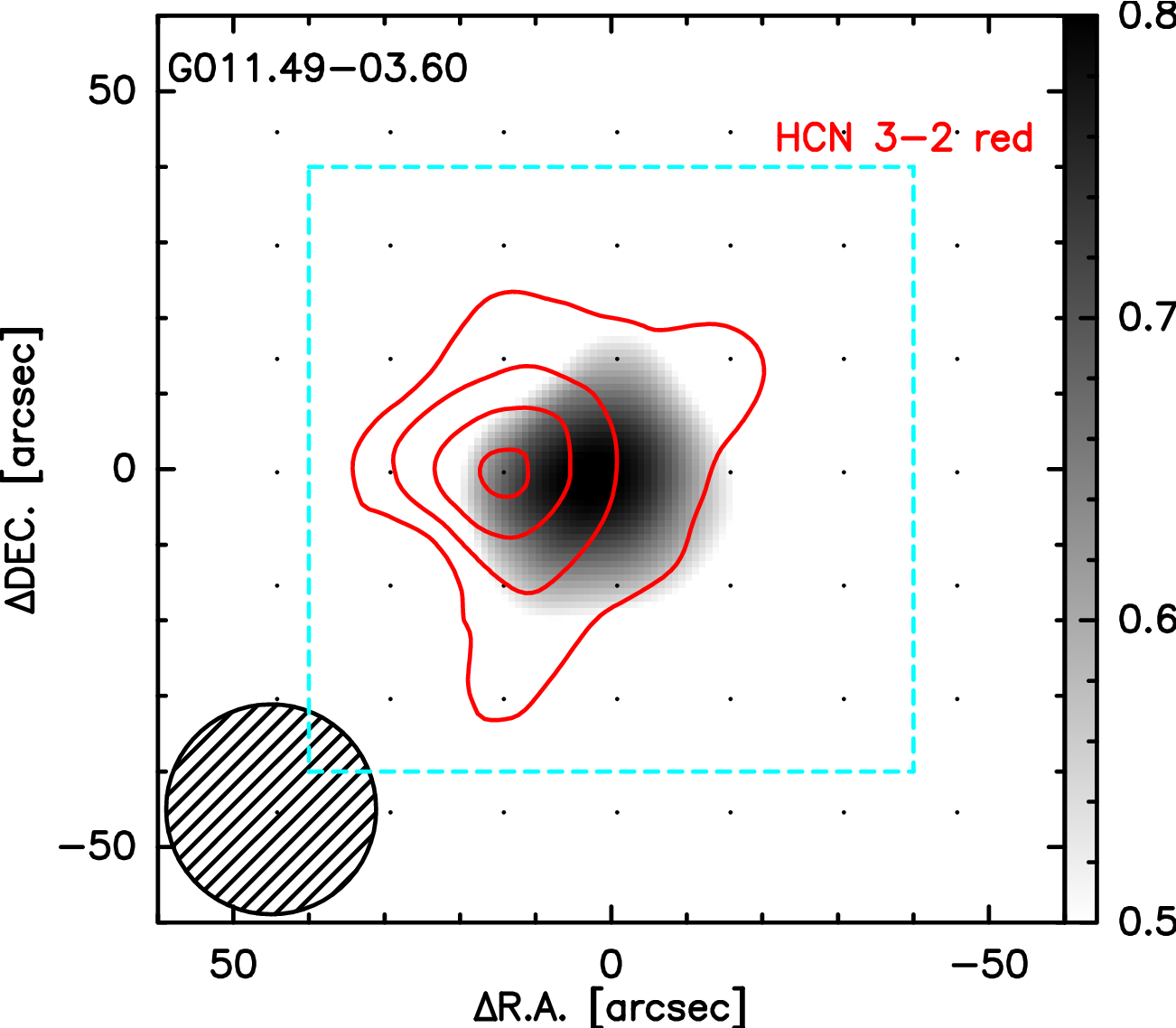} 
 \renewcommand{\thefigure}{3}
\addtocounter{figure}{-1}
\label{fig3}
\caption{Continued.}
\end{figure*}

\begin{figure*}
\includegraphics[width=0.5\textwidth]{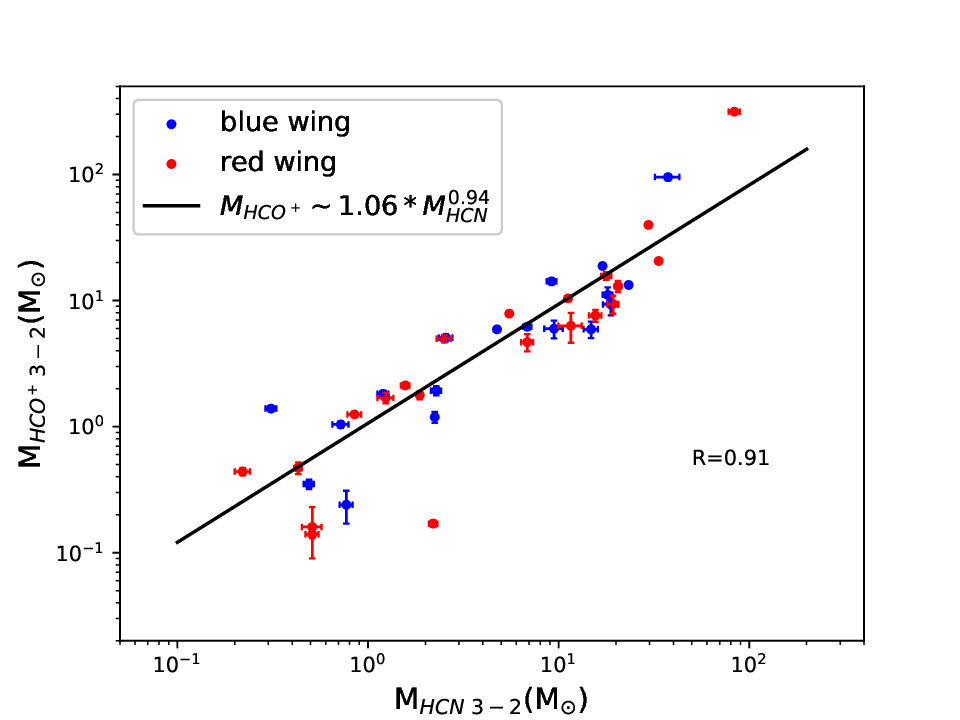}
 \includegraphics[width=0.5\textwidth]{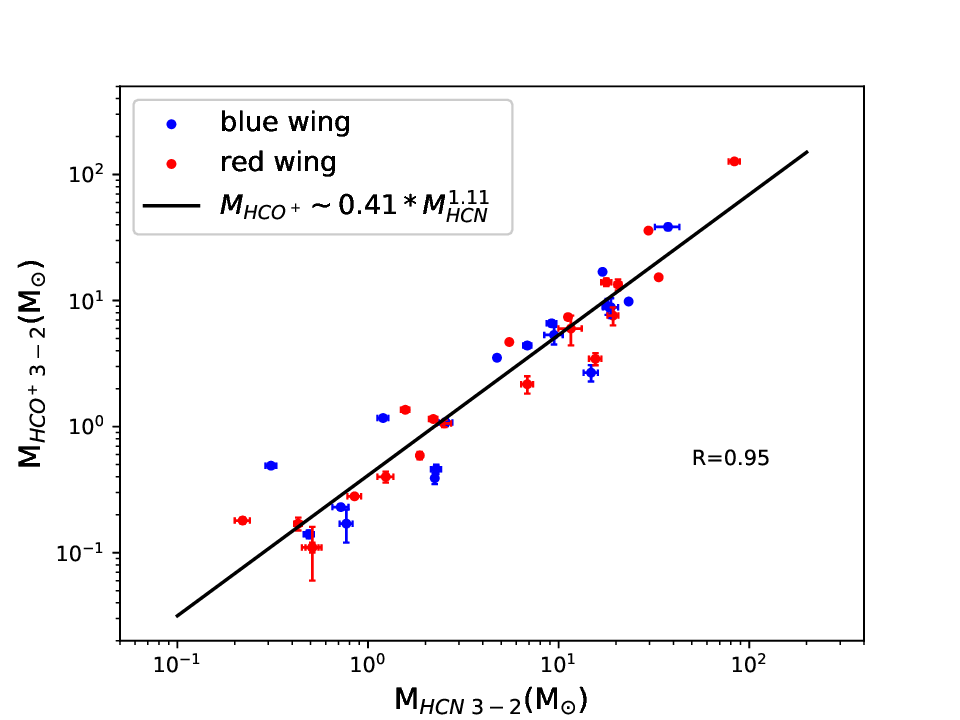}
\renewcommand{\thefigure}{4}
\flushleft
\caption{The relationship between the outflow mass traced by HCO$^+$ 3-2 and HCN 3-2 for 33 sources. In the left, outflow mass traced by HCO$^+$ 3-2 is obtained with fixed HCO$^+$ abundance, while  it is unfixed to derive outflow mass traced by HCO$^+$ 3-2 in the right. The solid lines represent the fitting results.}
\label{fig4}   
\end{figure*}

\begin{figure*}
\includegraphics[width=0.5\textwidth]{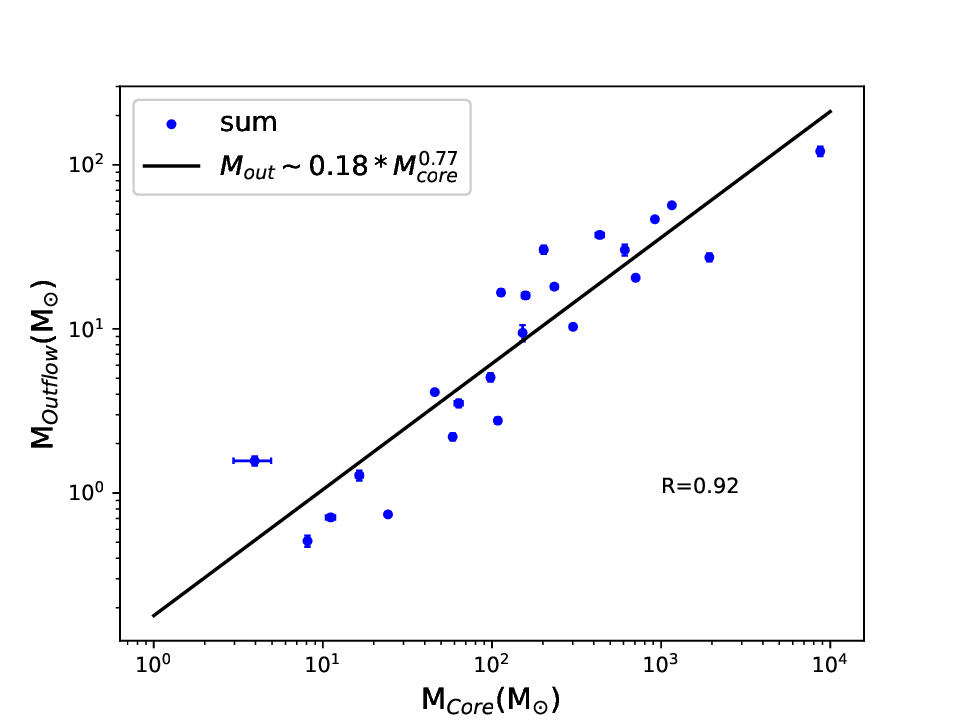}
 \includegraphics[width=0.5\textwidth]{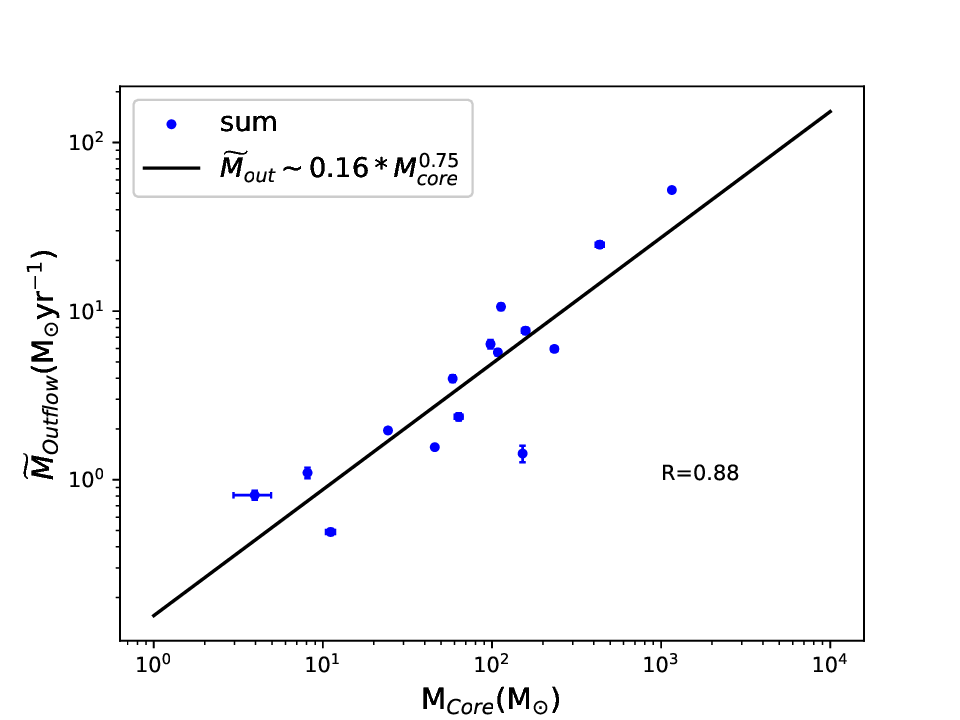}
 \renewcommand{\thefigure}{5}
\caption{ The left panel presents the outflow mass versus the cloud core mass for 23 sources. The right panel presents the mass lose rate versus the cloud core mass for 15 sources. The solid lines are obtained through linear regression using the least squares method.}
\label{fig5} 
\end{figure*}
\end{CJK*}
\end{document}